\documentclass[onecolumn ,aps,prd,a4paper,superscriptaddress,tightenlines,noeprint,10pt,floatfix,nobibnotes,notitlepage]{revtex4-1}
\RequirePackage{float}
\usepackage[normalem]{ulem}
\usepackage{graphicx,epsf}
\usepackage{subcaption}
\usepackage[utf8]{inputenc}
\usepackage[USenglish]{babel}
\usepackage[T1]{fontenc}
\usepackage[toc,page,header]{appendix}
\usepackage{minitoc}
\usepackage{tabularx}
\usepackage{cancel}
\usepackage{color}
\usepackage[bookmarksnumbered,hypertexnames=false,colorlinks,colorlinks=true,linkcolor=blue,urlcolor=black,citecolor=blue,anchorcolor=green]{hyperref}
\usepackage{bbm,braket,microtype,mathrsfs,amsmath,amssymb,color,amsthm,mathtools,fullpage,enumitem,ae,aecompl,bm,thm-restate}
\usepackage{physics}
\usepackage{tikz}
\usetikzlibrary{quantikz2}
\usepackage{lipsum}
\usepackage{nameref}
\usepackage{color}
\usepackage{booktabs, caption, palatino}
\usepackage[capitalize]{cleveref}
\usepackage{silence}

\newcommand{\ba}{\begin{eqnarray}}
\newcommand{\ea}{\end{eqnarray}}

\renewcommand{\cal}{\mathcal}
\newcommand{\ot}{\otimes}
\renewcommand{\tr}[1]{\rm{Tr}\left[#1\right]}
\newcommand{\pur}{\rm Pur\,}
\newcommand{\exv}[2]{\langle #1\rangle_{#2}}

\usepackage{stackengine}
\stackMath

\begin{document}

\title{Probes of chaos over the Clifford group and approach to Haar values}

\author{Stefano Cusumano}
\email[Corresponding author: ]{ste.cusumano@gmail.com}
\affiliation{Dipartimento di Fisica `Ettore Pancini', Universit\`a degli Studi di Napoli Federico II, Via Cintia 80126,  Napoli, Italy}
\affiliation{INFN, Sezione di Napoli, Italy}

\author{Gianluca Esposito}
\affiliation{INFN, Sezione di Napoli, Italy}
\affiliation{Scuola Superiore Meridionale, , Largo San Marcellino 10, 80138 Napoli, Italy}

\author{Alioscia Hamma}
\affiliation{Dipartimento di Fisica `Ettore Pancini', Universit\`a degli Studi di Napoli Federico II, Via Cintia 80126,  Napoli, Italy}
\affiliation{INFN, Sezione di Napoli, Italy}
\affiliation{Scuola Superiore Meridionale, , Largo San Marcellino 10, 80138 Napoli, Italy}

\begin{abstract}
Chaotic behavior of quantum systems can be characterized by the adherence of the expectation values of given probes to moments of the Haar distribution. In this work, we analyze the behavior of several probes of chaos using  a technique known as Isospectral Twirling \cite{10.21468/SciPostPhys.10.3.076}. This consists in fixing the spectrum of the Hamiltonian and picking its eigenvectors at random. Here, we study the transition from stabilizer bases to random bases according to the Haar measure by T-doped random quantum circuits. We then compute the average value of the probes over ensembles of random spectra from Random Matrix Theory, the Gaussian Diagonal Ensemble and the Gaussian Unitary Ensemble, associated with non-chaotic and chaotic behavior respectively. We also study the behavior of such probes over the Toric Code Hamiltonian.
\end{abstract}

\maketitle

\tableofcontents

\section{\label{sec:introduction}Introduction}

The origins of Random Matrix Theory (RMT) dates back to the works by Wigner and Dyson~\cite{Wigner_1951,Wigner_1955,Wigner_1955_II,Wigner_1958,dyson_statistical_1962,dyson_statistical_1962_II,dyson_statistical_1962_III}, who studied the spectral properties of complex many body systems. They found that the salient long-time equilibrium properties of such systems could be studied by matrices, describing the Hamiltonian of many-body systems, whose elements are distributed randomly according to some probability distribution. From its first applications to many-body physics, RMT has developed through the decades, finding applications in several fields of physics, from solid state physics, with applications on disordered solids~\cite{GorkovEliashberg1965JETP,Efetov1983AdvPhys,AltlandGefen1993PRL,AltlandGefen1995PRB,AltlandGefenMontambaux1996PRL} and quantum dots~\cite{Iida1990PRL,Iida1990AnnPhys,Pluhar1995AnnPhys,Pluhar1994PRL}, to many-body quantum field theory~\cite{Calogero1969_2191,Calogero1969_2197,Sutherland1971_JMP246,Sutherland1971_JMP251,Sutherland1971_PRA4,Sutherland1972_PRA5,SimonsLeeAltshuler1993_PRL}, quantum chromodynamics~\cite{Kalkreuter1995PRD,Kalkreuter1996NPBPS,Verbaarschot1994PRL} and quantum gravity~\cite{Abdalla1994_2DGravity}.

Most importantly to us, RMT is a fundamental tool for the study of quantum chaos and the approach to equilibrium of complex quantum systems.
Indeed, the choice of the probability distribution has a fundamental impact on the probability distribution of the eigenvalues~\cite{livan_introduction_2018,mehta_random_1991,haake_quantum_2001} and eigenvectors~\cite{IZRAILEV1987250,PhysRevA.42.1013,Zyczkowski_1990,zyczkowski_eigenvector_1991,PhysRevB.111.054301} of such matrices, which thus happen to describe different dynamical features of the complex systems they represent.   
Families of random spectra associated with different probability distributions, such as the Gaussian Diagonal Ensemble (GDE) and the Gaussian Unitary Ensemble (GUE), are associated with integrable and chaotic behavior of the corresponding dynamics, at least in a single symmetry sector~\cite{berry_level_1977}.

Chaotic dynamics, while well defined at classical level via the Lyapunov exponents~\cite{lyapunov_general_1992} and the butterfly effect~\cite{DeterministicNonperiodicFlow}, remains elusive at quantum level still today~\cite{MaldacenaShenkerStanford2016,Srednicki1994,ChotorlishviliUgulava2010,HaakeWiedemannZyczkowski1992,MankoVilelaMendes2000,WeinsteinLloydTsallis2002,gu2024simulatingquantumchaoschaos}.
Indeed, quantum chaos is most often associated to quantum systems exhibiting the same features of quantized classical chaotic systems. One such feature is the already mentioned statistics of the energy spectrum.
Several probes have been proposed and utilized: those based on collective dynamical features, such as OTOCs~\cite{LarkinOvchinnikov1969,ShenkerStanford2014,ShenkerStanford2014MultipleShocks,ShenkerStanford2015Stringy} and Loschmidt echo~\cite{PhysRev.80.580,Peres1984,Jalabert2001,CucchiettiPastawskiJalabert2004,Goussev2008} and those based on information-theoretical quantities, such as entanglement entropy~\cite{HaydenLeungWinter2006,WangGhoseSandersHu2004,VidmarRigol2017} and mutual information~\cite{DingHaydenWalter2016,HosurQiRobertsYoshida2016,TouilDeffner2020} or the entanglement spectrum\cite{PhysRevLett.112.240501,shaffer_irreversibility_2014,PhysRevB.96.020408,PhysRevA.109.L040401}. Another set of probes of chaos is obtained by looking at another specific feature of quantum systems, quantum coherence. By looking at the quantum coherence between different eigenstates of the system, it is indeed possible to define further probes of chaos, such as the norm of coherence and the Wigner-Yanase-Dyson skew information.

In search of probes  that can detect chaos in an unambiguous way \cite{e23081073,10.21468/SciPostPhys.10.3.076} we see that one can find such signatures either in the spectrum or its eigenvectors. The former connection is obtained through the aforementioned level statistics and spectral form factors \cite{PhysRevE.55.4067,PhysRevLett.80.1808,kunz_probability_1999,venuti2025integrabilitychaosfractalanalysis}. Complex or chaotic behavior of Hamiltonian through their eigenvectors is typically considered in approaches towards thermalization such as ETH \cite{PhysRevA.81.022113,doi:10.1142/9789814704090_0008,venuti2019ergodicityeigenstatethermalizationfoundations,de_palma_necessity_2015,PhysRevLett.130.140402,deutsch_eigenstate_2018,1jzy-sk9r}. 

In this work, we go beyond the paradigm set in \cite{e23081073,10.21468/SciPostPhys.10.3.076}, where it is studied the behavior in ensembles of Hamiltonian with completely random eigenvectors - according to the Haar measure - and with a fixed spectrum. As such, these results only depend on spectral properties and are indeed related to spectral form factors only. However, instead of picking eigenvectors at random from the Haar measure over the unitary group, one may consider other ensembles: for instance, one can consider ensembles of stabilizer states. Although very entangled, they are intrinsically simple as they can be efficiently simulated on a classical computer \cite{Dennis2002,BravyiKitaev1998,KitaevToricCode2003,VeitchMousavianGottesmanEmerson2014,PhysRevLett.132.080402,PhysRevA.109.022429,Leone2024learningtdoped}. One can then also study the transition towards more and more complex eigenstates through the model of T-doped quantum circuit \cite{PhysRevLett.112.240501,shaffer_irreversibility_2014,p8dn-glcw}. Specifically, we will be dealing with averages over the Clifford group and averages over T-doped Clifford circuits~\cite{Leone2021quantumchaosis}.

Choosing eigenvalues and eigenvectors of the Hamiltonian from two independent ensembles, i.e. RMT for the spectra and the algebraic theory of the Clifford group for the eigenvectors, allows one to consider how much simulations using Clifford resources can be used to simulate chaotic Hamiltonian, or, conversely, which probes of chaos are sensible to the spectrum alone, rather than on the joint eigenvalues and eigenvector distributions. Notice that this is an important point when going beyond the Haar average, as the decoupling of eigenvalues and eigenvectors is automatic only in this case, see also Sec.~\ref{sec:probes_and_isospectral}.
Concretely, we consider Hamiltonians of the form $H=V^\dagger DV$, where $D$ is diagonal in the computational basis and fixes the spectrum, while $V$ is drawn from the Clifford group, or from an ensemble of T-doped Clifford circuits, and fixes the eigenbasis. The corresponding evolution $U(t)=V^\dagger e^{-iDt}V$ is not in general a Clifford operation: the Clifford (or T-doped Clifford) restriction concerns only the unitary $V$ used to select the eigenbasis, while the operator $e^{-iDt}$ is in general non-Clifford. For stabilizer Hamiltonians, $D$ is a sum of mutually commuting Pauli strings, and $e^{-iDt}$ is thus a product of Pauli rotations, so that the whole evolution could be implemented by a circuit with a number of gates polynomial in the number of qubits~\cite{Bullock2004}. In contrast, for a generic diagonal $D$, implementing $e^{-iDt}$ requires a number of gates growing exponentially with the number of qubits.
Moreover, systems where the Hamiltonian eigenvalues are picked at random, while the eigenvectors are subject to some kind of algebraic constraint can be used in the study of the transition from  the Eigenstate Thermalization Hypothesis to the Many-Body Localized phase~\cite{nandkishore_many-body_2015,RevModPhys.91.021001}. In this sense, imposing the eigenvectors to be stabilizer states can be considered an extreme example of such algebraic constraints. More generally, there is wide interest in the study of algebraic and spectral properties of random and T-doped Clifford circuits, especially in connection with quantum information properties such as entanglement and non-stabilizerness and associated phase transitions~\cite{PhysRevLett.128.050402,oliviero_transitions_2021,viscardi2025interplayentanglementstructuresstabilizer,Iannotti2025entanglement,1jzy-sk9r,magni2025anticoncentrationstatedesigndoped,PhysRevB.111.054301}.

A fundamental concept in this analysis is the one of $k$-design. Shortly, the concept of $k$-design measures the adherence of a given unitary group, equipped with a probability measure, to the Haar distribution. In practice, the higher the $k$, the more the given group and probability measure will be able to mimic the full unitary group. The concept of $k$-design has applications in several fields of quantum physics~\cite{PashayanBartlettGross2020}, such as sampling~\cite{BoulandFeffermanNirkheVazirani2019,HangleiterBermejoVegaSchwarzEisert2018}, cryptography~\cite{DiVincenzoLeungTerhal2002} and tomography~\cite{HuangKuengPreskill2020}, and is intimately connected with quantum chaos~\cite{roberts_chaos_2017,magni2025anticoncentrationstatedesigndoped}. While the Clifford group together with the uniform probability distribution forms a 3-design~\cite{zhu2016cliffordgroupfailsgracefully}, non-stabilizerness has been recently found responsible for the transition of the Clifford group to form an arbitrary $k$-design~\cite{Leone2021quantumchaosis}. Namely, the introduction of non-Clifford gates in random Clifford circuits allows one to pick up random states which are closer to be Haar random distributed, that is, they are sampled according to a $k$-design.

By studying the behavior of spectral averaged probes of chaos when the isospectral twirling~\cite{10.21468/SciPostPhys.10.3.076} is performed over different unitary groups, we aim at several points. First, we want to uncover the role of non-stabilizerness in the transition from a group forming a 3-design, into one forming an arbitrary $k$-design. Moreover, as the adherence to high moments of the Haar distribution is a signal of chaotic behavior, by studying how a 3-design approaches a $k$-design we study also the transition to chaos. Finally, by the use of stabilizer Hamiltonian, we will also highlight how the introduction of non-stabilizerness by T-doping leads the Clifford average to approach the Haar average. In particular, we will consider two situations: a stabilizer Hamiltonian which is diagonal in the computational basis, whose spectrum is randomly sampled from an ensemble of RMT. We will also consider a model whose spectrum is fixed. Our choice falls on the Toric code Hamiltonian, as this model presents both a highly degenerate spectrum, and thus an extremely non-chaotic dynamics, and a set of stabilizer eigenstates.

We start by summarizing the main results of our work in Sec.~\ref{sec:main_results}. In order to pursue our program, we will have to introduce several technical tools in Sec.~\ref{sec:tools}. This will include group averaging (Sec.~\ref{sec:group_averages}), the isospectral twirling of probes of chaos (Sec.~\ref{sec:probes_and_isospectral}), RMT and averages over families of random spectra (Sec.~\ref{sec:RMT}). In Sec.~\ref{sec:probes_of_chaos} we will introduce the most significant probes of chaos for our analysis, together with the results associated with them. Finally, in Sec.~\ref{sec:conclusions} we draw our conclusions and give some outlook for future investigations.

\section{Main results\label{sec:main_results}}
Let us first summarize the main results and findings of this work.
First of all, we provide closed analytical expressions for several probes of chaos averaged over the Clifford group and over T-doped Clifford circuits. This allows us to identify the main dynamical features of so averaged probes, namely, their equilibrium value and their approach to such values, highlighting the differences between the averages over the Clifford group and the full Unitary group, and thus, the role that non-stabilizerness plays in Hamiltonians.

Generally speaking, we find that the probes of chaos we analyze are of two kinds: those sensible to non-stabilizerness and those which are not. This sensitivity is displayed by their dependence on the T-doping.   Table~\ref{tab:probes} shows a summary of such results for the large temporal limit. We find that scrambling measures such as the Loschmidt Echo and the Out-of-Time-Order-Correlators do depend on the stabilizerness of the eigenvectors, showing a factor $d={\rm dim} (\mathcal{H})$ of difference between the average over the Clifford group and the full unitary group. On the other hand, entropic and coherence measures are not sensible to the stabilizerness (or lack thereof) of the eigenvectors, and both random Clifford Hamiltonians or Hamiltonians with Haar-random eigenvectors yield the same long-time values.  This is  not surprising, as the Clifford group is able to generate maximal entanglement, and thus maximal decoherence. Remarkably, no probe ever shows a difference between the two sets of eigenvectors at short times. This is important as it would seem that fast scrambling behavior can be realized by Hamiltonians with stabilizer eigenvectors, which is relevant, e.g., for the scrambling behavior of black holes \cite{PhysRevLett.132.080402, PhysRevA.109.022429}.

As mentioned in Sec.~\ref{sec:introduction} we average our probes over two different spectral ensembles, the Gaussian Diagonal Ensemble and the Gaussian Unitary Ensemble, associated with integrable and chaotic dynamics respectively. We find that the probes averaged over the GDE show no particular sensibility over the group average, but the asymptotic value. On the contrary, when the spectra are taken from the GUE, the characteristic oscillations of this ensemble are suppressed when the average is taken over the Clifford group. Similarly, when analyzing the Toric Code Hamiltonian, we find that also such integrable model shows the same sensibility to the group average, that is, scrambling measures reach asymptotic values which discriminate between the averages.

\begin{figure}[!ht]
\centering
\begin{subfigure}{0.49\textwidth}
\centering
\includegraphics[width=\textwidth]{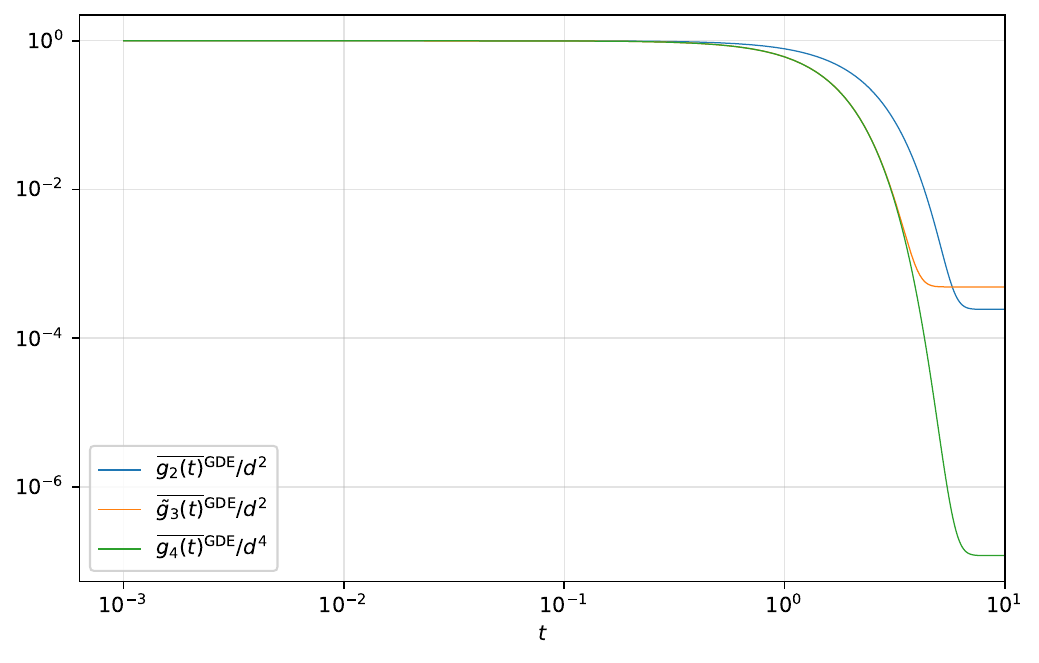}
\caption{}
\label{fig:compare_GDE}
\end{subfigure}
\begin{subfigure}{0.49\textwidth}
\centering
\includegraphics[width=\textwidth]{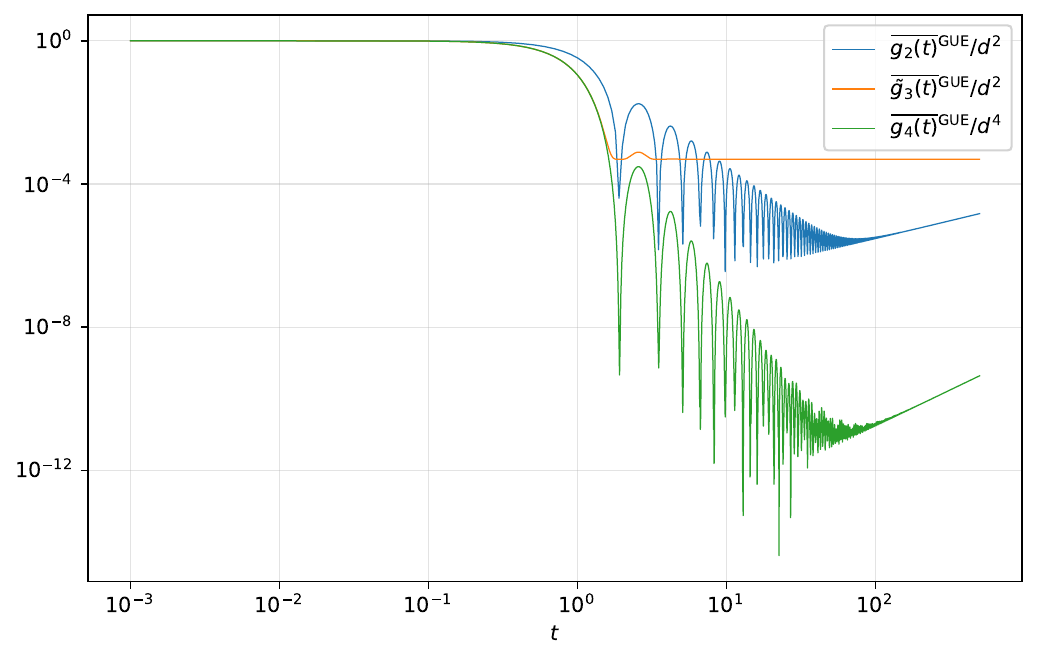}
\caption{}
\label{fig:compare_GUE}
\end{subfigure}
\caption{Comparison of $g_2(t)$, $\tilde{g}_3(t)$ and $g_4(t)$ averaged over the GDE (panel~\subref{fig:compare_GDE}) and the GUE (panel~\subref{fig:compare_GUE}) for $d=2^{16}$. One notices in both cases how the Clifford spectral form factor $\tilde{g}_3(t)$ shares the same equilibrium value of $g_2(t)$. Moreover, one notices the suppression of oscillations in the case of GUE.}
\label{fig:g3tilde_compare}
\end{figure}

Our analysis also allows us highlight some technical features of averages over the Clifford group. We find that, as the Clifford group is a 3-design, the 4-th order average taken over this group does not depend on the four point spectral form factor $g_4(t)$, but rather to a modified version $\tilde{g}_3(t)$. This Clifford spectral form factor does not depend on four energy values, but rather on three, and present features in between the ordinary two point and the four point spectral form factors, see also Fig.~\ref{fig:g3tilde_compare}. Indeed, it shares the same equilibrium value of $g_2(t)$, while it presents the same oscillations of $g_4(t)$, though suppressed by a factor $d^2$. This peculiar behavior is due to the projection on the Clifford commutant when taking the average over the Clifford group. Moreover, this function is responsible for the different behavior of the probes when averaged over the Clifford group.

Furthermore, we find that, in order to meaningfully average over the Clifford group, one has to impose the averaged operator to be stabilizer, that is, to possess a complete set of stabilizer eigenvectors. This is because the Clifford average does depend on the initial eigenvectors of the operator to be averaged. As a consequence, averaging a generic operator would introduce non-Clifford resources in the average, spoiling and biasing the analysis. It is for this reason that we choose to average unitary operators generated by stabilizer Hamiltonians. As a consequence, we compute the components of the unitary operators so generated over both the Clifford and Unitary commutants.

Furthermore, in order to compute the T-doped average we further analyze the formula derived in~\cite{Leone2021quantumchaosis}. We diagonalize the matrix $\Xi$ there derived (see Sec.~\ref{sec:group_averages}), which describes the effects of doping, computing its eigenvalues and eigenvectors. We find that two of these eigenvectors are proportional to projectors on the symmetric and anti-symmetric representations of the permutation group $S_4$. The effect of doping is to interpolate between the Clifford and Haar values of the averages, by progressively reducing the contributions from the Clifford group and increasing the contributions from the Unitary group, until, in the limit of infinite doping layers, the average reaches the Haar value.
We observe that in general, consistently with the literature, one needs an amount of doping layers of order $\mathcal{O}(\log d)$ in order to recover the Haar values of the averages.

\begin{table}[!ht]
\begin{tabular}{|c|c|c|}
\toprule
Probe&Clifford Equilibrium value&Haar Equilibrium Value\\
\midrule
Loschmidt Echo&$\mathcal{O}(d^{-1})$&$\mathcal{O}(d^{-2})$\\
${\rm OTOC}_4$&$\mathcal{O}(d^{-1})$&$\mathcal{O}(d^{-2})$\\
Tripartite Mutual Information&$\mathcal{O}(\log d^{-1})$&$\mathcal{O}(\log d^{-1})$\\
Entanglement Entropy&$\mathcal{O}(\log \sqrt{d})$&$\mathcal{O}(\log\sqrt{d})$\\
Coherence&$1-\mathcal{O}(d^{-1})$&$1-\mathcal{O}(d^{-1})$\\
WYD Skew Information&$1-\mathcal{O}(d^{-1})$&$1-\mathcal{O}(d^{-1})$\\
\bottomrule
\end{tabular}
\caption{Summary of the equilibrium values of probes of chaos averaged over the Clifford and Unitary group.}
\label{tab:probes}
\end{table}

\section{\label{sec:tools}Tools and methods}

As mentioned in the Introduction, there is still no clear definition of quantum chaos. Indeed, while classical chaos can be easily defined via Lyapunov exponents and the butterfly effect, the quantum scenario is not as straightforward. As quantum dynamics is unitary, that is, preserves distances, a concept of Lyapunov exponent can hardly be defined~\cite{HaakeWiedemannZyczkowski1992,CucchiettiPastawskiJalabert2004}. Thus, historically, the strategy to define quantum chaos has been to quantize classical systems known to possess chaotic behavior, and define quantum chaos in terms of the salient features of such dynamics~\cite{haake_quantum_2001}.
While this salient feature has been recognized in the statistical properties of the energy spectrum of the system, a measure of all energy correlation of a complex quantum system is clearly impossible to achieve.

This has led to the definition of a plethora of probes and diagnostics for quantum chaos. Some of these are sort of quantum equivalents of the butterfly effect, being connected with the scrambling of correlations in the system, like OTOCs and Loschmidt echo. Others are connected with quantum information properties of the system, while others are connected with the coherence of the system in a given basis.
While all the probes of chaos will be discussed in detail in Sec.~\ref{sec:probes_of_chaos}, here we want to introduce some general features of such probes, in order to explain better the tools and methods we are going to use.

The main idea is that the dynamics of a quantum system is coded into its energy spectrum, i.e. the eigenvalues of the Hamiltonian of the system $H$, and by the unitary operator $V=\exp[-iHt]$ generated by the Hamiltonian. As computation of many-body interacting quantum systems' exact spectrum is not feasible as soon as the size of the system increases, in order to study large systems one can rely on RMT. As we are going to see in Sec.~\ref{sec:RMT}, RMT deals itself with the spectra of matrices whose entries are random numbers distributed according to some probability distribution.
Depending on these probability distributions, the spectrum of the matrix will be different, and thus the dynamical features of the system it describes. Specifically, there are probability distributions that are known to lead to random spectra giving rise to chaotic dynamics, while others are known to lead to spectra describing integrable dynamics.

In defining probes of chaos, one wishes to find quantities which are sensible to chaotic features of the dynamics. As these features are dictated by the spectrum of the system, one would love to make these probes independent on the eigenvectors of the system, i.e. average over all possible dynamics with a given spectra. This allows one to discern effects which are due to the spectrum alone. When one averages further over spectra from a given families associated with chaotic and non-chaotic behavior, one can finally single out  features which are exclusive of chaotic systems.

The first average corresponds to what is called {\it isospectral twirling}. This is indeed an Haar average over a set of unitaries, so that the spectrum of the averaged operator is left unchanged. The isospectral twirling of order $2k$ of a unitary operator $V$ is defined as:
\ba
\label{eq:isospectral_twirling_definition}
\mathcal{R}_{\cal G}^{(2k)}=\int d\mu_{\cal G}\, G^{\dag 2k}V^{\ot k,k}G^{2k}
\ea
where $\mathcal{G}$ is the unitary group over which the average is performed, and we have introduced the notation $V^{\ot k,k}=V^{\ot k}\ot V^{\dag \ot k}$.

In~\cite{10.21468/SciPostPhys.10.3.076} the isospectral twirling has been applied to systems with random spectra, averaging over the whole unitary group $\mathcal{U}$. In this work we want to take a step further in the analysis initiated there, by performing the isospectral twirling with {\it Clifford} operations and {\it T-doped} Clifford circuits. The main difference between the averages performed over these group and the one over the unitary group is expressed by the concept of $k$-design. The concept of $k$-design expresses how much a given set of unitary operators, equipped with a given probability distribution, can mimic the properties of the whole unitary group. The Clifford group, together with the uniform probability measure, is known to form a 3-design, that is, any average over the Clifford group, such as the one in Eq.~\eqref{eq:isospectral_twirling_definition}, will be equal to the one performed over the unitary group for $2k\leq3$.

As the probes of chaos that we are going to examine will all depend on the $k=2$ isospectral twirling, our first result will consist in observing differences in the probes of chaos when the averages are performed over different groups. Moreover, we will also perform averages over T-doped
Clifford circuits. The role of T-doping is to introduce in the system a quantum resource known as non-stabilizerness, often dubbed magic. This information theoretic resource is responsible, together with entanglement, of quantum advantage in quantum information processing. This is intimately connected with the ability of a quantum circuit of sampling over the Haar distribution, that is, the capacity of a quantum circuit to reproduce a given $k$-design. Specifically, as studied in~\cite{Leone2021quantumchaosis}, the effect of T-doping is to make the Clifford group closer to the unitary group. Specifically, when a sufficient amount of doping is introduced in a Clifford circuit, this becomes a 4-design. Thus, by studying probes of chaos averaged over T-doped Clifford circuits we will be able to observe a {\it transition} to chaotic behavior.

In order to observe this transition, we will need to specify some Hamiltonian models whose eigenvectors are stabilizer states. This choice is necessary in order to observe a proper transition, as the choice of a random basis according to some distribution, e.g. Thomas-Porter, would introduce some uncontrolled amount of non-stabilizerness into the analysis, spoiling the results. This will lead us to consider a specific family of Hamiltonian, known as Stabilizer Hamiltonian. The main property of these Hamiltonian is to be generated by a limited amount of commuting Pauli operators, and to possess stabilizer eigenstates.

As the matter of this work is rather technical, in the following we will introduce all the tools necessary to our analysis, starting with the Haar measure tools and the computation of averages over the unitary groups of our interest in Sec.~\ref{sec:group_averages}. Then, in Sec.~\ref{sec:probes_and_isospectral} we will introduce the main features of the probes of chaos we are going to consider, their expression in terms of isospectral twirling, their dependence on the spectral form factors and the computation of their group average.
In Sec.~\ref{sec:RMT} we introduce RMT and the ensemble average over given families of random spectra. Finally, in Sec.~\ref{sec:physical_models} we introduce our physical models, and compute their spectral form factors.

\subsection{\label{sec:group_averages}Group averages}

In statistical physics one is often interested of averaging given quantities over a given set of states, i.e. average the expectation value of an operator over multiple possible states. When one is interested in the averaging over a random state picked uniformly at random from the Hilbert space, one has to use the Haar measure. The Haar measure formalizes the concept of picking random unitary matrices uniformly.

While we point the reader to other references for a thorough treatment of the Haar measure tools~\cite{Mele2024introductiontohaar,watrous_theory_2018,bengtsson_geometry_2007}, here we limit ourselves to introduce the basic concepts and notions that will be used throughout the paper. This will also allow us to set up some notation.

\subsubsection{\label{sec:haar_average}Haar average, moments of operators and \texorpdfstring{$k$}{}-designs}

The Haar measure $\mu_\mathcal{G}$ over a unitary group $\mathcal{G}$ is the unique probability measure which is both left and right invariant over the group $\mathcal{G}$. Formally, for any integrable function $f$ and $V\in\mathcal{G}$ one has:

\ba
\int_{\mathcal{G}}d\mu_{\mathcal{G}}\,f(G)=\int_{\mathcal{G}}d\mu_{\mathcal{G}}\,f(VG)=\int_{\mathcal{G}}d\mu_{\mathcal{G}}\,f(GV)
\ea

The main object we will be interested with in this work is the $k$-th moment operator of an operator $O$ with respect to a group $\mathcal{G}$. Indicating with $\mathcal{L}(\mathbb{C}^d)$ the space of linear operators over $\mathcal{C}^d$, the $k$-th moment operator of $\mathcal{O}\in\mathcal{L}((\mathbb{C}^d)^{\ot k})$ is given by:
\ba
\mathcal{R}_{\cal G}^{(k)}(O)=\int_{\mathcal{G}}d\mu_{\cal G}\,G^{\dag\ot k}OG^{\ot k}
\ea

In order to compute such object, one needs to introduce the commutant of order $k$ of a group $\mathcal{G}$, $\rm{Comm}(\mathcal{G},k)$. This is defined as:
\ba
\rm{Comm}(\mathcal{G},k)=\{O\in\mathcal{L}((\mathbb{C}^d)^{\ot k}):[O,G^{\ot k}]=0\,\forall G\in\mathcal{G}\}
\ea
The $k$-th moment of an operator $\mathcal{R}_{\cal G}^{(k)}(O)$ can be proven to be an element of $\rm{Comm}(\mathcal{G},k)$. As the latter is a vector space, one just needs to find an orthogonal basis $\{F^{(i)}_{\cal G}\}_{i=1}^{\dim(\rm{Comm}(\mathcal{G}),k)}$ of the commutant and then write the moment of the operator $O$ according to the Hilbert Schmidt product:
\ba
\label{eq:moment_operator_expansion}
\mathcal{R}_{\cal G}^{(k)}(O)=\sum_{i=1}^{\dim(\rm{Comm}(\cal{G},k))}\Tr[F_{\cal G}^{(i)}O]F_{\cal G}^{(i)}
\ea

When the average is performed over the unitary group $\mathcal{U}$, one talks about Haar $k$ moments. Connected to this, there is one more important concept to introduce for our work, the one of unitary $k$-design. Let us consider a set of unitaries $\mathcal{V}=\{V\}$, and a probability distribution $\nu$ associated with its elements. The distribution $\nu$ is said to form a unitary $k$-design if:
\ba
\int_{\mathcal{V}}d\nu\, V^{\dag\ot k}OV^{\ot k}=\int_{\cal U}d\mu_{\cal U}\,V^{\dag\ot k}OV^{\ot k},\quad\forall O\in\mathcal{L}((\mathbb{C}^d)^{\ot k})
\ea

With all of this in mind, let us move further to the computation of group averages over the unitary and Clifford groups, before introducing the concept of T-doping.

The computation of moments of operator over the unitary group can be performed rather straightforwardly knowing  that the set of permutations belonging to the permutation group $S_k$ form a base of the commutant of the unitary group. Indicating with $T_\pi$ the permutation operator corresponding to the permutation $\pi\in S_k$, one has from Eq.~\eqref{eq:moment_operator_expansion}:
\ba
\mathcal{R}_{\cal U}^{(k)}(O)=\sum_{\pi\in S_k}c_\pi(O)T_\pi
\ea
where one can show that the coefficients $c_\pi(O)$ are solutions of the system of equations:
\ba
\Tr[T^\dag_{\sigma}O]=\sum_{\pi\in S_k} c_\pi(O)\Tr[T^\dag_\sigma T_\pi]
\ea
The solution can be expressed as:
\ba
\label{eq:c_pi_sol}
c_\pi(O)=\sum_{\sigma}W_{\pi\sigma}\Tr[T_\sigma O]
\ea
Eq.~\eqref{eq:c_pi_sol}, together with the properties of permutation operators, allows us to write the $k-$th moment of the operator $O$ as:
\ba
\label{eq:haar_moments}
\mathcal{R}_{\cal U}^{(k)}=\sum_{\pi,\sigma\in S_k}W_{\pi\sigma}\Tr[T_\sigma O]T_\pi
\ea
The reader can refer to App.~\ref{app:permutations} for details and properties on permutation operators that will be repeatedly used throughout this manuscript.
The coefficients $W_{\pi\sigma}$ are known as  Weingarten functions. They are related to the characters of the irreducible representations of the group $S_k$, and indeed can be expressed in terms of these, see App.~\ref{app:weingarten}. However, there is a simpler method to compute them. Indeed, one can define the Gram matrix $\Omega$, defined as:
\ba
\Omega_{\pi\sigma}=\Tr\left[T_{\pi}T_{\sigma}^\dag\right].
\ea
Then one can easily compute the Weingarten functions as:
\ba
W_{\pi\sigma}=(\Omega^{-1})_{\pi\sigma}.
\ea
So that the final expression for the $k$-th moment operator is given by
\ba
\label{eq:matrix_expression_haar}
\mathcal{R}_\mathcal{U}^{(k)}(O)=\vec{c}^{\,T}(O)\Omega^{-1}\vec{T}=\vec{c}^{\,T}(O)W\vec{T}
\ea
where the vector $\vec{c}(O)$ is a column vector having as components the traces $c_\pi(O)=\tr{T_\pi O}$ and $\vec{T}$ is a column vector having as components the permutation operators $T_\sigma$. Naturally, $W$ is a matrix having as components the Weingarten functions $W_{\pi\sigma}$. As we are going to see, a similar matrix expression to the one in Eq.~\eqref{eq:matrix_expression_haar} can be written also for the Clifford and T-doped averages.

\subsubsection{Clifford group and Clifford 4-th moments}

Let us now discuss and introduce the Clifford group, its main characteristics and its importance in quantum information theory, before turning the attention to averaging over this group. 
The importance of the Clifford group in quantum computation can be hardly underestimated, as it generates the family of states known as stabilizer states. Their importance has its origins in the seminal works by Gottesman~\cite{Gottesman1998_Heisenberg,Gottesman1998FaultTolerant}, where it was shown that this class of states, widely used in quantum error correction~\cite{613213,PhysRevLett.79.953,PhysRevA.52.R2493,steane_multiple-particle_1996,PhysRevLett.77.198,BravyiKitaev1998,PhysRevA.54.4741}, could efficiently be represented on classical computers. The implication of this is that one needs a larger family of states in order to perform quantum computation unfeasible for classical devices.

The Clifford group is defined formally as the normalizer of the Pauli group. The N-qubit Pauli group $\mathcal{P}_N$ is the group whose elements include all Pauli strings, i.e. operators of the form $P=P_1\ot P_2\ot\cdots\ot P_N$ with $P_i\in\{I,X,Y,Z\}$ multiplied by a phase in the set $\phi=\{\pm1,\pm i\}$. The normalizer of a group is the set of operations which maps the group into itself, that is:
\ba
\mathcal{C}=\{C\in\mathcal{L}((\mathbb{C}^2)^{\ot N}):C^\dag PC=P'\in\mathcal{P}_N\}
\ea
Clifford operations are unitary operations, and form themselves a group. The set of generators of the Clifford group includes only three elements. These are the two single qubit gates, the Hadamard gate $H$ and the Phase gate $P$, and the two-qubit CNOT gate $C_X$:
\ba
H=\frac{1}{\sqrt{2}}\begin{pmatrix}
1&1\\
1&-1
\end{pmatrix},\quad P=\begin{pmatrix}
1&0\\
0&i
\end{pmatrix},\quad C_X=\begin{pmatrix}
1&0&0&0\\
0&1&0&0\\
0&0&0&1\\
0&0&1&0
\end{pmatrix}
\ea
Any Clifford unitary $C$ can be realized by combinations of these three elementary gates. Stabilizer states are those states that can be realized, starting from the all-zero string $\ket{0}^{\otimes N}$, by Clifford unitaries. In other words, the set of pure stabilizer states $\rm{PSTAB}$ can be characterized as:
\ba
\rm{PSTAB}=\left\{\ket{\phi}\in\mathcal{H}:\ket{\phi}=C\ket{0}^{\ot N}\right\}
\ea
As the Clifford group is a unitary group, it is possible to perform averages over it, and thus define a corresponding moment of an operator:
\ba
\mathcal{R}_{\cal C}^{(k)}(O)=\int d\mu_{\cal C}\,C^{\dag \ot k}OC^{\ot k}
\ea
The Clifford group has been shown to be a 3-design (almost a 4-design~\cite{zhu2016cliffordgroupfailsgracefully}). In our computation of average values of probes of chaos we will be interested in quantities depending on the fourth moment of a unitary operator $U$. As a consequence, we need to evaluate the moment operator for the Clifford group, as we cannot use the expression in Eq.~\eqref{eq:haar_moments}. The expression for this average has already been computed~\cite{PhysRevLett.121.170502,zhu2016cliffordgroupfailsgracefully,Leone2021quantumchaosis,bittel2025completetheorycliffordcommutant}, and it reads:
\ba
\label{eq:clifford_average}
\mathcal{R}_{\mathcal{C}}^{(4)}(O)=\int_{\cal{C}}d\mu_{\cal C}\,C^{\dag\ot 4}OC^{\ot4}=\sum_{\pi,\sigma\in S_4}W^+_{\pi\sigma}\Tr\left[OQT_\pi\right]QT_{\sigma}+W^-_{\pi\sigma}\Tr\left[OQ^\perp T_\pi\right]Q^\perp T_{\sigma}
\ea
In this expression the $W_{\pi}^{\pm}$ are the {\it generalized Weingarten functions}, which similarly to the Haar scenario can be computed using the group characters, see App.~\ref{app:weingarten}, or via the inverse of generalized Gram matrices:
\ba
W^{\pm}_{\pi\sigma}=(\Omega^{\pm})^{-1}_{\pi\sigma},\quad(\Omega^+)_{\pi\sigma}=\Tr[T_{\pi\sigma}Q],\quad(\Omega^-)_{\pi\sigma}=\Tr[T_{\pi\sigma}Q^{\perp}]
\ea
The operator $Q$ is defined as:
\ba
Q=d^{-2}\sum_{P\in\mathbb{P}_N}P^{\ot4}
\ea
where $d=2^N$ is the dimension of the N-qubit Hilbert space and $\mathbb{P}_N$ indicates the set of all N-qubit Pauli strings without the phases. The operator $Q$ is an orthogonal projector, so that also $Q^{\perp}=1-Q$ is an orthogonal projector. Moreover, $Q$ commutes with all permutation operators.

The Clifford fourth moment of an operator can also be expressed in matrix notation as:
\ba
\label{eq:matrix_exp_clifford}
\mathcal{R}_{\mathcal{C}}^{(4)}(O)=\vec{q}^{\,T} W^+ Q\vec{T}+\vec{q}_{\perp}^{\,T} W^- Q^{\perp}\vec{T}
\ea
where $\vec{q}_\pi=\tr{T_\pi QO}$ and similarly for $\vec{q}_\perp$. Further details on the calculation of the generalized Gram matrices can be found in App.~\ref{app:weingarten}.

\subsubsection{T-doping and the transition to \texorpdfstring{$k$}{}-design}

\begin{figure}[!ht]
\centering
\begin{subfigure}{0.49\linewidth}
\centering
\begin{quantikz}
\lstick{$\ket{0}$}&\gate[5]{C}&&\gate[5]{C}&&\gate[5]{C}&\\
\lstick{$\vdots$}&&&&&&\\
\lstick{$\ket{0}$}&&&&&&\\
\lstick{$\vdots$}&&&&&&\\
\lstick{$\ket{0}$}&&&&&&
\end{quantikz}
\caption{Quantum circuit with only Clifford layers.}
\label{fig:clifford_circuit}
\end{subfigure}
\begin{subfigure}{0.49\linewidth}
\centering
\begin{quantikz}
\lstick{$\ket{0}$}&\gate[5]{C}&\gate[5]{T}&\gate[5]{C}&\gate[5]{T}&\gate[5]{C}&\\
\lstick{$\vdots$}&&&&&&\\
\lstick{$\ket{0}$}&&&&&&\\
\lstick{$\vdots$}&&&&&&\\
\lstick{$\ket{0}$}&&&&&&
\end{quantikz}
\caption{T-doped Clifford circuit.}
\label{fig:t_doped_circuit}
\end{subfigure}
\caption{Illustration of Clifford~\ref{fig:clifford_circuit} and T-doped~\ref{fig:t_doped_circuit} circuits.}
\label{fig:quantum_circuits}
\end{figure}
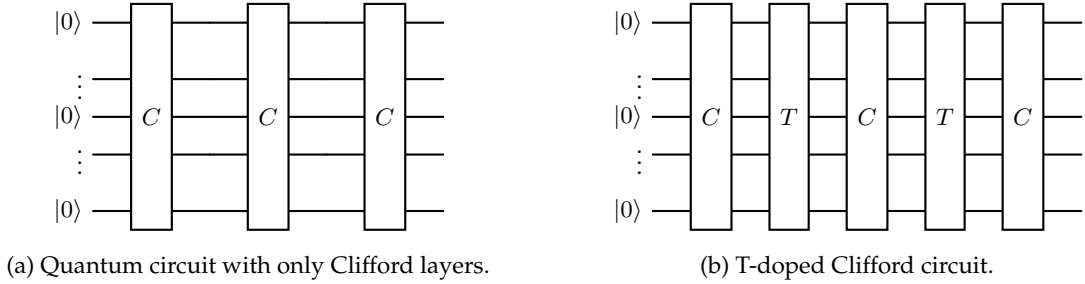

As already mentioned, the Clifford group equipped with the uniform probability measure forms a 3-design. However, the simple addition of another gate to the set of generators of the Clifford group is able to transform the Clifford group into the full unitary group. This gate can be any gate of the form $\Theta=\dyad{0}+e^{-i\theta}\dyad{1}$ with $\theta\neq n\pi/2$ with $n\in\mathbb{N}$, but the most common choice is the $T$ gate, corresponding to $\theta=\pi/4$. The simple addition of the $T$-gate allows one to exit the Clifford orbit and start exploring the whole Hilbert space.
The action of the T-gate consists in the injection into the system of the quantum resource called non-stabilizerness~\cite{turkeshi_magic_2025}. This resource, roughly speaking, measures the distance of a quantum state from the stabilizer polytope~\cite{VeitchMousavianGottesmanEmerson2014}, has been found to be connected with several physical phenomena, among which quantum state learning~\cite{Leone2024learningtdoped,PhysRevA.109.022429,PhysRevLett.132.080402,PhysRevA.106.062434}, relativistic quantum information~\cite{4brj-cl26,9ph7-cyzh,yang2025analytictoolsharvestingmagic}, the AdS/CFT correspondence~\cite{cao2025gravitationalbackreactionmagical,Cepollaro_2024,rz86-47h3}, many-body physics~\cite{10.21468/SciPostPhys.15.4.131,y9r6-dx7p,Iannotti2025entanglement,viscardi2025interplayentanglementstructuresstabilizer,cepollaro2025stabilizerentropysubspaces,Zhou2020SingleT} and quantum foundations~\cite{cusumano2025nonstabilizernessviolationschshinequalities,macedo2025witnessingmagicbellinequalities,haug2025efficientwitnessingtestingmagic}, just to cite a few.
The repeated application of random Clifford unitaries, interspersed with layers containing a single $T$-gate, as shown in Fig.~\ref{fig:quantum_circuits}, allows one to create arbitrary $k$-designs~\cite{Leone2021quantumchaosis,loio2026quantumstatedesignsmagic}.

As the addition of even a single $T$-gate to the circuit is sufficient to bring one out of the Clifford orbit, one is interested in what happens to the moment of an operator $O$. It turns out~\cite{Leone2021quantumchaosis} that one can write an explicit expression for the 4-th moment of an operator averaged over the T-doped Clifford circuits. This expression depends explicitly on both the phase $\theta$ of the gate use for doping and on the number of layers, reading:
\ba
\label{eq:k_doped_average}
\mathcal{R}_{\cal{C}_k}^{(4)}(O)=\sum_{\pi,\sigma\in S_4}\left[\left((\Xi^k)_{\pi\sigma}Q+\Gamma_{\pi\sigma}^{(k)}\right)t_{\pi}(O)+\delta_{\pi\sigma}b_\pi(O)\right]T_\sigma
\ea
The matrices $\Xi$ and $\Gamma^{(k)}$ contain the information about the doping, as they depend on the doping gate $\Theta$. The matrix $\Xi$ has elements:
\ba
\Xi_{\pi\sigma}=\sum_{\tau\in S_4}W_{\pi\tau}^+\Tr\left[T_\sigma \Theta^{\ot4}Q\Theta^{\dag\ot4}QT_\tau\right]-W_{\pi\tau}^-\Tr\left[T_\sigma \Theta^{\ot4}Q^\perp \Theta^{\dag\ot4}QT_\tau\right]
\ea
where $\Theta$ represents a circuit layer with a single $\Theta$ gate applied to any of the qubit.
The matrix $\Gamma_{\pi\sigma}^{(k)}$ is defined as:
\ba
\Gamma_{\pi\sigma}^{(k)}&=&\sum_{\tau\in S_4}\Lambda_{\pi\tau}\sum_{i=0}^{k-1}(\Xi^i)_{\tau\sigma}\\
\Lambda_{\pi\tau}&=&\sum_{\sigma\in S_4}W^-_{\pi\sigma}\Tr\left[T_\tau \Theta^{\ot4}Q\Theta^{\dag\ot4}Q^\perp T_\sigma\right].
\ea
Finally, all the information about the operator $O$ is encoded into the spectral form factors:
\ba
t_\pi(O)&=&\sum_{\sigma\in S_4}\left[W^+_{\pi\sigma}\Tr\left[OQT_\sigma\right]-W^-_{\pi\sigma}\Tr\left[OQ^\perp T_\sigma\right]\right]\\
b_\pi(O)&=&\sum_{\sigma\in S_4}W^-_{\pi\sigma}\Tr\left[OQ^\perp T_{\sigma}\right]
\ea
Just as for the case of Haar and Clifford averages, also in this case one can write the expression in matrix form:
\ba
\label{eq:matrix_exp_doped}
\mathcal{R}_{\cal{C}_k}^{(4)}(O)=\vec{t}^{\,T}\Xi^k Q\vec{T}+\vec{t}^{\,T}\Gamma^{(k)}\vec{T}+\vec{b}^{\,T}\cdot\vec{T}
\ea
where the vector $\vec{t}$ has for components the coefficients $t_\pi$ and similarly $\vec{b}$. While details on the calculation of all the matrices involved in Eq.~\eqref{eq:k_doped_average} can be found in App.~\ref{app:k_doped_average}, here we limit ourselves to illustrate their main features.
 
 The expression in Eq.~\eqref{eq:matrix_exp_doped} can be simplified by diagonalizing the matrix $\Xi$ as:
 \ba
 \Xi=VD_{\Xi}V^{T}
 \ea
 where the matrix $D_{\Xi}$ is diagonal and $V$ is a unitary matrix. One can then write:
\ba
\vec{t}^{\,T}\Xi^kQ\vec{T}=\vec{t}\,VD_\Xi^kV^{T}Q\vec{T}=\vec{v}^{\,T} D_\Xi^kQ\vec{T}_V
\ea
where we have defined $\vec{v}=V\vec{t}$ and $\vec{T}_V=V^{T}\vec{T}$. The main advantage of this rewriting lies in the fact that the matrix $D_\Xi$ has only 6 entries different from zero, corresponding to the non null eigenvalues of $\Xi$.
In a similar way one obtains:
\ba
\vec{t}^{\,T}\Gamma^{(k)}\vec{T}=\vec{t}^{\, T}\Lambda(\sum_{i=0}^{k-1}\Xi^i)\vec{T}=\vec{t}^{\, T}\Lambda V(\sum_{i=0}^{k-1}D_\Xi^i)V^{T}\vec{T}=\vec{\lambda}^{\,T}(1-D_\Xi^k)^{-1}\vec{T}_V
\ea
where this time we have defined:
\ba
\vec{\lambda}=V^{T}\Lambda^T\vec{t}.
\ea
As the matrix $\Xi$ only has six non zero eigenvalues, only the corresponding eigenvectors will contribute to the calculation. These are reported in App.~\ref{app:k_doped_average}, and most noticeably some of them are just projectors onto irreducible representations of the group $S_4$.

With all these substitutions in mind, the expression for the 4-th moment of an operator $O$ over a T-doped circuit with $k$ doping layers is:
\ba
\mathcal{R}_{\mathcal{C}_T}^{(4)}=\vec{v}^{\,T}(O) D_\Xi^kQ\vec{T}_V+\vec{\lambda}^{\,T}(O)(1-D_\Xi^k)^{-1}\vec{T}_V+\vec{b}^{\,T}(O)\cdot\vec{T}
\ea

This exhausts what we had to say regarding the group averages we are going to use throughout the paper. 
In Sec.~\ref{sec:probes_and_isospectral} we are going to introduce probes of chaos, and write them in terms of the isospectral twirling.

\subsection{\label{sec:probes_and_isospectral}Probes of chaos and isospectral twirling}
Let us now illustrate the general features of the probes of chaos we are going to study. 
All the probes of chaos $\mathcal{P}$ we are going to consider in this work consist in the expectation value of the product of two or more observables $A_i(t)$, which get evolved via the unitary operator $V=\exp[-iHt]$:
\ba
\mathcal{P}(\rho)=\Tr[A_1(t)A_2(t)\dots A_k(t)\rho]
\ea
where $A_i(t)=V^\dag A(0)V$ are time evolved operators and $\rho$ is the state of the system. This state is usually assumed to be a thermal one, i.e. $\rho\propto e^{-\beta H}$. In this work we will always work in the limit $\beta\rightarrow 0$, i.e. in the infinite temperature limit, so that $\rho=\mathbb{I}/d$, so that the probes of chaos will take the form:
\ba
\mathcal{P}=\mathcal{P}(\mathbb{I}/d)=d^{-1}\Tr[A_1(t)A_2(t)\dots A_k(t)]
\ea
One can then apply the swap trick (see App.~\ref{app:permutations}) and linearize this expectation value of a product of operators, where the price to pay is to consider multiple copies of the Hilbert space. We will have:
\ba
\mathcal{P}=\Tr[T_{\cal P}(A_1\ot A_2\ot\dots\ot A_k)V^{\ot k,k}]
\ea
where $T_{\cal P}$ represents a permutation operator and the permutation belongs to $S_k$. This probes of chaos will then be averaged over given groups by averaging over all possible unitary evolution $V$ with a given spectrum. As already clarified, this is done by mean of the isospectral twirling. We will indicate the value of a given probe $\mathcal{P}$ averaged over the group $\mathcal{G}$ as:
\ba
\langle\mathcal{P}\rangle_{\cal G}=\Tr[T_{\cal P}(A_1\ot A_2\ot\dots\ot A_k)\mathcal{R}_{\cal G}^{(2k)}(V^{\ot k,k})]
\ea

As the $\mathcal{R}_{\cal G}^{(2k)}$ are proportional either to $\vec{T}$ or $Q\vec{T}$, we will also define vectors of traces of the form:
\ba
\vec{\mathcal{P}}=\Tr[T_{\cal P}(A_1\ot A_2\ot\dots\ot A_k)\vec{T}],\qquad \vec{\mathcal{P}}^Q=\Tr[T_{\cal P}(A_1\ot A_2\ot\dots\ot A_k)Q\vec{T}]
\ea
This vectors will allow us to write the averages of probes of chaos in a very convenient and compact form exploiting the expressions in Eqs.~(\ref{eq:matrix_expression_haar},~\ref{eq:matrix_exp_clifford},~\ref{eq:matrix_exp_doped}).

Thus, the calculation of the isospectral twirling of a probe of chaos $\mathcal{P}$ can be split into two parts. The first part, which is common to all probes, is to compute the spectral dependence of the moment operator $\mathcal{R}_{\cal G}^{(4)}(V^{\ot2,2})$, encoded in the vectors $\vec{c}$,$\vec{q}$ and their combinations.
The second, different for each probe, is to compute the associated vector $\vec{\mathcal{P}}$. The first part will be treated now, as it will also allows us to introduce the reasons for using RMT and averaging over families of spectra. The second part will instead be treated individually for each probe of chaos studied in Sec.~\ref{sec:probes_of_chaos}.

The vector $\vec{c}$ defined in Sec.~\ref{sec:haar_average} has for components some traces involving the operator to be averaged. When this operator is a unitary $V$, these traces become proportional to the {\it spectral form factors} $g_k$ of the operator $V$, a ubiquitous quantity in the study of quantum chaos~\cite{PhysRevE.55.4067,PhysRevLett.80.1808,PhysRevE.60.3949,kunz_probability_1999,n7rj-gwwj,PhysRevD.98.086026,venuti2025integrabilitychaosfractalanalysis} and other fields~\cite{caceres_spectral_2022,PhysRevLett.131.151602,PhysRevResearch.2.043403,garcia-garcia_exact_2018}.

Let us consider as first example the computation of the vector $\vec{c}$ for the isospectral twirling of order $k=1$:
\ba
\mathcal{R}_{\cal U}^{(2)}(V^{\ot1,1})
\ea
In this case we need to compute two components of the vector $\vec{c}$, corresponding to the two permutations of $S_2$, the identity $I$ and the swap $T_{(12)}$. These reads:
\ba
\vec{c}_I=\Tr[IV^{\ot1,1}]=|\Tr[V]|^2=g_2(t),\qquad\vec{c}_{T_{(12)}}=\Tr[T_{(12)}V^{\ot1,1}]=\Tr[VV^\dag]=\Tr[\mathbb{I}_d]=d
\ea
The first component is proportional to the two point spectral form factor of $V$. We can exploit the spectral decomposition of $V$ to write the function $g_2$ explicitly. As $V=\exp[-iHt]$, we can write it in diagonal form as:
\ba
V=\sum_{i}e^{-iE_it}\dyad{E_i}
\ea
where the $E_i$ are the eigenvalues of $H$ and $\ket{E_i}$ the associated eigenvectors. Using this expression we can write:
\ba
g_2(t)=|\Tr[V]|^2=\sum_{i,j}e^{-i(E_i-E_j)t}
\ea
As we are going to see in Sec.~\ref{sec:RMT}, these spectral form factors will be later averaged over families of random spectra associated with integrable or chaotic dynamics.

The computation of the vector $\vec{c}$ for the isospectral twirling of order $k=2$ includes also traces proportional to the three and four points spectral form factors $g_3(t)$ and $g_4(t)$. While we leave the details of the computation in App.~\ref{app:spectral_functions}, here we only report the expressions of the three and four point spectral form factors:
\ba
g_3(t)&=&\sum_{i,j,k}e^{-i(2E_i-E_j-E_k)t}\\
g_4(t)&=&\sum_{i,j,k,\ell}e^{-i(E_i+E_j-E_k-E_\ell)t}
\ea

Let us then turn the attention on the components of the vector $\vec{q}$. This vector has for components traces of the form $\Tr[T_\pi QV^{\ot2,2}]$. As shown in App.~\ref{app:spectral_functions}, these traces depend on the Clifford spectral form factors $\tilde{g}_k$ there defined. These generalized spectral form factors, just as the usual spectral form factors, depend on the spectrum of $V$. However, in addition to this, they also depend on the expectation value of the eigenstates of $V$ on the Pauli strings $P$. Let us consider for instance the component of $\vec{q}$ associated with the identity permutation. One obtains:
\ba
\label{eq:pauli_correlation_function}
\Tr[IQV^{\ot2,2}]=d^{-2}\sum_P\sum_{i,j,k,\ell}e^{-i(E_i+E_j-E_k-E_\ell)t}\bra{E_i}P\ket{E_i}\bra{E_j}P\ket{E_j}\bra{E_k}P\ket{E_k}\bra{E_\ell}P\ket{E_\ell}
\ea
As one can see, the evaluation of the Clifford spectral form factors depend on the initial eigenstates of $V$. Indeed, the values $\bra{E_i}P\ket{E_i}$ serve as a sort of initial condition on the Clifford average: this is because, by averaging over the Clifford group, one is lead to explore the Clifford orbit of a given operator, rather than the whole set of possible unitary operator.
As shown in App.~\ref{app:spectral_functions}, it is in general not possible to go much further than the expression in Eq.~\eqref{eq:pauli_correlation_function}, which leads us to define the Clifford spectral form factors $\tilde{g}_k(t)$ in App.~\ref{app:spectral_functions}. However, some simplifications will be possible when dealing with stabilizer Hamiltonian: these can be written as $H=CDC$, with $C\in\mathcal{C}$ a unitary belonging to the Clifford group.  Because of this, when performing the average of $C$ over the Clifford group,  one will explore only and exactly the Clifford unitaries, recovering the separability between the spectral and eigenvectors averages.

Indeed, as shown in Sec.~\ref{sec:stabilizer_hamiltonian}, the Clifford spectral form factors in Eq.~\eqref{eq:pauli_correlation_function} will reduce to the usual spectral functions $g_k(t)$, the only exception being the one obtained from the identity permutation. While this might seem a coincidence, it is actually consistent with the fact that the Clifford group forms a 3-design, so that only spectral form factors containing more than three energy values will differ. Indeed, we will see that the spectral form factor stemming from the identity permutation, corresponding to $g_4(t)$ in the unitary case, will be different from the ones obtained for the average over the unitary group. It will be dubbed $\tilde{g}_3(t)$, since, in contrast with the full unitary case, it will only depend on three energy values.

In the following sections we will introduce RMT and the ensemble of spectra over which our spectral form factors will be averaged. Moreover, we will also introduce the other average of interest in this work, the long time average.

\subsection{\label{sec:RMT}Random matrix theory and spectral averages}
As already mentioned in Sec.~\ref{sec:introduction}, the origins of RMT date back to Wigner, who first tried to make prediction on the emission spectra of complex nuclei in neutron scattering experiments.
The idea was that, in absence of a precise and reliable physical model to describe the nuclei,one could still predict properties of the emission spectrum by mean of statistical properties that the spectra of such systems should possess.
The way to build this random spectra is to consider the complex quantum system under examination to possess an Hamiltonian described by a random matrix, and then to average this matrix over the probability distribution from which its elements are sampled.

In this work we will consider two ensembles. The first one will be the Gaussian Diagonal Ensemble (GDE), made out of real diagonal matrices whose entries are distributed according to a Gaussian probability distribution centered in zero with unit width. The probability distribution of the spectrum of such matrices reads:
\ba
P_{\rm GDE}\left(\{E_i\}\right)\propto\left(\frac{2}{\pi}\right)^{\frac{1}{2}}\exp\left[-2\sum_i E_i^2\right]
\ea
where $P\{E_i\}=P(E_1,E_2,\cdots,E_d)$ indicates the probability of obtaining a given spectrum. This distribution of the energy eigenvalues characterize integrable systems~\cite{riser_power_2021}.

The other ensemble we will be dealing with is the Gaussian Unitary Ensemble (GUE). This ensemble includes complex Hermitian matrices, whose elements are sampled according to a Gaussian distribution centered in zero having width $d^{-1/2}$. The corresponding probability distribution for the spectrum of these matrices is:
\ba
P_{\rm GUE}\left(\{E_i\}\right)\propto\exp\left[-\frac{d}{2}\sum_i E_i^2\right]\prod_{i<j}|E_i-E_j|^2
\ea
Besides the different normalization in the exponential term, the main difference between $P_{\rm GDE}$ and $P_{\rm GUE}$ is the presence in the latter of the level repulsion term $\prod_{i<j}|E_i-E_j|^2$, which is characteristic of chaotic systems.

Let us now compute the spectral average of the two point spectral form factor $g_2(t)$ for the GDE and the GUE, leaving the other $g_k(t)$ for App.~\ref{app:spectral_averages}.

The definition of the spectral average is:
\ba
\nonumber
\overline{g_2(t)}^{\rm E}&=&\sum_{i,j=1}^d\int dE_1\,dE_2\,\cdots dE_de^{-i(E_i-E_j)t}P_{\rm E}(\{E_k\})\\
\nonumber
&=&\sum_{i,j=1}^d\int dE_i\,dE_j\,e^{-i(E_i-E_j)t}\int d\{E_{/ij}\}P_{\rm E}(\{E_k\})
\ea
where $d\{E_{/ij}\}$ stands for the product of all the $dE_k$ but $dE_i,dE_j$. The first thing to note is that $E_i,E_j$ are dummy variables. So, when in the sum $i=j$ the result is automatically 1 because of the normalization of the probability distribution $P_E$, and the sum reduces to the counting $i\neq j$.
Let us also define the n-point marginal probability distribution $\rho_{\rm E}^{(n)}(E_1,E_2,\cdots,E_n)$ as:
\ba
\label{eq:n_marginal_prob_def}
\rho_{\rm E}^{(n)}(E_1,E_2,\cdots,E_n)=\int dE_{n+1}\cdots dE_d P_{\rm E}(\{E_k\})
\ea
Using $\rho_{\rm E}^{(n)}(E_1,E_2,\cdots,E_n)$ one can rewrite the spectral average as:
\ba
\overline{g_2(t)}^{\rm E}&=&d+d(d-1)\int dE_1\,dE_2\,e^{-i(E_1-E_2)t}\rho_{\rm E}^{(2)}(E_1,E_2)
\ea
We are now ready to compute the averages over the GDE and GUE.
First of all, let us notice that the GDE distribution, missing the level repulsion terms, factorizes:
\ba
P_{\rm GDE}(\{E_i\})=\prod_{i=1}^d P_{\rm GDE}(E_i),\qquad P_{\rm GDE}(E_i)=\left(\frac{2}{\pi}\right)^{\frac{1}{2}}e^{-2E_i^2}
\ea
this makes the calculation of $\overline{g_2(t)}^{\rm GDE}$ extremely straightforward:
\ba
\overline{g_2(t)}^{\rm GDE}&=&d+d(d-1)\left(\left(\frac{2}{\pi}\right)\int dE_1\,dE_2\,e^{-i(E_1-E_2)t}e^{-2E_1^2}e^{-2E_2^2}\right)\\
&=&d+d(d-1)e^{-\frac{t^2}{4}}
\ea

Let us now turn to a way less trivial example, the average of $g_2(t)$ over the GUE. Thanks to Eq.~\eqref{eq:n_marginal_prob_def}, we only need to compute $\rho_{\rm GUE}^{(n)}(E_1,\cdots,E_n)$. This can be expressed as~\cite{dyson_statistical_1962,dyson_statistical_1962_II,dyson_statistical_1962_III,PhysRevE.55.4067,PhysRevD.98.086026,mehta_random_1991,haake_quantum_2001}:
\ba
\label{eq:n_marginal_prob_formula}
\rho_{\rm GUE}^{(n)}(E_1,\cdots,E_n)=\frac{(d-n)!}{d!}\det[K]
\ea
where $K$ is a matrix whose elements, in the large $d$ limit, read:
\ba
K_{ij}=\delta_{ij}\frac{d}{2\pi}\sqrt{4-E_i^2}+(1-\delta_{ij})\frac{\sin(d(E_i-E_j))}{\pi(E_i-E_j)}
\ea
The two point marginal probability distribution can then be easily computed as:
\ba
\rho_{\rm GUE}^{(2)}(E_1,E_2)=\frac{1}{d(d-1)}\left(\frac{d^2}{4\pi^2}\rho(E_1)\rho(E_2)-\frac{\sin^2(d(E_1-E_2))}{\pi(E_1-E_2)^2}\right)
\ea
Finally enforcing the box approximation~\cite{Cotler2017Chaos}, one obtains:
\ba
\overline{g_2(t)}^{\rm GUE}=d+d^2r_1^2(t)-r_2(t)
\ea
where $r_1(t)=J_1(2t)/t$ with $J_1(x)$ being the Bessel function of the first kind and $r_2=\theta(2d-t)\left(1-\frac{t}{2d}\right)$ where $\theta(x)$ is a function equal to $1$ for $x>0$ and equal to $0$ for $x<0$.

\begin{figure}
\centering
\begin{subfigure}{0.49\textwidth}
\centering
\includegraphics[width=\linewidth]{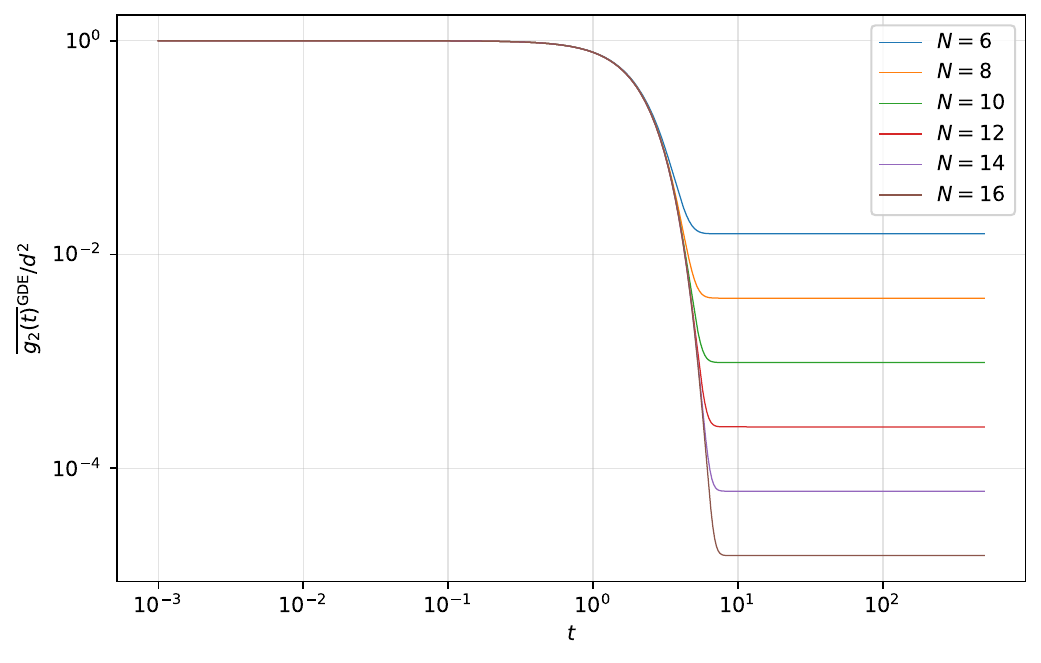}
\caption{Log-Log plot of the normalized $\overline{g_2(t)}^{GDE}/d^2$}
\label{fig:g2_GDE}
\end{subfigure}
\begin{subfigure}{0.49\textwidth}
\centering
\includegraphics[width=\linewidth]{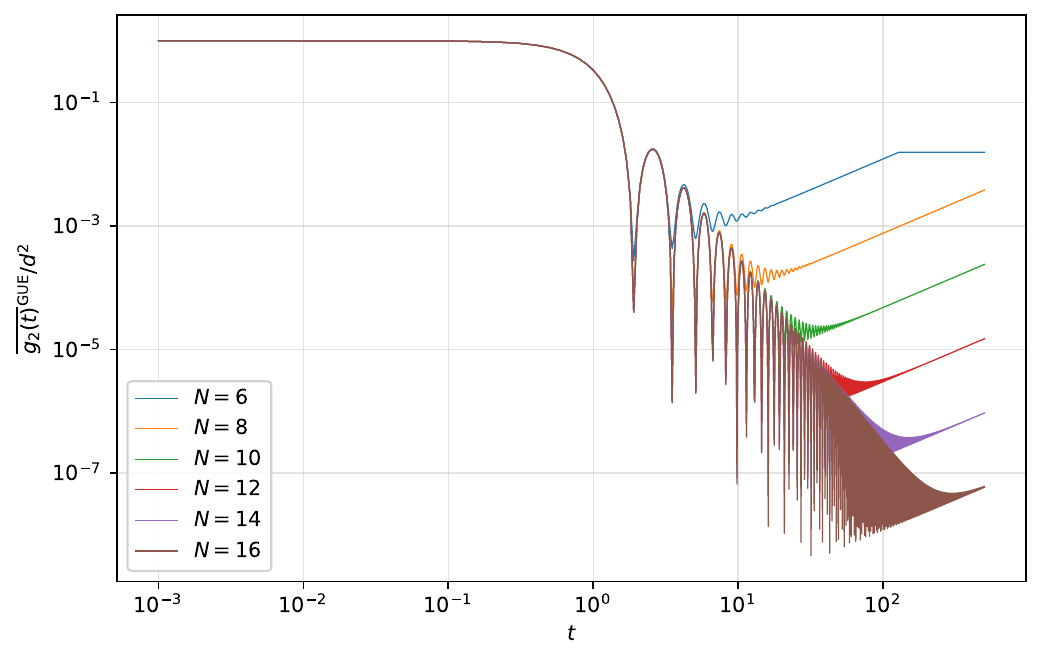}
\caption{Log-Log plot of the normalized $\overline{g_2(t)}^{\rm GUE}/d^2$}
\label{fig:g2_GUE}
\end{subfigure}
\caption{Plot of the normalized versions of $\overline{g_2(t)}^{\rm GDE}$ (~\ref{fig:g2_GDE}) and $\overline{g_2(t)}^{\rm GUE}$ (~\ref{fig:g2_GUE}) for $d=2^{N}$.}
\label{fig:g2}
\end{figure}

Let us also briefly review how to compute the envelope curves of the averaged spectral form factors over the GUE and the long time behavior.

As shown in~\cite{Cotler2017Chaos,10.21468/SciPostPhys.10.3.076}, in order to compute the envelope curves one has to first compute the large $d$ limit, and then suppress all the oscillating terms.

One can consider for instance the two point-spectral form factor $g_2(t)$ averaged over the GDE and GUE. One has:
\ba
\overline{g_2(t)}^{\rm GUE}&=&d + (d r_1(t))^2-dr_2(t)
\ea
Taking the large $d$ limit and suppressing any oscillating part results in:
\ba
\overline{g_2(t)}^{\rm GUE}&\approx &d+d^2 \pi^{-1} t^{-3}+d\theta(t-2d)\left(1-\frac{t}{2d}\right)
\ea
From the envelope function it is possible to compute some dynamical features of the spectral form factor. Namely, it is possible to compute the equilibration time $t_{\rm eq}$ after which the function reaches its long time value. To do this one has to impose that the time independent part of the envelope function equals the time dependent part. For instance, considering $\overline{g_2(t)}^{\rm GUE}$, neglecting the $r_2(t)$ part, one has to solve the equation:
\ba
d\approx \frac{d^2}{\pi t^3}\Rightarrow d\approx \pi t^3\Rightarrow t_{\rm eq}=\mathcal{O}(d^{1/3})
\ea
that is, $\overline{g_2(t)}^{\rm GUE}$ equilibrates to its long time value after a time of order $d^{1/3}$.
As for the long time limit, one can easily compute it from the envelope curves as:
\ba
\lim_{t\rightarrow\infty}\overline{g_2(t)}^{\rm GUE}=d
\ea
while for the GDE, as there are no oscillating pieces, one can simply compute the long time limit as:
\ba
\lim_{t\rightarrow\infty}\overline{g_2(t)}^{\rm GDE}&=&\lim_{t\rightarrow\infty}d+d(d - 1) e^{-t^2/4}=d
\ea
while the equilibration time can be computed the same way as done for $\overline{g_2(t)}$.
One can proceed the same way for all the other spectral form factors, obtaining for $g_4(t)$:
\ba
\overline{g_4(t)}^{\rm GUE}&\approx&\frac{d^4}{\pi^{2}t^{6}}+d^2\Big[2-\frac{31}{8\pi t^3}-\frac{8r_2(2t)-16r_2(t)+r_2(t)/\sqrt{2}}{\pi^{3/2}t^{5/2}}-4r_2(t)+2r_2^2(t)+\frac{r_2^2(t)}{\pi^2t^2}\Big]
\ea
to which correspond a dip time of order $\mathcal{O}(d^{1/6})$ and an equilibration time of order $\mathcal{O}(d)$. The long time value is easily seen to be
\ba
\lim_{t\rightarrow\infty}\overline{g_4(t)}^{GUE}=2d^2
\ea
For $\overline{\tilde{g}_3(t)}^{\rm GUE}$ one gets:
\ba
\overline{\tilde{g}_3(t)}^{\rm GUE}\approx 2d+d^2\left[\frac{1}{\pi^2 t^6}-\frac{4r_2(t)}{\pi^2t^3}-\frac{2r_2(t)}{\pi^3 t^4}+\frac{2r_2^2(t)}{d}+\frac{r_2^2(t)}{\pi^2t^2d}\right]
\ea
One notices from this envelope function that the Clifford spectral form factor $\tilde{g}_3(t)$ presents features in between $g_2(t)$ and $g_4(t)$. Just as $g_2(t)$, $\tilde{g}_3(t)$ has an initial value of $d^2$, and then quickly decays to a value $\mathcal{O}(d)$ in a time $\mathcal{O}(1)$. On the other hand, $\tilde{g}_3(t)$ inherits its oscillatory features from $g_4(t)$. However, this oscillatory features are suppressed by a factor $d^2$ compared to $g_4(t)$, so that $\tilde{g}_3(t)$ after a short and small revival around a time $\mathcal{O}(d^{1/6})$ effectively equilibrates to its equilibrium value of $\mathcal{O}(d)$. The salient features of $g_2(t),\tilde{g}_3(t),g_4(t)$ are summarized in Table~\ref{tab:envelopes}.

\begin{table}[!ht]
\begin{tabular}{|c|c|c|}
\toprule
$\overline{g_2(t)}^{\rm GUE}$&$\overline{\tilde{g}_3(t)}^{\rm GUE}$&$\overline{g_4(t)}^{\rm GUE}$\\
\midrule
\begin{tabular}{|c|c|}
\toprule
Value&Time\\
\midrule
$\mathcal{O}(d)$&$\mathcal{O}(1)$\\
$\mathcal{O}(d^{3/2})$&$\mathcal{O}(d^{1/6})$\\
$\mathcal{O}(d)$&$\mathcal{O}(d^{1/3})$\\
$\mathcal{O}(d^{1/2})$&$\mathcal{O}(d^{1/2})$\\
$\mathcal{O}(d)$&$\mathcal{O}(d)$\\
\bottomrule
\end{tabular}&
\begin{tabular}{|c|c|}
\toprule
Value&Time\\
\midrule
$\mathcal{O}(2d)$&$\mathcal{O}(1)$\\
$\mathcal{O}(3d)$&$\mathcal{O}(d^{1/6})$\\
$\mathcal{O}(2d)$&$\mathcal{O}(d^{1/3})$\\
$\mathcal{O}(2d)$&$\mathcal{O}(d^{1/2})$\\
$\mathcal{O}(2d)$&$\mathcal{O}(d)$\\
\bottomrule
\end{tabular}&
\begin{tabular}{|c|c|}
\toprule
Value&Time\\
\midrule
$\mathcal{O}(d^2)$&$\mathcal{O}(1)$\\
$\mathcal{O}(d^3)$&$\mathcal{O}(d^{1/6})$\\
$\mathcal{O}(d^2)$&$\mathcal{O}(d^{1/3})$\\
$\mathcal{O}(d)$&$\mathcal{O}(d^{1/2})$\\
$\mathcal{O}(d^2)$&$\mathcal{O}(d)$\\
\bottomrule
\end{tabular}\\
\bottomrule
\end{tabular}
\caption{Summary of the behavior of the envelope functions }
\label{tab:envelopes}
\end{table}

\subsection{\label{sec:physical_models}Physical models}

As we have seen, the Clifford spectral form factors depend on the expectation value of the eigenstates of $V$ over the Pauli strings. For this technical reason we need to specify an initial Hamiltonian model, in contrast with~\cite{10.21468/SciPostPhys.10.3.076}, where the initial condition is not needed.

The choice we make is to consider stabilizer Hamiltonian. The first reason for this choice is technical, as all eigenstates of a stabilizer Hamiltonian are stabilizer states, which have a simple $\pm1$ expectation value over Pauli strings, simplifying the computation of the Clifford spectral form factors.
Besides this technical reason, there is also a more profound reason to choose Hamiltonian models having stabilizer eigenstates. The most general random Hamiltonian has its eigenvectors sampled from the Thomas-Porter distribution. These vectors are uniformly distributed on a $d-1$ dimensional sphere, and in general are not stabilizer. Thus, using a random Hamiltonian, via its random eigenstates, we would be introducing non-stabilizerness in the system, spoiling the possibility of obtaining information on the role of non-stabilizerness in the transition from integrable to chaotic behavior.

In the following sections we are going to illustrate the main features of stabilizer Hamiltonian and their eigenstates. We will first show that, because all the quantities we are going to study are invariant under Clifford rotation, we can choose a stabilizer Hamiltonian whose eigenstates are the computational basis states.
Then, we will focus on a specific instance of stabilizer Hamiltonian, the Toric code Hamiltonian.

\subsubsection{\label{sec:stabilizer_hamiltonian}Stabilizer Hamiltonians}
A stabilizer Hamiltonian is a quantum many‑body Hamiltonian constructed from a set of mutually commuting Pauli strings. Among the many properties of these Hamiltonian, the (degenerate) eigenspace corresponding to the eigenvalue $+1$ is the ground states of such Hamiltonian, offering the possibility of error protected quantum computation~\cite{LidarBrunQEC2013,TerhalQECMemories2015,BrownColloquium2016} in the context of topological error correcting codes~\cite{Dennis2002,Poulin2005Subsystem,Fowler2012}, such as the Toric code~\cite{BravyiKitaev1998,KitaevToricCode2003} and the Color code~\cite{BombinMartinDelgado2006,Bombin2015GaugeColorCode}. In these codes, errors correspond to higher energy levels of the Hamiltonian, which are thus penalized. Stabilizer Hamiltonians have also been studied in other contexts, such as thermodynamics~\cite{temme2015faststabilizerhamiltoniansthermalize,PhysRevB.77.064302} and statistical mechanics~\cite{alicki_statistical_2007,bravyi_no-go_2009,PhysRevLett.123.230503,PhysRevA.83.042330,10.21468/SciPostPhys.6.4.041}. 

The most generic N-qubit stabilizer Hamiltonian can be written as:
\ba
\label{eq:stab_ham}
H_{\rm stab}=\sum_{P\in\langle\mathcal{P}_{\rm ab}\rangle}\omega_PP
\ea
where $\langle\mathcal{P}_{\rm ab}\rangle$ indicates the set of generators of an Abelian subgroup of the Pauli group $\mathcal{P}_N$. A (trivial) example of such Hamiltonian is the one of $N$ non-interacting spins, each with transition frequency $\omega_i$:
\ba
\label{eq:comp_basis_ham}
H_{\rm cb}=\sum_{i=1}^N\omega_i Z_i
\ea
The Hamiltonian in Eq.~\eqref{eq:comp_basis_ham} can be easily seen to have eigenstates coinciding with the computational basis states. Moreover, it is easy to see that any Hamiltonian of the form in Eq.~\eqref{eq:stab_ham} can be brought back to the computational basis one by means of Clifford operations. This implies that we can study the Hamiltonian in Eq.~\eqref{eq:comp_basis_ham} without loss of generality.

The Pauli strings appearing in a stabilizer Hamiltonian, being the generators of an Abelian group, all commute with each other. Moreover, the product of all the elements of an Abelian group is always equal to the identity, unless there is a unique elements of order 2 in the group~\cite{7f616d19-7bad-3244-9920-e3bd00bd46f6,Gorowski2014714}, which is never the case unless $N=1$. In formulas we have:
\ba
\prod_{P\in\mathcal{P}_{\rm ab}}P=\mathbb{I},\qquad [P,P']=0\;\forall P\in\langle\mathcal{P}_{\rm ab}\rangle
\ea
Because of these properties, the unitary operator generated by a stabilizer Hamiltonian can always be brought in the form:
\ba
V_{\rm stab}=\exp[-iH_{\rm stab}t]=\prod_{P\in\langle\mathcal{P}_{\rm ab}\rangle}\exp[-i\omega_PPt]=\prod_{P\in\langle\mathcal{P}_{\rm ab}\rangle}\left(\cos(\omega_P t)\mathbb{I}-i\sin(\omega_Pt) P\right)
\ea
Let us finally notice that the energy eigenvalues $E_i$ of any $H_{\rm stab}$ can be written in terms of the coefficient $\omega_P$ as:
\ba
E_i=\vec{e}_i\cdot\vec{\omega}_P
\ea
where $\vec{\omega}_P=(\omega_1,\omega_2,\cdots,\omega_N)$ is an N-dimensional vector with components being the parameters $\omega_P$, while $\vec{e}_i=(e_i^1,e_i^2,\cdots,e_i^N)$ is another N-dimensional vector whose elements $e_i^j=\pm1$. As there are $2^N=d$ such vectors, one can see that mapping is valid.

The corresponding unitary operator $V_{\rm stab}$ allows for some simplifications in the computation of the Clifford spectral form factors. To see this, let us compute only a couple of traces here, leaving the rest for App.~\ref{app:stab_ham_spectral_functions}. Specifically, let us show the component $\vec{q}_I$ of the vector $\vec{q}$ for the case of the computational basis Hamiltonian, as this is the only one giving a spectral form factor not identical to one of the $g_k(t)$ found for the Haar case. In order to perform this trace, one has to express the Pauli strings as:
\ba
P=\mathbf{X}^{\vec{x}}\mathbf{Z}^{\vec{z}}
\ea
where $\mathbf{X}=X_1\ot X_2\ot\dots\ot X_N$ and similarly $\mathbf{Z}=Z_1\ot Z_2\ot\dots\ot Z_N$. The vectors $\vec{x},\vec{z}$ are $N$ components vectors from $\{0,1\}^{\times N}$.
Moreover, the action of $\mathbf{Z}^{\vec{z}}$ on a computational basis state $\ket{i}$ is given by:
\ba
\mathbf{Z}^{\vec{z}}\ket{i}=(-1)^{\vec{z}\cdot\vec{i}}\ket{i}
\ea
where $\vec{i}$ is an $N$ components vector corresponding to the binary string $\ket{i}$. Finally, we will make use of the following property:
\ba
\sum_{\vec{z}\in\{0,1\}^{\times N}}(-1)^{\vec{z}\cdot\vec{i}}=d\delta_{\vec{i},\vec{0}}
\ea
Let us finally turn to the computation of $\vec{q}_I$. We have:
\ba
\nonumber
&&\vec{q}_I=\Tr[IQV_{\rm cb}^{\ot2,2}]=d^{-2}\sum_P\left|\Tr[V_{\rm cb}P]\right|^4\\
\nonumber
&&=d^{-2}\sum_P\sum_{i,j,k,\ell}e^{-i(E_i+E_j-E_k-E_\ell)t}\Tr[P\dyad{E_i}]\Tr[P\dyad{E_j}]\Tr[P\dyad{E_k}]\Tr[P\dyad{E_\ell}]\\
\nonumber
&&=d^{-2}\sum_{\vec{z},\vec{x}}\sum_{i,j,k,\ell}e^{-i(E_i+E_j-E_k-E_\ell)t}\Tr[\mathbf{X}^{\vec{x}}\mathbf{Z}^{\vec{z}}\dyad{E_i}]\Tr[\mathbf{X}^{\vec{x}}\mathbf{Z}^{\vec{z}}\dyad{E_j}]\Tr[\mathbf{X}^{\vec{x}}\mathbf{Z}^{\vec{z}}\dyad{E_k}]\Tr[\mathbf{X}^{\vec{x}}\mathbf{Z}^{\vec{z}}\dyad{E_\ell}]\\
\nonumber
&&=d^{-2} \sum_{\vec{x},\vec{z}}\sum_{i,j,k,\ell} e^{-i(E_i+E_j-E_k-E_\ell)t}(-1)^{\vec{z}\cdot(\vec{i}\oplus\vec{j}\oplus\vec{k}\oplus\vec{\ell})}\delta_{\vec{x},\vec{0}}\\
\nonumber
&&=d^{-1}\sum_{i,j,k,\ell} e^{-i(E_i+E_j-E_k-E_\ell)t}\delta_{(\vec{i}\oplus\vec{j}\oplus\vec{k}\oplus\vec{\ell}),\vec{0}}\\
&&=d^{-1}\sum_{i,j,k} e^{-i(E_i+E_j-E_k-E_{i\oplus j\oplus k})t}=\tilde{g}^{\rm cb}_3(t)
\ea
As one notices, $\vec{q}_I$ depends on only three indices, and that is why we dubbed the corresponding spectral form factor as $\tilde{g}^{\rm cp}_3(t)$. Moreover, because of the equivalence between $H_{\rm cp}$ and any $H_{\rm stab}$, this result holds true also for the latter class of Hamiltonians. Finally, we highlight once again that this is consistent with the Clifford group being a 3-design: the permutation $I$, whose corresponding term should give rise to the four point spectral form factor as for $\vec{c}_I$, gives instead rise to a modified three point spectral form factor.

Before turning to a specific example of stabilizer Hamiltonians, namely the Toric Code Hamiltonian, let us also show the general mechanism under which all the other traces in $\vec{q}$ reduce to some $g_k(t)$ with $k\leq3$. Let us thus consider the permutation $T_{(12)}$. For the case of a stabilizer Hamiltonian, the corresponding component of $\vec{q}$ reads:
\ba
\vec{q}_{T_{(12)}}=\Tr[T_{(12)}QV_{\rm stab}^{\ot2,2}]=d^{-2}\sum_P\Tr[PV_{\rm stab}PV_{\rm stab}]\Tr[PV_{\rm stab}^\dag]^2
\ea The only thing to note in order to compute this trace is that, because of the term $\Tr[PV_{\rm stab}^\dag]$, only the strings $P$ belonging to the Abelian subgroup generated by the strings in $H_{\rm stab}$ will contribute to the sum. Thus one gets:
\ba
\nonumber
\vec{q}_{T_{(12)}}&=&d^{-2}\sum_P\Tr[PV_{\rm stab}PV_{\rm stab}]\Tr[PV_{\rm stab}^\dag]^2
=d^{-2}\sum_{P\in\mathcal{P}_{\rm ab}}\Tr[PV_{\rm stab}PV_{\rm stab}]\Tr[PV_{\rm stab}^\dag]^2\\
\nonumber
&=&d^{-2}\sum_{P\in\mathcal{P}_{\rm ab}}\Tr[V_{\rm stab}^2]\Tr[PV_{\rm stab}^\dag]^2=d^{-2}\Tr[V_{\rm stab}^2]\sum_{P}\Tr[PV_{\rm stab}^\dag]^2\\
\nonumber
&=&d^{-1}\Tr[V_{\rm stab}^2]\Tr[T_{(12)}V_{\rm stab}^{\dag\ot2}]=d^{-1}g_2(2t)
\ea
As shown in App.~\ref{app:stab_ham_spectral_functions} one can proceed in a similar way for the other permutations, finding the results summarized in Table~\ref{table:stab_ham_q}.

\begin{table}[!ht]
\begin{tabular}{|c|c|}
\toprule
$T_\pi$&$\tr{T_\pi QV_{\rm cb}^{\ot 2,2}}$\\
\midrule
$I$&$\tilde{g}_{3}^{\rm cb}(t)$\\
$T_{(ij)}/\{T_{(12)},T_{(34)}\}$&$d$\\
$T_{(12)}$&$d^{-1}g_2(2t)$\\
$T_{(34)}$&$d^{-1} g_2(2t)$\\
$T_{(ij)(k\ell)}$&$\tilde{g}_3^{\rm cb}(t)$\\
$T_{(ijk)}$&$1$ \\
$T_{(ijk\ell)}$ crossing&$d^{-1} g_2(2t)$\\
$T_{(ijk\ell)}$ non-crossing&$d$\\
\bottomrule
\end{tabular}
\caption{Value of the components of $\vec{q}$ for stabilizer Hamiltonians. By crossing 4-cycles we mean permutations that lead to alternating $U$ and $U^\dag$ sequences (up to the cyclicity of the trace), and by non-crossing we refer to permutations that do not do that.}
\label{table:stab_ham_q}
\end{table}

\subsubsection{The Toric Code}

\begin{figure}
\begin{tikzpicture}[scale=1]
    \def\n{5}
    \def\s{1}

    \foreach \i in {0,...,\n} {
        \foreach \j in {0,...,\n} {
            \fill (\i*\s,\j*\s) circle (2pt);
        }
    }

    \foreach \i in {0,...,\n} {
        \foreach \j in {0,...,\n} {
            \ifnum\i<\n
                \draw (\i*\s,\j*\s) -- ({(\i+1)*\s},\j*\s);
            \fi
            \ifnum\j<\n
                \draw (\i*\s,\j*\s) -- (\i*\s,{(\j+1)*\s});
            \fi
        }
    }


    \draw[red,very thick] (1*\s,2*\s) -- (2*\s,2*\s);
    \draw[red,very thick] (2*\s,2*\s) -- (2*\s,3*\s);
    \draw[red,very thick] (2*\s,3*\s) -- (1*\s,3*\s);
    \draw[red,very thick] (1*\s,3*\s) -- (1*\s,2*\s);

    \node at (1.5*\s, 2.5*\s) {$B_f$};


    \ifnum 4<\n
        \draw[blue,very thick] (4*\s,4*\s) -- (5*\s,4*\s);
    \fi

    \ifnum 4<\n
        \draw[blue,very thick] (4*\s,4*\s) -- (4*\s,5*\s);
    \fi

    \ifnum 4<\n
        \draw[blue,very thick] (4*\s,4*\s) -- (4*\s,3*\s);
    \fi

    \ifnum 4<\n
        \draw[blue,very thick] (4*\s,4*\s) -- (3*\s,4*\s);
    \fi

    \node[blue,anchor=south west] at (4*\s,4*\s) {$A_v$};
\end{tikzpicture}
\caption{The Toric code. A qubit lives on each of the $2N^2$ edges of the lattice. For each vertex $v$ and facet $f$ one defines respectively a vertex $A_v$ and facet $B_f$ operator respectively. Notice that the lattice has the topology of a torus, i.e. one has periodic boundary conditions.}
\label{fig:toric_code}
\end{figure}
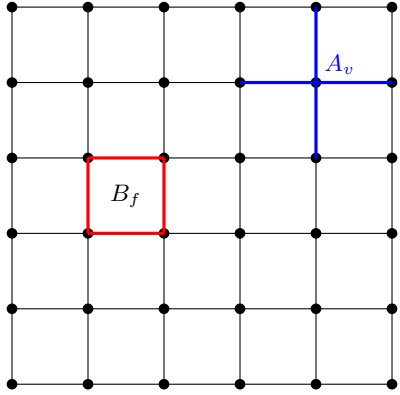

Let us now illustrate the Toric code. The Toric Code, introduced by Kitaev~\cite{KitaevToricCode2003,KITAEV20062}, represents a prototypical example of a stabilizer code. In the Toric Code, $2N^2$ qubits are placed on the edges of an $N\times N$ lattice, see Fig.~\ref{fig:toric_code}. Introducing periodic boundary conditions, the topology of the lattice corresponds to the one of a torus. Indeed, the Toric Code represents also the first example of topological quantum computation, which has then been developed and extended to more complicated scenarios. Let us now go back to our problem, and introduce the Toric Code Hamiltonian. As we said, a qubit is associated to each of the $2N^2$ edges of the lattice. One then defines the Hamiltonian of the Toric code as:
\ba
H_{\rm Tor}=-J\sum_{v=1}^{N^2}A_v-J\sum_{f=1}^{N^2}B_f
\ea
where the $A_v$ are the vertex operators and the $B_f$ are the facet operators, see again Fig.~\ref{fig:toric_code}. The vertex and facet operators are defined as:
\ba
A_v=\bigotimes_{e\sim v}X_e,\qquad B_f=\bigotimes_{e\in\partial f}Z_e,
\ea
where $e\sim v$ means that the edge $e$ has a vertex in $v$ and $e\in\partial f$ means that the edge $e$ belongs to the boundary of the facet $f$.
The vertex and facet operators all commute with each other:
\ba
\comm{A_v}{A_{v'}}=\comm{B_f}{B_{f'}}=\comm{A_v}{B_f}=0
\ea
Moreover, as the vertex and facet operators are generators of an Abelian group, it holds:
\ba
\label{eq:toric_prod_rule}
\prod_v A_v=\prod_f B_f=I\Rightarrow\prod_{v,f}A_vB_f=I
\ea
Because of these relations, one can write the unitary operator $V_{\rm Tor}$ generated by $H_{\rm Tor}$ as:
\ba
V_{\rm Tor}=\exp\left[-iH_{\rm Tor}t\right]=\prod_{v,f=1}^{N^2}e^{+iJA_vt}e^{+iJB_ft}=\prod_{v,f=1}^{N^2}(\cos JtI+i\sin JtA_v)(\cos JtI+i\sin JtB_f)
\ea
The Toric Code Hamiltonian is a sum of commuting Pauli strings generating an Abelian group: since all terms in the sum commute, it is exactly solvable and trivially integrable in this sense. Moreover, the spectrum of the Toric Hamiltonian is 4-fold degenerate with equally spaced levels. In App.~\ref{app:stab_ham_spectral_functions} we compute all the spectral form factors of the Toric code exactly, which will serve as a benchmark for comparisons with chaotic dynamics.

\begin{figure}[!th]
\centering
\begin{subfigure}{0.33\textwidth}
\centering
\includegraphics[width=\linewidth]{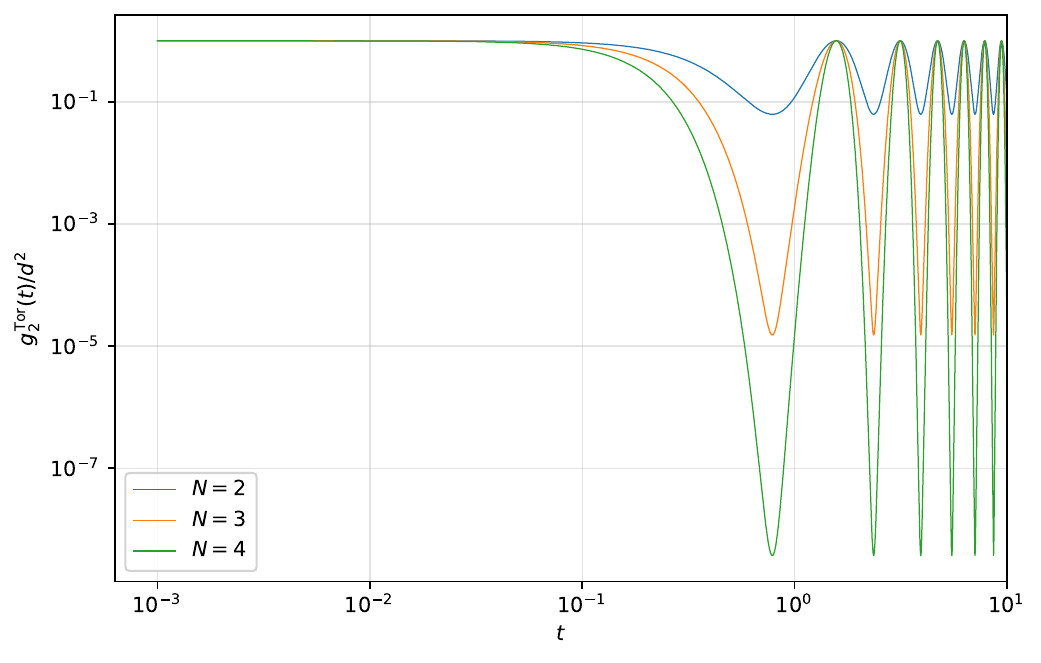}
\caption{Log-log plot of $g_2^{\rm Tor}(t)/d^2$.}
\label{fig:g2Tor}
\end{subfigure}
\begin{subfigure}{0.33\textwidth}
\centering
\includegraphics[width=\linewidth]{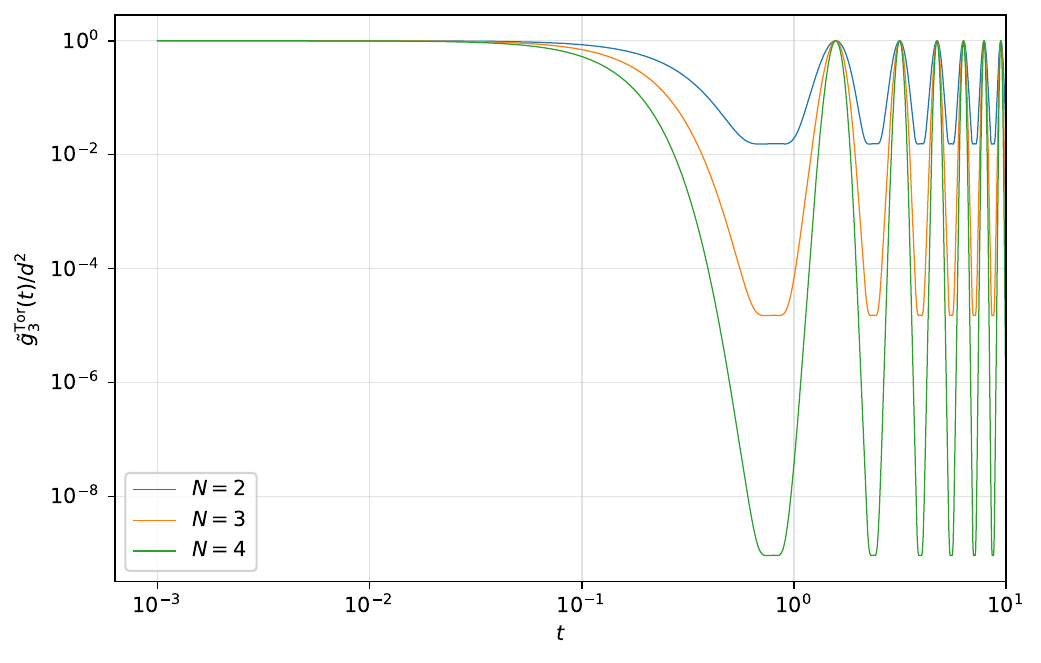}
\caption{Log-log plot of $\tilde{g}_3^{\rm Tor}(t)/d^2$.}
\label{fig:g3tilde_Tor}
\end{subfigure}
\begin{subfigure}{0.33\textwidth}
\centering
\includegraphics[width=\linewidth]{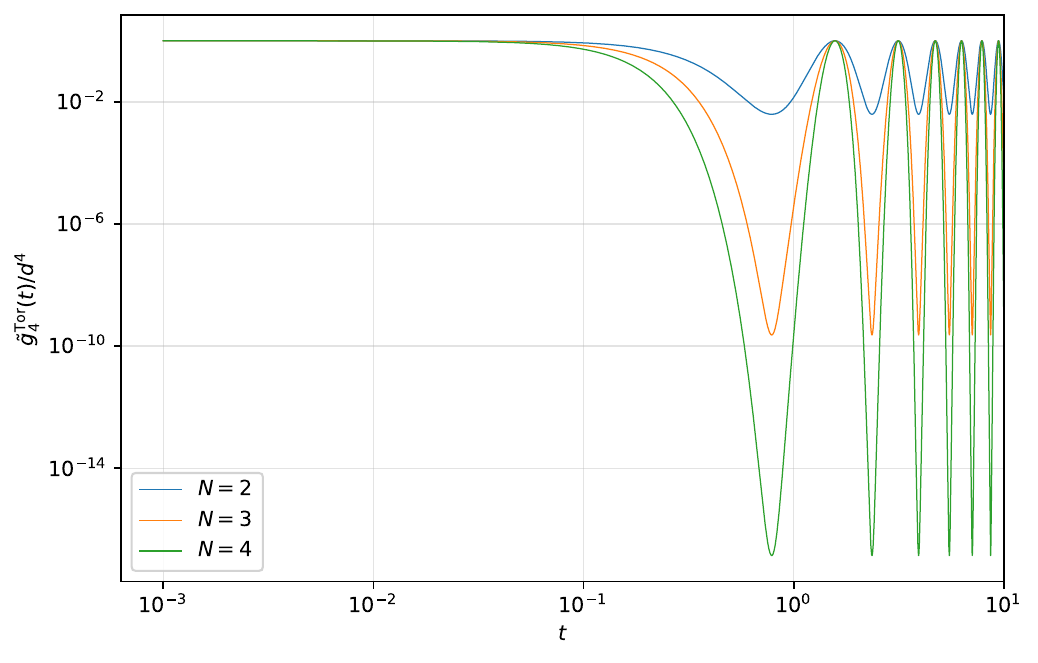}
\caption{Log-log plot of $g_4^{\rm Tor}(t)/d^4$.}
\label{fig:g4Tor}
\end{subfigure}
\caption{Plot of the functions $g_2^{\rm Tor}(t)$ (panel~\subref{fig:g2Tor}), $\tilde{g}_3(t)$ (panel~\ref{fig:g3tilde_Tor}) and $g_4^{\rm Tor}(t)$ (panel~\subref{fig:g4Tor}) for different lattice size $N$. One can once again observe how the behavior of $\tilde{g}_3(t)$ is in between the ones of $g_2(t)$ and $g_4(t)$.}
\label{fig:g24Tor}
\end{figure}

\section{\label{sec:probes_of_chaos}Probes of chaos}
Let us finally turn to the object of this paper, namely probes of chaos. In this section we are going to illustrate the results for the two main probes that exhibit significant differences between the Clifford and Haar averages: Loschmidt echo of the second kind and the OTOC.
We then illustrate one example of probe that does not depend on the group average, the Tripartite Mutual Information, in order to show how the average happens to not depend on the group. We leave most details and all the other probes, including the entanglement entropy, the norm of coherence and the WYD skew information, to the appendix sections. Finally, let us stress here that all the plots shown in this work come from exact analytical expressions.

\subsection{Loschmidt echo\label{sec:loschmidt}}

The Loschmidt paradox dates back to the nineteenth century, when Johann Loschmidt posed a very important question: where does irreversibility come from? Indeed, if the microscopic dynamics of the constituents of a gas is reversible, why is it not possible to invert all particles velocity and ``unscramble'' the gas?

Quantum Loschmidt echo was then introduced at the end of the twentieth century in the context of quantum chaos~\cite{Peres1984,Gorin2006,Richter2002,Andersen2006} and decoherence~\cite{Jalabert2001,Pastawski2000}, in an attempt to solve the absence of a quantum definition of Lyapunov exponents and other classical indicators of chaotic behavior.
It has then been applied successfully in contexts such as quantum information theory~\cite{Emerson2003,Prosen2007}, dynamical phase transitions~\cite{Heyl2013,Heyl2018} and localization~\cite{TorresHerrera2015,Znidaric2008}.

The main idea behind the Loschmidt echo is that if a quantum system is chaotic, then reverting the dynamics of the system with an imperfect Hamiltonian should lead back to an initial state which is way different from the original one. On the contrary, if the dynamics is integrable, small perturbations of the Hamiltonian used to reverse the dynamics should not lead to large differences in the initial state.

The Loschmidt echo of the first kind $\mathcal{L}_1(V)$ is defined as the squared modulus of the overlap between an initial state $\ket{\psi}$ and its evolved counterpart $V\ket{\psi}$, so that:
\ba
\label{eq:loschmidt_first_kind_def}
\mathcal{L}_1(V)=|\mel{\psi}{V}{\psi}|^2=|\tr{V\dyad{\psi}}|^2
\ea
The Loschmidt echo of the first kind is a measure of the reversibility~\cite{Pastawski2001PhysRep} of the dynamics described by the operator $V$, and it can be used to detect dynamical phase transitions~\cite{NiuWang2023PRA,VanhalaOjanen2023PRResearch} and revivals~\cite{JafariJohannesson2017PRL,Pastawski1995PRL}.
One can easily linearize Eq.~\eqref{eq:loschmidt_first_kind_def} with the usual swap trick, and then take the average of $V$ with respect to the appropriate group:
\ba
\nonumber
\mathcal{L}_1(V)&=&|\tr{V\dyad{\psi}}|^2=\tr{V\dyad{\psi}}\tr{V^\dag\dyad{\psi}}=\tr{(V\ot V^\dag)\dyad{\psi}^{\ot2}}\\
&&\Rightarrow \langle\mathcal{L}_1(V)\rangle_{\cal G}=\tr{\mathcal{R}^{(2)}_{\cal G}(V^{\ot1,1})\dyad{\psi}^{\ot2}}
\ea
As the Clifford group is a 3-design, one can immediately see that no difference can be detected between the averages on Clifford and t-doped Clifford through $\mathcal{L}_1$, as it only depends on the second moment of $V$. Indeed, it was already found in~\cite{10.21468/SciPostPhys.10.3.076} that the Loschmidt echo of the first kind is less efficient than the its second kind counterpart in detecting chaotic behavior, as it only depends on the two point spectral form factor $g_2(t)$.

\begin{figure*}[!th]
\centering
\begin{subfigure}{0.49\textwidth}
\centering
\includegraphics[width=\textwidth]{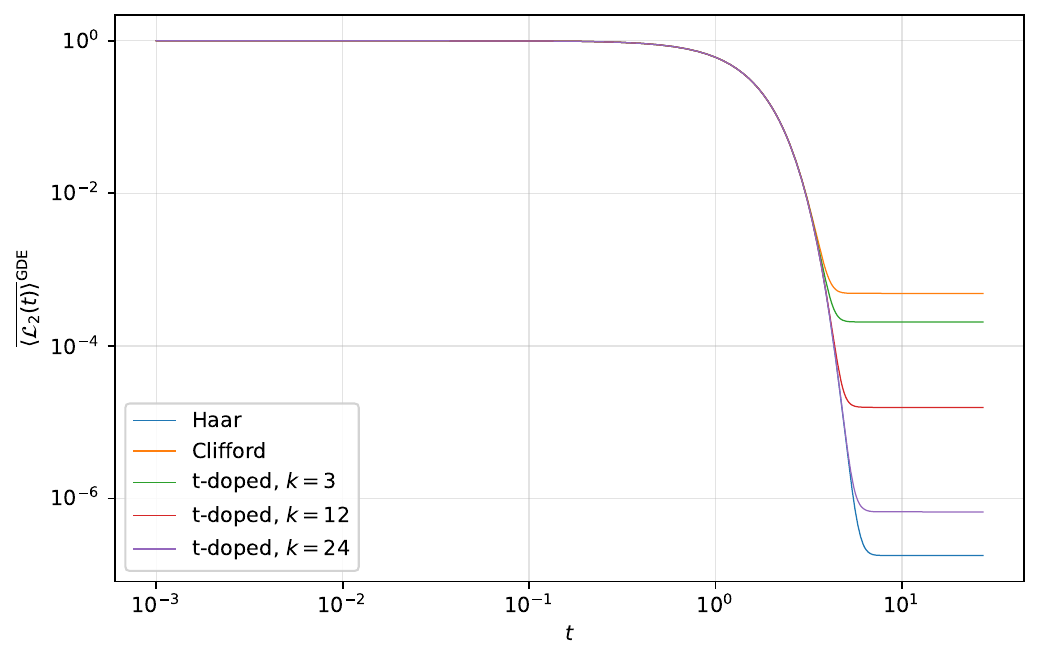}
\caption{Log-log plot of $\langle\mathcal{L}_2\rangle_{\cal G}$ averaged over the GDE.}
\label{fig:loschmidt_GDE}
\end{subfigure}
\begin{subfigure}{0.49\textwidth}
\centering
\includegraphics[width=\textwidth]{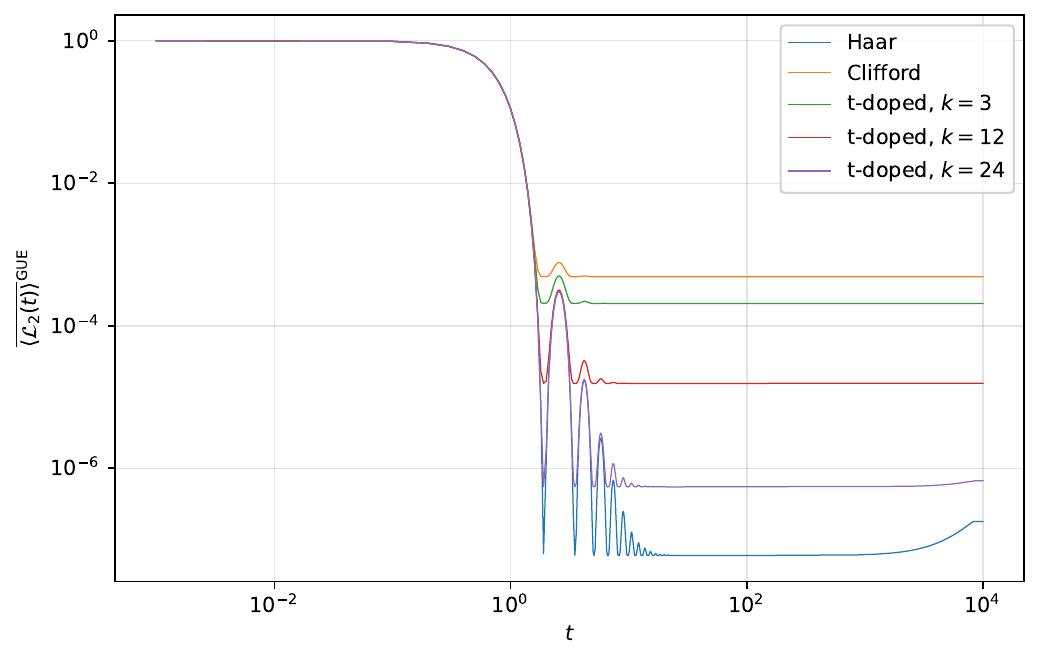}
\caption{Log-log plot of $\langle\mathcal{L}_2\rangle_{\cal G}$ averaged over the GUE.}
\label{fig:loschmidt_GUE}
\end{subfigure}\\
\begin{subfigure}{0.49\textwidth}
\centering
\includegraphics[width=\textwidth]{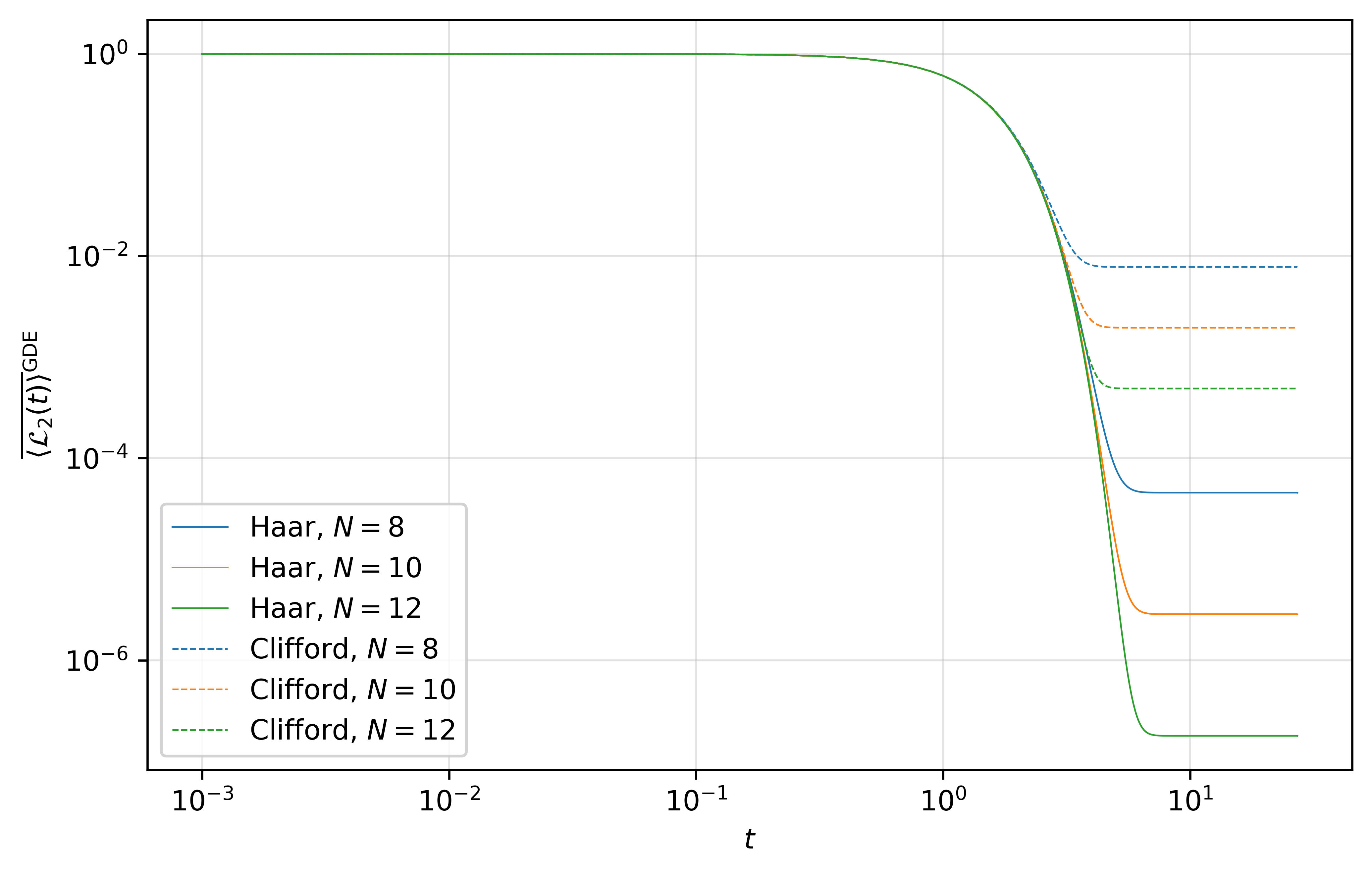}
\caption{Log-log plot of $\langle\mathcal{L}_2\rangle_{\cal G}$ averaged over the GDE for different Hilbert space dimension.}
\label{fig:loschmidt_GDE_dimension}
\end{subfigure}
\begin{subfigure}{0.49\textwidth}
\centering
\includegraphics[width=\textwidth]{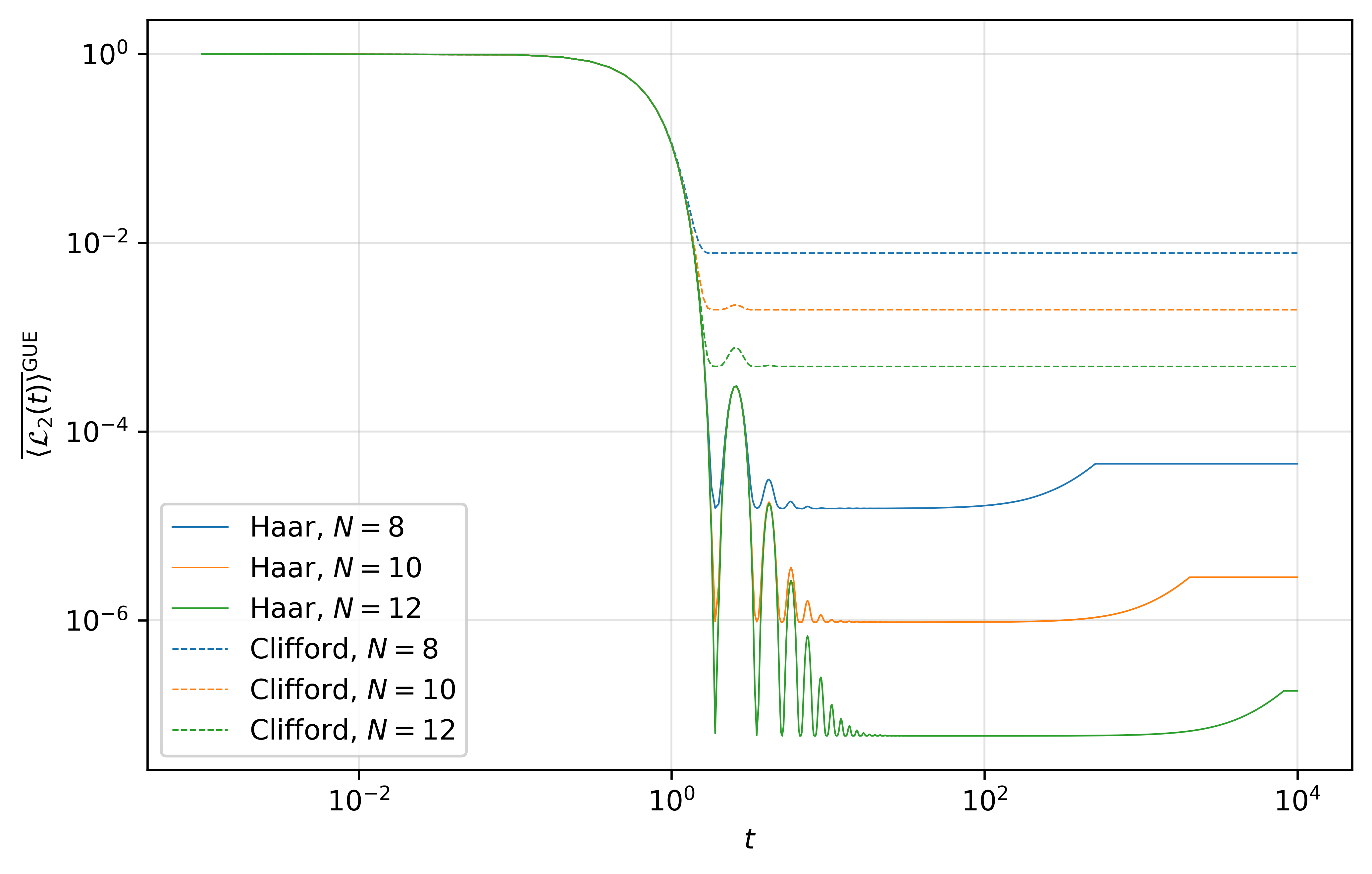}
\caption{Log-log plot of $\langle\mathcal{L}_2\rangle_{\cal G}$ averaged over the GUE for different Hilbert space dimension.}
\label{fig:loschmidt_GUE_dimension}
\end{subfigure}
\caption{Plot of the Loschmidt echo of the second kind $\overline{\langle\mathcal{L}_2\rangle}^{\rm E}_{\mathcal{G}}$ with $E=\{GDE,GUE\}$ in panel~\subref{fig:loschmidt_GDE} and~\subref{fig:loschmidt_GUE} respectively. The dimension is $d=2^{12}$. One notices that the Haar and Clifford averaged echoes are indeed very different, as they reach different asymptotic values, $\mathcal{O}(d^{-2})$ and $\mathcal{O}(d^{-1})$ respectively. Moreover, one notices how doping makes the Clifford average closer and closer to the Haar values. As for the time to reach the asymptotic value, one notices from panel~\subref{fig:loschmidt_GDE_dimension} no difference between the Clifford and the Haar average under the GDE, as both approaches their long time value in a time $\mathcal{O}(\sqrt{\log d})$. On the contrary, as shown in panel~\subref{fig:loschmidt_GUE_dimension} the presence of doping influences the equilibration time in the case of the GUE, which ranges from the $\mathcal{O}(d^{1/3})$ for the Clifford average, to $\mathcal{O}(d)$ for the Haar case.}
\label{fig:loschmidt_GDE_GUE}
\end{figure*}

The Loschmidt echo of the second kind is defined as the Hilbert-Schmidt scalar product between the unitary evolution $V$ and a perturbed backward evolution $V^{'\dag}$. It is interesting to consider the case of perturbations of the generator of $V$, the Hermitian operator $H$, such that the spectrum is left unperturbed, i.e. $\rm{Sp}[H]=\rm{Sp}[H+\delta H]$, so that $V=\exp[-iHt]$ and $V^{'\dag}=\exp[+i(H+\delta H)t]$, so that:
\ba
\mathcal{L}_2(t)=d^{-2}\left|\tr{e^{+i(H+\delta H)t}e^{-iHt}}\right|^2
\ea
Under the hypothesis of the spectrum being left unchanged, one can write $H+\delta H=A^\dag HA$ with $A\in\mathcal{U}$ being a unitary operator close to the identity. In this case one has $e^{i(H+\delta H)t}=e^{iA^\dag HAt}=A^\dag e^{iHt}A$, and with the help of the swap trick one gets:
\ba
\nonumber
\left|\tr{e^{+i(H+\delta H)t}e^{-iHt}}\right|^2&=&\tr{A^\dag V^\dag A V}\tr{A^\dag V^\dag A V}=\tr{T_{(12)(34)}(A^\dag\ot A\ot A^\dag\ot A)(V^\dag\ot V\ot V^\dag\ot V)}\\
&=&\tr{T_{(14)(23)}A^{\ot(\dag,1\dag,1)}V^{\ot2,2}}
\ea
where we indicate $A^{\ot(\dag,1\dag,1)}=A^\dag\ot A\ot A^\dag\ot A$. Taking the isospectral twirling one obtains:
\ba
\langle\mathcal{L}_2(t)\rangle_{\cal G}=\tr{T_{(14)(23)}A^{\ot(\dag,1\dag,1)}\mathcal{R}_{\cal G}^{(4)}(V)}
\ea
From this formula, one can immediately see that the Loschmidt echo of the second kind distinguishes between Unitary, Clifford and t-doped Clifford scenarios due to the fact  that the Clifford group fails to be a 4-design. Indeed, assuming the operator $A$ to be a Pauli string, one obtains for the averages over the Haar and Clifford groups:
\ba
\label{eq:loschmidt_haar}
&&\langle\mathcal{L}_2(t)\rangle_{\cal U}=\frac{d^{4}
  + \bigl(g_2(2t)-4g_{2}(t)+g_4(t)-9\bigr)d^{2}-2\Re\bigl[g_3(t)\bigr]d-6\bigl(g_{2}(2t)-4g_2(t)+g_{4}(t)\bigr)}{d^2(d^{4}-10d^{2}+9)}\\
\label{eq:loschmidt_clifford}
&&\langle\mathcal{L}_2(t)\rangle_{\cal C}=\frac{\left(\tilde{g}_3-1\right)}{(d^2-1)}
\ea
while the formula for the average over T-doped circuits is long and cumbersome, and is thus reported in App.~\ref{app:loschmidt}.

First of all, let us notice how the average over the Clifford group does not depend on $c_4(t)$, compared to the average over the unitary group. This has consequences on the long time value of the Loschmidt echo, as one notices from the plots in Fig.~\ref{fig:loschmidt_GDE_GUE}.

As one can see, the Clifford and Haar averages reach very different asymptotic values for the case of the GDE.  This is because $g_4(t)$ reaches an asymptotic value of order $\mathcal{O}(d^2)$, while $\tilde{g}_3(t)$ has an asymptotic value of order $\mathcal{O}(d)$. The immediate consequence of this, as can be easily calculated from Eqs.~(\ref{eq:loschmidt_haar},~\ref{eq:loschmidt_clifford}), is that the asymptotic value of $\langle\mathcal{L}_2(t)\rangle_{\cal U}$ is of order $\mathcal{O}(d^{-2})$, while $\langle\mathcal{L}_2(t)\rangle_{\cal C}$ is $\mathcal{O}(d^{-1})$. 

Thus, contrary to the Haar scenario, averaging over the Clifford group allows the system to retain some memory of its initial state. As doping layers are mixed into the average the asymptotic value starts approaching the Haar value, thus leading the system to retain less and less memory of its initial state.

As for the asymptotic equilibration time, one notices that averaging over the Clifford or unitary group makes little difference in the GDE case, while it has a strong impact in the GUE case, causing the system to equilibrate in a time $\mathcal{O}(d^{1/3})$, rather than $\mathcal{O}(d^2)$.

\begin{figure}[!th]
\begin{subfigure}{0.33\textwidth}
\centering
\includegraphics[width=\linewidth]{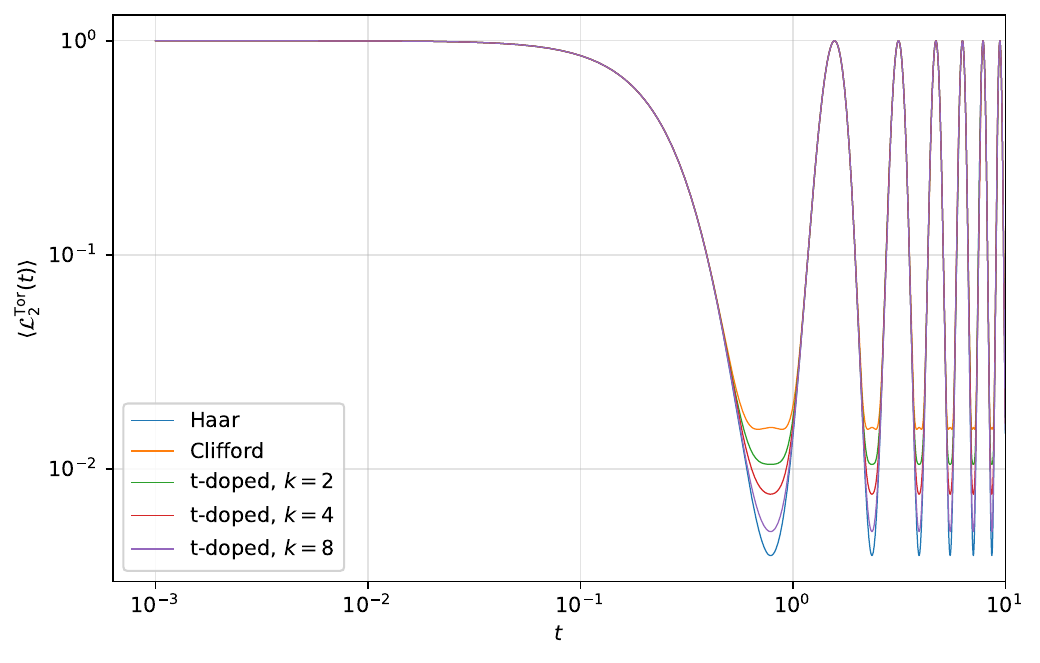}
\caption{$N=2$}
\label{fig:loschmidt_Tor_N2}
\end{subfigure}
\begin{subfigure}{0.33\textwidth}
\centering
\includegraphics[width=\linewidth]{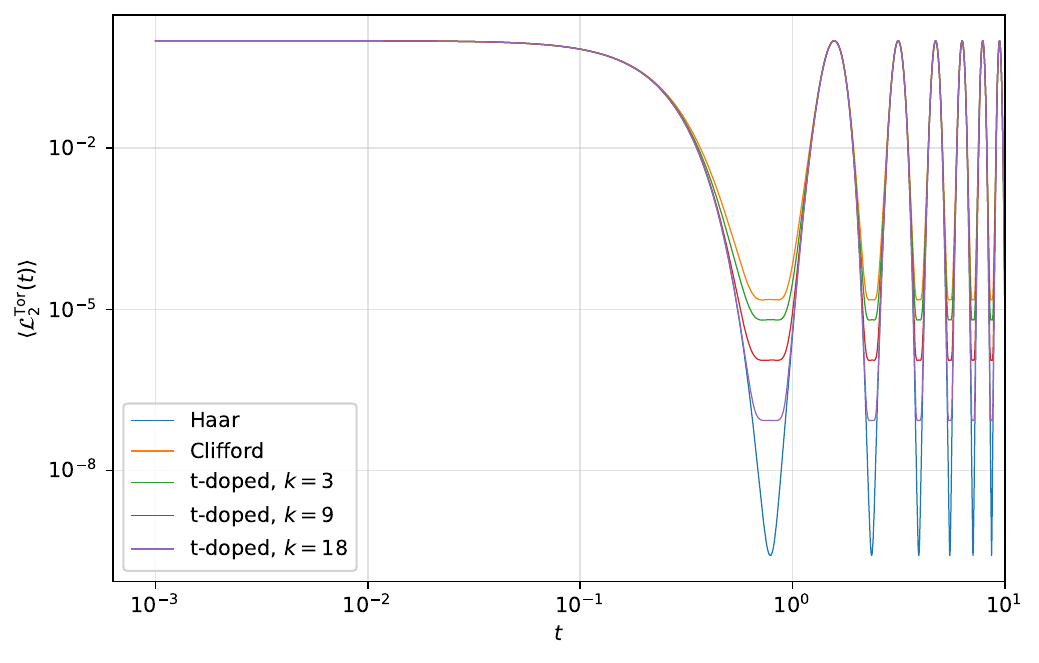}
\caption{$N=3$}
\label{fig:loschmidt_Tor_N3}
\end{subfigure}
\begin{subfigure}{0.33\textwidth}
\centering
\includegraphics[width=\linewidth]{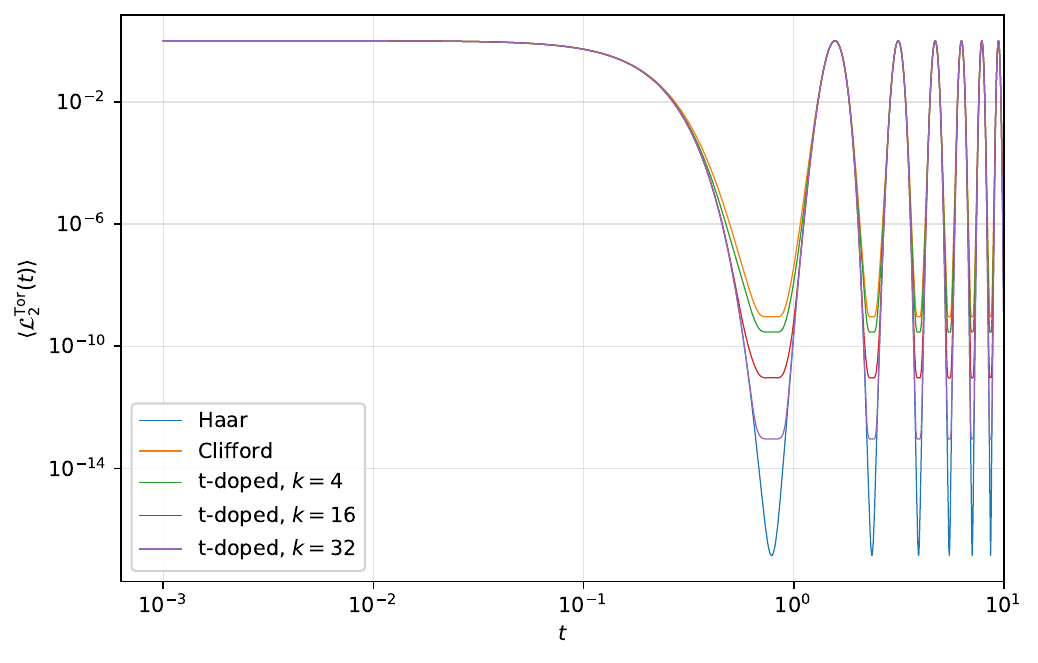}
\caption{$N=4$}
\label{fig:loschmidt_Tor_N4}
\end{subfigure}
\caption{Plot of the Loschmidt echo for the Toric Code for different lattice size. One notices a behavior consistent with our finding over random spectra, including a different minimum when the average is taken over the Clifford or Unitary group. This difference goes with the dimensionality of the system, as one notices from the different plots for different lattice dimension. Notice also that the equilibration time is the same independently of the dimension of the system, as the timescale for the Toric code are dictated by the parameter $J$.}
\label{fig:Loschmidt_Tor}
\end{figure}

The Toric Code gives results compatible with this picture. As shown in Fig.~\ref{fig:Loschmidt_Tor}, the minimum value reached by $\mathcal{L}_2^{\rm Tor}$ is of order $\mathcal{O}(d^{-1})$ when the average is taken over the Clifford group, while it is $\mathcal{O}(d^{-2})$ in the Haar case.

\subsection{Out of Time Order Correlators (OTOCs)\label{sec:otoc}}
Out-of-Time-Order-Correlators~\cite{Garcia-Mata:2023,XuSwingle2022} (OTOCs) are the quantum equivalent of the butterfly effect, as they measure how much an initially localized operator spreads over the system during unitary evolution.
OTOCs namely describe the average evolution of Heisenberg operators at different times, and thus they are a widely used measure of scrambling and information spreading.

They were originally conceived for the study of  superconductors~\cite{LarkinOvchinnikov1969}, and later found application in connecting black holes horizon geometry and chaos~\cite{ShenkerStanford2014,ShenkerStanford2014MultipleShocks,ShenkerStanford2015Stringy}, allowing to show that black holes are  fast scramblers~\cite{SekinoSusskind2008}. Moreover, OTOC have also applications in strongly correlated systems~\cite{Kitaev2015Talk,rz86-47h3} and information scrambling~\cite{MaldacenaShenkerStanford2016,Swingle2018,e23081073,vardhan2025freemutualinformationhigherpoint}, and due to their wide applicability found much attention in recent times.

Given 2k non overlapping operators $A_\ell,B_\ell$, $\ell=1,\dots,k$, the 4k-point OTOC is defined as:
\ba
\rm{OTOC}_{4k}(t)=d^{-1}\tr{V^\dag A_1^\dag VB_1^\dag\dots V^\dag A_k^\dag VB_k^\dag V^\dag A_1 VB_1\dots V^\dag A_kVB_k}
\ea
One can easily see that the trace can always be written in linearized form through an appropriate permutation operator from the order 4k symmetric group $S_{4k}$. As we are going to be interested in $\rm{OTOC}_4$, the formula in this case reads:
\ba
\nonumber
\rm{OTOC}_4&=&d^{-1}\tr{V^\dag A^\dag VB^\dag V^\dag A VB}=d^{-1}\tr{T_{(1234)}(V^\dag\ot V\ot V^\dag\ot V)(A^\dag\ot B^\dag\ot A\ot B)}\\
\qquad&=&d^{-1}\tr{T_{(1423)}(A^{\ot1,1}\ot B^{\dag\ot1,1})V^{\ot2,2}}
\ea
Thus after averaging one has:
\ba
\langle \rm{OTOC}_4\rangle_{\cal G}=d^{-1}\tr{T_{(1423)}(A^{\ot1,1}\ot B^{\dag\ot1,1})\mathcal{R}_{\cal G}^{(4)}(V)}
\ea
Assuming the operators $A,B$ to be non overlapping Pauli strings, one can compute the expressions of the group averaged ${\rm OTOC}_4$ as:
\ba
\langle {\rm OTOC}_4\rangle_{\cal U}&=&\frac{-6\Re\bigl[g_{3}(t)\bigr]+\bigl(9+g_{2}(2t)-4g_{2}(t)+g_{4}(t)\bigr)d-d^{3}}{(d-1)(d+1)(d-3)(d+3)}\\
\langle {\rm OTOC}_4\rangle_{\cal C}&=&\frac{d\left(4-2g_{2}(2t)+(-3+\tilde{g}_{3}(t))d^{2}\right)}{(d-1)(d+1)(d-2)(d+2)}
\ea
where once again we leave the formula for the T-doped average to the appendix.

\begin{figure*}
\centering
\begin{subfigure}[t]{0.49\textwidth}
\centering
\includegraphics[width=\linewidth]{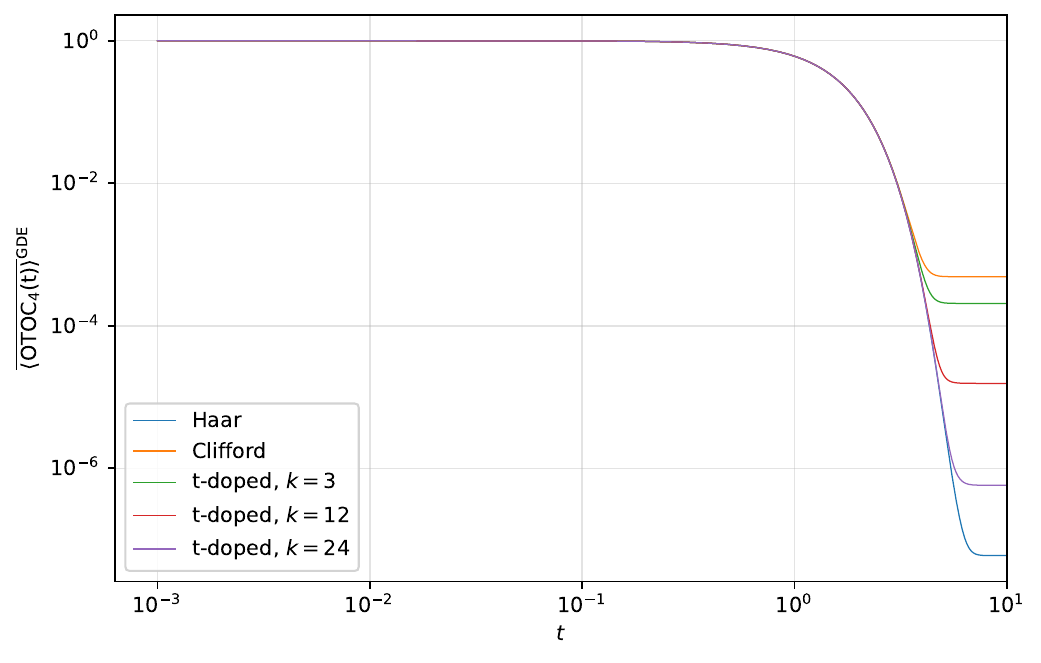}
\caption{Log-log plot of the $\rm{OTOC}_4$ averaged over the GDE.}
\label{fig:otoc4_GDE}
\end{subfigure}
\begin{subfigure}[t]{0.49\textwidth}
\centering
\includegraphics[width=\linewidth]{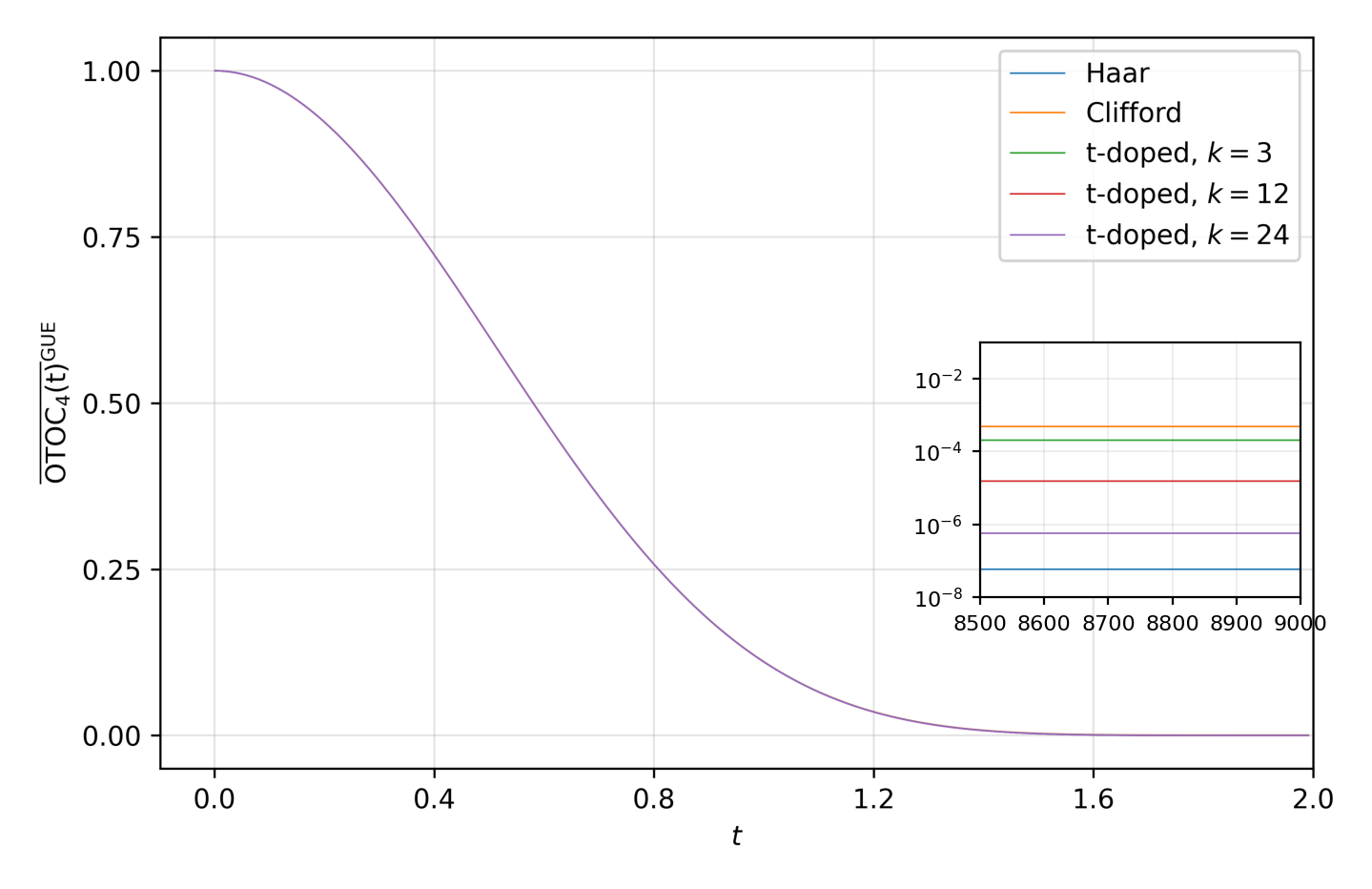}
\caption{Linear plot of the $\langle\rm{OTOC}_4\rangle_{\cal G}$ averaged over the GUE. The inset show the asymptotic value in logarithmic scale for the various averages.}
\label{fig:otoc4_GUE}
\end{subfigure}
\caption{Plot of the $\rm{OTOC}_4$ averaged over the GDE (~\subref{fig:otoc4_GDE})and the GUE(~\subref{fig:otoc4_GUE}) for $d=2^{12}$. One notices that averaging over the Clifford or Unitary group only influences the asymptotic value in both cases, but not the equilibration time. For both spectral families the asymptotic value of the Haar average is $\mathcal{O}(d^{-2})$, while for the Clifford average it scales as $\mathcal{O}(d^{-1})$.
}
\label{fig:otoc4}
\end{figure*}

The curves for the $\rm{OTOC}_4$ are shown in Fig.~\ref{fig:otoc4}. The average over GDE in Fig.~\ref{fig:otoc4_GDE} shows how averaging over the Clifford group only changes the asymptotic value reached by the OTOC, but not the equilibration time. Indeed, similarly to Loschmidt echo, the asymptotic value for the GDE averaged OTOC is $\mathcal{O}(d^{-1})$ for the Clifford average, while the value of the Haar average scales as $\mathcal{O}(d^{-2})$. The equilibration time for the GDE scales $\mathcal{O}(\sqrt{\log d})$. Regarding GUE, the plot in Fig.~\ref{fig:otoc4_GUE} is not in logarithmic scale, as the GUE averaged OTOC over the unitary group becomes negative at times $\mathcal{O}(d^{1/2})$, in contrast with the Clifford one, which is never negative, before reaching the asymptotic value at times $\mathcal{O}(d)$. Nonetheless, as shown in the inset of Fig.~\ref{fig:otoc4_GUE}, the asymptotic values scale like $\mathcal{O}(d^{-2})$ and $\mathcal{O}(d^{-1})$ for the Haar and Clifford group respectively. Overall, these results are very similar to the ones obtained for the Loschmidt echo, as these two probes are connected~\cite{Cotler2017Chaos}.

\begin{figure}[!th]
\begin{subfigure}{0.33\textwidth}
\centering
\includegraphics[width=\linewidth]{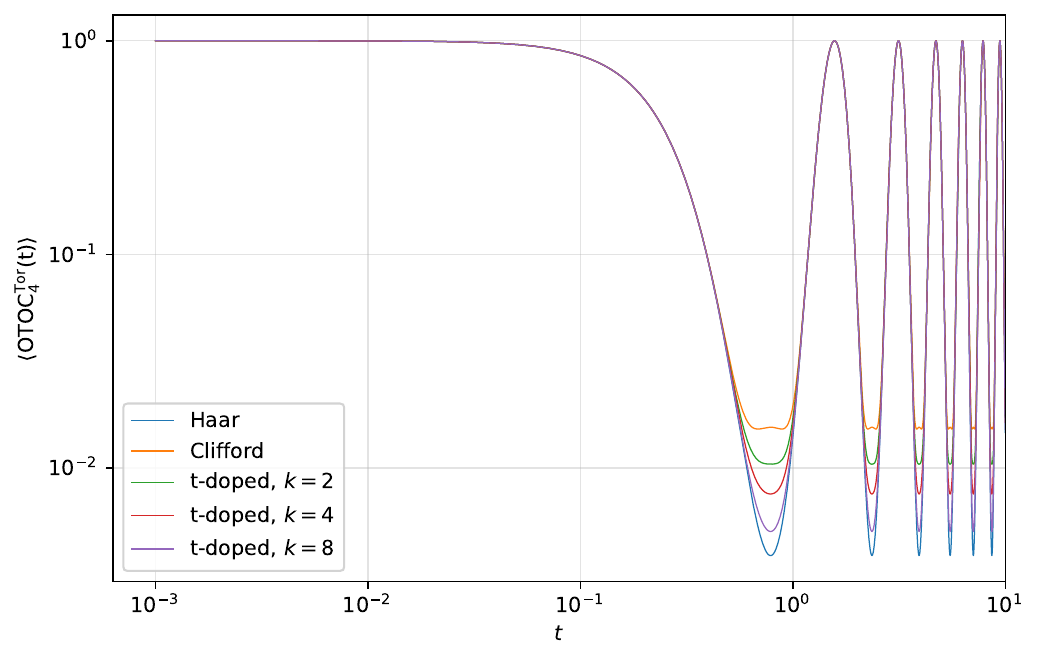}
\caption{$N=2$}
\label{fig:OTOC4_Tor_N2}
\end{subfigure}
\begin{subfigure}{0.33\textwidth}
\centering
\includegraphics[width=\linewidth]{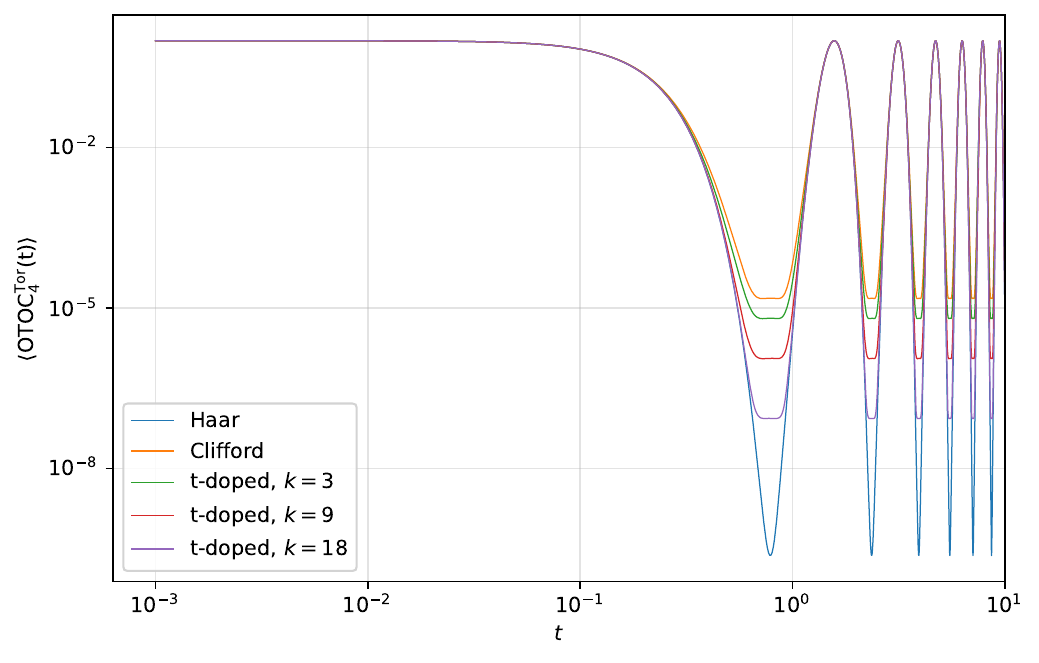}
\caption{$N=3$}
\label{fig:OTOC4_Tor_N3}
\end{subfigure}
\begin{subfigure}{0.33\textwidth}
\centering
\includegraphics[width=\linewidth]{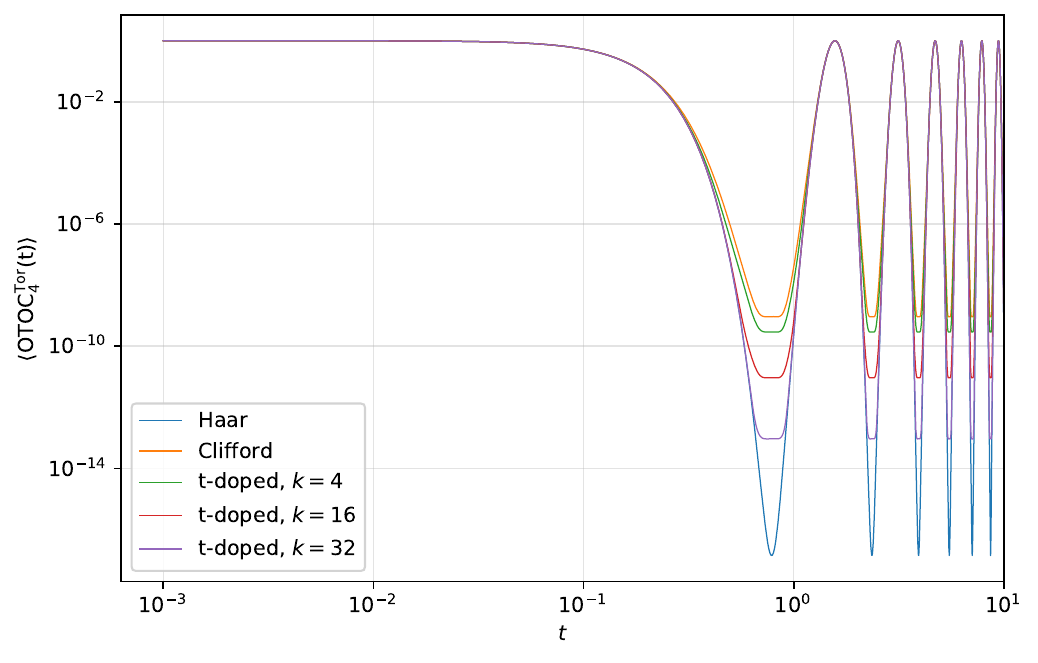}
\caption{$N=4$}
\label{fig:OTOC4_Tor_N4}
\end{subfigure}
\caption{Plot of the $\rm{OTOC}_4$ for the Toric code for different lattice size. Similarly to the Loschmidt echo, one can once again notice a different minimum when the average is taken over the Clifford ($\mathcal{O}(d^{-1})$) or the Unitary ($\mathcal{O}(d^{-2})$) group.}
\label{fig:OTOC4_Tor}
\end{figure}

\subsection{Tripartite Mutual Information\label{sec:tripartite_info}}
Another quantity from the family of entropic functionals which is deeply connected to quantum chaos is tripartite mutual information (TMI). Tripartite mutual information stems from classical information theory, where its classical precursor can be individuated in the interaction information~\cite{1057469}. The interest in this quantity from the quantum community has mainly originated from the study of quantum capacity of noisy quantum channels~\cite{PhysRevA.55.1613,PhysRevA.68.062323,PhysRevA.70.012307}.

The tripartite mutual information plays a central role in holography nowadays~\cite{Hayden2011_HolographicMI,Headrick2010_Renyi}, quantum gravity~\cite{SekinoSusskind2008,YoshidaKitaev2017_HaydenPreskill,PhysRevA.106.062434,PhysRevA.109.022429} and quantum chaos~\cite{HosurQiRobertsYoshida2016,DingHaydenWalter2016}, as it allows one to study how information is spread through different parts (subsystems) of a quantum system, and how much this information correlates the subsystems.

Consider for instance a quantum system whose state lives on a bipartite quantum system $\mathcal{H}_A\ot\mathcal{H}_B$. After some evolution given by a quantum channel, the system is mapped into a new state, living on the Hilbert space $\mathcal{H}_C\ot\mathcal{H}_D$. A scrambling channel is one for which any action performed on a part of the input state, say $A$, cannot be detected by measurement on a subsystem of the output state, say $C$. A quantum channel inducing scrambling on the quantum systems shall then be characterized by a low value of the tripartite mutual information, defined as:
\ba
I_3(A:C:D)=I(A:C)+I(A:D)-I(A:CD).
\ea
where $I(A:C)=S(A)+S(C)-S(A,C)$ is the bipartite mutual information.
The TMI is an entropic functional, and indeed can be written in terms of quantum entropies as:
\ba
I_3(A:C:D)=S(C)+S(D)-S(AC)-S(AD)
\ea
This expression can indeed be put in terms of an expectation value~\cite{e23081073} as:
\ba
I_{3_{(2)}}\leq\log d+\log\tr{T_{(13)(24)}\mathcal{R}_{\cal G}^{(4)}(V)T_{(12)}^{(C)\ot2}}+\log\tr{T_{(13)(24)}\mathcal{R}_{\cal G}^{(4)}(V)T_{(12)}^{(C)}\ot T_{(12)}^{(D)}}
\ea
where we assumed $A=C$ and $B=D$, and by $I_{3_{(2)}}$ indicates that we are using the $\alpha=2$ Rényi Entropy to define the mutual information. Notice that, since the Rényi entropies involve a logarithm, we can only average the argument in order to obtain an upper bound exploiting the Jensen's inequality.

\begin{figure}[!th]
\begin{subfigure}{0.49\textwidth}
\centering
\includegraphics[width=\textwidth]{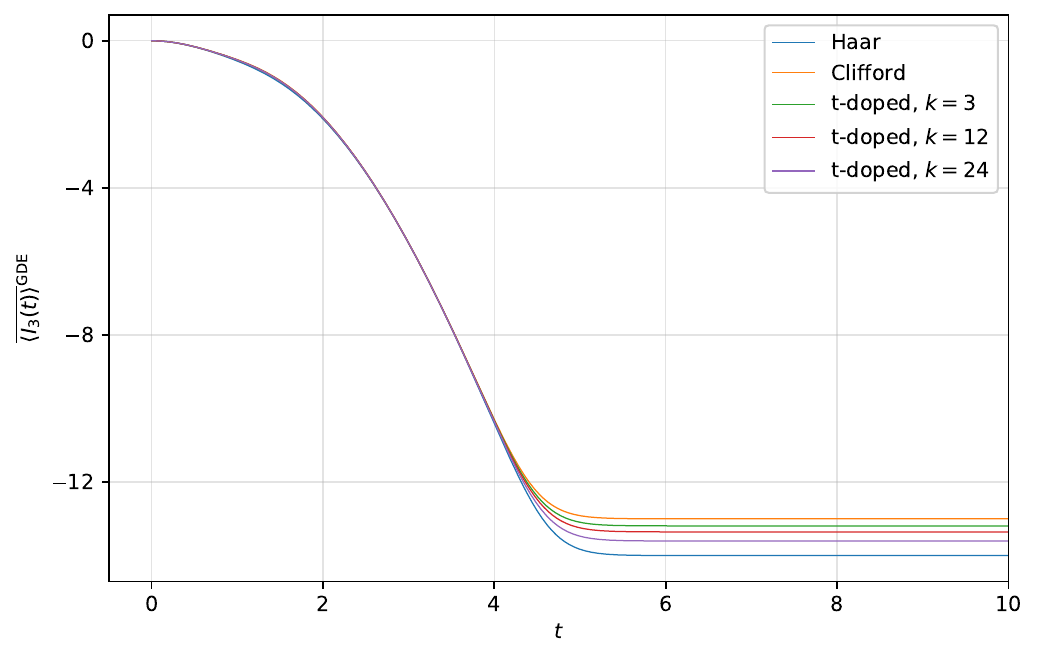}
\caption{Plot of $\overline{\langle I_3\rangle}^{GDE}$}
\label{fig:tripartite_GDE}
\end{subfigure}
\begin{subfigure}{0.49\textwidth}
\centering
\includegraphics[width=\textwidth]{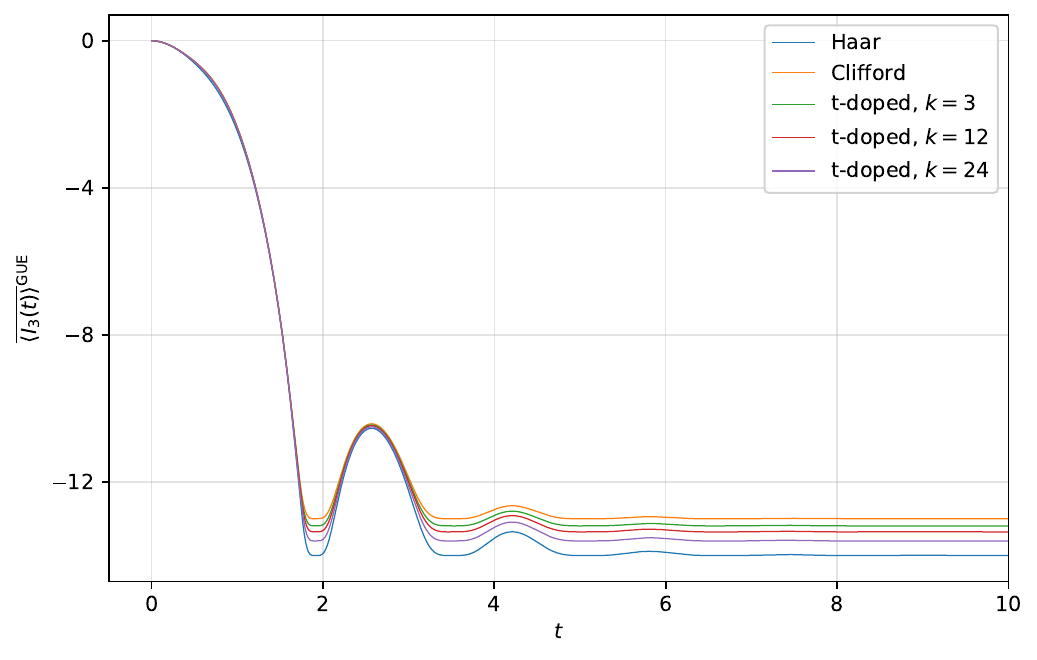}
\caption{Plot of $\overline{\langle I_3\rangle}^{GUE}$}
\label{fig:tripartite_GUE}
\end{subfigure}
\caption{Plot of the bound on the tripartite info $I_{3_{(2)}}$ averaged over the GDE (~\ref{fig:tripartite_GDE}) and the GUE(~\ref{fig:tripartite_GUE}) for $d=2^{12}$. Averaging over different groups does not influence the qualitative behavior of the bound, that is, it does not influence the equilibration time, nor the presence of oscillations for the GUE. The main difference between the two averages is seen in the asymptotic value, though this difference would be washed out by increasing the dimension $d$.}
\label{fig:tripartite_info}
\end{figure}
We can then dub the arguments of the logarithms as:
\ba
\left\langle I_{3_{(2)}}^{C^2}\right\rangle_{\cal G}&=&\tr{T_{(13)(24)}\mathcal{R}_{\cal G}^{(4)}(V)T_{(12)}^{(C)\ot2}}\\
\left\langle I_{3_{(2)}}^{CD}\right\rangle_{\cal G}&=&\tr{T_{(13)(24)}\mathcal{R}_{\cal G}^{(4)}(V)T_{(12)}^{(C)}\ot T_{(12)}^{(D)}}
\ea

One can compute the Unitary and Clifford averages of the two arguments of the logarithm as:
\ba
\nonumber
\left\langle I_{3_{(2)}}^{C^2}\right\rangle_{\cal U}&=&\frac{6\Re[g_{3}(t)]+\big(g_{2}(2t)-4 g_{2}(t)+g_{4}(t)\big)(d^{2}-d)-6d\Re[g_{3}(t)]-18d^{3}+2d^{5}}{(d-3) d (d+1)(d+3)}\\
\nonumber
\left\langle I_{3_{(2)}}^{CD}\right\rangle_{\cal U}&=&\frac{3\big(g_{2}(2t)-4g_{2}(t)+g_{4}(t)\big)-(3 g_{2}(2t)-12 g_{2}(t)+2\Re[g_3(t)]+3 g_{4}(t))d+2\Re[g_3(t)]d^{2}-18d^{3}+2 d^{5}}{(d-3)d(d+1)(d+3)}
\\
\nonumber
\left\langle I_{3_{(2)}}^{C^2}\right\rangle_{\cal C}&=&\frac{-2g_{2}(2t)(-1+d)+d^{2}\left(-6+\tilde{g}_3(t)(-1+d)+2(-1+d)d\right)}{(1+d)(d^{2}-4)}\\
\nonumber
\left\langle I_{3_{(2)}}^{CD}\right\rangle_{\cal C}&=&\frac{d\left(g_{2}(2t)(-1+d)+2\left(-2+\tilde{g}_3(t)-(2+\tilde{g}_3(t))d+d^{3}\right)\right)}{(1+d)(d^{2}-4)}
\ea

\begin{figure}[!ht]
\begin{subfigure}{0.33\textwidth}
\centering
\includegraphics[width=\linewidth]{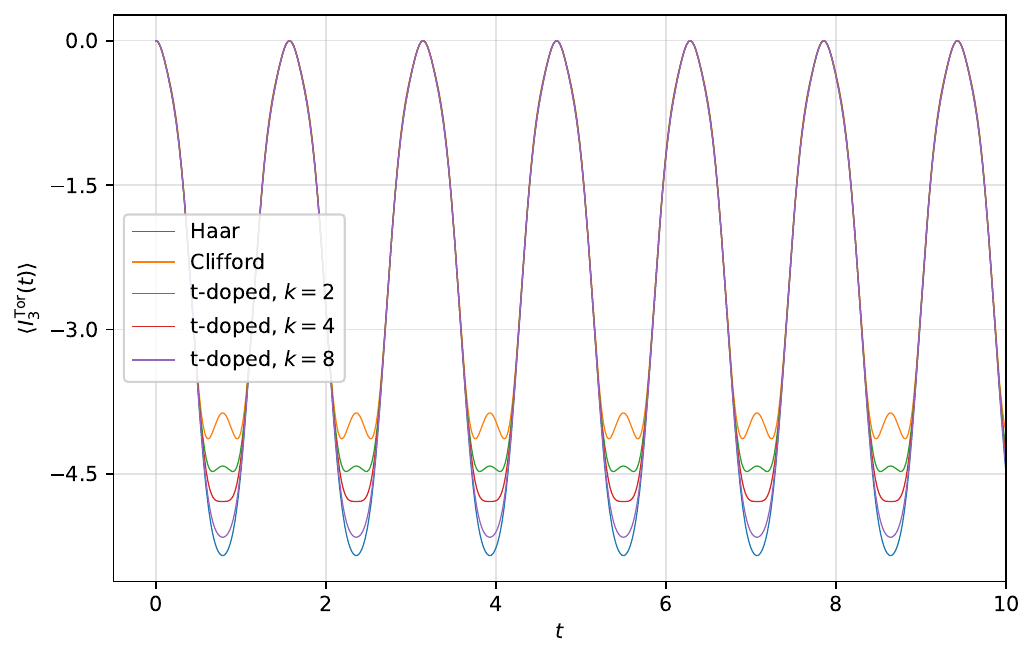}
\caption{$N=2$}
\label{fig:tripartite_Tor_N2}
\end{subfigure}
\begin{subfigure}{0.33\textwidth}
\centering
\includegraphics[width=\linewidth]{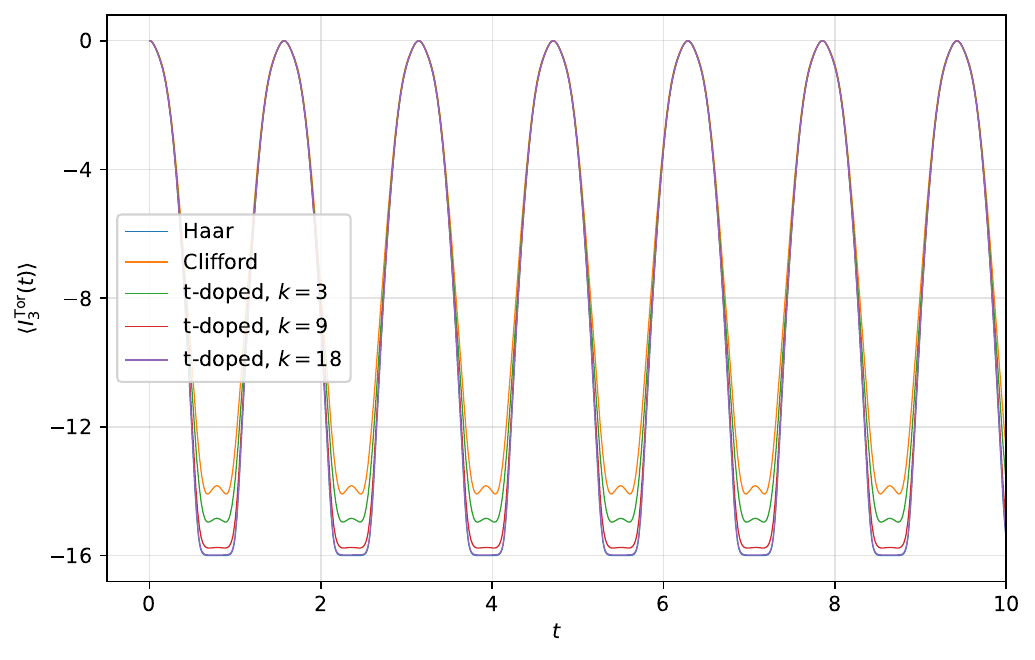}
\caption{$N=3$}
\label{fig:tripartite_Tor_N3}
\end{subfigure}
\begin{subfigure}{0.33\textwidth}
\centering
\includegraphics[width=\linewidth]{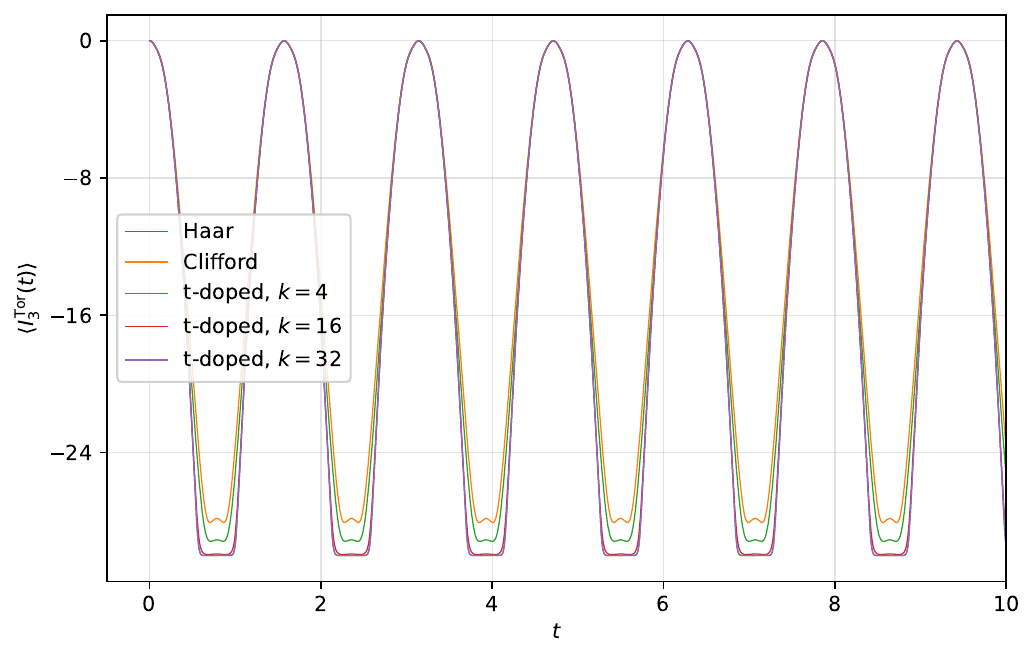}
\caption{$N=4$}
\label{fig:tripartite_Tor_N4}
\end{subfigure}
\caption{Tripartite Mutual Information for the Toric Code over different lattice size.}
\label{fig:tripartite_Tor}
\end{figure}

From the plots in Fig.~\ref{fig:tripartite_info}, one would be tempted to say that the bound on the tripartite information is influenced by whether the average is taken over the Clifford or unitary group. However, the asymptotic value of the argument of the logarithm for the Haar case scales as $g_4(t)/d^2$, while for the Clifford case scales as $\tilde{g}_3(t)$. As $g_4(t)$ scales as $d^2$, while $\tilde{g}_3(t)$ scales as $d$, one can see that the asymptotic value will still be $\mathcal{O}(\log d)$, and thus the observed difference is only a finite size effect. This can be noticed by looking at the corresponding plots for the Toric Code, Fig.~\ref{fig:tripartite_Tor}. In those plots one notices that as the dimensionality of the system increases, the difference between the Clifford and Haar averages become less and less relevant. This behavior is consistent with the Clifford group being able to generate maximal entanglement, so one should not expect to observe differences between the Clifford and the Haar averages with entropic measures.

\section{\label{sec:conclusions}Conclusions and outlook}
Let us first summarize the content of this work. In the first part of the paper we have written down all the main technical ingredients needed for our analysis. This included writing down the expressions for all the group averages. In particular, in order to write a manageable formula for the T-doped average, we had to diagonalize the matrix $\Xi$, find its eigenvectors, which we found to be connected to the projectors onto various irreps of the symmetric group. Then we reported all the techniques needed to manage probes of chaos, before introducing RMT ensembles and the spectral averages. We then introduced our Hamiltonian models, needed in order to obtain an initial condition for the Clifford averages. This has led us to compute some modified spectral form factors, which we dubbed Clifford spectral form factors. Indeed, we computed the explicit expression for one of this Clifford spectral form factors, $\tilde{g}_3(t)$, and computed its spectral average over the GDE and GUE ensembles. Moreover, we also computed its envelope curve and the asymptotic behavior.

Thanks to our method, it has been possible to decouple the features of the dynamics depending on the average over the eigenstates from those depending on the spectral distribution: indeed, by averaging over the Clifford group one does not average over eigenvectors belonging to the Thomas-Porter distribution, which are just uniform vectors, but rather over stabilizer states.
We studied several probes of chaos from different families: from scrambling diagnostics such as the Loschmidt echo and the OTOC, to entropic functionals such as the R\'enyi entropy and the tripartite mutual information and coherence measures such as the norm of coherence and the WYD skew information.
For the case of scrambling measures, we observed that averaging over the Clifford group leads to variations in the asymptotic value reached by such probes and in the equilibration times.

On the contrary, entropic measures have shown no sensibility to the group over which the average is taken. This is indeed very consistent, as entropic measures mostly depend on entanglement, which can be efficiently generated via Clifford operations.

Finally, coherence measures show an initial dependence on the ensemble over which the average is taken, which is however washed away in the long time limit.

An interesting common feature of all the probes of chaos when averaged over the Clifford group, is their dependence on the eigenstates of the initial Hamiltonian. This feature is indeed due to the dependence of the Clifford spectral form factor on the expectation value of the eigenstates of the Hamiltonian over Pauli strings. This is on the one hand very interesting, as one could possibly compute the eigenstates of spin models, and then compute the evolution of given probes or fluctuation when averaged over different groups. On the other hand, this pushes the need for an extensive study of the eigenvector distributions of random matrix ensembles.

Besides our findings, this work also indicates us some possible directions to investigate in the future. First of all, it would be of great interest to extend this analysis to higher moments of the Haar distribution. This would require computing the spectral averages of the corresponding spectral functions~\cite{PhysRevD.98.086026} and computing explicit formulas for the Clifford average and the T-doped average~\cite{bittel2025completetheorycliffordcommutant}.
Such a research endeavor would also require extending the web of probes of chaos to ones involving higher moments of the Haar distribution, beyond higher order OTOCs.
Moreover, investigating the difference between higher moments of the  Haar and Clifford groups could have applications in quantum cryptography and information processing, where one is interested in the amount of resources needed in order to sample from a given moment of the Haar distribution.

It would be also of interest to investigate the average over eigenvectors of random matrix ensembles, and compute them for the case of the Clifford spectral form factors. This could have interesting applications in random matrix theory, and especially in the assessment of quantum resources need for the simulation of physical systems. Examples of this could be the use of the techniques of this paper to analyze ensembles of random circuits, such as matchgate circuits and brickwall architectures.

\section*{Acknowledgements}
SC and AH acknowledge  support from the
National Quantum Science and Technology Institute (NQSTI), PNRR MUR project PE0000023-
NQSTI. SC and GE thank Jovan Odavi\'c for useful comments and discussions.

\bibliographystyle{SciPost_bibstyle}
\bibliography{isospectral_twirling_t_doped.bib}

@article{Bullock2004,
  author        = {Bullock, Stephen S. and Markov, Igor L.},
  title         = {Asymptotically optimal circuits for arbitrary $n$-qubit diagonal computations},
  journal       = {Quantum Information and Computation},
  volume        = {4},
  number        = {1},
  pages         = {27--47},
  year          = {2004},
  doi= {10.26421/QIC4.1-3},
  url= {https://www.rintonpress.com/journals/doi/QIC4.1-3.html}
}

@article{PhysRevLett.128.050402,
  title = {Stabilizer R\'enyi Entropy},
  author = {Leone, Lorenzo and Oliviero, Salvatore F. E. and Hamma, Alioscia},
  journal = {Phys. Rev. Lett.},
  volume = {128},
  issue = {5},
  pages = {050402},
  numpages = {5},
  year = {2022},
  month = {Feb},
  publisher = {American Physical Society},
  doi = {10.1103/PhysRevLett.128.050402},
  url = {https://link.aps.org/doi/10.1103/PhysRevLett.128.050402}
}

@article{berry_level_1977,
	title = {Level clustering in the regular spectrum},
	volume = {356},
	issn = {0080-4630},
	url = {https://doi.org/10.1098/rspa.1977.0140},
	doi = {10.1098/rspa.1977.0140},
	number = {1686},
	journal = {Proceedings of the Royal Society of London. A. Mathematical and Physical Sciences},
	author = {Berry, Michael Victor and Tabor, M.},
	month = sep,
	year = {1977},
	note = {\_eprint: https://royalsocietypublishing.org/rspa/article-pdf/356/1686/375/62525/rspa.1977.0140.pdf},
	pages = {375--394},
}

@article{7f616d19-7bad-3244-9920-e3bd00bd46f6,
 ISSN = {0003486X, 19398980},
 URL = {http://www.jstor.org/stable/1967333},
 author = {G. A. Miller},
 journal = {Annals of Mathematics},
 number = {4},
 pages = {188--190},
 publisher = {[Annals of Mathematics, Trustees of Princeton University on Behalf of the Annals of Mathematics, Mathematics Department, Princeton University]},
 title = {A New Proof of the Generalized Wilson's Theorem},
 urldate = {2026-07-20},
 volume = {4},
 year = {1903}
}

@article{Gorowski2014714,
author = {Górowski, Jan and Łomnicki, Adam},
doi = {10.2478/aupcsm-2014-0001},
url = {https://doi.org/10.2478/aupcsm-2014-0001},
title = {Simple proofs of some generalizations of the Wilson’s theorem},
journal = {Annales Universitatis Paedagogicae Cracoviensis. Studia Mathematica},
number = {1},
volume = {13},
year = {2014},
pages = {7--14}
}

@article{oliviero_transitions_2021,
	title = {Transitions in entanglement complexity in random quantum circuits by measurements},
	volume = {418},
	issn = {0375-9601},
	url = {https://www.sciencedirect.com/science/article/pii/S0375960121005855},
	doi = {https://doi.org/10.1016/j.physleta.2021.127721},
	journal = {Physics Letters A},
	author = {Oliviero, Salvatore F. E. and Leone, Lorenzo and Hamma, Alioscia},
	year = {2021},
	pages = {127721},
}

@article{RevModPhys.91.021001,
  title = {Colloquium: Many-body localization, thermalization, and entanglement},
  author = {Abanin, Dmitry A. and Altman, Ehud and Bloch, Immanuel and Serbyn, Maksym},
  journal = {Rev. Mod. Phys.},
  volume = {91},
  issue = {2},
  pages = {021001},
  numpages = {26},
  year = {2019},
  month = {May},
  publisher = {American Physical Society},
  doi = {10.1103/RevModPhys.91.021001},
  url = {https://link.aps.org/doi/10.1103/RevModPhys.91.021001}
}

@article{nandkishore_many-body_2015,
	title = {Many-{Body} {Localization} and {Thermalization} in {Quantum} {Statistical} {Mechanics}},
	volume = {6},
	issn = {1947-5462},
	url = {https://www.annualreviews.org/content/journals/10.1146/annurev-conmatphys-031214-014726},
	doi = {https://doi.org/10.1146/annurev-conmatphys-031214-014726},
	number = {Volume 6, 2015},
	journal = {Annual Review of Condensed Matter Physics},
	publisher = {Annual Reviews},
	author = {Nandkishore, Rahul and Huse, David A.},
	year = {2015},
	note = {Type: Journal Article},
	pages = {15--38},
}

@article{1jzy-sk9r,
  title = {Anticoncentration and Nonstabilizerness Spreading under Ergodic Quantum Dynamics},
  author = {Tirrito, Emanuele and Turkeshi, Xhek and Sierant, Piotr},
  journal = {Phys. Rev. Lett.},
  volume = {135},
  issue = {22},
  pages = {220401},
  numpages = {9},
  year = {2025},
  month = {Nov},
  publisher = {American Physical Society},
  doi = {10.1103/1jzy-sk9r},
  url = {https://link.aps.org/doi/10.1103/1jzy-sk9r}
}

@misc{magni2025anticoncentrationstatedesigndoped,
      title={Anticoncentration and State Design of Doped Real Clifford Circuits and Tensor Networks}, 
      author={Beatrice Magni and Markus Heinrich and Lorenzo Leone and Xhek Turkeshi},
      year={2025},
      eprint={2512.15880},
      archivePrefix={arXiv},
      primaryClass={quant-ph},
      url={https://arxiv.org/abs/2512.15880}, 
}

@article{PhysRevB.111.054301,
  title = {Pauli spectrum and nonstabilizerness of typical quantum many-body states},
  author = {Turkeshi, Xhek and Dymarsky, Anatoly and Sierant, Piotr},
  journal = {Phys. Rev. B},
  volume = {111},
  issue = {5},
  pages = {054301},
  numpages = {12},
  year = {2025},
  month = {Feb},
  publisher = {American Physical Society},
  doi = {10.1103/PhysRevB.111.054301},
  url = {https://link.aps.org/doi/10.1103/PhysRevB.111.054301}
}

@article{turkeshi_magic_2025,
	title = {Magic spreading in random quantum circuits},
	volume = {16},
	issn = {2041-1723},
	url = {https://doi.org/10.1038/s41467-025-57704-x},
	doi = {10.1038/s41467-025-57704-x},
	number = {1},
	journal = {Nature Communications},
	author = {Turkeshi, Xhek and Tirrito, Emanuele and Sierant, Piotr},
	month = mar,
	year = {2025},
	pages = {2575},
}

@article{p8dn-glcw,
  title = {Anticoncentration in Clifford Circuits and Beyond: From Random Tensor Networks to Pseudomagic States},
  author = {Magni, Beatrice and Christopoulos, Alexios and De Luca, Andrea and Turkeshi, Xhek},
  journal = {Phys. Rev. X},
  volume = {15},
  issue = {3},
  pages = {031071},
  numpages = {16},
  year = {2025},
  month = {Sep},
  publisher = {American Physical Society},
  doi = {10.1103/p8dn-glcw},
  url = {https://link.aps.org/doi/10.1103/p8dn-glcw}
}

@misc{vardhan2025freemutualinformationhigherpoint,
      title={Free mutual information and higher-point OTOCs}, 
      author={Shreya Vardhan and Jinzhao Wang},
      year={2025},
      eprint={2509.13406},
      archivePrefix={arXiv},
      primaryClass={quant-ph},
      url={https://arxiv.org/abs/2509.13406}, 
}

@article{deutsch_eigenstate_2018,
	title = {Eigenstate thermalization hypothesis},
	volume = {81},
	url = {https://doi.org/10.1088/1361-6633/aac9f1},
	doi = {10.1088/1361-6633/aac9f1},
	number = {8},
	journal = {Reports on Progress in Physics},
	publisher = {IOP Publishing},
	author = {Deutsch, Joshua M},
	month = jul,
	year = {2018},
	pages = {082001},
}

@article{PhysRevLett.130.140402,
  title = {Non-Abelian Eigenstate Thermalization Hypothesis},
  author = {Murthy, Chaitanya and Babakhani, Arman and Iniguez, Fernando and Srednicki, Mark and Yunger Halpern, Nicole},
  journal = {Phys. Rev. Lett.},
  volume = {130},
  issue = {14},
  pages = {140402},
  numpages = {8},
  year = {2023},
  month = {Apr},
  publisher = {American Physical Society},
  doi = {10.1103/PhysRevLett.130.140402},
  url = {https://link.aps.org/doi/10.1103/PhysRevLett.130.140402}
}

@article{de_palma_necessity_2015,
	title = {Necessity of {Eigenstate} {Thermalization}},
	volume = {115},
	url = {https://link.aps.org/doi/10.1103/PhysRevLett.115.220401},
	doi = {10.1103/PhysRevLett.115.220401},
	number = {22},
	journal = {Physical Review Letters},
	publisher = {American Physical Society},
	author = {De Palma, Giacomo and Serafini, Alessio and Giovannetti, Vittorio and Cramer, Marcus},
	month = nov,
	year = {2015},
	pages = {220401},
}

@misc{venuti2019ergodicityeigenstatethermalizationfoundations,
      title={Ergodicity, eigenstate thermalization, and the foundations of statistical mechanics in quantum and classical systems}, 
      author={Lorenzo Campos Venuti and Lawrence Liu},
      year={2019},
      eprint={1904.02336},
      archivePrefix={arXiv},
      primaryClass={cond-mat.stat-mech},
      url={https://arxiv.org/abs/1904.02336}, 
}

@inbook{doi:10.1142/9789814704090_0008,
author = {L. Campos Venuti},
title = {Theory of Temporal Fluctuations in Isolated Quantum Systems},
booktitle = {Quantum Criticality in Condensed Matter},
chapter = {},
publisher={World Scientific},
year={2015},
pages = {203-219},
doi = {10.1142/9789814704090_0008},
URL = {\url{https://www.worldscientific.com/doi/abs/10.1142/9789814704090_0008}},
eprint = {\url{https://www.worldscientific.com/doi/pdf/10.1142/9789814704090_0008}}
}

@article{PhysRevA.81.022113,
  title = {Unitary equilibrations: Probability distribution of the Loschmidt echo},
  author = {Campos Venuti, Lorenzo and Zanardi, Paolo},
  journal = {Phys. Rev. A},
  volume = {81},
  issue = {2},
  pages = {022113},
  numpages = {14},
  year = {2010},
  month = {Feb},
  publisher = {American Physical Society},
  doi = {10.1103/PhysRevA.81.022113},
  url = {https://link.aps.org/doi/10.1103/PhysRevA.81.022113}
}

@article{PhysRevB.96.020408,
  title = {Entanglement complexity in quantum many-body dynamics, thermalization, and localization},
  author = {Yang, Zhi-Cheng and Hamma, Alioscia and Giampaolo, Salvatore M. and Mucciolo, Eduardo R. and Chamon, Claudio},
  journal = {Phys. Rev. B},
  volume = {96},
  issue = {2},
  pages = {020408},
  numpages = {5},
  year = {2017},
  month = {Jul},
  publisher = {American Physical Society},
  doi = {10.1103/PhysRevB.96.020408},
  url = {https://link.aps.org/doi/10.1103/PhysRevB.96.020408}
}

@article{shaffer_irreversibility_2014,
	title = {Irreversibility and entanglement spectrum statistics in quantum circuits},
	volume = {2014},
	url = {https://doi.org/10.1088/1742-5468/2014/12/P12007},
	doi = {10.1088/1742-5468/2014/12/P12007},
	number = {12},
	journal = {Journal of Statistical Mechanics: Theory and Experiment},
	publisher = {IOP Publishing and SISSA},
	author = {Shaffer, Daniel and Chamon, Claudio and Hamma, Alioscia and Mucciolo, Eduardo R},
	month = dec,
	year = {2014},
	pages = {P12007},
}

@article{PhysRevLett.112.240501,
  title = {Emergent Irreversibility and Entanglement Spectrum Statistics},
  author = {Chamon, Claudio and Hamma, Alioscia and Mucciolo, Eduardo R.},
  journal = {Phys. Rev. Lett.},
  volume = {112},
  issue = {24},
  pages = {240501},
  numpages = {5},
  year = {2014},
  month = {Jun},
  publisher = {American Physical Society},
  doi = {10.1103/PhysRevLett.112.240501},
  url = {https://link.aps.org/doi/10.1103/PhysRevLett.112.240501}
}

@article{PhysRevA.109.L040401,
  title = {Quantifying nonstabilizerness through entanglement spectrum flatness},
  author = {Tirrito, Emanuele and Tarabunga, Poetri Sonya and Lami, Gugliemo and Chanda, Titas and Leone, Lorenzo and Oliviero, Salvatore F. E. and Dalmonte, Marcello and Collura, Mario and Hamma, Alioscia},
  journal = {Phys. Rev. A},
  volume = {109},
  issue = {4},
  pages = {L040401},
  numpages = {6},
  year = {2024},
  month = {Apr},
  publisher = {American Physical Society},
  doi = {10.1103/PhysRevA.109.L040401},
  url = {https://link.aps.org/doi/10.1103/PhysRevA.109.L040401}
}

@article{LuoEntanglement2005,
  author  = {S. Luo},
  title   = {Quantum Fisher information and entanglement},
  journal = {Letters in Mathematical Physics},
  volume  = {71},
  pages   = {1--11},
  year    = {2005}
}

@article{FuruichiYanagiKuriyama2004,
  author  = {S. Furuichi and K. Yanagi and K. Kuriyama},
  title   = {Fundamental properties of the Wigner–Yanase–Dyson information},
  journal = {Journal of Mathematical Physics},
  volume  = {45},
  number  = {12},
  pages   = {4868--4877},
  year    = {2004}
}

@article{Furuichi2006,
  author  = {S. Furuichi},
  title   = {On generalized entropies and related topics in quantum information theory},
  journal = {Journal of Mathematical Physics},
  volume  = {47},
  number  = {2},
  year    = {2006}
}

@article{Luo2004,
  author  = {S. Luo},
  title   = {Wigner–Yanase skew information and uncertainty relations},
  journal = {Physics Letters A},
  volume  = {346},
  pages   = {393--398},
  year    = {2005}
}

@article{LuoZhang2004,
  author  = {S. Luo and Z. Zhang},
  title   = {On skew information and uncertainty relations},
  journal = {Journal of Statistical Physics},
  volume  = {114},
  pages   = {1557--1576},
  year    = {2004}
}

@article{Furuichi2008,
  author  = {S. Furuichi},
  title   = {Information quantities, uncertainty relations, and trace inequalities},
  journal = {Journal of Mathematical Physics},
  volume  = {50},
  number  = {1},
  year    = {2009}
}

@article{Luo2003,
  author  = {S. Luo},
  title   = {Wigner–Yanase skew information and uncertainty relations},
  journal = {Physics Letters A},
  volume  = {320},
  pages   = {73--78},
  year    = {2003}
}

@article{GibiliscoIsola2005,
  author  = {Paolo Gibilisco and Tommaso Isola},
  title   = {Wigner–Yanase information on quantum state space: The geometric approach},
  journal = {Journal of Mathematical Physics},
  volume  = {44},
  pages   = {3752--3762},
  year    = {2003}
}

@article{LuoFuruichi2005,
  author  = {S. Luo and S. Furuichi},
  title   = {Wigner–Yanase skew information and Fisher information},
  journal = {Reports on Mathematical Physics},
  volume  = {56},
  number  = {3},
  pages   = {439--447},
  year    = {2005}
}

@incollection{WignerYanase1997,
  author    = {E. P. Wigner and M. M. Yanase},
  title     = {Information contents of distributions},
  booktitle = {Part I: Particles and Fields. Part II: Foundations of Quantum Mechanics},
  pages     = {452--460},
  publisher = {Springer Berlin Heidelberg},
  year      = {1997},
  doi       = {10.1007/978-3-662-09203-3_48}
}

@incollection{Lieb2002,
  author    = {Elliott H. Lieb},
  title     = {Convex Trace Functions and the Wigner-Yanase-Dyson Conjecture},
  booktitle = {Selected Papers of {E}. {H}. Lieb},
  publisher = {Springer Berlin Heidelberg},
  address   = {Berlin, Heidelberg},
  year      = {2002},
  isbn      = {978-3-642-55925-9},
  doi       = {10.1007/978-3-642-55925-9_13}
}

@article{Yanagi2010,
  author    = {K. Yanagi},
  title     = {Uncertainty relation on Wigner--Yanase--Dyson skew information},
  journal   = {Journal of Mathematical Analysis and Applications},
  volume    = {365},
  number    = {1},
  pages     = {12--18},
  year      = {2010},
  doi       = {10.1016/j.jmaa.2009.09.060}
}

@article{YoshidaKitaev2017_HaydenPreskill,
  author    = {Yoshida, Beni and Kitaev, Alexei},
  title     = {Efficient decoding for the Hayden--Preskill protocol},
  journal   = {arXiv preprint},
  year      = {2017},
  eprint    = {1710.03363},
  archivePrefix = {arXiv},
  primaryClass  = {quant-ph}
}

@article{Headrick2010_Renyi,
  author    = {Headrick, Matthew},
  title     = {Entanglement R{\'e}nyi entropies in holography},
  journal   = {Physical Review D},
  volume    = {82},
  number    = {12},
  pages     = {126010},
  year      = {2010},
  doi       = {10.1103/PhysRevD.82.126010}
}

@article{Hayden2011_HolographicMI,
  author    = {Hayden, Patrick and Headrick, Matthew and Maloney, Alexander},
  title     = {Holographic mutual information},
  journal   = {Journal of High Energy Physics},
  volume    = {2011},
  number    = {01},
  pages     = {120},
  year      = {2011},
  doi       = {10.1007/JHEP01(2011)120}
}

@article{PhysRevA.70.012307,
  title = {Capacity of nonlinear bosonic systems},
  author = {Giovannetti, Vittorio and Lloyd, Seth and Maccone, Lorenzo},
  journal = {Phys. Rev. A},
  volume = {70},
  issue = {1},
  pages = {012307},
  numpages = {6},
  year = {2004},
  month = {Jul},
  publisher = {American Physical Society},
  doi = {10.1103/PhysRevA.70.012307},
  url = {https://link.aps.org/doi/10.1103/PhysRevA.70.012307}
}

@article{PhysRevA.68.062323,
  title = {Broadband channel capacities},
  author = {Giovannetti, Vittorio and Lloyd, Seth and Maccone, Lorenzo and Shor, Peter W.},
  journal = {Phys. Rev. A},
  volume = {68},
  issue = {6},
  pages = {062323},
  numpages = {11},
  year = {2003},
  month = {Dec},
  publisher = {American Physical Society},
  doi = {10.1103/PhysRevA.68.062323},
  url = {https://link.aps.org/doi/10.1103/PhysRevA.68.062323}
}

@article{PhysRevA.55.1613,
  title = {Capacity of the noisy quantum channel},
  author = {Lloyd, Seth},
  journal = {Phys. Rev. A},
  volume = {55},
  issue = {3},
  pages = {1613--1622},
  numpages = {0},
  year = {1997},
  month = {Mar},
  publisher = {American Physical Society},
  doi = {10.1103/PhysRevA.55.1613},
  url = {https://link.aps.org/doi/10.1103/PhysRevA.55.1613}
}

@ARTICLE{1057469,
  author={McGill, W.},
  journal={Transactions of the IRE Professional Group on Information Theory}, 
  title={Multivariate information transmission}, 
  year={1954},
  volume={4},
  number={4},
  pages={93-111},
  doi={10.1109/TIT.1954.1057469}}

@article{PhysRevA.97.053811,
  title = {Interferometric modulation of quantum cascade interactions},
  author = {Cusumano, Stefano and Mari, Andrea and Giovannetti, Vittorio},
  journal = {Phys. Rev. A},
  volume = {97},
  issue = {5},
  pages = {053811},
  numpages = {11},
  year = {2018},
  month = {May},
  publisher = {American Physical Society},
  doi = {10.1103/PhysRevA.97.053811},
  url = {https://link.aps.org/doi/10.1103/PhysRevA.97.053811}
}

@article{PhysRevA.95.053838,
  title = {Interferometric quantum cascade systems},
  author = {Cusumano, Stefano and Mari, Andrea and Giovannetti, Vittorio},
  journal = {Phys. Rev. A},
  volume = {95},
  issue = {5},
  pages = {053838},
  numpages = {15},
  year = {2017},
  month = {May},
  publisher = {American Physical Society},
  doi = {10.1103/PhysRevA.95.053838},
  url = {https://link.aps.org/doi/10.1103/PhysRevA.95.053838}
}

@article{PhysRevA.98.032119,
  title = {Entropy production and asymptotic factorization via thermalization: A collisional model approach},
  author = {Cusumano, Stefano and Cavina, Vasco and Keck, Maximilian and De Pasquale, Antonella and Giovannetti, Vittorio},
  journal = {Phys. Rev. A},
  volume = {98},
  issue = {3},
  pages = {032119},
  numpages = {9},
  year = {2018},
  month = {Sep},
  publisher = {American Physical Society},
  doi = {10.1103/PhysRevA.98.032119},
  url = {https://link.aps.org/doi/10.1103/PhysRevA.98.032119}
}

@article{PhysRevResearch.3.023214,
  title = {Quantum coherence as a signature of chaos},
  author = {Anand, Namit and Styliaris, Georgios and Kumari, Meenu and Zanardi, Paolo},
  journal = {Phys. Rev. Res.},
  volume = {3},
  issue = {2},
  pages = {023214},
  numpages = {25},
  year = {2021},
  month = {Jun},
  publisher = {American Physical Society},
  doi = {10.1103/PhysRevResearch.3.023214},
  url = {https://link.aps.org/doi/10.1103/PhysRevResearch.3.023214}
}

@article{Korzekwa2016_WorkExtraction,
  author    = {Korzekwa, Kamil and Lostaglio, Matteo and Oppenheim, Jonathan and Jennings, David},
  title     = {The extraction of work from quantum coherence},
  journal   = {New Journal of Physics},
  volume    = {18},
  number    = {2},
  pages     = {023045},
  year      = {2016},
  doi       = {10.1088/1367-2630/18/2/023045}
}

@incollection{Campaioli2018_QuantumBatteries,
  author    = {Campaioli, Federico and Pollock, Felix A. and Vinjanampathy, Sai},
  title     = {Quantum Batteries},
  booktitle = {Fundamental Theories of Physics},
  publisher = {Springer International Publishing},
  pages     = {207--225},
  year      = {2018},
  doi       = {10.1007/978-3-319-99046-0_8}
}

@article{AlickiFannes2013_EntanglementBoost,
  author    = {Alicki, Robert and Fannes, Mark},
  title     = {Entanglement boost for extractable work from ensembles of quantum batteries},
  journal   = {Physical Review E},
  volume    = {87},
  number    = {4},
  pages     = {042123},
  year      = {2013},
  doi       = {10.1103/PhysRevE.87.042123}
}

@article{Andolina2019_ExtractableWork,
  author    = {Andolina, Gian Marcello and Keck, Michael and Mari, Andrea
                and Campisi, Michele and Giovannetti, Vittorio and Polini, Marco},
  title     = {Extractable work, the role of correlations, and asymptotic freedom in quantum batteries},
  journal   = {Physical Review Letters},
  volume    = {122},
  number    = {4},
  pages     = {047702},
  year      = {2019},
  doi       = {10.1103/PhysRevLett.122.047702}
}

@article{GarciaPintos2020_Fluctuations,
  author    = {Garc{\'i}a-Pintos, Luis Pedro and Hamma, Alioscia and del Campo, Adolfo},
  title     = {Fluctuations in extractable work bound the charging power of quantum batteries},
  journal   = {Physical Review Letters},
  volume    = {125},
  number    = {4},
  pages     = {040601},
  year      = {2020},
  doi       = {10.1103/PhysRevLett.125.040601}
}

@article{PhysRevLett.127.028901,
  title = {Comment on ``Fluctuations in Extractable Work Bound the Charging Power of Quantum Batteries''},
  author = {Cusumano, Stefano and Rudnicki, \L{}ukasz},
  journal = {Phys. Rev. Lett.},
  volume = {127},
  issue = {2},
  pages = {028901},
  numpages = {2},
  year = {2021},
  month = {Jul},
  publisher = {American Physical Society},
  doi = {10.1103/PhysRevLett.127.028901},
  url = {https://link.aps.org/doi/10.1103/PhysRevLett.127.028901}
}

@article{Styliaris2019_PRB,
  author    = {Styliaris, Georgios and Anand, N. and Campos Venuti, Lorenzo and Zanardi, Paolo},
  title     = {Quantum coherence and the localization transition},
  journal   = {Physical Review B},
  volume    = {100},
  number    = {22},
  pages     = {224204},
  year      = {2019},
  doi       = {10.1103/PhysRevB.100.224204}
}

@article{StyliarisZanardi2019_PRL,
  author    = {Styliaris, Georgios and Zanardi, Paolo},
  title     = {Quantifying the incompatibility of quantum measurements relative to a basis},
  journal   = {Physical Review Letters},
  volume    = {123},
  number    = {7},
  pages     = {070401},
  year      = {2019},
  doi       = {10.1103/PhysRevLett.123.070401}
}

@unpublished{StyliarisAnandZanardi2020_arXiv,
  author    = {Styliaris, Georgios and Anand, N. and Zanardi, Paolo},
  title     = {Information scrambling over bipartitions: Equilibration, entropy production, and typicality},
  year      = {2020},
  eprint    = {2007.08570},
  archivePrefix = {arXiv},
  primaryClass  = {quant-ph},
  note      = {arXiv preprint}
}

@article{ZanardiPaunkovic2006_PRE,
  author    = {Zanardi, Paolo and Paunkovi{\'c}, Nikola},
  title     = {Ground state overlap and quantum phase transitions},
  journal   = {Physical Review E},
  volume    = {74},
  number    = {3},
  pages     = {031123},
  year      = {2006},
  doi       = {10.1103/PhysRevE.74.031123}
}

@article{ZanardiStyliarisVenuti2017_CGP,
  author    = {Zanardi, Paolo and Styliaris, Georgios and Campos Venuti, Lorenzo},
  title     = {Coherence-generating power of quantum unitary maps and beyond},
  journal   = {Physical Review A},
  volume    = {95},
  number    = {5},
  pages     = {052306},
  year      = {2017},
  doi       = {10.1103/PhysRevA.95.052306}
}

@article{ZanardiStyliarisVenuti2017_Unital,
  author    = {Zanardi, Paolo and Styliaris, Georgios and Campos Venuti, Lorenzo},
  title     = {Measures of coherence-generating power for quantum unital operations},
  journal   = {Physical Review A},
  volume    = {95},
  number    = {5},
  pages     = {052307},
  year      = {2017},
  doi       = {10.1103/PhysRevA.95.052307}
}

@article{ZanardiVenuti2018_Grassmannian,
  author    = {Zanardi, Paolo and Venuti, Lorenzo Campos},
  title     = {Quantum coherence generating power, maximally abelian subalgebras, and Grassmannian geometry},
  journal   = {Journal of Mathematical Physics},
  volume    = {59},
  number    = {1},
  pages     = {012203},
  year      = {2018},
  doi       = {10.1063/1.4997146}
}

@article{StreltsovAdessoPlenio2017_Colloquium,
  author    = {Streltsov, Alexander and Adesso, Gerardo and Plenio, Martin B.},
  title     = {Colloquium: Quantum coherence as a resource},
  journal   = {Reviews of Modern Physics},
  volume    = {89},
  number    = {4},
  pages     = {041003},
  year      = {2017},
  doi       = {10.1103/RevModPhys.89.041003}
}

@article{WinterYang2016,
  author    = {Winter, Andreas and Yang, Dong},
  title     = {Operational resource theory of coherence},
  journal   = {Physical Review Letters},
  volume    = {116},
  number    = {12},
  pages     = {120404},
  year      = {2016},
  doi       = {10.1103/PhysRevLett.116.120404}
}

@article{MarvianSpekkens2016,
  author    = {Marvian, Iman and Spekkens, Robert W.},
  title     = {How to quantify coherence: Distinguishing speakable and unspeakable notions},
  journal   = {Physical Review A},
  volume    = {94},
  number    = {5},
  pages     = {052324},
  year      = {2016},
  doi       = {10.1103/PhysRevA.94.052324}
}

@article{Du2015,
  author    = {Du, Shuanping and Bai, Zhaoqi and Guo, Yu},
  title     = {Conditions for coherence transformations under incoherent operations},
  journal   = {Physical Review A},
  volume    = {91},
  number    = {5},
  pages     = {052120},
  year      = {2015},
  doi       = {10.1103/PhysRevA.91.052120}
}

@article{Streltsov2015,
  author    = {Streltsov, Alexander and Singh, Uttam and Dhar, Himadri Shekhar and Bera, Manabendra Nath and Adesso, Gerardo},
  title     = {Measuring quantum coherence with entanglement},
  journal   = {Physical Review Letters},
  volume    = {115},
  number    = {2},
  pages     = {020403},
  year      = {2015},
  doi       = {10.1103/PhysRevLett.115.020403}
}

@article{ChitambarHsieh2016,
  author    = {Chitambar, Eric and Hsieh, Min-Hsiu},
  title     = {Relating the resource theories of entanglement and quantum coherence},
  journal   = {Physical Review Letters},
  volume    = {117},
  number    = {2},
  pages     = {020402},
  year      = {2016},
  doi       = {10.1103/PhysRevLett.117.020402}
}

@article{Zeh1970,
  author    = {Zeh, H. Dieter},
  title     = {On the interpretation of measurement in quantum theory},
  journal   = {Foundations of Physics},
  volume    = {1},
  number    = {1},
  pages     = {69--76},
  year      = {1970},
  doi       = {10.1007/BF00708656},
  note      = {Reprinted in later collections}
}

@article{Zurek1981,
  author    = {Zurek, Wojciech Hubert},
  title     = {Pointer basis of quantum apparatus: Into what mixture does the wave packet collapse?},
  journal   = {Physical Review D},
  volume    = {24},
  number    = {6},
  pages     = {1516--1525},
  year      = {1981},
  doi       = {10.1103/PhysRevD.24.1516}
}

@article{Zurek1982,
  author    = {Zurek, Wojciech Hubert},
  title     = {Environment-induced superselection rules},
  journal   = {Physical Review D},
  volume    = {26},
  number    = {8},
  pages     = {1862--1880},
  year      = {1982},
  doi       = {10.1103/PhysRevD.26.1862}
}

@article{JoosZeh1985,
  author    = {Joos, E. and Zeh, H. Dieter},
  title     = {The emergence of classical properties through interaction with the environment},
  journal   = {Zeitschrift f{\"u}r Physik B Condensed Matter},
  volume    = {59},
  number    = {2},
  pages     = {223--243},
  year      = {1985},
  doi       = {10.1007/BF01725541}
}

@article{PhysRevLett.113.140401,
  title = {Quantifying Coherence},
  author = {Baumgratz, T. and Cramer, M. and Plenio, M. B.},
  journal = {Phys. Rev. Lett.},
  volume = {113},
  issue = {14},
  pages = {140401},
  numpages = {5},
  year = {2014},
  month = {Sep},
  publisher = {American Physical Society},
  doi = {10.1103/PhysRevLett.113.140401},
  url = {https://link.aps.org/doi/10.1103/PhysRevLett.113.140401}
}

@article{doi:10.1142/S0219887824400103,
author = {Iannotti, Daniele and Hamma, Alioscia},
title = {Geometric methods in quantum information and entanglement variational principle},
journal = {International Journal of Geometric Methods in Modern Physics},
volume = {21},
number = {10},
pages = {2440010},
year = {2024},
doi = {10.1142/S0219887824400103},
URL = { https://doi.org/10.1142/S0219887824400103},
eprint = { https://doi.org/10.1142/S0219887824400103}
}

@article{Srednicki1993,
  author = {M. Srednicki},
  title = {Entropy and area},
  journal = {Physical Review Letters},
  volume = {71},
  pages = {666},
  year = {1993}
}

@article{Vidal2003,
  author = {G. Vidal and J. I. Latorre and E. Rico and A. Kitaev},
  title = {Entanglement in quantum critical phenomena},
  journal = {Physical Review Letters},
  volume = {90},
  pages = {227902},
  year = {2003},
}

@article{CalabreseCardy2004,
  author = {P. Calabrese and J. Cardy},
  title = {Entanglement entropy and quantum field theory},
  journal = {Journal of Statistical Mechanics: Theory and Experiment},
  year = {2004},
  pages = {P06002}
}

@article{RyuTakayanagi2006,
  author = {S. Ryu and T. Takayanagi},
  title = {Holographic derivation of entanglement entropy from the anti–de Sitter space/conformal field theory correspondence},
  journal = {Physical Review Letters},
  volume = {96},
  pages = {181602},
  year = {2006}
}

@article{Page1993,
  author    = {D. N. Page},
  title     = {Average entropy of a subsystem},
  journal   = {Physical Review Letters},
  volume    = {71},
  pages     = {1291},
  year      = {1993},
  doi       = {10.1103/PhysRevLett.71.1291}
}

@article{ZanardiZalkaFaoro2000,
  author    = {P. Zanardi and C. Zalka and L. Faoro},
  title     = {Entangling power of quantum evolutions},
  journal   = {Physical Review A},
  volume    = {62},
  pages     = {030301},
  year      = {2000},
  doi       = {10.1103/PhysRevA.62.030301}
}

@article{Zanardi2001,
  author    = {P. Zanardi},
  title     = {Entanglement of quantum evolutions},
  journal   = {Physical Review A},
  volume    = {63},
  pages     = {040304},
  year      = {2001},
  doi       = {10.1103/PhysRevA.63.040304}
}

@article{KumariGhose2019,
  author    = {M. Kumari and S. Ghose},
  title     = {Untangling entanglement and chaos},
  journal   = {Physical Review A},
  volume    = {99},
  pages     = {042311},
  year      = {2019},
  doi       = {10.1103/PhysRevA.99.042311}
}

@article{MezeiStanford2017,
  author    = {M. Mezei and D. Stanford},
  title     = {On entanglement spreading in chaotic systems},
  journal   = {Journal of High Energy Physics},
  volume    = {2017},
  number    = {5},
  pages     = {065},
  year      = {2017},
  doi       = {10.1007/JHEP05(2017)065}
}

@article{YouGu2018,
  author    = {Y.-Z. You and Y. Gu},
  title     = {Entanglement features of random Hamiltonian dynamics},
  journal   = {Physical Review B},
  volume    = {98},
  pages     = {014309},
  year      = {2018},
  doi       = {10.1103/PhysRevB.98.014309}
}

@ARTICLE{Garcia-Mata:2023,
AUTHOR = {García-Mata, I.  and Jalabert, R. A. and Wisniacki, D. Ariel},
TITLE   = {{O}ut-of-time-order correlations and quantum chaos},
YEAR    = {2023},
JOURNAL = {Scholarpedia},
VOLUME  = {18},
NUMBER  = {4},
PAGES   = {55237},
DOI     = {10.4249/scholarpedia.55237},
NOTE    = {revision \#204529}
}

@article{SekinoSusskind2008,
  author    = {Y. Sekino and L. Susskind},
  title     = {Fast scramblers},
  journal   = {Journal of High Energy Physics},
  volume    = {2008},
  number    = {10},
  pages     = {065},
  year      = {2008},
  doi       = {10.1088/1126-6708/2008/10/065}
}

@misc{Kitaev2015Talk,
  author       = {A. Kitaev},
  title        = {A simple model of quantum holography},
  note         = {Talk given at KITP Program: Entanglement in Strongly Correlated Quantum Matter},
  year         = {2015},
  howpublished = {KITP seminar},
  url          = {http://online.kitp.ucsb.edu/online/entangled15/},
}

@article{Swingle2018,
  author    = {B. Swingle},
  title     = {Unscrambling the physics of out-of-time-order correlators},
  journal   = {Nature Physics},
  volume    = {14},
  pages     = {988},
  year      = {2018},
  doi       = {10.1038/s41567-018-0295-5}
}

@misc{XuSwingle2022,
  author    = {S. Xu and B. Swingle},
  title     = {Scrambling dynamics and out-of-time ordered correlators in quantum many-body systems: a tutorial},
  year      = {2022},
  eprint    = {2202.07060},
  archivePrefix = {arXiv},
  primaryClass  = {quant-ph}
}

@article{ShenkerStanford2014,
  author    = {S. H. Shenker and D. Stanford},
  title     = {Black holes and the butterfly effect},
  journal   = {Journal of High Energy Physics},
  volume    = {2014},
  number    = {3},
  pages     = {67},
  year      = {2014},
  doi       = {10.1007/JHEP03(2014)067}
}

@article{ShenkerStanford2014MultipleShocks,
  author    = {S. H. Shenker and D. Stanford},
  title     = {Multiple shocks},
  journal   = {Journal of High Energy Physics},
  volume    = {2014},
  number    = {12},
  pages     = {46},
  year      = {2014},
  doi       = {10.1007/JHEP12(2014)046}
}

@article{ShenkerStanford2015Stringy,
  author    = {S. H. Shenker and D. Stanford},
  title     = {Stringy effects in scrambling},
  journal   = {Journal of High Energy Physics},
  volume    = {2015},
  number    = {5},
  pages     = {132},
  year      = {2015},
  doi       = {10.1007/JHEP05(2015)132}
}

@article{Pastawski2001PhysRep,
  title        = {Quantum interference phenomena in the Loschmidt echo},
  author       = {Pastawski, Horacio M. and Levstein, Patricia R. and Usaj, Gonzalo and Raya, Jorge and Hirschinger, Jana},
  journal      = {Physics Reports},
  volume       = {355},
  number       = {1},
  pages        = {35--85},
  year         = {2001},
  doi          = {10.1016/S0370-1573(01)00025-4},
  note         = {Comprehensive review of Loschmidt echo}
}

@article{Pastawski1995PRL,
  title        = {Quantum Chaos: An Experimentalist’s Approach via Nuclear Magnetic Resonance},
  author       = {Pastawski, Horacio M. and Levstein, Patricia R. and Usaj, Gonzalo},
  journal      = {Physical Review Letters},
  volume       = {75},
  pages        = {4310--4313},
  year         = {1995},
  doi          = {10.1103/PhysRevLett.75.4310}
}

@article{VanhalaOjanen2023PRResearch,
  title        = {Theory of the Loschmidt echo and dynamical quantum phase transitions in disordered Fermi systems},
  author       = {Vanhala, Tuomas I. and Ojanen, Teemu},
  journal      = {Physical Review Research},
  volume       = {5},
  pages        = {033178},
  year         = {2023},
  doi          = {10.1103/PhysRevResearch.5.033178}
}

@article{NiuWang2023PRA,
  title        = {Excited state quantum phase transition and Loschmidt echo spectra in a spinor Bose–Einstein condensate},
  author       = {Niu, Zhen-Xia and Wang, Qian},
  journal      = {Physical Review A},
  volume       = {107},
  pages        = {033307},
  year         = {2023},
  doi          = {10.1103/PhysRevA.107.033307}
}

@article{JafariJohannesson2017PRL,
  title        = {Loschmidt Echo Revivals: Critical and Noncritical},
  author       = {Jafari, R. and Johannesson, Henrik},
  journal      = {Physical Review Letters},
  volume       = {118},
  pages        = {015701},
  year         = {2017},
  doi          = {10.1103/PhysRevLett.118.015701}
}

@article{Andersen2006,
  title = {Echo spectroscopy and quantum chaos},
  author = {Andersen, M. F. and Kaplan, A. and Grunzweig, T. and Davidson, N.},
  journal = {Physical Review Letters},
  volume = {97},
  number = {10},
  pages = {104102},
  year = {2006},
  doi = {10.1103/PhysRevLett.97.104102}
}

@article{Emerson2003,
  title = {Fidelity decay as an efficient indicator of quantum chaos},
  author = {Emerson, Joseph and Weinstein, Yaakov S. and Lloyd, Seth and Cory, David G.},
  journal = {Science},
  volume = {302},
  number = {5653},
  pages = {2098--2100},
  year = {2003},
  doi = {10.1126/science.1090790}
}

@article{Prosen2007,
  title = {Is the efficiency of classical simulations of quantum dynamics related to integrability?},
  author = {Prosen, Toma\v{z} and \v{Z}nidari\v{c}, Marko},
  journal = {Physical Review E},
  volume = {75},
  number = {1},
  pages = {015202},
  year = {2007},
  doi = {10.1103/PhysRevE.75.015202}
}

@article{Heyl2013,
  title = {Dynamical quantum phase transitions in the transverse-field Ising model},
  author = {Heyl, Markus and Polkovnikov, Anatoli and Kehrein, Stefan},
  journal = {Physical Review Letters},
  volume = {110},
  number = {13},
  pages = {135704},
  year = {2013},
  doi = {10.1103/PhysRevLett.110.135704}
}

@article{Heyl2018,
  title = {Dynamical quantum phase transitions: a review},
  author = {Heyl, Markus},
  journal = {Reports on Progress in Physics},
  volume = {81},
  number = {5},
  pages = {054001},
  year = {2018},
  doi = {10.1088/1361-6633/aaaf9a}
}

@article{TorresHerrera2015,
  title = {Dynamics at the many-body localization transition},
  author = {Torres-Herrera, E. J. and Santos, Lea F.},
  journal = {Physical Review B},
  volume = {92},
  number = {1},
  pages = {014208},
  year = {2015},
  doi = {10.1103/PhysRevB.92.014208}
}

@article{Znidaric2008,
  title = {Many-body localization in the Heisenberg XXZ magnet},
  author = {\v{Z}nidari\v{c}, Marko and Prosen, Toma\v{z} and Prelov\v{s}ek, Peter},
  journal = {Physical Review B},
  volume = {77},
  number = {6},
  pages = {064426},
  year = {2008},
  doi = {10.1103/PhysRevB.77.064426}
}

@article{Jalabert2001,
  title = {Environment-independent decoherence rate in classically chaotic systems},
  author = {Jalabert, Ra\'{u}l A. and Pastawski, Horacio M.},
  journal = {Physical Review Letters},
  volume = {86},
  number = {12},
  pages = {2490--2493},
  year = {2001},
  doi = {10.1103/PhysRevLett.86.2490}
}

@article{Gorin2006,
  title = {Dynamics of Loschmidt echoes and fidelity decay},
  author = {Gorin, Thomas and Prosen, Toma\v{z} and Seligman, Thomas H. and \v{Z}nidari\v{c}, Marko},
  journal = {Physics Reports},
  volume = {435},
  number = {2--5},
  pages = {33--156},
  year = {2006},
  doi = {10.1016/j.physrep.2006.09.003}
}

@article{Pastawski2000,
  title = {Decoherence as irreversibility in quantum mechanics},
  author = {Pastawski, Horacio M. and Levstein, Patricia R. and Usaj, Gonzalo and Raya, Jorge and Hirschinger, Jordi},
  journal = {Physica A},
  volume = {283},
  number = {1--2},
  pages = {166--170},
  year = {2000},
  doi = {10.1016/S0378-4371(00)00111-9}
}

@article{Richter2002,
  title = {Semiclassical theory of chaotic systems},
  author = {Richter, Klaus and Sieber, Martin},
  journal = {Physical Review Letters},
  volume = {89},
  number = {20},
  pages = {206801},
  year = {2002},
  doi = {10.1103/PhysRevLett.89.206801}
}

@article{Cotler2017Chaos,
  title        = {Chaos, complexity, and random matrices},
  author       = {Cotler, J. and Hunter-Jones, N. and Liu, J. and Yoshida, B.},
  journal      = {Journal of High Energy Physics},
  volume       = {2017},
  number       = {11},
  pages        = {48},
  year         = {2017},
  doi          = {10.1007/JHEP11(2017)048},
}

@article{riser_power_2021,
	title = {Power spectrum and form factor in random diagonal matrices and integrable billiards},
	volume = {425},
	issn = {0003-4916},
	url = {https://www.sciencedirect.com/science/article/pii/S0003491620303274},
	doi = {https://doi.org/10.1016/j.aop.2020.168393},
	journal = {Annals of Physics},
	author = {Riser, Roman and Kanzieper, Eugene},
	year = {2021},
	pages = {168393},
}

@article{PhysRevA.83.042330,
  title = {Local stabilizer codes in three dimensions without string logical operators},
  author = {Haah, Jeongwan},
  journal = {Phys. Rev. A},
  volume = {83},
  issue = {4},
  pages = {042330},
  numpages = {16},
  year = {2011},
  month = {Apr},
  publisher = {American Physical Society},
  doi = {10.1103/PhysRevA.83.042330},
  url = {https://link.aps.org/doi/10.1103/PhysRevA.83.042330}
}

@Article{10.21468/SciPostPhys.6.4.041,
	title={{Foliated fracton order from gauging subsystem symmetries}},
	author={Wilbur Shirley and Kevin Slagle and Xie Chen},
	journal={SciPost Phys.},
	volume={6},
	pages={041},
	year={2019},
	publisher={SciPost},
	doi={10.21468/SciPostPhys.6.4.041},
	url={https://scipost.org/10.21468/SciPostPhys.6.4.041},
}

@article{bravyi_no-go_2009,
	title = {A no-go theorem for a two-dimensional self-correcting quantum memory based on stabilizer codes},
	volume = {11},
	url = {https://doi.org/10.1088/1367-2630/11/4/043029},
	doi = {10.1088/1367-2630/11/4/043029},
	number = {4},
	journal = {New Journal of Physics},
	author = {Bravyi, Sergey and Terhal, Barbara},
	month = apr,
	year = {2009},
	pages = {043029},
}

@article{alicki_statistical_2007,
	title = {A statistical mechanics view on {Kitaev}'s proposal for quantum memories},
	volume = {40},
	url = {https://doi.org/10.1088/1751-8113/40/24/012},
	doi = {10.1088/1751-8113/40/24/012},
	number = {24},
	journal = {Journal of Physics A: Mathematical and Theoretical},
	author = {Alicki, R and Fannes, M and Horodecki, M},
	month = may,
	year = {2007},
	pages = {6451},
}

@article{PhysRevB.77.064302,
  title = {Autocorrelations and thermal fragility of anyonic loops in topologically quantum ordered systems},
  author = {Nussinov, Zohar and Ortiz, Gerardo},
  journal = {Phys. Rev. B},
  volume = {77},
  issue = {6},
  pages = {064302},
  numpages = {16},
  year = {2008},
  month = {Feb},
  publisher = {American Physical Society},
  doi = {10.1103/PhysRevB.77.064302},
  url = {https://link.aps.org/doi/10.1103/PhysRevB.77.064302}
}

@article{PhysRevLett.123.230503,
  title = {Universality Classes of Stabilizer Code Hamiltonians},
  author = {Weinstein, Zack and Ortiz, Gerardo and Nussinov, Zohar},
  journal = {Phys. Rev. Lett.},
  volume = {123},
  issue = {23},
  pages = {230503},
  numpages = {6},
  year = {2019},
  month = {Dec},
  publisher = {American Physical Society},
  doi = {10.1103/PhysRevLett.123.230503},
  url = {https://link.aps.org/doi/10.1103/PhysRevLett.123.230503}
}

@misc{temme2015faststabilizerhamiltoniansthermalize,
      title={How fast do stabilizer Hamiltonians thermalize?}, 
      author={Kristan Temme and Michael J. Kastoryano},
      year={2015},
      eprint={1505.07811},
      archivePrefix={arXiv},
      primaryClass={quant-ph},
      url={https://arxiv.org/abs/1505.07811}, 
}

@article{Zhou2020SingleT,
  author       = {Zhou, S. and Yang, Z. and Hamma, A. and Chamon, C.},
  title        = {Single T gate in a Clifford circuit drives transition to universal entanglement spectrum statistics},
  journal      = {SciPost Physics},
  volume       = {9},
  number       = {6},
  pages        = {87},
  year         = {2020},
  doi          = {10.21468/SciPostPhys.9.6.087}
}

@article{BrownColloquium2016,
  author  = {Brown, B. J. and Loss, Daniel and Pachos, Jiannis K. and Self, Chris N. and Wootton, J. R.},
  title   = {Colloquium: Quantum memories based on topological stability},
  journal = {Reviews of Modern Physics},
  volume  = {88},
  year    = {2016},
  doi     = {10.1103/RevModPhys.88.045005}
}

@article{BombinMartinDelgado2006,
  author  = {Bombin, H. and Martin-Delgado, M. A.},
  title   = {Topological quantum distillation},
  journal = {Physical Review Letters},
  volume  = {97},
  pages   = {180501},
  year    = {2006},
  doi     = {10.1103/PhysRevLett.97.180501}
}

@article{Bombin2015GaugeColorCode,
  author  = {Bombin, Hector},
  title   = {Gauge color codes: optimal transversal gates and gauge fixing in topological stabilizer codes},
  journal = {New Journal of Physics},
  volume  = {17},
  pages   = {083002},
  year    = {2015}
}

@article{Poulin2005Subsystem,
  author  = {Poulin, D.},
  title   = {Stabilizer Formalism for Operator Quantum Error Correction},
  journal = {Physical Review Letters},
  volume  = {95},
  year    = {2005},
  doi     = {10.1103/PhysRevLett.95.230504}
}

@article{KitaevToricCode2003,
  author  = {Kitaev, Alexei Y.},
  title   = {Fault-tolerant quantum computation by anyons},
  journal = {Annals of Physics},
  volume  = {303},
  pages   = {2--30},
  year    = {2003},
  doi     = {10.1016/S0003-4916(02)00018-0}
}

@article{Dennis2002,
  author  = {Dennis, E. and Kitaev, A. and Landahl, A. and Preskill, J.},
  title   = {Topological quantum memory},
  journal = {Journal of Mathematical Physics},
  volume  = {43},
  pages   = {4452--4505},
  year    = {2002},
  doi     = {10.1063/1.1499754}
}

@article{Fowler2012,
  author  = {Fowler, Austin G. and Mariantoni, Matteo and Martinis, John M. and Cleland, Andrew N.},
  title   = {Surface codes: Towards practical large-scale quantum computation},
  journal = {Physical Review A},
  volume  = {86},
  year    = {2012},
  doi     = {10.1103/PhysRevA.86.032324}
}

@article{BravyiKitaev1998,
  author  = {Bravyi, Sergey and Kitaev, Alexei},
  title   = {Quantum codes on a lattice with boundary},
  journal = {arXiv:quant-ph/9811052},
  year    = {1998}
}

@book{LidarBrunQEC2013,
  title        = {Quantum Error Correction},
  editor       = {Lidar, Daniel A. and Brun, Todd A.},
  year         = {2013},
  publisher    = {Cambridge University Press},
  address      = {Cambridge},
  isbn         = {9781107002173}
}

@article{TerhalQECMemories2015,
  author       = {Terhal, Barbara M.},
  title        = {Quantum error correction for quantum memories},
  journal      = {Reviews of Modern Physics},
  volume       = {87},
  number       = {2},
  pages        = {307--346},
  year         = {2015},
  publisher    = {American Physical Society},
  doi          = {10.1103/RevModPhys.87.307}
}

@article{garcia-garcia_exact_2018,
	title = {Exact moments of the {Sachdev}-{Ye}-{Kitaev} model up to order 1/{N2}},
	volume = {2018},
	issn = {1029-8479},
	url = {https://doi.org/10.1007/JHEP04(2018)146},
	doi = {10.1007/JHEP04(2018)146},
	number = {4},
	journal = {Journal of High Energy Physics},
	author = {García-García, Antonio M. and Jia, Yiyang and Verbaarschot, Jacobus J. M.},
	month = apr,
	year = {2018},
	pages = {146},
}

@article{PhysRevResearch.2.043403,
  title = {Statistics of the spectral form factor in the self-dual kicked Ising model},
  author = {Flack, Ana and Bertini, Bruno and Prosen, Toma\ifmmode \check{z}\else \v{z}\fi{}},
  journal = {Phys. Rev. Res.},
  volume = {2},
  issue = {4},
  pages = {043403},
  numpages = {14},
  year = {2020},
  month = {Dec},
  publisher = {American Physical Society},
  doi = {10.1103/PhysRevResearch.2.043403},
  url = {https://link.aps.org/doi/10.1103/PhysRevResearch.2.043403}
}

@article{PhysRevLett.131.151602,
  title = {Supersymmetric Spectral Form Factor and Euclidean Black Holes},
  author = {Choi, Sunjin and Kim, Seok and Song, Jaewon},
  journal = {Phys. Rev. Lett.},
  volume = {131},
  issue = {15},
  pages = {151602},
  numpages = {5},
  year = {2023},
  month = {Oct},
  publisher = {American Physical Society},
  doi = {10.1103/PhysRevLett.131.151602},
  url = {https://link.aps.org/doi/10.1103/PhysRevLett.131.151602}
}

@article{caceres_spectral_2022,
	title = {Spectral form factor in sparse {SYK} models},
	volume = {2022},
	issn = {1029-8479},
	url = {https://doi.org/10.1007/JHEP08(2022)236},
	doi = {10.1007/JHEP08(2022)236},
	number = {8},
	journal = {Journal of High Energy Physics},
	author = {Cáceres, Elena and Misobuchi, Anderson and Raz, Amir},
	month = aug,
	year = {2022},
	pages = {236},
}

@article{PhysRevD.98.086026,
  title = {Spectral form factors and late time quantum chaos},
  author = {Liu, Junyu},
  journal = {Phys. Rev. D},
  volume = {98},
  issue = {8},
  pages = {086026},
  numpages = {28},
  year = {2018},
  month = {Oct},
  publisher = {American Physical Society},
  doi = {10.1103/PhysRevD.98.086026},
  url = {https://link.aps.org/doi/10.1103/PhysRevD.98.086026}
}

@article{n7rj-gwwj,
  title = {Statistics of the random matrix spectral form factor},
  author = {Altland, Alex and Divi, Francisco and Micklitz, Tobias and Pappalardi, Silvia and Rezaei, Maedeh},
  journal = {Phys. Rev. Res.},
  volume = {7},
  issue = {3},
  pages = {033138},
  numpages = {15},
  year = {2025},
  month = {Aug},
  publisher = {American Physical Society},
  doi = {10.1103/n7rj-gwwj},
  url = {https://link.aps.org/doi/10.1103/n7rj-gwwj}
}

@article{kunz_probability_1999,
	title = {The probability distribution of the spectral form factor in random matrix theory},
	volume = {32},
	url = {https://doi.org/10.1088/0305-4470/32/11/011},
	doi = {10.1088/0305-4470/32/11/011},
	number = {11},
	journal = {Journal of Physics A: Mathematical and General},
	author = {Kunz, Hervé},
	month = mar,
	year = {1999},
	pages = {2171},
}

@article{PhysRevE.60.3949,
  title = {Ergodic properties of a generic nonintegrable quantum many-body system in the thermodynamic limit},
  author = {Prosen, Toma\ifmmode \check{z}\else \v{z}\fi{}},
  journal = {Phys. Rev. E},
  volume = {60},
  issue = {4},
  pages = {3949--3968},
  numpages = {0},
  year = {1999},
  month = {Oct},
  publisher = {American Physical Society},
  doi = {10.1103/PhysRevE.60.3949},
  url = {https://link.aps.org/doi/10.1103/PhysRevE.60.3949}
}

@article{PhysRevLett.80.1808,
  title = {Time Evolution of a Quantum Many-Body System: Transition from Integrability to Ergodicity in the Thermodynamic Limit},
  author = {Prosen, Toma\ifmmode \check{z}\else \v{z}\fi{}},
  journal = {Phys. Rev. Lett.},
  volume = {80},
  issue = {9},
  pages = {1808--1811},
  numpages = {0},
  year = {1998},
  month = {Mar},
  publisher = {American Physical Society},
  doi = {10.1103/PhysRevLett.80.1808},
  url = {https://link.aps.org/doi/10.1103/PhysRevLett.80.1808}
}

@article{PhysRevE.55.4067,
  title = {Spectral form factor in a random matrix theory},
  author = {Br\'ezin, E. and Hikami, S.},
  journal = {Phys. Rev. E},
  volume = {55},
  issue = {4},
  pages = {4067--4083},
  numpages = {0},
  year = {1997},
  month = {Apr},
  publisher = {American Physical Society},
  doi = {10.1103/PhysRevE.55.4067},
  url = {https://link.aps.org/doi/10.1103/PhysRevE.55.4067}
}

@misc{venuti2025integrabilitychaosfractalanalysis,
      title={Integrability and Chaos via fractal analysis of Spectral Form Factors: Gaussian approximations and exact results}, 
      author={Lorenzo Campos Venuti and Jovan Odavić and Alioscia Hamma},
      year={2025},
      eprint={2505.05199},
      archivePrefix={arXiv},
      primaryClass={quant-ph},
      url={https://arxiv.org/abs/2505.05199}, 
}

@misc{loio2026quantumstatedesignsmagic,
      title={Quantum State Designs via Magic Teleportation}, 
      author={Hugo Lóio and Guglielmo Lami and Lorenzo Leone and Max McGinley and Xhek Turkeshi and Jacopo De Nardis},
      year={2026},
      eprint={2510.13950},
      archivePrefix={arXiv},
      primaryClass={quant-ph},
      url={https://arxiv.org/abs/2510.13950}, 
}

@misc{macedo2025witnessingmagicbellinequalities,
      title={Witnessing Magic with Bell inequalities}, 
      author={Rafael A. Macedo and Patrick Andriolo and Santiago Zamora and Davide Poderini and Rafael Chaves},
      year={2025},
      eprint={2503.18734},
      archivePrefix={arXiv},
      primaryClass={quant-ph},
      url={https://arxiv.org/abs/2503.18734}, 
}

@misc{haug2025efficientwitnessingtestingmagic,
      title={Efficient witnessing and testing of magic in mixed quantum states}, 
      author={Tobias Haug and Poetri Sonya Tarabunga},
      year={2025},
      eprint={2504.18098},
      archivePrefix={arXiv},
      primaryClass={quant-ph},
      url={https://arxiv.org/abs/2504.18098}, 
}

@misc{cepollaro2025stabilizerentropysubspaces,
      title={Stabilizer Entropy of Subspaces}, 
      author={Simone Cepollaro and Gianluca Cuffaro and Matthew B. Weiss and Stefano Cusumano and Alioscia Hamma and Seth Lloyd},
      year={2025},
      eprint={2512.23013},
      archivePrefix={arXiv},
      primaryClass={quant-ph},
      url={https://arxiv.org/abs/2512.23013}, 
}

@misc{viscardi2025interplayentanglementstructuresstabilizer,
      title={Interplay of entanglement structures and stabilizer entropy in spin models}, 
      author={Michele Viscardi and Marcello Dalmonte and Alioscia Hamma and Emanuele Tirrito},
      year={2025},
      eprint={2503.08620},
      archivePrefix={arXiv},
      primaryClass={quant-ph},
      url={https://arxiv.org/abs/2503.08620}, 
}

@article{Iannotti2025entanglement,
  doi = {10.22331/q-2025-07-21-1797},
  url = {https://doi.org/10.22331/q-2025-07-21-1797},
  title = {Entanglement and {S}tabilizer entropies of random bipartite pure quantum states},
  author = {Iannotti, Daniele and Esposito, Gianluca and Campos Venuti, Lorenzo and Hamma, Alioscia},
  journal = {{Quantum}},
  issn = {2521-327X},
  publisher = {{Verein zur F{\"{o}}rderung des Open Access Publizierens in den Quantenwissenschaften}},
  volume = {9},
  pages = {1797},
  month = jul,
  year = {2025}
}

@article{y9r6-dx7p,
  title = {Stabilizer entropy in nonintegrable quantum evolutions},
  author = {Odavi\ifmmode \acute{c}\else \'{c}\fi{}, J. and Viscardi, M. and Hamma, A.},
  journal = {Phys. Rev. B},
  volume = {112},
  issue = {10},
  pages = {104301},
  numpages = {14},
  year = {2025},
  month = {Sep},
  publisher = {American Physical Society},
  doi = {10.1103/y9r6-dx7p},
  url = {https://link.aps.org/doi/10.1103/y9r6-dx7p}
}

@Article{10.21468/SciPostPhys.15.4.131,
	title={{Complexity of frustration: A new source of non-local non-stabilizerness}},
	author={Jovan Odavić and Tobias Haug and Gianpaolo Torre and Alioscia Hamma and Fabio Franchini and Salvatore Marco Giampaolo},
	journal={SciPost Phys.},
	volume={15},
	pages={131},
	year={2023},
	publisher={SciPost},
	doi={10.21468/SciPostPhys.15.4.131},
	url={https://scipost.org/10.21468/SciPostPhys.15.4.131},
}

@misc{cusumano2025nonstabilizernessviolationschshinequalities,
      title={Non-stabilizerness and violations of CHSH inequalities}, 
      author={Stefano Cusumano and Lorenzo Campos Venuti and Simone Cepollaro and Gianluca Esposito and Daniele Iannotti and Barbara Jasser and Jovan Odavi\' c and Michele Viscardi and Alioscia Hamma},
      year={2025},
      eprint={2504.03351},
      archivePrefix={arXiv},
      primaryClass={quant-ph},
      url={https://arxiv.org/abs/2504.03351}, 
}

@article{9ph7-cyzh,
  title = {Harvesting magic from the vacuum},
  author = {Nystr\"om, Ron and Pranzini, Nicola and Keski-Vakkuri, Esko},
  journal = {Phys. Rev. Res.},
  volume = {7},
  issue = {3},
  pages = {033085},
  numpages = {5},
  year = {2025},
  month = {Jul},
  publisher = {American Physical Society},
  doi = {10.1103/9ph7-cyzh},
  url = {https://link.aps.org/doi/10.1103/9ph7-cyzh}
}

@misc{yang2025analytictoolsharvestingmagic,
      title={Analytic Tools for Harvesting Magic Resource in Curved Spacetime}, 
      author={Jiayue Yang and Dyuman Bhattacharya and Ming Zhang and Robert B. Mann},
      year={2025},
      eprint={2508.16466},
      archivePrefix={arXiv},
      primaryClass={quant-ph},
      url={https://arxiv.org/abs/2508.16466}, 
}

@article{Cepollaro_2024,
   title={Stabilizer entropy of quantum tetrahedra},
   volume={109},
   ISSN={2470-0029},
   url={http://dx.doi.org/10.1103/PhysRevD.109.126008},
   DOI={10.1103/physrevd.109.126008},
   number={12},
   journal={Physical Review D},
   publisher={American Physical Society (APS)},
   author={Cepollaro, Simone and Chirco, Goffredo and Cuffaro, Gianluca and Esposito, Gianluca and Hamma, Alioscia},
   year={2024},
   month=jun }

@misc{cao2025gravitationalbackreactionmagical,
      title={Gravitational back-reaction is magical}, 
      author={ChunJun Cao and Gong Cheng and Alioscia Hamma and Lorenzo Leone and William Munizzi and Savatore F. E. Oliviero},
      year={2025},
      eprint={2403.07056},
      archivePrefix={arXiv},
      primaryClass={hep-th},
      url={https://arxiv.org/abs/2403.07056}, 
}

@article{rz86-47h3,
  title = {Stabilizer entropy and entanglement complexity in the Sachdev-Ye-Kitaev model},
  author = {Jasser, Barbara and Odavi\ifmmode \acute{c}\else \'{c}\fi{}, Jovan and Hamma, Alioscia},
  journal = {Phys. Rev. B},
  volume = {112},
  issue = {17},
  pages = {174204},
  numpages = {16},
  year = {2025},
  month = {Nov},
  publisher = {American Physical Society},
  doi = {10.1103/rz86-47h3},
  url = {https://link.aps.org/doi/10.1103/rz86-47h3}
}

@article{4brj-cl26,
  title = {Harvesting stabilizer entropy and nonlocality from a quantum field},
  author = {Cepollaro, Simone and Cusumano, Stefano and Hamma, Alioscia and Lo Giudice, Giorgio and Odavi\ifmmode \acute{c}\else \'{c}\fi{}, Jovan},
  journal = {Phys. Rev. D},
  volume = {112},
  issue = {10},
  pages = {105012},
  numpages = {17},
  year = {2025},
  month = {Nov},
  publisher = {American Physical Society},
  doi = {10.1103/4brj-cl26},
  url = {https://link.aps.org/doi/10.1103/4brj-cl26}
}

@article{PhysRevA.106.062434,
  title = {Retrieving information from a black hole using quantum machine learning},
  author = {Leone, Lorenzo and Oliviero, Salvatore F. E. and Piemontese, Stefano and True, Sarah and Hamma, Alioscia},
  journal = {Phys. Rev. A},
  volume = {106},
  issue = {6},
  pages = {062434},
  numpages = {9},
  year = {2022},
  month = {Dec},
  publisher = {American Physical Society},
  doi = {10.1103/PhysRevA.106.062434},
  url = {https://link.aps.org/doi/10.1103/PhysRevA.106.062434}
}

@article{PhysRevLett.132.080402,
  title = {Unscrambling Quantum Information with Clifford Decoders},
  author = {Oliviero, Salvatore F. E. and Leone, Lorenzo and Lloyd, Seth and Hamma, Alioscia},
  journal = {Phys. Rev. Lett.},
  volume = {132},
  issue = {8},
  pages = {080402},
  numpages = {6},
  year = {2024},
  month = {Feb},
  publisher = {American Physical Society},
  doi = {10.1103/PhysRevLett.132.080402},
  url = {https://link.aps.org/doi/10.1103/PhysRevLett.132.080402}
}

@article{PhysRevA.109.022429,
  title = {Learning efficient decoders for quasichaotic quantum scramblers},
  author = {Leone, Lorenzo and Oliviero, Salvatore F. E. and Lloyd, Seth and Hamma, Alioscia},
  journal = {Phys. Rev. A},
  volume = {109},
  issue = {2},
  pages = {022429},
  numpages = {29},
  year = {2024},
  month = {Feb},
  publisher = {American Physical Society},
  doi = {10.1103/PhysRevA.109.022429},
  url = {https://link.aps.org/doi/10.1103/PhysRevA.109.022429}
}

@article{Leone2024learningtdoped,
  doi = {10.22331/q-2024-05-27-1361},
  url = {https://doi.org/10.22331/q-2024-05-27-1361},
  title = {Learning t-doped stabilizer states},
  author = {Leone, Lorenzo and Oliviero, Salvatore F. E. and Hamma, Alioscia},
  journal = {{Quantum}},
  issn = {2521-327X},
  publisher = {{Verein zur F{\"{o}}rderung des Open Access Publizierens in den Quantenwissenschaften}},
  volume = {8},
  pages = {1361},
  month = may,
  year = {2024}
}

@article{PhysRevA.54.4741,
  title = {Simple quantum error-correcting codes},
  author = {Steane, A. M.},
  journal = {Phys. Rev. A},
  volume = {54},
  issue = {6},
  pages = {4741--4751},
  numpages = {0},
  year = {1996},
  month = {Dec},
  publisher = {American Physical Society},
  doi = {10.1103/PhysRevA.54.4741},
  url = {https://link.aps.org/doi/10.1103/PhysRevA.54.4741}
}

@article{PhysRevLett.77.198,
  title = {Perfect Quantum Error Correcting Code},
  author = {Laflamme, Raymond and Miquel, Cesar and Paz, Juan Pablo and Zurek, Wojciech Hubert},
  journal = {Phys. Rev. Lett.},
  volume = {77},
  issue = {1},
  pages = {198--201},
  numpages = {0},
  year = {1996},
  month = {Jul},
  publisher = {American Physical Society},
  doi = {10.1103/PhysRevLett.77.198},
  url = {https://link.aps.org/doi/10.1103/PhysRevLett.77.198}
}

@article{steane_multiple-particle_1996,
	title = {Multiple-particle interference and quantum error correction},
	volume = {452},
	issn = {1364-5021},
	url = {https://doi.org/10.1098/rspa.1996.0136},
	doi = {10.1098/rspa.1996.0136},
	number = {1954},
	journal = {Proceedings of the Royal Society A: Mathematical, Physical and Engineering Sciences},
	author = {Steane, Andrew},
	month = nov,
	year = {1996},
	note = {\_eprint: https://royalsocietypublishing.org/rspa/article-pdf/452/1954/2551/998878/rspa.1996.0136.pdf},
	pages = {2551--2577},
}

@article{PhysRevA.52.R2493,
  title = {Scheme for reducing decoherence in quantum computer memory},
  author = {Shor, Peter W.},
  journal = {Phys. Rev. A},
  volume = {52},
  issue = {4},
  pages = {R2493--R2496},
  numpages = {0},
  year = {1995},
  month = {Oct},
  publisher = {American Physical Society},
  doi = {10.1103/PhysRevA.52.R2493},
  url = {https://link.aps.org/doi/10.1103/PhysRevA.52.R2493}
}

@article{PhysRevLett.79.953,
  title = {A Nonadditive Quantum Code},
  author = {Rains, Eric M. and Hardin, R. H. and Shor, Peter W. and Sloane, N. J. A.},
  journal = {Phys. Rev. Lett.},
  volume = {79},
  issue = {5},
  pages = {953--954},
  numpages = {0},
  year = {1997},
  month = {Aug},
  publisher = {American Physical Society},
  doi = {10.1103/PhysRevLett.79.953},
  url = {https://link.aps.org/doi/10.1103/PhysRevLett.79.953}
}

@INPROCEEDINGS{613213,
  author={Calderbank, A.R. and Rains, E.M. and Shor, P.W. and Sloane, N.J.A.},
  booktitle={Proceedings of IEEE International Symposium on Information Theory}, 
  title={Quantum error correction via codes over GF(4)}, 
  year={1997},
  volume={},
  number={},
  pages={292-},
  doi={10.1109/ISIT.1997.613213}}

@article{Gottesman1998FaultTolerant,
  title        = {Theory of Fault-Tolerant Quantum Computation},
  author       = {Gottesman, Daniel},
  journal      = {Physical Review A},
  volume       = {57},
  number       = {1},
  pages        = {127--137},
  year         = {1998},
  doi          = {10.1103/PhysRevA.57.127},
  url          = {https://doi.org/10.1103/PhysRevA.57.127},
  archivePrefix= {arXiv},
  eprint       = {quant-ph/9702029},
  primaryClass = {quant-ph}
}

@article{Gottesman1998_Heisenberg,
  author    = {Gottesman, Daniel},
  title     = {The Heisenberg representation of quantum computers},
  journal   = {arXiv preprint},
  year      = {1998},
  eprint    = {quant-ph/9807006},
  archivePrefix = {arXiv}
}

@article{VeitchMousavianGottesmanEmerson2014,
  author    = {Veitch, Victor and Mousavian, Seyed Ali Hamed and Gottesman, Daniel and Emerson, Joseph},
  title     = {The Resource Theory of Stabilizer Computation},
  journal   = {New Journal of Physics},
  volume    = {16},
  number    = {1},
  pages     = {013009},
  year      = {2014},
  doi       = {10.1088/1367-2630/16/1/013009},
  eprint    = {1307.7171},
  archivePrefix = {arXiv},
  primaryClass  = {quant-ph}
}

@article{DiVincenzoLeungTerhal2002,
  author    = {DiVincenzo, David P. and Leung, Debbie W. and Terhal, Barbara M.},
  title     = {Quantum data hiding},
  journal   = {IEEE Transactions on Information Theory},
  volume    = {48},
  number    = {3},
  pages     = {580--598},
  year      = {2002},
  doi       = {10.1109/18.985948}
}

@article{HangleiterBermejoVegaSchwarzEisert2018,
  author    = {Hangleiter, D. and Bermejo-Vega, J. and Schwarz, M. and Eisert, J.},
  title     = {Anticoncentration theorems for schemes showing a quantum speedup},
  journal   = {Quantum},
  volume    = {2},
  pages     = {65},
  year      = {2018},
  doi       = {10.22331/q-2018-05-22-65}
}

@article{PashayanBartlettGross2020,
  author    = {Pashayan, H. and Bartlett, S. D. and Gross, D.},
  title     = {From estimation of quantum probabilities to simulation of quantum circuits},
  journal   = {Quantum},
  volume    = {4},
  pages     = {223},
  year      = {2020},
  doi       = {10.22331/q-2020-01-13-223}
}

@article{BoulandFeffermanNirkheVazirani2019,
  author    = {Bouland, Adam and Fefferman, Bill and Nirkhe, Chinmay and Vazirani, Umesh},
  title     = {On the complexity and verification of quantum random circuit sampling},
  journal   = {Nature Physics},
  volume    = {15},
  pages     = {159--163},
  year      = {2019},
  doi       = {10.1038/s41567-018-0318-2}
}

@article{HuangKuengPreskill2020,
  author    = {Huang, H.-Y. and Kueng, R. and Preskill, J.},
  title     = {Predicting many properties of a quantum system from very few measurements},
  journal   = {Nature Physics},
  volume    = {16},
  pages     = {1050--1057},
  year      = {2020},
  doi       = {10.1038/s41567-020-0932-7}
}

@article{HosurQiRobertsYoshida2016,
  author    = {Hosur, P. and Qi, X.-L. and Roberts, D. A. and Yoshida, B.},
  title     = {Chaos in quantum channels},
  journal   = {Journal of High Energy Physics},
  volume    = {2016},
  number    = {2},
  pages     = {4},
  year      = {2016},
  doi       = {10.1007/JHEP02(2016)004}
}

@article{TouilDeffner2020,
  author    = {Touil, A. and Deffner, S.},
  title     = {Quantum scrambling and the growth of mutual information},
  journal   = {Quantum Science and Technology},
  volume    = {5},
  number    = {3},
  pages     = {035005},
  year      = {2020},
  doi       = {10.1088/2058-9565/ab8ebb}
}

@article{DingHaydenWalter2016,
  author    = {Ding, D. and Hayden, P. and Walter, M.},
  title     = {Conditional mutual information of bipartite unitaries and scrambling},
  journal   = {Journal of High Energy Physics},
  volume    = {2016},
  number    = {12},
  pages     = {145},
  year      = {2016},
  doi       = {10.1007/JHEP12(2016)145}
}

@article{WangGhoseSandersHu2004,
  author    = {Wang, X. and Ghose, S. and Sanders, B. C. and Hu, B.},
  title     = {Entanglement as a signature of quantum chaos},
  journal   = {Physical Review E},
  volume    = {70},
  pages     = {016217},
  year      = {2004},
  doi       = {10.1103/PhysRevE.70.016217}
}

@article{VidmarRigol2017,
  author    = {Vidmar, L. and Rigol, M.},
  title     = {Entanglement entropy of eigenstates of quantum chaotic Hamiltonians},
  journal   = {Physical Review Letters},
  volume    = {119},
  pages     = {220603},
  year      = {2017},
  doi       = {10.1103/PhysRevLett.119.220603}
}

@article{HaydenLeungWinter2006,
  author    = {Hayden, Patrick and Leung, Debbie W. and Winter, Andreas},
  title     = {Aspects of generic entanglement},
  journal   = {Communications in Mathematical Physics},
  volume    = {265},
  number    = {1},
  pages     = {95--117},
  year      = {2006},
  doi       = {10.1007/s00220-006-1535-6}
}

@article{PhysRev.80.580,
  title = {Spin Echoes},
  author = {Hahn, E. L.},
  journal = {Phys. Rev.},
  volume = {80},
  issue = {4},
  pages = {580--594},
  numpages = {0},
  year = {1950},
  month = {Nov},
  publisher = {American Physical Society},
  doi = {10.1103/PhysRev.80.580},
  url = {https://link.aps.org/doi/10.1103/PhysRev.80.580}
}

@article{Peres1984,
  author    = {Peres, A.},
  title     = {Stability of quantum motion in chaotic and regular systems},
  journal   = {Physical Review A},
  volume    = {30},
  pages     = {1610--1615},
  year      = {1984},
  doi       = {10.1103/PhysRevA.30.1610}
}

@article{CucchiettiPastawskiJalabert2004,
  author    = {Cucchietti, F. M. and Pastawski, H. M. and Jalabert, R. A.},
  title     = {Universality of the Lyapunov regime for the Loschmidt echo},
  journal   = {Physical Review B},
  volume    = {70},
  pages     = {035311},
  year      = {2004},
  doi       = {10.1103/PhysRevB.70.035311}
}

@article{Goussev2008,
  author    = {Goussev, A. and Waltner, D. and Richter, K. and Jalabert, R. A.},
  title     = {Loschmidt echo for local perturbations: non-monotonic cross-over from the Fermi-golden-rule to the escape-rate regime},
  journal   = {New Journal of Physics},
  volume    = {10},
  pages     = {093010},
  year      = {2008},
  doi       = {10.1088/1367-2630/10/9/093010}
}

@article{LarkinOvchinnikov1969,
  author    = {Larkin, A. I. and Ovchinnikov, Yu. N.},
  title     = {Quasiclassical Method in the Theory of Superconductivity},
  journal   = {Soviet Physics JETP},
  volume    = {28},
  number    = {6},
  pages     = {1200--1205},
  year      = {1969},
  note      = {Original: Zh. Eksp. Teor. Fiz. 55, 2262 (1968)}
}

@article{MaldacenaShenkerStanford2016,
  author    = {Maldacena, Juan and Shenker, Stephen H. and Stanford, Douglas},
  title     = {A bound on chaos},
  journal   = {Journal of High Energy Physics},
  year      = {2016},
  volume    = {2016},
  number    = {8},
  pages     = {106},
  doi       = {10.1007/JHEP08(2016)106}
}

@article{Srednicki1994,
  author    = {Srednicki, Mark},
  title     = {Chaos and quantum thermalization},
  journal   = {Physical Review E},
  volume    = {50},
  pages     = {888},
  year      = {1994},
  doi       = {10.1103/PhysRevE.50.888}
}

@article{ChotorlishviliUgulava2010,
  author    = {Chotorlishvili, L. and Ugulava, A.},
  title     = {Quantum chaos and its kinetic stage of evolution},
  journal   = {Physica D: Nonlinear Phenomena},
  volume    = {239},
  number    = {3--4},
  pages     = {103},
  year      = {2010},
  doi       = {10.1016/j.physd.2009.08.017}
}

@article{HaakeWiedemannZyczkowski1992,
  author    = {Haake, Fritz and Wiedemann, H. and Zyczkowski, K.},
  title     = {Lyapunov exponents from quantum dynamics},
  journal   = {Annalen der Physik},
  volume    = {504},
  number    = {7},
  pages     = {531},
  year      = {1992},
  doi       = {10.1002/andp.19925040706}
}

@article{MankoVilelaMendes2000,
  author    = {Man'ko, V. I. and Vilela Mendes, R.},
  title     = {Lyapunov exponent in quantum mechanics: A phase-space approach},
  journal   = {Physica D: Nonlinear Phenomena},
  volume    = {145},
  number    = {3--4},
  pages     = {330--348},
  year      = {2000},
  doi       = {10.1016/s0167-2789(00)00117-2}
}

@article{WeinsteinLloydTsallis2002,
  author    = {Weinstein, Y. S. and Lloyd, S. and Tsallis, C.},
  title     = {Border between regular and chaotic quantum dynamics},
  journal   = {Physical Review Letters},
  volume    = {89},
  pages     = {214101},
  year      = {2002},
  doi       = {10.1103/PhysRevLett.89.214101}
}

@misc{gu2024simulatingquantumchaoschaos,
      title={Simulating quantum chaos without chaos}, 
      author={Andi Gu and Yihui Quek and Susanne Yelin and Jens Eisert and Lorenzo Leone},
      year={2024},
      eprint={2410.18196},
      archivePrefix={arXiv},
      primaryClass={quant-ph},
      url={https://arxiv.org/abs/2410.18196}, 
}

@Article{e23081073,
AUTHOR = {Leone, Lorenzo and Oliviero, Salvatore F. E. and Hamma, Alioscia},
TITLE = {Isospectral Twirling and Quantum Chaos},
JOURNAL = {Entropy},
VOLUME = {23},
YEAR = {2021},
NUMBER = {8},
ARTICLE-NUMBER = {1073},
URL = {https://www.mdpi.com/1099-4300/23/8/1073},
PubMedID = {34441214},
ISSN = {1099-4300},
DOI = {10.3390/e23081073}
}

@book{Abdalla1994_2DGravity,
  author    = {Abdalla, E. and Abdalla, M. C. B. and Dalmazi, D. and Zadra, A.},
  title     = {2D-Gravity in Non-Critical Strings: Discrete and Continuum Approaches},
  series    = {Lecture Notes in Physics Monographs},
  volume    = {20},
  publisher = {Springer},
  address   = {Berlin, Heidelberg},
  year      = {1994},
}

@article{Verbaarschot1994PRL,
  author  = {Verbaarschot, J. J. M.},
  title   = {Spectrum of the QCD Dirac operator and chiral random matrix theory},
  journal = {Physical Review Letters},
  volume  = {72},
  pages   = {2531--2533},
  year    = {1994},
  doi     = {10.1103/PhysRevLett.72.2531}
}

@inproceedings{Kalkreuter1996NPBPS,
  author    = {Kalkreuter, T.},
  title     = {Study of Lanczos Methods for Wilson Fermions},
  booktitle = {Nuclear Physics B (Proceedings Supplements)},
  volume    = {49},
  pages     = {168},
  year      = {1996}
}

@article{Kalkreuter1995PRD,
  author  = {Kalkreuter, T.},
  title   = {Spectrum of the Dirac operator and multigrid algorithm with dynamical staggered fermions},
  journal = {Physical Review D},
  volume  = {51},
  pages   = {1305--1313},
  year    = {1995},
  doi     = {10.1103/PhysRevD.51.1305}
}

@article{Calogero1969_2191,
  author  = {Calogero, F.},
  title   = {Solution of a Three-Body Problem in One Dimension},
  journal = {Journal of Mathematical Physics},
  volume  = {10},
  number  = {12},
  pages   = {2191--2196},
  year    = {1969},
  doi     = {10.1063/1.1664820}
}

@article{Calogero1969_2197,
  author  = {Calogero, F.},
  title   = {Ground State of a One-Dimensional N-Body System},
  journal = {Journal of Mathematical Physics},
  volume  = {10},
  pages   = {2197--2200},
  year    = {1969},
  doi     = {10.1063/1.1664821}
}

@article{Sutherland1971_JMP246,
  author  = {Sutherland, B.},
  title   = {Quantum Many-Body Problem in One Dimension: Ground State},
  journal = {Journal of Mathematical Physics},
  volume  = {12},
  pages   = {246--250},
  year    = {1971},
  doi     = {10.1063/1.1665584}
}

@article{Sutherland1971_JMP251,
  author  = {Sutherland, B.},
  title   = {Quantum Many-Body Problem in One Dimension: Thermodynamics},
  journal = {Journal of Mathematical Physics},
  volume  = {12},
  pages   = {251--256},
  year    = {1971},
  doi     = {10.1063/1.1665585}
}

@article{Sutherland1971_PRA4,
  author  = {Sutherland, B.},
  title   = {Exact Results for a Quantum Many-Body Problem in One Dimension},
  journal = {Physical Review A},
  volume  = {4},
  pages   = {2019--2021},
  year    = {1971},
  doi     = {10.1103/PhysRevA.4.2019}
}

@article{Sutherland1972_PRA5,
  author  = {Sutherland, B.},
  title   = {Exact Results for a Quantum Many-Body Problem in One Dimension. II},
  journal = {Physical Review A},
  volume  = {5},
  pages   = {1372--1376},
  year    = {1972},
  doi     = {10.1103/PhysRevA.5.1372}
}

@article{SimonsLeeAltshuler1993_PRL,
  author  = {Simons, B. D. and Lee, P. A. and Altshuler, B. L.},
  title   = {Exact Description of Spectral Correlators by a Quantum One-Dimensional Model with Inverse-Square Interaction},
  journal = {Physical Review Letters},
  volume  = {70},
  number  = {26},
  pages   = {4122--4125},
  year    = {1993},
  doi     = {10.1103/PhysRevLett.70.4122}
}

@article{AltlandGefenMontambaux1996PRL,
  author  = {Altland, Alexander and Gefen, Yuval and Montambaux, Gilles},
  title   = {What is the Thouless Energy for Ballistic Systems?},
  journal = {Physical Review Letters},
  volume  = {76},
  number  = {7},
  pages   = {1130--1133},
  year    = {1996},
  doi     = {10.1103/PhysRevLett.76.1130}
}

@article{AltlandGefen1995PRB,
  author  = {Altland, Alexander and Gefen, Yuval},
  title   = {Spectral statistics of nondiffusive disordered electron systems: A comprehensive approach},
  journal = {Physical Review B},
  volume  = {51},
  number  = {16},
  pages   = {10671--10690},
  year    = {1995},
  doi     = {10.1103/PhysRevB.51.10671}
}

@article{AltlandGefen1993PRL,
  author  = {Altland, Alexander and Gefen, Yuval},
  title   = {Spectral statistics in nondiffusive regimes},
  journal = {Physical Review Letters},
  volume  = {71},
  number  = {20},
  pages   = {3339--3342},
  year    = {1993},
  doi     = {10.1103/PhysRevLett.71.3339}
}

@article{Efetov1983AdvPhys,
  author  = {Efetov, K. B.},
  title   = {Supersymmetry and Theory of Disordered Metals},
  journal = {Advances in Physics},
  volume  = {32},
  number  = {1},
  pages   = {53--127},
  year    = {1983},
  doi     = {10.1080/00018738300101531}
}

@article{GorkovEliashberg1965JETP,
  author  = {Gor'kov, L. P. and Eliashberg, G. M.},
  title   = {Minute Metallic Particles in an Electromagnetic Field},
  journal = {Soviet Physics JETP},
  volume  = {21},
  pages   = {940--947},
  year    = {1965}
}

@article{Pluhar1994PRL,
  title   = {Suppression of Weak Localization Due to Magnetic Flux in Few-Channel Ballistic Microstructures},
  author  = {Pluhař, Z. and Weidenmüller, H. A. and Zuk, J. A. and Lewenkopf, C. H.},
  journal = {Physical Review Letters},
  volume  = {73},
  pages   = {2115--2119},
  year    = {1994},
  doi     = {10.1103/PhysRevLett.73.2115}
}

@article{Pluhar1995AnnPhys,
  title   = {Crossover from Orthogonal to Unitary Symmetry for Ballistic Electron Transport in Chaotic Microstructures},
  author  = {Pluhar, Z. and Weidenm{\"u}ller, H. A. and Zuk, J. A. and Lewenkopf, C. H. and Wegner, F. J.},
  journal = {Annals of Physics},
  volume  = {243},
  pages   = {1--64},
  year    = {1995},
  doi     = {10.1006/aphy.1995.1089}
}

@article{Iida1990PRL,
  title    = {Wave propagation through disordered media and universal conductance fluctuations},
  author   = {Iida, S. and Weidenm{\"u}ller, H. A. and Zuk, J. A.},
 journal = {Physical Review Letters},
  volume   = {64},
  number   = {5},
  pages    = {583--586},
  doi      = {10.1103/PhysRevLett.64.583},
  url      = {https://link.aps.org/doi/10.1103/PhysRevLett.64.583},
  year     = {1990}
}

@article{Iida1990AnnPhys,
  title    = {Statistical scattering theory, the supersymmetry method and universal conductance fluctuations},
  author   = {Iida, S. and Weidenm{\"u}ller, H. A. and Zuk, J. A.},
  journal= {Annals of Physics},
  volume   = {200},
  number   = {2},
  pages    = {219--270},
  doi      = {10.1016/0003-4916(90)90275-S},
  year     = {1990}
}

@article{roberts_chaos_2017,
	title = {Chaos and complexity by design},
	volume = {2017},
	issn = {1029-8479},
	url = {https://doi.org/10.1007/JHEP04(2017)121},
	doi = {10.1007/JHEP04(2017)121},
	number = {4},
	journal = {Journal of High Energy Physics},
	author = {Roberts, Daniel A. and Yoshida, Beni},
	month = apr,
	year = {2017},
	pages = {121},
}

@article{KITAEV20062,
title = {Anyons in an exactly solved model and beyond},
journal = {Annals of Physics},
volume = {321},
number = {1},
pages = {2-111},
year = {2006},
note = {January Special Issue},
issn = {0003-4916},
doi = {https://doi.org/10.1016/j.aop.2005.10.005},
url = {https://www.sciencedirect.com/science/article/pii/S0003491605002381},
author = {Alexei Kitaev}
}

@Article{10.21468/SciPostPhys.10.3.076,
	title={{Random matrix theory of the isospectral twirling}},
	author={Salvatore F. E. Oliviero and Lorenzo Leone and Francesco Caravelli and Alioscia Hamma},
	journal={SciPost Phys.},
	volume={10},
	pages={076},
	year={2021},
	publisher={SciPost},
	doi={10.21468/SciPostPhys.10.3.076},
	url={https://scipost.org/10.21468/SciPostPhys.10.3.076},
}

@book{sagan_symmetric_2001,
	series = {Graduate {Texts} in {Mathematics}},
	title = {The {Symmetric} {Group}: {Representations}, {Combinatorial} {Algorithms}, and {Symmetric} {Functions}},
	isbn = {978-0-387-95067-9},
	url = {https://books.google.it/books?id=dmrnR48_x38C},
	publisher = {Springer New York},
	author = {Sagan, B.},
	year = {2001},
	lccn = {00040042},
}

@misc{westrich2011youngsnaturalrepresentationsmathcals4,
      title={Young's Natural Representations of $\mathcal{S}_4$}, 
      author={Quinton Westrich},
      year={2011},
      eprint={1112.0687},
      archivePrefix={arXiv},
      primaryClass={math.RT},
      url={https://arxiv.org/abs/1112.0687}, 
}

@misc{bittel2025completetheorycliffordcommutant,
      title={A complete theory of the Clifford commutant}, 
      author={Lennart Bittel and Jens Eisert and Lorenzo Leone and Antonio A. Mele and Salvatore F. E. Oliviero},
      year={2025},
      eprint={2504.12263},
      archivePrefix={arXiv},
      primaryClass={quant-ph},
      url={https://arxiv.org/abs/2504.12263}, 
}

@article{Leone2021quantumchaosis,
  doi = {10.22331/q-2021-05-04-453},
  url = {https://doi.org/10.22331/q-2021-05-04-453},
  title = {Quantum {C}haos is {Q}uantum},
  author = {Leone, Lorenzo and Oliviero, Salvatore F. E. and Zhou, You and Hamma, Alioscia},
  journal = {{Quantum}},
  issn = {2521-327X},
  publisher = {{Verein zur F{\"{o}}rderung des Open Access Publizierens in den Quantenwissenschaften}},
  volume = {5},
  pages = {453},
  month = may,
  year = {2021}
}

@article{PhysRevLett.121.170502,
  title = {Recovering Quantum Gates from Few Average Gate Fidelities},
  author = {Roth, I. and Kueng, R. and Kimmel, S. and Liu, Y.-K. and Gross, D. and Eisert, J. and Kliesch, M.},
  journal = {Phys. Rev. Lett.},
  volume = {121},
  issue = {17},
  pages = {170502},
  numpages = {8},
  year = {2018},
  month = {Oct},
  publisher = {American Physical Society},
  doi = {10.1103/PhysRevLett.121.170502},
  url = {https://link.aps.org/doi/10.1103/PhysRevLett.121.170502}
}

@misc{zhu2016cliffordgroupfailsgracefully,
      title={The Clifford group fails gracefully to be a unitary 4-design}, 
      author={Huangjun Zhu and Richard Kueng and Markus Grassl and David Gross},
      year={2016},
      eprint={1609.08172},
      archivePrefix={arXiv},
      primaryClass={quant-ph},
      url={https://arxiv.org/abs/1609.08172}, 
}

@book{bengtsson_geometry_2007,
	title = {Geometry of {Quantum} {States}: {An} {Introduction} to {Quantum} {Entanglement}},
	isbn = {978-1-139-45346-2},
	url = {https://books.google.it/books?id=aA4vXMbuOTUC},
	publisher = {Cambridge University Press},
	author = {Bengtsson, I. and Zyczkowski, K.},
	year = {2007},
}

@book{watrous_theory_2018,
	title = {The {Theory} of {Quantum} {Information}},
	isbn = {978-1-316-85312-2},
	url = {https://books.google.it/books?id=jQFWDwAAQBAJ},
	publisher = {Cambridge University Press},
	author = {Watrous, J.},
	year = {2018},
}

@article{Mele2024introductiontohaar,
  doi = {10.22331/q-2024-05-08-1340},
  url = {https://doi.org/10.22331/q-2024-05-08-1340},
  title = {Introduction to {H}aar {M}easure {T}ools in {Q}uantum {I}nformation: {A} {B}eginner's {T}utorial},
  author = {Mele, Antonio Anna},
  journal = {{Quantum}},
  issn = {2521-327X},
  publisher = {{Verein zur F{\"{o}}rderung des Open Access Publizierens in den Quantenwissenschaften}},
  volume = {8},
  pages = {1340},
  month = may,
  year = {2024}
}

@article {DeterministicNonperiodicFlow,
      author = "Edward N.  Lorenz",
      title = "Deterministic Nonperiodic Flow",
      journal = "Journal of Atmospheric Sciences",
      year = "1963",
      publisher = "American Meteorological Society",
      address = "Boston MA, USA",
      volume = "20",
      number = "2",
      doi = "10.1175/1520-0469(1963)020<0130:DNF>2.0.CO;2",
      pages=      "130 - 141",
      url = "https://journals.ametsoc.org/view/journals/atsc/20/2/1520-0469_1963_020_0130_dnf_2_0_co_2.xml"
}

@book{lyapunov_general_1992,
	series = {Control {Theory} and {Applications} {Series}},
	title = {General {Problem} of the {Stability} {Of} {Motion}},
	isbn = {978-0-7484-0062-1},
	url = {https://books.google.it/books?id=4tmAvU3_SCoC},
	publisher = {Taylor \& Francis},
	author = {Lyapunov, A.M.},
	year = {1992},
	lccn = {92032800},
}

@article{zyczkowski_eigenvector_1991,
	title = {Eigenvector statistics for the transitions from the orthogonal to the unitary ensemble},
	volume = {82},
	issn = {1431-584X},
	url = {https://doi.org/10.1007/BF01324340},
	doi = {10.1007/BF01324340},
	number = {2},
	journal = {Zeitschrift für Physik B Condensed Matter},
	author = {Życzkowski, Karol and Lenz, Georg},
	month = jun,
	year = {1991},
	pages = {299--303},
}

@article{Zyczkowski_1990,
doi = {10.1088/0305-4470/23/20/005},
url = {https://doi.org/10.1088/0305-4470/23/20/005},
year = {1990},
month = {oct},
publisher = {},
volume = {23},
number = {20},
pages = {4427},
author = {K Zyczkowski},
title = {Indicators of quantum chaos based on eigenvector statistics},
journal = {Journal of Physics A: Mathematical and General}
}

@article{PhysRevA.42.1013,
  title = {Random-matrix theory and eigenmodes of dynamical systems},
  author = {Haake, Fritz and \ifmmode \dot{Z}\else \.{Z}\fi{}yczkowski, Karol},
  journal = {Phys. Rev. A},
  volume = {42},
  issue = {2},
  pages = {1013--1016},
  numpages = {0},
  year = {1990},
  month = {Jul},
  publisher = {American Physical Society},
  doi = {10.1103/PhysRevA.42.1013},
  url = {https://link.aps.org/doi/10.1103/PhysRevA.42.1013}
}

@article{IZRAILEV1987250,
title = {Chaotic stucture of eigenfunctions in systems with maximal quantum chaos},
journal = {Physics Letters A},
volume = {125},
number = {5},
pages = {250-252},
year = {1987},
issn = {0375-9601},
doi = {https://doi.org/10.1016/0375-9601(87)90203-9},
url = {https://www.sciencedirect.com/science/article/pii/0375960187902039},
author = {F.M. Izrailev}
}

@book{haake_quantum_2001,
	series = {Physics and astronomy online library},
	title = {Quantum {Signatures} of {Chaos}},
	isbn = {978-3-540-67723-9},
	url = {https://books.google.it/books?id=Orv0BXoorFEC},
	publisher = {Springer},
	author = {Haake, F.},
	year = {2001},
	lccn = {92004599},
}

@book{mehta_random_1991,
	title = {Random {Matrices}},
	isbn = {978-0-12-488051-1},
	url = {https://books.google.it/books?id=-sloQgAACAAJ},
	publisher = {Academic Press},
	author = {Mehta, M.L.},
	year = {1991},
	lccn = {90000257},
}

@book{livan_introduction_2018,
	series = {{SpringerBriefs} in {Mathematical} {Physics}},
	title = {Introduction to {Random} {Matrices}: {Theory} and {Practice}},
	isbn = {978-3-319-70885-0},
	url = {https://books.google.it/books?id=-lJHDwAAQBAJ},
	publisher = {Springer International Publishing},
	author = {Livan, G. and Novaes, M. and Vivo, P.},
	year = {2018},
}

@article{dyson_statistical_1962_III,
	title = {Statistical {Theory} of the {Energy} {Levels} of {Complex} {Systems}. {III}},
	volume = {3},
	issn = {0022-2488},
	url = {https://doi.org/10.1063/1.1703775},
	doi = {10.1063/1.1703775},
	number = {1},
	journal = {Journal of Mathematical Physics},
	author = {Dyson, Freeman J.},
	month = jan,
	year = {1962},
	pages = {166--175}
}

@article{dyson_statistical_1962_II,
	title = {Statistical {Theory} of the {Energy} {Levels} of {Complex} {Systems}. {II}},
	volume = {3},
	issn = {0022-2488},
	url = {https://doi.org/10.1063/1.1703774},
	doi = {10.1063/1.1703774},
	number = {1},
	journal = {Journal of Mathematical Physics},
	author = {Dyson, Freeman J.},
	month = jan,
	year = {1962},
	pages = {157--165}
}

@article{dyson_statistical_1962,
	title = {Statistical {Theory} of the {Energy} {Levels} of {Complex} {Systems}. {I}},
	volume = {3},
	issn = {0022-2488},
	url = {https://doi.org/10.1063/1.1703773},
	doi = {10.1063/1.1703773},
	number = {1},
	journal = {Journal of Mathematical Physics},
	author = {Dyson, Freeman J.},
	month = jan,
	year = {1962},
	pages = {140--156}
}

@article{Wigner_1958,
 ISSN = {0003486X, 19398980},
 URL = {http://www.jstor.org/stable/1970008},
 author = {Eugene P. Wigner},
 journal = {Annals of Mathematics},
 number = {2},
 pages = {325--327},
 publisher = {[Annals of Mathematics, Trustees of Princeton University on Behalf of the Annals of Mathematics, Mathematics Department, Princeton University]},
 title = {On the Distribution of the Roots of Certain Symmetric Matrices},
 urldate = {2026-02-08},
 volume = {67},
 year = {1958}
}

@article{Wigner_1955_II,
 ISSN = {0003486X, 19398980},
 URL = {http://www.jstor.org/stable/1969956},
 author = {Eugene P. Wigner},
 journal = {Annals of Mathematics},
 number = {2},
 pages = {203--207},
 publisher = {[Annals of Mathematics, Trustees of Princeton University on Behalf of the Annals of Mathematics, Mathematics Department, Princeton University]},
 title = {Characteristics Vectors of Bordered Matrices with Infinite Dimensions II},
 urldate = {2026-02-08},
 volume = {65},
 year = {1957}
}

@article{Wigner_1955,
 ISSN = {0003486X, 19398980},
 URL = {http://www.jstor.org/stable/1970079},
 author = {Eugene P. Wigner},
 journal = {Annals of Mathematics},
 number = {3},
 pages = {548--564},
 publisher = {[Annals of Mathematics, Trustees of Princeton University on Behalf of the Annals of Mathematics, Mathematics Department, Princeton University]},
 title = {Characteristic Vectors of Bordered Matrices With Infinite Dimensions},
 urldate = {2026-02-08},
 volume = {62},
 year = {1955}
}

@article{Wigner_1951,
title={On the statistical distribution of the widths and spacings of nuclear resonance levels}, 
volume={47}, 
DOI={10.1017/S0305004100027237}, 
number={4}, 
journal={Mathematical Proceedings of the Cambridge Philosophical Society}, 
author={Wigner, Eugene P.}, 
year={1951}, 
pages={790–798}}

\appendix

\section{\label{app:permutations}Permutation operators properties}

The symmetric group $S_k$ is the group whose elements permute the order of $k$ elements of a set. For instance the group $S_2$ is the group of operations exchanging the order of two elements. As such, it contains only two elements, the identity $I$ and the swap $T_{(12)}$, as these are the only possible exchanges one can do over two elements. The symmetric group $S_k$ has cardinality equal to $k!$. Each permutation $\pi\in S_k$ exchange the ordering of some elements of set of $k$ elements. A faithful representation of permutations acting over a $d$ dimensional complex space is given by the set of $d$ dimensional orthogonal matrices. In the following, and in all main text, we will always indicate such operators as $T_\pi$, understanding the dimensionality of the representation. Because of the properties of the symmetric group, one has the following properties for the permutation operators:
\ba
T_{\pi^{-1}}=T_{\pi}^\dag\\
T_{\pi\sigma}=T_\pi T_\sigma
\ea

Let us now also explain briefly the cyclic notation of permutations. Let us explain this starting with the simplest example, the symmetric group $S_2$ and the swap operator $T_{(12)}$ corresponding to the permutation $(12)$. This notation tells us that we need to move the object in position 1 into position 2, and the object in position 2 into position 1. With this in mind, let us now move to a more complicated example. Let us consider the symmetric group $S_4$. Let us consider three permutations from this set, $(13), (123),(1234)$. Remember that this permutations act on a set of 4 objects. The first permutation corresponds to moving the object in position 1 into position 3, leaving 2 and 4 untouched. Permutation $(123)$  corresponds to moving the object 1 into position 2, the object 2 into position 3, and the object 3 into position 1, i.e. to cyclically permute them. Similarly, $(1234)$ indicates that we must move object 1 into position 2 and so on until we move the object 4 into position 1. More generally, the permutation belonging to $S_k$ indicated with $(i\dots j)(k\dots\ell)(m\dots n)$ stands for permutating cyclically all the indices $i\dots j$ and so on, leaving untouched the unmentioned indices.

Let us now explain how the permutation operators can be used to express traces over multiple copies of an Hilbert space. Let us start by showing the so-called swap trick:
\ba
\label{eq:swap_trick}
\Tr[AB]=\Tr[T_{(12)}A\ot B]
\ea
where $T_{(12)}$ is the permutation operator exchanging the two copies of the Hilbert space, namely the swap operator. It acts on state defined over the tensor product of two copies of an Hilbert space $\mathcal{H}^{\ot2}$ having orthonormal basis $\{\ket{i}\}$ as:
\ba
T_{(12)}\ket{i}\ot\ket{j}=\ket{j}\ot\ket{i}\Rightarrow
T_{(12)}=\sum_{ij}\dyad{ji}{ij}
\ea
Expanding the operators $A,B$ in terms of the basis $\{\ket{i}\}$, on the one hand one has:
\ba
\nonumber
\Tr[AB]&=&\Tr[\sum_{i,j,k,\ell}A_{ij}B_{k\ell}\dyad{i}{j}\dyad{k}{\ell}]\\
\nonumber
&=&\Tr[\sum_{i,j,\ell}A_{ij}B_{j\ell}\dyad{i}{\ell}]\\
&=&\sum_{i,j}A_{ij}B_{ji}.
\ea
On the other hand:
\ba
\nonumber
\Tr[T_{(12)}A\ot B]&=&\Tr[\sum_{m,n}\dyad{nm}{mn}\sum_{i,j,k,\ell}A_{ij}B_{k\ell}\dyad{ik}{j\ell}]\\
\nonumber
&=&\Tr[\sum_{m,n}\sum_{j,\ell}A_{mj}B_{n\ell}\dyad{nm}{j\ell}]\\
&=&\sum_{j,\ell}A_{\ell j}B_{j\ell}
\ea
from which~\eqref{eq:swap_trick} follows.
Exploiting
\ba
T_{(1 k\,k-1\,\dots2)}\ket{i_1}\ot\cdots\ot\ket{i_k}=\ket{i_2}\ot\cdots\ket{i_k}\ot\ket{i_1}
\ea
one can show the same way that:
\ba
\Tr[A_1\cdots A_k]=\Tr[T_{(1\,k\,k-1\dots\,2)}(A_1\ot\cdots\ot A_k)]
\ea

\section{\label{app:weingarten}Weingarten functions}

\subsection{Gram matrices computation\label{app:gram_matrices}}
In this section we are going to compute the Gram matrices and their (pseudo)inverse.

Let us start with the simplest Gram matrix, the one for the Haar averages. Its elements are defined as:
\ba
\Omega_{\pi\sigma}=\Tr[T_{\pi}T_\sigma]
\ea
Let us first notice that the product $T_\pi T_\sigma=T_{\pi\sigma}$, so that we effectively need to compute traces of single permutation operators. Let us do explicitly the calculation for the case of second and fourth moments, that is, when permutations are picked from the groups $S_2$ and $S_4$ respectively. The group $S_2$ comprises only two elements, the identity and the swap. Using the properties of permutation operators in App.~\ref{app:permutations}, we can compute:
\ba
\Tr[I]=\Tr[I\mathbb{I}_d^{\ot2}]=\Tr[\mathbb{I}_d]^2=d^2\\
\Tr[T_{(12)}]=\Tr[T_{(12)}\mathbb{I}_d^{\ot2}]=\Tr[\mathbb{I}_d^2]=d
\ea
where $\mathbb{I}_d$ is the identity operator on a $d$-dimensional vector space. We can finally write the corresponding Gram matrix and its inverse as:
\ba
\Omega=\begin{pmatrix}
d^2&d\\
d&d^2
\end{pmatrix},\quad\Omega^{-1}=(d^2-1)^{-1}\begin{pmatrix}
1&-d^{-1}\\
-d^{-1}&1
\end{pmatrix}
\ea
The Gram matrix for the 4-th Haar moment can be computed in a similar way. Indeed, the traces to compute for each conjugacy class of the permutations of $S_4$ are:
\ba
\Tr[I]&=&\Tr[I\mathbb{I}_d^{\ot4}]=\Tr[\mathbb{I}]^4=d^4\\
\Tr[T_{(ij)}]&=&\Tr[T_{(ij)}\mathbb{I}_d^{\ot4}]=\Tr[\mathbb{I}^2]\Tr[\mathbb{I}]^2=d^3\\
\Tr[T_{(ijk)}]&=&\Tr[T_{(ijk)}\mathbb{I}_d^{\ot4}]=\Tr[\mathbb{I}]\Tr[\mathbb{I}^3]=d^2\\
\Tr[T_{(ijk\ell)}]&=&\Tr[T_{(ijk\ell)}\mathbb{I}_d^{\ot4}]=\Tr[\mathbb{I}^4]=d\\
\Tr[T_{(ij)(k\ell)}]&=&\Tr[T_{(ij)(k\ell)}\mathbb{I}_d^{\ot4}]=\Tr[\mathbb{I}^2]^2=d^2
\ea
With these traces one can write the Gram matrix for $S_4$ and its inverse.

Let us now turn our attention to the generalized Gram matrices $\Omega^{\pm}$, whose matrix elements reads:
\ba
\Omega^+_{\pi\sigma}=\Tr[T_\pi T_\sigma Q],\qquad\Omega^-_{\pi\sigma}=\Tr[T_\pi T_\sigma Q^\perp]=\Omega_{\pi\sigma}-\Omega^+_{\pi\sigma}
\ea
Once again, we simply need to compute these traces for each conjugacy class, obtaining:
\ba
\Tr\left[IQ\right]&=&\frac{1}{d^2}\sum_P\Tr\left[IP^{\ot4}\right]=\frac{1}{d^2}\sum_P\Tr\left[P\right]^4=d^2\\
\Tr\left[T_{(ij)}Q\right]&=&\frac{1}{d^2}\sum_P\Tr\left[T_{(ij)}P^{\ot4}\right]=\frac{1}{d^2}\sum_P\Tr\left[P^2\right]\Tr\left[P\right]^2=d\\
\Tr\left[T_{(ij)(k\ell)}Q\right]&=&\frac{1}{d^2}\sum_P\Tr\left[T_{(ij)(k\ell)}P^{\ot4}\right]=\frac{1}{d^2}\sum_P\Tr\left[P^2\right]^2=d^2\\
\Tr\left[T_{(ijk)}Q\right]&=&\frac{1}{d^2}\sum_P\Tr\left[T_{(ijk)}P^{\ot4}\right]=\frac{1}{d^2}\sum_P\Tr\left[P\right]^2=1\\
\Tr\left[T_{(ijk\ell)}Q\right]&=&\frac{1}{d^2}\sum_P\Tr\left[T_{(ijk\ell)}P^{\ot4}\right]=\frac{1}{d^2}\sum_P\Tr\left[P^4\right]=d
\ea
To obtain these results one needs to use few basic properties of Pauli strings:
\ba
\Tr[P]=d\delta_{P,I}\\
P^2=I\;\;\forall P\\
\Tr[PP']=d\delta_{P,P'}
\ea
These traces allow one to compute the generalized Gram matrices and their inverse, obtaining the generalized Weingarten functions.

\subsection{Weingarten functions and group characters}
It is possible to compute the Weingarten functions also from the group characters of the symmetric group. In this section we briefly report some results in this regard.

The expression of the Weingarten functions $W_{\pi\sigma}$ in terms of the characters of the irreps of the symmetric  group $S_4$ is given by:
\ba
W_\pi=\sum_{\lambda}\frac{d_\lambda^2}{(4!)^2}\frac{\chi^\lambda(\pi)}{D_\lambda}
\ea
where the sum runs over all the irreducible representations $\lambda$ of $S_4$, $d_\lambda$ is the dimension of the corresponding irreps, $\chi^\lambda(\pi)$ are the characters and $D_\lambda$ is the $d$ dimensional projector onto the irrep $\lambda$:
\ba
D_\lambda=\Tr[\Pi_\lambda],\quad\Pi_\lambda=\frac{d_\lambda}{4!}\sum_{\pi\in S_4}\chi^\lambda(\pi)T_\pi
\ea
The elements of the permutation group $S_4$ can be divided into five conjugacy classes, while the group has five irreducible representations. In Tab.~\ref{table:s4characters} the characters of the irreps of $S_4$  summarized, while details on the computation can be found in appropriate references~\cite{westrich2011youngsnaturalrepresentationsmathcals4,sagan_symmetric_2001}.

\begin{table}[!ht]
\centering
\begin{tabular}{|c|c|c|c|c|c|c|}
\toprule
Conjugacy class/Irrep&$\chi^\lambda(I)$&$\chi^\lambda(T_{(ij)})$&$\chi^\lambda(T_{(ij)(k\ell)})$&$\chi^\lambda(T_{(ijk)})$&$\chi^\lambda(T_{ijk\ell})$&$d_\lambda$\\
\midrule
$\lambda_1$&$1$&$-1$&$1$&$1$&$-1$&$1$\\
$\lambda_2$&$3$&$-1$&$-1$&$0$&$1$&$3$\\
$\lambda_3$&$2$&$0$&$2$&$-1$&$0$&$2$\\
$\lambda_4$&$3$&$1$&$-1$&$0$&$-1$&$3$\\
$\lambda_5$&$1$&$1$&$1$&$1$&$1$&$1$\\
\bottomrule
\end{tabular}
\caption{Character and dimension of each irrep for each conjugacy class of the symmetric group $S_4$.}
\label{table:s4characters}
\end{table}

Also the generalized Weingarten functions $W_\pi^\pm$ can be expressed in terms of the characters of $S_4$. Their expression is:
\ba
W_\pi^\pm=\sum_\lambda\frac{d_\lambda^2}{(4!)^2}\frac{\chi^\lambda(\pi)}{D^\pm_\lambda}
\ea
where the $D^\pm_\lambda$ are defined as:
\ba
D^+_\lambda=\Tr\left[Q\Pi_\lambda\right]\qquad D_\lambda^-=\Tr\left[Q^\perp \Pi_\lambda\right]
\ea

Then we need to compute the values of $D_\lambda$, $D_\lambda^+$ and $D_\lambda=D_\lambda-D_\lambda^+$. We can start by writing the expressions for the projectors onto the irreps $\Pi_\lambda$:
\ba
\Pi_{\lambda_1}&=&\frac{1}{24}\left\{I-\sum_{i,j}T_{(ij)}+\sum_{i,j,k,\ell}T_{(ij)(k\ell)}+\sum_{i,j,k}T_{(ijk)}-\sum_{i,j,k,\ell}T_{(ijk\ell)}\right\}\\
\Pi_{\lambda_2}&=&\frac{1}{8}\left\{3I-\sum_{i,j}T_{(ij)}-\sum_{i,j,k,\ell}T_{(ij)(k\ell)}+\sum_{i,j,k,\ell}T_{(ijk\ell)}\right\}\\
\Pi_{\lambda_3}&=&\frac{1}{12}\left\{2I+2\sum_{i,j,k,\ell}T_{(ij)(k\ell)}-\sum_{i,j,k}T_{(ijk)}\right\}\\
\Pi_{\lambda_4}&=&\frac{1}{8}\left\{3I+\sum_{i,j}T_{(ij)}-\sum_{i,j,k,\ell}T_{(ij)(k\ell)}-\sum_{i,j,k,\ell}T_{(ijk\ell)}\right\}\\
\Pi_{\lambda_5}&=&\frac{1}{24}\left\{I+\sum_{i,j}T_{(ij)}+\sum_{i,j,k,\ell}T_{(ij)(k\ell)}+\sum_{i,j,k}T_{(ijk)}+\sum_{i,j,k,\ell}T_{(ijk\ell)}\right\}=\Pi_{\rm sym}\\
\ea

From these, one can easily compute $D_\lambda$ and $D^\pm_\lambda$, as the computation of the traces involved is the same as the one needed for the Gram matrices.
\begin{table}[!ht]
\centering
\begin{tabular}{|c|c|c|c|}
\toprule
Irrep&$D_\lambda^+$&$D_\lambda^-$&$D_\lambda$\\
\midrule
$\lambda_1$&$\frac{(d-2)(d-1)}{6}$&$\frac{(d-4)(d-2)(d-1)(d+1)}{24}$&$\frac{d(d-1)(d-2)(d-3)}{24}$\\
$\lambda_2$&$0$&$\frac{(d-2)(d-1)d(d+1)}{8}$&$\frac{d(d+1)(d-2)(d-1)}{8}$\\
$\lambda_3$&$\frac{(d+1)(d-1)}{3}$&$\frac{(d-2)(d-1)(d+1)(d+2)}{12}$&$\frac{d^2(d-1)(d+1)}{12}$\\
$\lambda_4$&$0$&$\frac{(d-1)d(d+1)(d+2)}{8}$&$\frac{d(d+1)(d+2)(d-1)}{8}$\\
$\lambda_5$&$\frac{(d+2)(d+1)}{6}$&$\frac{(d-1)(d+1)(d+2)(d+4)}{24}$&$\frac{d(d+1)(d+2)(d+3)}{24}$\\
\bottomrule
\end{tabular}
\caption{Values of $D_\lambda^\pm$ and  $D_\lambda$ for each irrep.}
\label{table:dlambda}
\end{table}

The values of all the $D_\lambda$ are reported in Tab.~\ref{table:dlambda}. Using these values it is then possible to compute the generalized Weingarten functions. Let us start from the $W^+_\pi$, which can be rewritten as:
\ba
W^+_\pi=\sum_{\lambda_1,\lambda_5}\frac{d_\lambda^2}{(4!)^2}\frac{\chi^\lambda(\pi)}{D_\lambda^+}=\frac{1}{(4!)^2}\frac{2}{d(d+1)(d-1)}\left[d(\chi^{\lambda_1}(\pi)+\chi^{\lambda_5}(\pi))+(\chi^{\lambda_1}(\pi)-\chi^{\lambda_5}(\pi))\right]
\ea
Using this formula it is easy to compute the coefficients $W^+_\pi$. With the same procedure, one can also write an equivalent (though not as short and simple) formula for the $W^-_\pi$:
\ba
\nonumber
W_\pi^-=&&\frac{1}{4!}\left[d^2(d-1)(d+1)(d+2)(d-2)(d+6)(d-6)\right]^{-1}\\
\nonumber
\Bigg[&&d(d-2)(d+2)(d+6)\chi^{\lambda_1}(\pi)+(d+1)(d+2)(d-6)(d+6)\chi^{\lambda_2}(\pi)\\
\nonumber
&&+(d-2)(d+2)(d-6)(d+6)\chi^{\lambda_3}(\pi)+d(d-2)(d-6)(d+6)\chi^{\lambda_4}(\pi)\\
&&+d(d-2)(d+2)(d-6)\chi^{\lambda_5}(\pi)\Bigg]
\ea
Once again, it suffices to plug the values of the characters into the formula and get the $W^-_\pi$ reported in Tab.~\ref{table:weingarten}

\begin{table}[!ht]
\centering
\begin{tabular}{|c|c|c|}
\toprule
Conjugacy class&$W_\pi^+$&$W_\pi^-$\\
\midrule
$I$&$\frac{1}{4!}\frac{1}{(d+2)(d-2)}$&$\frac{1}{4!}\frac{3(3d+10)}{(d-1)(d+1)(d-2)(d+2)(d-4)}$\\
$T_{(ij)}$&$-\frac{1}{4!}\frac{3}{2(d+1)(d-1)(d+2)(d-2)}$&$\frac{1}{4!}\frac{d-8}{d(d-1)(d+1)(d-2)(d+4)}$\\
$T_{(ij)(k\ell)}$&$\frac{1}{4!}\frac{1}{(d+2)(d-2)}$&$\frac{1}{4!}\frac{1}{(d-1)(d+1)(d+2)(d+4)}$\\
$T_{(ijk)}$&$\frac{1}{4!}\frac{d^2+8}{(d+1)(d-1)(d+2)(d-2)}$&$-\frac{1}{4!}\frac{6}{(d-1)(d+1)(d+2)(d-2)(d+4)}$\\
$T_{(ijk\ell)}$&$-\frac{1}{4!}\frac{3d}{2d(d+1)(d-1)(d+2)(d-2)}$&$\frac{1}{4!}\frac{d^2+2d+16}{d(d-1)(d+1)(d-2)(d+2)(d+4)}$\\
\bottomrule
\end{tabular}
\caption{Generalized Weingarten functions $W_\pi^\pm$.}
\label{table:weingarten}
\end{table}

\subsection{Example: computation of the Haar second moment of a unitary operator \texorpdfstring{$V$}{V}\label{app:second_moment}}
Let us then consider as working example the calculation of the isospectral twirling of a unitary operator of the form $V^{\ot k,k}$ for $k=1$, i.e. the calculation of $\hat{\cal{R}}^{(2)}(V^{\ot1,1})$. One has that the symmetric group of order 2 has only two elements, $S_2=\{I,T_{(12)}\}$ where $I$ is the identity permutation and $T_{(12)}$ is the swap operator. Thus, to compute $\hat{\mathcal{R}}_{\cal U}^{(2)}(V^{\ot1,1})$ we need to put together the Gram matrix computed in App.~\ref{app:gram_matrices}, and the components of the vector $\vec{c}$, which have already been shown in Sec.~\ref{sec:probes_and_isospectral} to be proportional to the two point spectral form factor $g_2(t)$. Let us start by computing the inverse of the Gram matrix:
\ba
\Omega=\begin{pmatrix}
d^2&d\\
d&d^2
\end{pmatrix}\Rightarrow\Omega^{-1}=(d^2-1)^{-1}\begin{pmatrix}
1&-d^{-1}\\
-d^{-1}&1
\end{pmatrix}
\ea
We then need to compute the corresponding traces for the operator $V^{\ot1,1}$. This has already been done in Sec.~\ref{sec:probes_and_isospectral}, the result being:
\ba
\Tr\left[IU^{\ot1,1}\right]&=&g_2(t)\\
\Tr\left[T_{(12)}U^{\ot1,1}\right]&=&=d
\ea
where
\ba
g_2(t)=\sum_{j,k=1}^de^{-i(E_j-E_k)t}
\ea
is the two point spectral form factor. Inserting all these results into the expression for $\hat{\cal{R}}^{(2)}(V^{\ot1,1})$ one obtains:
\ba
\nonumber
\hat{\cal{R}}^{(2)}(V^{\ot1,1})&=&\vec{c}^T\Omega^{-1}\vec{T}=(d^2-1)^{-1}\begin{pmatrix}
d&g_2(t)
\end{pmatrix}\begin{pmatrix}
1&-d^{-1}\\
-d^{-1}&1
\end{pmatrix}\begin{pmatrix}
I\\
T_{(12)}
\end{pmatrix}\\
\nonumber
&=&\frac{g_2(t)}{d^2-1}I-\frac{g_2(t)}{d(d^2-1)}T_{(12)}-\frac{1}{(d^2-1)}I+\frac{d}{(d^2-1)}T_{(12)}\\
&=&\frac{g_2(t)-1}{d^2-1}I+\frac{d^2-g_2(t)}{d(d^2-1)}T_{12}
\ea

\section{\label{app:k_doped_average}Computation of T-doped averages}

\subsection{The \texorpdfstring{$\Xi$}{} Matrix}
\subsubsection{Computation of the \texorpdfstring{$\Xi_{\pi\sigma}$}{}}
In order to compute the matrix elements $\Xi_{\pi\sigma}$ we start by rewriting the expression as:
\ba
\nonumber
\Xi_{\pi\sigma}&=&\sum_{\tau\in S_4}W_{\pi\tau}^+\Tr\left[T_\sigma \Theta^{\ot4}Q\Theta^{\dag\ot4}QT_\tau\right]-W_{\pi\tau}^-\Tr\left[T_\sigma \Theta^{\ot4}Q^\perp \Theta^{\dag\ot4}QT_\tau\right]\\
\nonumber
&=&\sum_{\tau\in S_4}W_{\pi\tau}^+\Tr\left[T_\sigma \Theta^{\ot4}Q\Theta^{\dag\ot4}QT_\tau\right]-W_{\pi\tau}^-\Tr\left[T_\sigma \Theta^{\ot4}\left(1-Q\right) \Theta^{\dag\ot4}QT_\tau\right]\\
&=&\sum_{\tau\in S_4}\left(W_{\pi\tau}^++W_{\pi\tau}^-\right)\Tr\left[T_\sigma \Theta^{\ot4}Q\Theta^{\dag\ot4}QT_\tau\right]-W_{\pi\tau}^-\Tr\left[T_\sigma QT_\tau\right]
\ea
Let us then define the two matrices $K^{(1)}_{\tau\sigma}$ and $K_{\tau\sigma}^{(2)}$:
\ba
K^{(1)}_{\tau\sigma}&=&\Tr\left[T_\sigma QT_\tau\right]\\
K^{(2)}_{\tau\sigma}&=&\Tr\left[T_\sigma \Theta^{\ot4}Q\Theta^{\dag\ot4}QT_\tau\right]
\ea
One notices immediately that the matrix elements $K_{\tau\sigma}^{(1)}$ are just the matrix elements of the generalized Gram matrix $\Omega^+$. 

We are left to evaluate the argument of the trace in the elements of $\Theta_{\tau\sigma}^{(2)}$. The calculation can be performed essentially the same way as done in Sec.B4 of~\cite{Leone2021quantumchaosis}, and is reported here for completeness. As already mentioned, the operator $\Theta$ acts non trivially on a single-qubit with a $\Theta$ gate. This qubit can be assumed to be the first without loss of generality.
One can then compute the action of this gate on Pauli operators, obtaining:
\ba
\Theta X\Theta^\dag&=&\cos\theta X+\sin\theta Y\\
\Theta Y\Theta^\dag&=&\cos\theta Y-\sin\theta X\\
\Theta Z\Theta^\dag&=&Z\\
\Theta \mathbb{I}\Theta^\dag&=&\mathbb{I}
\ea
Consider then the following permutation operator $R\in S_{4N}$ acting on a ket vector as:
\ba
R(\ket{i_1i_2\cdots i_N})^{\ot4}=\ket{i_1}^{\ot4}\otimes\ket{i_2\cdots i_N}^{\ot4}.
\ea
The effect of the permutation operator $R$ is to reorder the elements of the 4-fold tensor product such that the elements corresponding to the 4 copies of the first qubit all stand at the beginning of the string, leaving the others unchanged.
Then we need to check the effect of the operator $R$ on the operators $\Theta^{\ot4}$, $Q$ and $T_{\tau\sigma}$.
The effect on the theta gate is immediately seen to be:
\ba
\Theta_R=R\Theta^{\ot4}R^{-1}=\Theta_1^{\ot4}\otimes\left(\otimes_{i=2}^N\mathbb{I}_i\right)^{\ot4}.
\ea
The effect of $R$ on a permutation operator has been shown to be:
\ba
T_{\sigma}^R=RT_{\sigma}R^{-1}=T_\sigma^{(2)}\otimes T_{\sigma}^{(d/2)}
\ea
where $T_\sigma^{(2)}$ and $T_\sigma^{(d/2)}$ are permutation operators acting on the copies of the first qubit and on the remaining $N-1$ qubit respectively.

Finally, the effect of $R$ on the projector $Q$ is:
\ba
\nonumber
Q_R=RQR^{-1}&=&\frac{1}{d^2}\sum_{P\in\mathbb{P}_N}RPR^{-1}=\frac{1}{4}\frac{1}{(d/2)^2}\sum_{P\in\mathbb{P}_{N-1}}\left(\mathbb{I}^{\ot4}+X^{\ot4}+Y^{\ot4}+Z^{\ot4}\right)\otimes P_{(d/2)}\\
\nonumber
&=&\frac{\left(\mathbb{I}^{\ot4}+X^{\ot4}+Y^{\ot4}+Z^{\ot4}\right)}{4}\ot\frac{1}{(d/2)^2}\sum_{P\in\mathbb{P}_{N-1}}=\frac{\left(\mathbb{I}^{\ot4}+X^{\ot4}+Y^{\ot4}+Z^{\ot4}\right)}{4}\ot Q_{(d/2)}\\
&=&\frac{Q_\mathbb{I}+Q_X+Q_Y+Q_Z}{4}
\ea
where we have defined:
\ba
Q_{(d/2)}=\frac{1}{(d/2)^2}\sum_{P\in\mathbb{P}_{N-1}}P_{(d/2)},\qquad Q_A=A^{\ot4}\otimes Q_{(d/2)}
\ea

We now have all the ingredients to manipulate the expression of $K_{\tau\sigma}^{(2)}$:
\ba
\nonumber
K_{\tau\sigma}^{(2)}=&&\Tr\left[T_\sigma \Theta^{\ot4}Q\Theta^{\dag\ot4}QT_\tau\right]=\tr{RT_{\tau\sigma}R^{-1}R\Theta^{\ot4}R^{-1}RQR^{-1}R\Theta^{\dag\ot4}R^{-1}RQR^{-1}}\\
\nonumber
&&=\frac{1}{16}\tr{T_{\tau\sigma}^R\Theta_R(Q_\mathbb{I}+Q_X+Q_Y+Q_Z)\Theta_R^\dag(Q_\mathbb{I}+Q_X+Q_Y+Q_Z)}\\
\nonumber
&&=\frac{1}{16}\left[\tr{T_{\tau\sigma}^R(Q_\mathbb{I}+Q_Z)(Q_\mathbb{I}+Q_X+Q_Y+Q_Z)}+\tr{T_{\tau\sigma}^R\Theta_R(Q_X+Q_Y)\Theta_R^\dag(Q_\mathbb{I}+Q_X+Q_Y+Q_Z)}\right]\\
\nonumber
&&=\frac{1}{2}\tr{T_{\tau\sigma} Q}+\frac{1}{16}\tr{T_{\tau\sigma}^R(Q_{\Theta X\Theta^\dag}+Q_{\Theta Y\Theta^\dag})(Q_\mathbb{I}+Q_X+Q_Y+Q_Z)}\\
\nonumber
&&=\frac{1}{2}\tr{T_{\tau\sigma} Q}+\frac{1}{8}\tr{T_{\tau\sigma}^R(Q_{\Theta X\Theta^\dag}+Q_{\Theta Y\Theta^\dag}+Q_{\Theta X\Theta^\dag X}+Q_{\Theta X\Theta^\dag Y})}\\
&&=\frac{1}{2}\tr{T_{\tau\sigma} Q}+\frac{1}{8}\tr{T_{\tau\sigma}^R\tilde{Q}_R}
\ea
where we have defined:
\ba
\tilde{Q}_R=Q_{\Theta X\Theta^\dag}+Q_{\Theta Y\Theta^\dag}+Q_{\Theta X\Theta^\dag X}+Q_{\Theta X\Theta^\dag Y}.
\ea
Let us then show shortly how to compute the second term, considering for instance the summand proportional to $Q_{\Theta X\Theta^\dag}$. As the two permutations are factorized, and as so are also the operators, one can simply compute each trace separately and then multiply, i.e.:
\ba
\tr{T_{\tau\sigma}^RQ_{\Theta X\Theta^\dag}}=\tr{T_{\tau\sigma}^{(2)}(\Theta X\Theta^\dag)^{\ot4}}\tr{T_{\tau\sigma}^{(d/2)}Q_{(d/2)}}
\ea
Consider for instance the case $\tau\sigma=I$, then:
\ba
\tr{I(\Theta X\Theta^\dag)^{\ot4}}=\tr{\Theta X\Theta^\dag}^4=\tr{\cos\theta X+\sin\theta Y}^4=0
\ea
In a similar fashion:
\ba
&&\tr{I(\Theta Y\Theta^\dag)^{\ot4}}=0\\
&&\tr{I(\Theta X\Theta^\dag X)^{\ot4}}=\tr{(\cos\theta X+\sin\theta Y)X}^4=\tr{\cos\theta\mathbb{I}}^4=16\cos^4\theta\\
&&\tr{I(\Theta X\Theta^\dag Y)^{\ot4}}=\tr{(\cos\theta X+\sin\theta Y)Y}^4=\tr{\sin\theta\mathbb{I}}^4=16\sin^4\theta,
\ea
so that:
\ba
\tr{I\tilde{Q}}=16(\cos^4\theta+\sin^4\theta)\tr{Q_{(d/2)}}=4(\cos^4\theta+\sin^4\theta)d^2.
\ea
The other permutations can be computed in a similar fashion, and the results are reported in Table~\ref{table:theta2}.
\begin{table}[!ht]
\begin{tabular}{|c|c|c|c|}
\toprule
$T_{\tau\sigma}$&$\tr{T_{\tau\sigma}^{(2)}\sum f(   \Theta)}$&$\tr{T_{\tau\sigma}^R\tilde{Q}_R}$&Multiplicity\\
\midrule
$I$&$16(\cos^4\theta+\sin^4\theta)$&$4(\cos^4\theta+\sin^4\theta)d^2$&$1$\\
$T_{(ij)}$&$8\cos^22\theta$&$4\cos^22\theta d$&$6$\\
$T_{(ij)(kl)}$&$4(3+\cos4\theta)$&$(3+\cos4\theta)d^2$&$3$\\
$T_{(ijk)}$&$4\cos4\theta$&$4\cos4\theta$&$8$\\
$T_{(ijkl)}$&$8\cos^22\theta$&$4\cos^22\theta d$&$6$\\
\bottomrule
\end{tabular}
\caption{Values of $\tr{T_{\tau\sigma}^R\tilde{Q}_R}$ for all the permutations}
\label{table:theta2}
\end{table}

\subsubsection{Diagonalization of \texorpdfstring{$\Xi$}{}}
As stated in the main text, the matrix $\Xi$ can be diagonalized. The results of the diagonalization are the eigenvalues and associated eigenvectors of matrix $\Xi$. These can simplify a lot all computations, since only 6 eigenvalues of $\Xi$ are non null. These are:
\ba
&&\xi_+=\frac{(7+\cos4\theta)d^2+3d(1-\cos4\theta)-8}{8(d^2-1)}\\
&&\xi_-=\frac{(7+\cos4\theta)d^2-3d(1-\cos4\theta)-8}{8(d^2-1)}\\
&&\xi_1=\xi_2=\xi_3=\xi_4=\frac{\xi_++\xi_-}{2}=\frac{(7+\cos4\theta)d^2-8}{8(d^2-1)}
\ea
The eigenvectors of $\Xi$ are given by the rows of the matrix $V^{T}$ diagonalizing the matrix. Here we report the corresponding combinations once the matrix $V^{T}$ is applied to either $\vec{T}$ or $\vec{t}$. The elements of $V^{T}\vec{T}$ associated with non null eigenvalues are:
\ba
&&T_+=\frac{1}{2\sqrt{6}}\left[I-\sum_{i,j}T_{(ij)}+\sum_{i,j,k}T_{(ijk)}-\sum_{i,j,k,\ell}T_{(ijk\ell)}+\sum_{i,j,k,\ell}T_{(ij)(k\ell)}\right]=\frac{4!}{2\sqrt{6}}\Pi_{\lambda_1}\\
&&T_-=\frac{4!}{2\sqrt{6}}\Pi_{\lambda_5}=\frac{4!}{2\sqrt{6}}\Pi_{\rm sym}\\
&&T_1=\frac{1}{2\sqrt{2}}\left[I-T_{(124)}-T_{(132)}-T_{(143)}-T_{(234)}+\sum_{i,j,k,\ell}T_{(ij)(k\ell)}\right]\\
&&T_2=\frac{1}{2\sqrt{2}}\left[T_{(13)}+T_{(24)}-T_{(14)}-T_{(23)}-T_{(1342)}-T_{(1243)}+T_{(1234)}+T_{(1432)}\right]\\
\nonumber
&&T_3=\frac{1}{2\sqrt{6}}\Big[2(T_{(12)}+T_{(34)})-(T_{(13)}+T_{(14)}+T_{(23)}+T_{(24)})+2(T_{(1423)}+T_{(1324)})\\
&&\qquad\qquad\qquad-(T_{(1234)}+T_{(1243)}+T_{(1342)}+T_{(1432)})\Big]\\
\nonumber
&&T_4=-\frac{1}{2\sqrt{6}}\Big[I+(T_{(124)}+T_{(132)}+T_{(143)}+T_{(234)})-2(T_{(123)}+T_{(134)}+T_{(142)}+T_{(243)})\\
&&\qquad\qquad\qquad\qquad+(T_{(12)(34)}+T_{(13)(24)}+T_{(14)(23)})\Big]
\ea
One can compute the same way the corresponding components of the vector $\vec{v}=V^{T}\vec{t}$, expressed in terms of the spectral form factors:
\ba
&&v_+=\frac{
g_{3}(t)-g_{4}(t)+g_{3}^{*}(t)
-g_{2}(2t)(-2+d)
+4g_{2}(t)(-2+d)
+(\tilde g_{3}-2d)(-3+d)d}
{2\sqrt{6}\,(-4+d)(-2+d)(-1+d)(1+d)}\\
&&v_-=\frac{
-g_{3}(t)-g_{4}(t)-g_{3}^{*}(t)
+g_{2}(2t)(2+d)
-4g_{2}(t)(2+d)
+d(3+d)\bigl(\tilde g_{3}+2d\bigr)}
{2\sqrt{6}\,(-1+d)(1+d)(2+d)(4+d)}\\
&&v_1=-\frac{g_4(t)+g_2(2t)-4g_2(t)-\bigl(\tilde{g}_3-3\bigr)d^{2}}
      {2\sqrt{2}\bigl(d^4-5d^2+4\bigr)}\\
&&v_2=\frac{d^3-dg_2(2t)}
     {4\sqrt{2}\bigl(d^4-5d^2+4\bigr)}\\
&&v_3=\frac{-4\bigl(g_3(t)+g_3^*(t)\bigr)
      +\bigl(g_{2}(2t)+8\,g_{2}(t)\bigr)d
      -d^{3}}
     {4\sqrt{6}\,\bigl(d^4-5d^2+4\bigr)}\\
\nonumber
&&v_4=\frac{g_2(2t)-4\,g_2(t)+g_4(t)
      -\bigl(-3+\tilde g_3\bigr)d^2}
     {2\sqrt{6}\,\bigl(d^4-5d^2+4\bigr)}
\ea
With all of this in mind one can finally write:
\ba
\vec{t}^T\Xi^kQ\vec{T}&=&\left[\xi_+^kv_+T_++\xi_-^kv_-T_-+\xi_4^k\sum_{i=1}^4v_iT_i\right]Q
\ea

\subsection{The Matrix \texorpdfstring{$\Gamma^{(k)}$}{}}

We remind that the matrix $\Gamma_{\pi\sigma}^{(k)}$ is defined as:
\ba
\Gamma_{\pi\sigma}^{(k)}=\sum_{\tau\in S_4}\Lambda_{\pi\tau}\sum_{i=0}^{k-1}(\Xi^i)_{\tau\sigma}
\ea

The matrix $\Lambda$ is defined as:
\ba
\Lambda_{\pi\tau}&=&\sum_{\sigma\in S_4}W^-_{\pi\sigma}\Tr\left[T_\tau \Theta^{\ot4}Q\Theta^{\dag\ot4}Q^\perp T_\sigma\right]
\ea
Let us rewrite the expression as:
\ba
&&\sum_{\sigma\in S_4}W^-_{\pi\sigma}\Tr\left[T_\tau \Theta^{\ot4}Q\Theta^{\dag\ot4}Q^\perp T_\sigma\right]=\sum_{\sigma\in S_4}W^-_{\pi\sigma}\Tr\left[T_\tau \Theta^{\ot4}Q\Theta^{\dag\ot4}(1-Q) T_\sigma\right]\\
&&=\sum_{\sigma\in S_4}W^-_{\pi\sigma}\Tr\left[T_\tau \Theta^{\ot4}Q\Theta^{\dag\ot4}T_\sigma\right]-W^-_{\pi\sigma}\Tr\left[T_\tau \Theta^{\ot4}Q\Theta^{\dag\ot4}Q T_\sigma\right]\\
&&=\sum_{\sigma\in S_4}W^-_{\pi\sigma}\Tr\left[T_\tau QT_\sigma\right]-W^-_{\pi\sigma}\Tr\left[T_\tau \Theta^{\ot4}Q\Theta^{\dag\ot4}Q T_\sigma\right]\\
&&=\sum_{\sigma\in S_4}W^-_{\pi\sigma}K_{\sigma\tau}^{(1)}-W^-_{\pi\sigma}K_{\sigma\tau}^{(2)}
\ea
Thus the matrix $\Lambda$ is expressible in terms of the matrices $K^{(1)}$ and $K^{(2)}$, which we have already computed.
Also, the computation of $\Lambda$ allows us to write the expressions of $\vec{\lambda}=\Lambda V^{T}\vec{t}$, which turns out to be simply $\vec{\lambda}=\frac{\sin^22\theta}{d^2-1}\vec{v}$.

We can then use the diagonal form of the matrix $\Xi=VD_\Xi V^{T}$ to write:
\ba
\sum_{i=0}^{k-1}\Xi^i=\sum_{i=0}^{k-1}(VD_\Xi V^{T})^i=V\left(\sum_{i=0}^{k-1}D^i\right)V^{T}
\ea
The last point to note is that all the eigenvalues of $\Xi$ are lower than 1, and that we are dealing with a geometric progression. We can define the matrix $\tilde{D}_\Xi^{(k)}$ as the diagonal matrix:
\ba
\tilde{D}^{(k)}=\begin{cases}
0&\rm{if }\xi_i=0\\
\frac{1-\xi_i^k}{1-\xi_i}&\rm{if }\xi_i\neq0
\end{cases}
\ea
where the $\{\xi_j\}$ are the eigenvalues of the matrix $\Xi$. Thus, we finally have:
\ba
\Gamma_{\pi\sigma}^{(k)}=\sum_{\tau,\rho\in S_4}\Lambda_{\pi\tau}V_{\tau\rho}\tilde{D}_{\rho\rho}^{(k)}V_{\rho\sigma}^\dag=\Lambda V\tilde{D}^{(k)}V^{T}
\ea
Inserting this into the expression of the fourth moment one gets:
\ba
\nonumber
\vec{t}^T\Gamma^{(k)}\vec{T}&=&\vec{t}^T\Lambda V\tilde{D}^{(k)}V^{T}\vec{T}=\vec{\lambda}^T\tilde{D}^{(k)}\vec{T}_V\\
\nonumber
&=&\frac{(1-\xi_+^k)}{1-\xi_+}\lambda_+T_++\frac{(1-\xi_-^k)}{1-\xi_-}\lambda_-T_-+\sum_{i=1}^4\frac{1-\xi_i^k}{1-\xi_i}\lambda_iT_i\\
&=&\frac{\sin^22\theta}{d^2-1}\left[\frac{(1-\xi_+^k)}{1-\xi_+}v_+T_++\frac{(1-\xi_-^k)}{1-\xi_-}v_-T_-+\frac{1-\xi_4^k}{1-\xi_4}\sum_{i=1}^4v_iT_i\right]
\ea

\section{\label{app:spectral_functions}Spectral form factors}
\subsection{The vector \texorpdfstring{$\vec{c}$}{}}
Let us now compute the components of the vector $\vec{c}$ for the order $k=2$ isospectral twirling. This corresponds to evaluate the traces:
\ba
\Tr[T_\pi V^{\ot 2,2}],\quad\forall\pi\in S_4
\ea
These traces can be evaluated according to the conjugacy classes of $S_4$.Let us start with the identity permutation :
\ba
\vec{c}_I=\Tr[IV^{\ot2,2}]=|\Tr[V]|^4=\sum_{i,j,k,\ell}e^{-i(E_i+E_j-E_k-E_\ell)t}=g_4(t)
\ea
Thus, the identity permutation introduces the four point spectral form factor $g_4(t)$. Similarly we can compute the trace corresponding to the permutation $T_{(12)}$:
\ba
\vec{c}_{T_{(12)}}=\Tr[T_{(12)}V^{\ot2,2}]=\Tr[V^2]\Tr[V^\dag]^2=\sum_{i,j,k}e^{-i(2E_i-E_j-E_k)t}=g_3(t)
\ea
Similarly $\vec{c}_{T_{(34)}}=g^*_3(t)$.
In a similar fashion one can compute all the other traces, summarized in Table~\ref{table:haar_spectral_functions}.

The component of $\vec{c}$ for all the other permutations are computed as:
\ba
\vec{c}_{T_{(ij)}}&=&\Tr[T_{(ij)}V^{\ot2,2}]=\Tr[VV^\dag]\left|Tr[V]\right|^2=d\sum_{i,j,}e^{-i(E_i-E_j)t}=dg_2(t)\\
\vec{c}_{T_{(ijk)}}&=&\Tr[T_{(ijk)}V^{\ot2,2}]=\Tr[VV^\dag V]\Tr[V^\dag]=\sum_{i,j}e^{-i(E_i-E_j)t}=g_2(t)\\
\vec{c}_{T_{(ijk\ell)}}&=&\Tr[T_{(ijk\ell)}V^{\ot2,2}]=\Tr[VV^\dag VV^\dag]=d\\
\vec{c}_{T_{(12)(34)}}&=&\Tr[T_{(12)(34)}V^{\ot2,2}]=\Tr[V^2]\Tr[V^{\dag2}]=g_2(2t)\\
\vec{c}_{T_{(ij)(k\ell)}}&=&\Tr[T_{(ij)(k\ell)}V^{\ot2,2}]=\Tr[VV^\dag]\Tr[VV^\dag]=d^2
\ea

\begin{table}[!ht]
\begin{tabular}{|c|c|}
\toprule
$T_\pi$&$\Tr[T_\pi V^{\ot2,2}]$\\
\midrule
$I$&$g_4(t)$\\
$T_{(12)}$&$g_3(t)$\\
$T_{(34)}$&$g^*_3(t)$\\
$T_{(ij)}$&$dg_2(t)$\\
$T_{(ijk)}$&$g_2(t)$\\
$T_{(ijk\ell)}$&$d$\\
$T_{(12)(34)}$&$g_2(2t)$\\
$T_{(ij)(k\ell)}$&$d^2$\\
\bottomrule
\end{tabular}
\caption{Components of the vector $\vec{c}$}
\label{table:haar_spectral_functions}
\end{table}

\subsection{The vector \texorpdfstring{$\vec{q}$}{}}
The computation of the components of the vector $\vec{q}$ is similar to the one of the vector $\vec{c}$, but not as simple.
Indeed, the components of the vector $\vec{q}$ are defined as:
\ba
\vec{q}_\pi=\Tr[T_\pi QV^{\ot2,2}]
\ea
To evaluate these traces, we once again write the unitary $V$ in its diagonal form.
Let us then consider a couple of instances, starting from identity permutation. In this case one gets:
\ba
\vec{q}_I=\Tr{QV^{\ot2,2}}=d^{-2}\sum_P|\Tr[UP]|^4=d^{-2}\sum_{i,j,k,\ell}e^{-i(E_i+E_j-E_k-E_\ell)t}\bra{E_i}P\ket{E_i}=\tilde{g}_4(t)
\ea
where we have defined the generalized 4 point spectral form factor $\tilde{g}_4(t)$. In contrast with $g_4(t)$, $\tilde{g}_4(t)$ depends on the expectation value of the energy eigenstates over the Pauli strings. This is also the reason that motivated us in considering stabilizer Hamiltonian as physical models.

In a similar manner one can compute the remaining traces, starting with $T_{(12)}$ and $T_{(ij)}$:
\ba
\nonumber
&&\vec{q}_{T_{(12)}}=\vec{q}^*_{T_{(34)}}=d^{-2}\sum_{P}\Tr[PVPV]\Tr[PV^\dag]^2\\
&&=\sum_{P}\sum_{i,j,k,\ell}e^{-i(E_i+E_j-E_k-E_\ell)}\bra{E_i}P\dyad{E_j}P\ket{E_i}\bra{E_k}P\ket{E_k}\bra{E_\ell}P\ket{E_\ell}=\tilde{g}_3(t)\\
\nonumber
&&\vec{q}_{T_{(ij)}}=d^{-2}\sum_{P}\Tr[PV^\dag PV]|\Tr[PV]|^2\\
&&=\sum_{P}\sum_{i,j,k,\ell}e^{-i(E_i+E_j-E_k-E_\ell)}\bra{E_i}P\dyad{E_k}P\ket{E_i}\bra{E_j}P\ket{E_j}\bra{E_\ell}P\ket{E_\ell}=\tilde{g}'_3(t)
\ea
As for the traces of the form $T_{(ijk)}^{(1,2)}$ standing for any 3-cycle permutation where either the first or second element is kept fixed, we get:
\ba
&&\vec{q}_{T_{(ijk)}^{(1,2)}}=\vec{q}^*_{T_{(ijk)}^{(3,4)}}=d^{-2}\sum_P\Tr[PVPV^\dag PV^\dag]\Tr[PV]\\
&&=d^{-2}\sum_P\sum_{i,j,k,\ell}e^{-i(E_i+E_j-E_k-E_\ell)t}\bra{E_i}P\ket{E_i}\bra{E_j}P\dyad{E_k}P\dyad{E_\ell}P\ket{E_j}=\tilde{g}_2(t)
\ea
Then, for permutations of the form $T_{(ijk\ell)}$ we get:
\ba
\nonumber
&&\vec{q}_{T_{(1342)}}=\vec{q}_{T_{(1243)}}=d^{-2}\sum_P\Tr[PVPV^\dag PVPV^\dag]\\
&&=d^{-2}\sum_P\sum_{i,j,k,\ell}e^{-i(E_i+E_j-E_k-E_\ell)t}\bra{E_\ell}P\dyad{E_i}P\dyad{E_k}P\dyad{E_j}P\ket{E_\ell}=\tilde{g}_1(t)\\
\nonumber
&&\vec{q}_{T_{(ijk\ell)}}=d^{-2}\sum_P\Tr[PVPVPV^\dag PV^\dag]\\
&&=d^{-2}\sum_P\sum_{i,j,k,\ell}e^{-i(E_i+E_j-E_k-E_\ell)t}\bra{E_\ell}P\dyad{E_i}P\dyad{E_j}P\dyad{E_k}P\ket{E_\ell}=\tilde{g}'_1(t)
\ea
Finally, double swaps $T_{(ij)(k\ell)}$ evaluate to:
\ba
\nonumber
&&\vec{q}_{T_{(12)(34)}}=d^{-2}\sum_P|\Tr[PVPV]|^2\\
&&=d^{-2}\sum_P\sum_{i,j,k,\ell}e^{-i(E_i+E_j-E_k-E_\ell)t}\bra{E_i}P\dyad{E_j}P\ket{E_i}\bra{E_k}P\dyad{E_\ell}P\ket{E_k}=\tilde{g}'_2(t)\\
\nonumber
&&\vec{q}_{T_{(ij)(k\ell)}}=d^{-2}\sum_P\Tr[PVPV^\dag]^2\\
&&=d^{-2}\sum_P\sum_{i,j,k,\ell}e^{-i(E_i+E_j-E_k-E_\ell)t}\bra{E_i}P\dyad{E_k}P\ket{E_i}\bra{E_j}P\dyad{E_\ell}P\ket{E_j}=\tilde{g}''_2(t)
\ea

The result for all possible permutations are summarized in Tab.~\ref{table:q_elements}.
\begin{table}[!ht]
\begin{tabular}{|c|c|}
\toprule
$T_{\pi}$&$\tr{T_\pi QV^{\ot2,2}}$\\
\midrule
$I$&$\tilde{g}_4(t)$\\
$T_{(12)}$&$\tilde{g}_3(t)$\\
$T_{(34)}$&$\tilde{g}^*_3(t)$\\
$T_{(ij)}$&$\tilde{g}'_3(t)$\\
$T_{(ijk)}^{(1,2)}$&$\tilde{g}_2(t)$\\
$T_{(ijk)}^{(3,4)}$&$\tilde{g}^*_2(t)$\\
$T_{(1342)},T_{(1243)}$&$\tilde{g}_1(t)$\\
$T_{(ijk\ell)}$&$\tilde{g}'_1(t)$\\
$T_{(12)(34)}$&$\tilde{g}'_2(t)$\\
$T_{(ij)(k\ell)}$&$\tilde{g}''_2(t)$\\
\bottomrule
\end{tabular}
\caption{The values of all the components of the vector $\vec{q}$.}
\label{table:q_elements}
\end{table}

\section{\label{app:spectral_averages}Spectral averages}

\subsection{The three point spectral form factor  \texorpdfstring{$g_3(t)$}{}}
Let us first note that in any computation only the real part of $g_3(t)$ will appear, so that we can limit ourselves to average only $\Re g_3(t)$.

First of all, we rewrite the expression of $g_3(t)$ as:
\ba
g_3(t)=\sum_{i,j.,k}e^{-i(2E_i-E_j-E_k)t}=d+\sum_{i\neq j}e^{-2i(E_i-E_j)t}+2e^{-i(E_i-E_j)t}+\sum_{i\neq j\neq k}e^{-i(2E_i-E_j-E_k)t}
\ea
Let us then write the expression for the average:
\ba
\nonumber
\overline{g_3(t)}^{E}&=&\sum_{i,j.,k}\int \{dE_k\}e^{-i(2E_i-E_j-E_k)t}P_E(\{E_k\})\\
\nonumber
&=&d+d(d-1)\int dE_1\,dE_2\,e^{-2i(E_1-E_2)t}\rho_E^{(2)}(E_1,E_2)+2d(d-1)\int dE_1\,dE_2\,e^{-i(E_1-E_2)t}\rho_E^{(2)}(E_1,E_2)\\
&&+d(d-1)(d-2)\int dE_1\,dE_2\,dE_3\,e^{-i(2E_1-E_2-E_3)t}\rho_E^{(3)}(E_1,E_2,E_3)
\ea
The first two terms have been already computed in the main text, while to compute the last one we need to compute the three point marginal probability distribution $\rho_{\rm GUE}^{(3)}(E_1,E_2,E_3)$ (the one for the GDE is trivial because of factorization). To do this we apply the expression in Eq.~\eqref{eq:n_marginal_prob_formula} to obtain:
\ba
\nonumber
\rho_{\rm GUE}^{(3)}(E_1,E_2,E_3)&=&\frac{1}{d(d-1)(d-2)}\Bigg[d^3\rho_{\rm GUE}^{(1)}(E_1)\rho_{\rm GUE}^{(1)}(E_2)\rho_{\rm GUE}^{(1)}(E_3)-d\sum_{i,j,k}\rho_{\rm GUE}^{(1)}(E_i)\left(\frac{\sin(d(E_j-E_k))}{\pi(E_j-E_k)}\right)^2\\
&&+2\left(\frac{\sin(d(E_1-E_2))}{\pi(E_1-E_2)}\right)\left(\frac{\sin(d(E_2-E_3))}{\pi(E_2-E_3)}\right)\left(\frac{\sin(d(E_1-E_3))}{\pi(E_1-E_3)}\right)\Bigg]
\ea
One can then use this expression to perform the integral
\ba
\nonumber
&&\int dE_1\,dE_2\,dE_3\,e^{-i(2E_1-E_2-E_3)t}\rho_E^{(3)}(E_1,E_2,E_3)\\
\nonumber
&=&\frac{1}{d(d-1)(d-2)}\Bigg\{\int dE_1\,dE_2\,dE_3\,e^{-i(2E_1-E_2-E_3)t}d^3\rho_{\rm GUE}^{(1)}(E_1)\rho_{\rm GUE}^{(1)}(E_2)\rho_{\rm GUE}^{(1)}(E_3)\\
\nonumber
&+&\int dE_1\,dE_2\,dE_3\,e^{-i(2E_1-E_2-E_3)t}d\sum_{i,j,k}\rho_{\rm GUE}^{(1)}(E_i)\left(\frac{\sin(d(E_j-E_k))}{\pi(E_j-E_k)}\right)^2\\
\nonumber
&+&2\int dE_1\,dE_2\,dE_3\,e^{-i(2E_1-E_2-E_3)t}\left(\frac{\sin(d(E_1-E_2))}{\pi(E_1-E_2)}\right)\left(\frac{\sin(d(E_2-E_3))}{\pi(E_2-E_3)}\right)\left(\frac{\sin(d(E_1-E_3))}{\pi(E_1-E_3)}\right)\Bigg\}\\
\ea
Before proceeding further and show the results of the integral, let us introduce the following functions, which will serve as building blocks for the next results:
\ba
r_1(t)=\frac{J_1(2t)}{t},\qquad
r_2(t)=\theta(2d-t)\left(1-\frac{t}{2d}\right),\qquad r_3(t)=\frac{\sin(\pi t/2)}{\pi t/2}
\ea
The result of the first integral is:
\ba
\int dE_1\,dE_2\,dE_3\,e^{-i(2E_1-E_2-E_3)t}d^3\rho_{\rm GUE}^{(1)}(E_1)\rho_{\rm GUE}^{(1)}(E_2)\rho_{\rm GUE}^{(1)}(E_3)=d^3r_1^2(t)r_1(2t)
\ea
The result of the second integral is:
\ba
&&\int dE_1\,dE_2\,dE_3\,e^{-i(2E_1-E_2-E_3)t}d\sum_{i,j,k}\rho_{\rm GUE}^{(1)}(E_i)\left(\frac{\sin(d(E_j-E_k))}{\pi(E_j-E_k)}\right)^2\\
&=&-d^2r_1(2t)r_2(t)r_3(2t)-2d^2r_1(t)r_2(2t)r_3(t)
\ea
Finally, the third term contribute as:
\ba
\nonumber
&&2\int dE_1\,dE_2\,dE_3\,e^{-i(2E_1-E_2-E_3)t}\left(\frac{\sin(d(E_1-E_2))}{\pi(E_1-E_2)}\right)\left(\frac{\sin(d(E_2-E_3))}{\pi(E_2-E_3)}\right)\left(\frac{\sin(d(E_1-E_3))}{\pi(E_1-E_3)}\right)=2dr_2(3t)\\
&&
\ea

Summing up all contributions we finally get:
\ba
\nonumber
\overline{g_3(t)}^{GUE}&=&d^3r_1^2(t)r_1(2t)-d^2r_1(2t)r_2(t)r_3(2t)-2d^2r_1(t)r_2(2t)r_3(t)+2dr_2(3t)\\
&&+d^2r_1^2(2t)-dr_2(2t)+2d^2r_1^2(t)-2dr_2(t)+d
\ea

\begin{figure}[!ht]
\centering
\begin{subfigure}{0.49\textwidth}
\centering
\includegraphics[width=\linewidth]{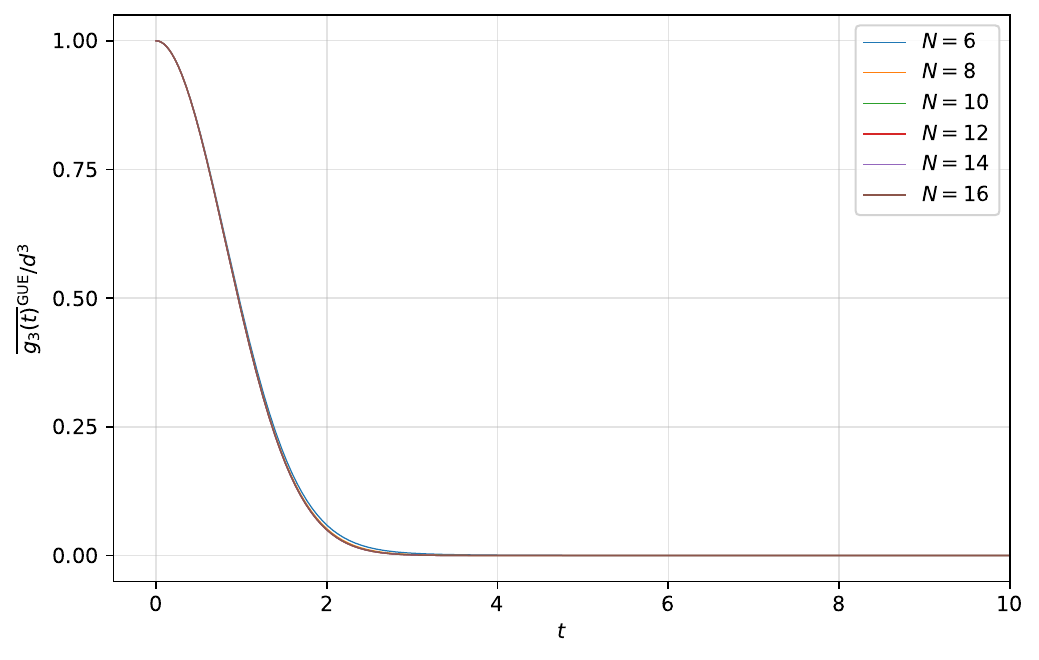}
\caption{Plot of the normalized $\overline{g_3(t)}^{GDE}/d^3$.}
\label{fig:g3_GDE}
\end{subfigure}
\begin{subfigure}{0.49\textwidth}
\centering
\includegraphics[width=\linewidth]{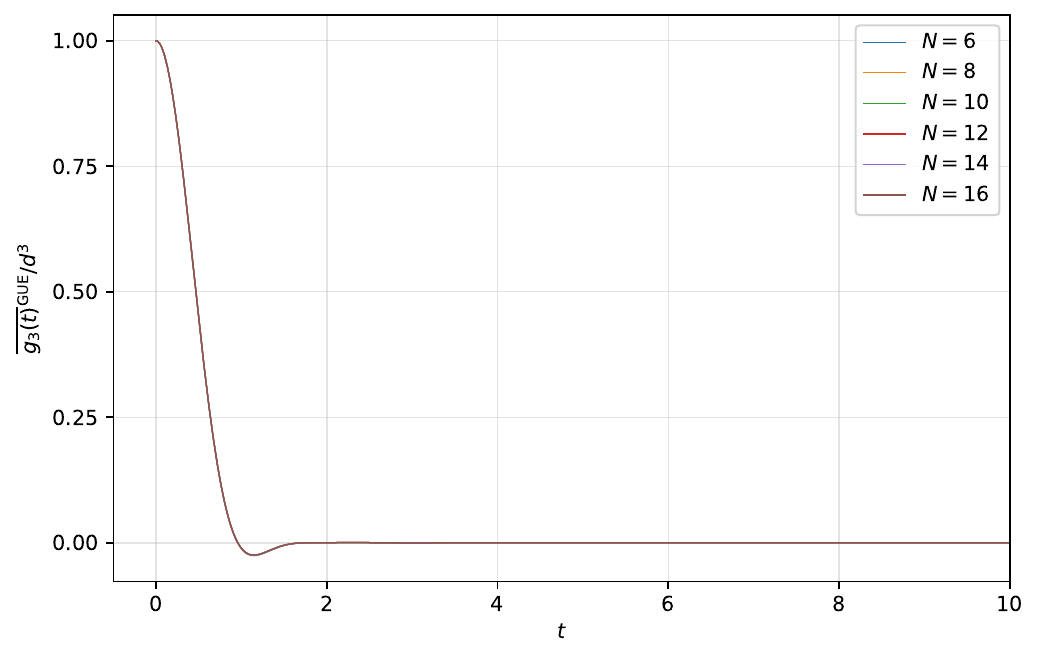}
\caption{Plot of the normalized $\overline{g_3(t)}^{\rm GUE}/d^3$.}
\label{fig:g3_GUE}
\end{subfigure}
\caption{Plot of the normalized versions of $\overline{g_3(t)}^{\rm GDE}$ (\subref{fig:g3_GDE}) and $\overline{g_3(t)}^{\rm GUE}$ (\subref{fig:g3_GUE}) for $d=2^N$. Notice how in both cases the curve essentially does not depend on the dimension and the asymptotic value is reached extremely quickly. The main difference between $\overline{g_3(t)}^{\rm GDE}$ and $\overline{g_3(t)}^{\rm GUE}$ is that the latter shortly becomes negative before reaching its asymptotic value.}
\label{fig:g3}
\end{figure}

As for the average over the GDE, once again this is trivial as the corresponding probability distribution factorizes. Indeed, exploiting this fact and the fact that $\{i,j,k\}$ are dummy indices, one obtains:
\ba
\overline{g_3(t)}^{GDE}=d+d(d-1)e^{-t^2}+2d(d-1)e^{-\frac{t^2}{4}}+d(d-1)(d-2)e^{-\frac{3t^2}{4}}
\ea
Both functions are plotted in Fig.~\ref{fig:g3}.

\subsection{The four point spectral form factor  \texorpdfstring{$g_4(t)$}{}}
Let us now compute the spectral averages of the four point spectral form factor $g_4(t)$. As usual, we first rewrite it in a more manageable shape as:
\ba
\nonumber
g_{4}(t)&=&\sum_{i\neq j \neq k \neq l} e^{-i(E_i+E_j-E_k-E_l)t}+\sum_{i\neq j \neq k}e^{-i(2E_i-E_j-E_k)t}+\sum_{i\neq j \neq k}e^{-i(E_i+E_j-2E_k)t}\\
&&+4(d-1)\sum_{i\neq j}e^{-i(E_i-E_j)t}+\sum_{i\neq j}e^{-i2(E_i - E_j)t}+2d(d-1)+d
\ea
At this point we note that all terms but the first have been computed in the computation of either $g_3(t)$ or $g_2(t)$, so that we can focus on it alone.

Once again, we need to compute the four point marginal probability distribution $\rho_{\rm GUE}^{(4)}(E_1,E_2,E_3,E_4)$ via the determinant of the matrix $K$, obtaining:
\ba
\nonumber
\rho_E^{(4)}(E_1,E_2,E_3,E_4)&=&\frac{1}{d(d-1)(d-2)(d-3)}\Big[d^4\rho_{\rm GUE}^{(1)}(E_1)\rho_{\rm GUE}^{(1)}(E_2)\rho_{\rm GUE}^{(1)}(E_3)\rho_{\rm GUE}^{(1)}(E_4)\\
\nonumber
&&-2\left(K_{13}K_{32}K_{34}K_{41}+K_{12}K_{23}K_{34}K_{41}+K_{12}K_{24}K_{43}K_{31}\right)\\
\nonumber
&&-d^2\sum_{i,j,k,\ell}\rho_{\rm GUE}^{(1)}(E_i)\rho_{\rm GUE}^{(1)}(E_j)K_{k\ell}^2+K_{12}^2K_{34}^2+K_{13}^2K_{24}^2+K_{14}^2K_{23}^2\\
&&+d\sum_i \rho_{\rm GUE}^{(1)}(E_i)K_{jk}K_{kj}K_{j\ell}\Big]
\ea
where for notational compactness we have used the matrix elements of $K$ to shorten the expressions.
Exploiting once more the fact that the indices $i,j,k,\ell$ are dummy, it turns out that we need to compute the integral:
\ba
\nonumber
&&d(d-1)(d-2)(d-3)\int dE_1\,dE_2\,dE_3\,dE_4\,e^{-i(E_1+E_2-E_3-E_4)}\rho_E^{(4)}(E_1,E_2,E_3,E_4)\\
\nonumber
&=&\int dE_1\,dE_2\,dE_3\,dE_4\,e^{-i(E_1+E_2-E_3-E_4)}d^4\rho_{\rm GUE}^{(1)}(E_1)\rho_{\rm GUE}^{(1)}(E_2)\rho_{\rm GUE}^{(1)}(E_3)\rho_{\rm GUE}^{(1)}(E_4)\\
\nonumber
&-&2\int dE_1\,dE_2\,dE_3\,dE_4\,e^{-i(E_1+E_2-E_3-E_4)}\left(K_{13}K_{32}K_{34}K_{41}+K_{12}K_{23}K_{34}K_{41}+K_{12}K_{24}K_{43}K_{31}\right)\\
\nonumber
&-&\int dE_1\,dE_2\,dE_3\,dE_4\,e^{-i(E_1+E_2-E_3-E_4)}d^2\sum_{i,j,k,\ell}\rho_{\rm GUE}^{(1)}(E_i)\rho_{\rm GUE}^{(1)}(E_j)K_{k\ell}^2\\
\nonumber
&+&\int dE_1\,dE_2\,dE_3\,dE_4\,e^{-i(E_1+E_2-E_3-E_4)}K_{12}^2K_{34}^2+K_{13}^2K_{24}^2+K_{14}^2K_{23}^2\\
\label{eq:g_4_integral_sum}
&+&\int dE_1\,dE_2\,dE_3\,dE_4\,e^{-i(E_1+E_2-E_3-E_4)}d\sum_i\rho_{\rm GUE}^{(1)}(E_i)K_{jk}K_{kj}K_{j\ell}
\ea
Let us now evaluate the integrals one by one. The first one is worth:
\ba
\int dE_1\,dE_2\,dE_3\,dE_4\,e^{-i(E_1+E_2-E_3-E_4)}d^4\rho_{\rm GUE}^{(1)}(E_1)\rho_{\rm GUE}^{(1)}(E_2)\rho_{\rm GUE}^{(1)}(E_3)\rho_{\rm GUE}^{(1)}(E_4)=d^4r_1^4(t)
\ea
Then we have:
\ba
\nonumber
&&-2\int dE_1\,dE_2\,dE_3\,dE_4\,e^{-i(E_1+E_2-E_3-E_4)}\left(K_{13}K_{32}K_{34}K_{41}+K_{12}K_{23}K_{34}K_{41}+K_{12}K_{24}K_{43}K_{31}\right)=-6dr_2(2t)\\
\ea
The third integral is worth:
\ba
\nonumber
-\int dE_1\,dE_2\,dE_3\,dE_4\,e^{-i(E_1+E_2-E_3-E_4)}d^2\sum_{i,j,k,\ell}\rho_{\rm GUE}^{(1)}(E_i)\rho_{\rm GUE}^{(1)}(E_j)K_{k\ell}^2=-2d^3r_1^2(t)r_2(t)r_3(2t)-4d^3r_1^2(t)r_2(t)\\
\ea
As for the fourth integral we get:
\ba
\int dE_1\,dE_2\,dE_3\,dE_4\,e^{-i(E_1+E_2-E_3-E_4)}K_{12}^2K_{34}^2+K_{13}^2K_{24}^2+K_{14}^2K_{23}^2=2d^2r_2^2(t)+d^2r_2^2(t)r_3^2(2t)
\ea
Finally, we can evaluate the last integral:
\ba
\int dE_1\,dE_2\,dE_3\,dE_4\,e^{-i(E_1+E_2-E_3-E_4)}d\sum_i\rho_{\rm GUE}^{(1)}(E_i)K_{jk}K_{kj}K_{j\ell}=8d^2r_1(t)r_2(t)r_3(t)
\ea
Putting everything together one gets:
\ba
\nonumber
&&\overline{g_4(t)}^{\rm GUE}=d^4r_1^4(t)-6dr_2(2t)-2d^3r_1^2(t)r_2(t)r_3(2t)-4d^3r_1^2(t)r_2(t)+2d^2r_2^2(t)+d^2r_2^2(t)r_3^2(2t)+8d^2r_1(t)r_2(t)r_3(t)\\
\nonumber
&&2\Re\left[d^3r_1^2(t)r_1(2t)-d^2r_1(2t)r_2(t)r_3(2t)-2d^2r_1(t)r_2(2t)r_3(t)+2dr_2(3t)+d^2r_1^2(2t)-dr_2(2t)+2d^2r_1^2(t)-2dr_2(t)\right]\\
&&+4(d-1)\left(d^2r_1^2(t)-dr_2(t)\right)+d^2r_1^2(2t)-dr_2(2t)+2d(d-1)+d
\ea

The average over the GDE is once again performed easily because of the factorization of the probability distribution, resulting into:
\ba
\nonumber
\overline{g_4(t)}^{\rm GDE}=d(2d-1)+4d(d-1)(d-1)e^{-t^{2}/4}+2d(d-1)(d-2)e^{-3t^{2}/4}+d(d-1)(d-2)(d-3)e^{-t^{2}/2}\\
\ea

\begin{figure}[!ht]
\begin{subfigure}{0.49\textwidth}
\centering
\includegraphics[width=\textwidth]{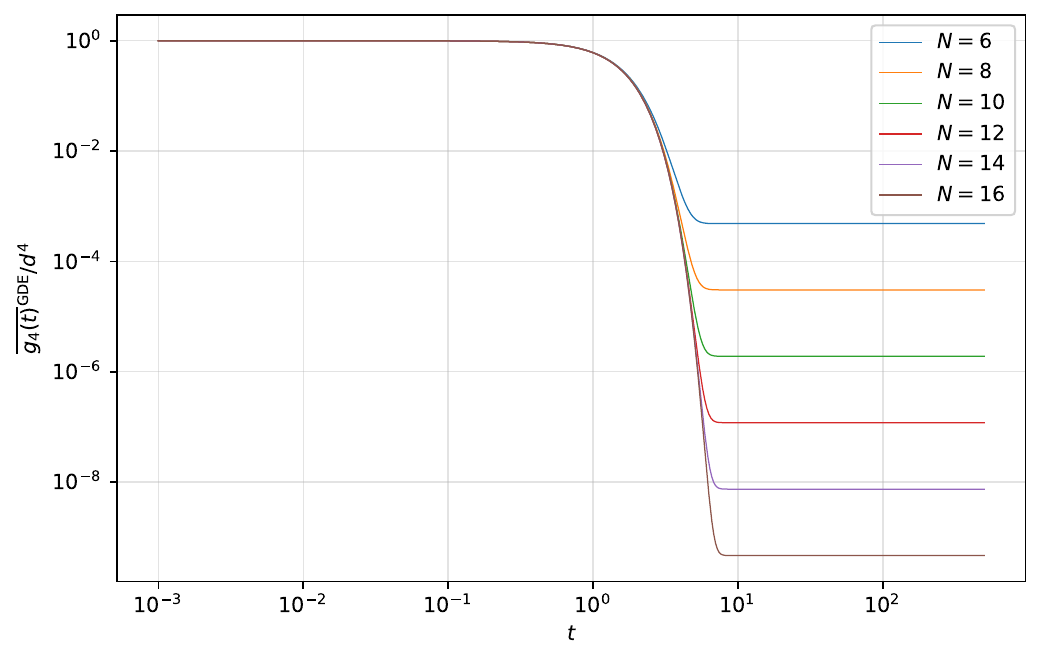}
\caption{Log-Log plot $\overline{g_4(t)}^{\rm GDE}/d^4$.}
\label{fig:g4GDE}
\end{subfigure}
\begin{subfigure}{0.49\textwidth}
\centering
\includegraphics[width=\textwidth]{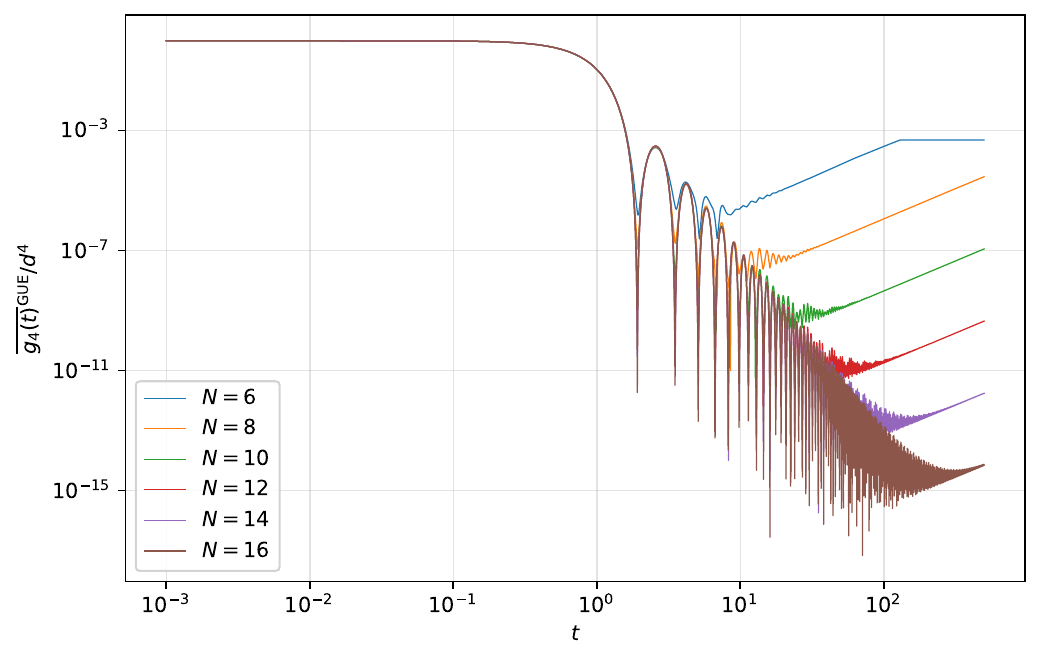}
\caption{Log-Log plot of $\overline{g_4(t)}^{\rm GUE}/d^4$.}
\label{fig:g4GUE}
\end{subfigure}
\caption{Plot of the normalized four points spectral form factor $g_4(t)/d^4$ averaged over the GDE (panel~\subref{fig:g4GDE}) and the GUE(panel~\subref{fig:g4GUE}) for several dimensions $d=2^N$. One can notice the characteristic oscillations and ramps of chaotic systems for $\overline{g_4(t)}^{GUE}$.}
\label{fig:g4}
\end{figure}

\subsection{The three point Clifford spectral form factor  \texorpdfstring{$\tilde{g}^{(\rm{cb})}_3(t)$}{}}
\begin{figure}[!ht]
\centering
\begin{subfigure}{0.49\textwidth}
\centering
\includegraphics[width=\textwidth]{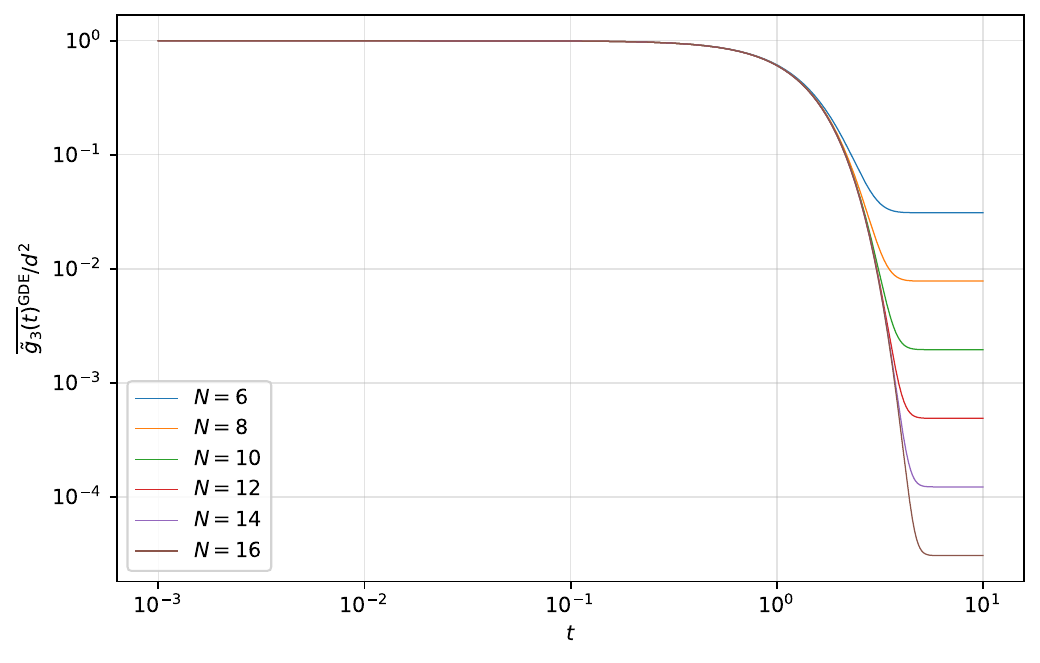}
\caption{Plot of $\overline{\tilde{g}_3(t)}^{\rm GDE}/d^2$.}
\label{fig:g3tildeGDE}
\end{subfigure}
\begin{subfigure}{0.49\textwidth}
\centering
\includegraphics[width=\textwidth]{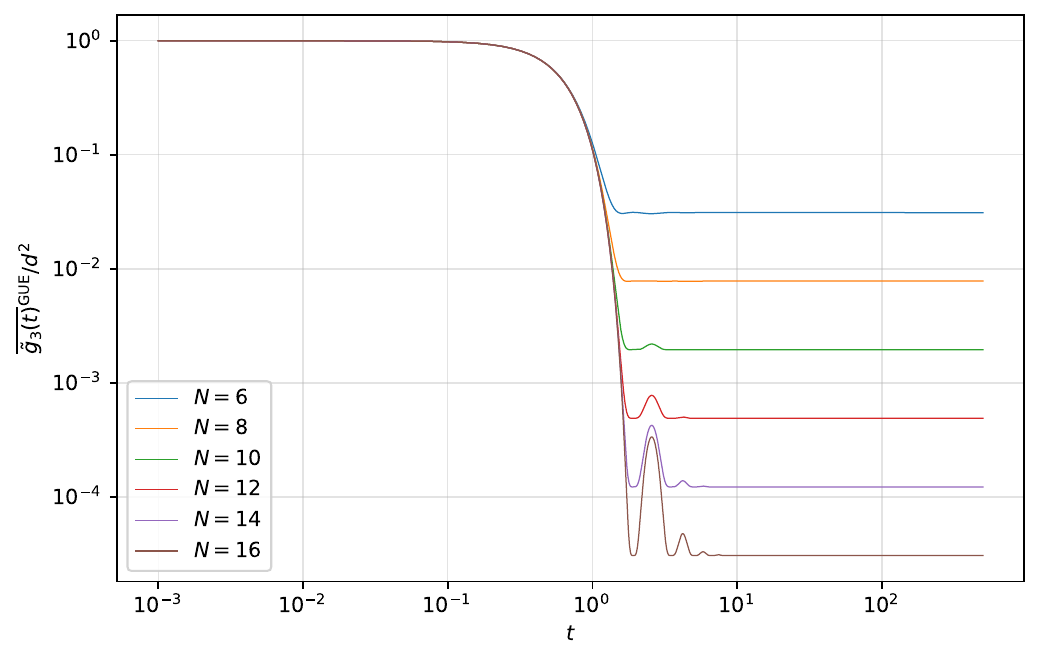}
\caption{Plot of $\overline{\tilde{g}_3(t)}^{\rm GUE}/d^2$.}
\label{fig:g3tildeGUE}
\end{subfigure}
\caption{Plot of the function $\tilde{g}_3(t)$ averaged over the GDE (panel~\subref{fig:g3tildeGDE}) and the GUE(panel~\subref{fig:g3tildeGUE}) for $d=2^N$.}
\label{fig:g3tilde}
\end{figure}

First of all, let us rewrite this spectral form factor as:
\ba
\nonumber
\tilde{g}^{(\rm{cb})}_3(t)&=&d^{-1}\sum_{i,j,k}e^{-i(E_i+E_j-E_k-E_{i\oplus j\oplus k})t}\\
\nonumber
&=&d^{-1}\Bigg[\sum_{i=j=k}e^{-i(E_i+E_j-E_k-E_{i\oplus j\oplus k})t}+\sum_{i=j,k}e^{-i(E_i+E_j-E_k-E_{i\oplus j\oplus k})t}+\sum_{i=k,j}e^{-i(E_i+E_j-E_k-E_{i\oplus j\oplus k})t}\\
\nonumber
&&+\sum_{k=j,i}e^{-i(E_i+E_j-E_k-E_{i\oplus j\oplus k})t}+\sum_{i\neq k\neq j}e^{-i(E_i+E_j-E_k-E_{i\oplus j\oplus k})t}\Bigg]\\
\nonumber
&=&d^{-1}\Bigg[\sum_{i=j=k}1+\sum_{i=j,k}e^{-i2(E_i-E_k)t}+\sum_{i=k,j}1+\sum_{k=j,i}1+\sum_{i\neq k\neq j}e^{-i(E_i+E_j-E_k-E_{i\oplus j\oplus k})t}\Bigg]\\
\nonumber
&=&d^{-1}\left[d+\sum_{i\neq k}e^{-i2(E_i-E_k)t}+2d(d-1)+\sum_{i\neq k\neq j}e^{-i(E_i+E_j-E_k-E_{i\oplus j\oplus k})t}\right]\\
\label{eq:g3_tilde_comp}
&=&1+2(d-1)+d^{-1}\sum_{i\neq k}e^{-i2(E_i-E_k)t}+d^{-1}\sum_{i\neq j\neq k}e^{-i(E_i+E_j-E_k-E_{i\oplus j\oplus k})t}
\ea
We already computed the average of all the terms above but the last one. Let us consider:
\ba
\nonumber
&&\overline{\sum_{i\neq j\neq k}e^{-i(E_i+E_j-E_k-E_{i\oplus j\oplus k})t}}^{\rm GUE}=\sum_{i\neq j\neq k}\int dE_1\cdots dE_d\,e^{-i(E_i+E_j-E_k-E_{i\oplus j\oplus k})t}P_{\rm GUE}(\{E_\alpha\})\\
\nonumber
&&=\sum_{i\neq j\neq k}\int dE_i\,dE_j\,dE_k\,dE_{i\oplus j\oplus k}\,e^{-i(E_i+E_j-E_k-E_{i\oplus j\oplus k})t}P_{\rm GUE}(\{E_{/i,j,k,i\oplus j\oplus k}\})\\
\label{eq:sum_clifford_g_3}
&&=d(d-1)(d-2)\int dE_1\,dE_2\,dE_3\,dE_{1\oplus 2\oplus 3}\rho_E^{(4)}(E_1,E_2,E_3,E_{1\oplus 2\oplus 3})
\ea
The integral in Eq.~\eqref{eq:sum_clifford_g_3} is just the same as the one in Eq.~\eqref{eq:g_4_integral_sum} with a missing $(d-3)$ factor. Thus, the final result is:
\ba
\overline{\tilde{g}_3(t)}^{\rm GDE}=1+2(d-1)+(d-1)e^{-t^2}+(d-1)(d-2)e^{-\frac{3 t^2}{4}}
\ea
for the GDE, while for the GUE one gets:
\ba
&&\overline{\tilde{g}_3(t)}^{\rm GUE}=1+2(d - 1)-r_2(2t)+
 dr_1^2(2 t)+\\
\nonumber
&+&\frac{(d^3r_1^4(t)-6r_2(2t)-2d^2r_1^2(t)r_2(t)r_3(2t)-4d^2r_1^2(t)r_2(t)+2dr_2^2(t)+dr_2^2(t)r_3^2(2t)+8dr_1(t)r_2(t)r_3(t))}{(d-3)}
\ea
Both functions are plotted in Fig.~\ref{fig:g3tilde}. It is important to highlight that the Clifford spectral form factor $\tilde{g}_3(t)$ has a functional dependence similar to $g_4(t)$, as it is the spectral form factor stemming from the identity permutation. However, due to the correlation between the indices in Eq.~\eqref{eq:g3_tilde_comp}, $\tilde{g}_3(t)$ is normalized to $d^2$, i.e. $\tilde{g}_3(0)/d^2=1$, and reaches an asymptotic value of order $\mathcal{O}(d)$, to be compared with the $d^2$ reached by $g_4(t)$.

\section{Hamiltonian models}

\subsection{\label{app:stab_ham_spectral_functions}Spectral form factors of a Stabilizer Hamiltonian}

In this section we want to compute the remaining traces corresponding to the components of the vector $\vec{q}$ for a stabilizer Hamiltonian.

For the permutation of the form $T_{(ij)}$ we get:
\ba
\nonumber
\vec{q}_{T_{(ij)}}&=&\Tr[T_{(ij)}QV_{\rm stab}^{\ot2,2}]=d^{-2}\sum_P\Tr[PV_{\rm stab}PV_{\rm stab}^\dag]\Tr[PV_{\rm stab}]\Tr[PV_{\rm stab}^\dag]\\
\nonumber
&=&d^{-2}\sum_{P\in\mathcal{P}_{\rm ab}}\Tr[PV_{\rm stab}PV_{\rm stab}^\dag]\Tr[PV_{\rm stab}]\Tr[PV_{\rm stab}^\dag]\\
\nonumber
&=&d^{-2}\sum_{P\in\mathcal{P}_{\rm ab}}\Tr[V_{\rm stab}V_{\rm stab}^\dag]\Tr[PV_{\rm stab}]\Tr[PV_{\rm stab}^\dag]\\
\nonumber
&=&d^{-1}\sum_{P\in\mathcal{P}_{\rm ab}}\Tr[PV_{\rm stab}]\Tr[PV_{\rm stab}^\dag]\\
&=&d^{-1}\sum_{P}\Tr[P^{\ot2}V_{\rm stab}^{\ot1,1}]=\Tr[T_{(12)}V_{\rm stab}^{\ot1,1}]=d
\ea

One can then compute the 3-cycles $T_{(ijk)}$ as:
\ba
\nonumber
\vec{q}_{T_{(ijk)}}&=&\Tr[T_{(ijk)}QV_{\rm stab}^{\ot2,2}]=d^{-2}\sum_P\Tr[PV_{\rm stab}PV_{\rm stab}^\dag PV_{\rm stab}]\Tr[PV_{\rm stab}^\dag]\\
\nonumber
&=&d^{-2}\sum_{P\in\mathcal{P}_{\rm ab}}\Tr[PV_{\rm stab}PV_{\rm stab}^\dag PV_{\rm stab}]\Tr[PV_{\rm stab}^\dag]\\
\nonumber
&=&d^{-2}\sum_{P\in\mathcal{P}_{\rm ab}}\Tr[PV_{\rm stab}]\Tr[PV_{\rm stab}^\dag]\\
\nonumber
&=&d^{-2}\sum_{P}\Tr[P^{\ot2}V_{\rm stab}^{\ot1,1}]\\
&=&d^{-1}\Tr[T_{(12)}V_{\rm stab}^{\ot1,1}]=1
\ea
As for 4-cycles leading to an alternate sequence of $V,V^\dag$ one has:
\ba
\nonumber
\vec{q}_{T_{(1342)}}&=&\Tr[T_{(1342)}QV_{\rm stab}^{\ot2,2}]=d^{-2}\sum_P\Tr[PV_{\rm stab}PV_{\rm stab}^\dag PV_{\rm stab}PV_{\rm stab}^\dag]\\
\nonumber
&=&d^{-2}\sum_P\Tr[V_{\rm stab}^\dag PV_{\rm stab}P PV_{\rm stab}PV_{\rm stab}^\dag]\\
\nonumber
&=&d^{-2}\sum_P\Tr[PV_{\rm stab}^2PV_{\rm stab}^{\dag2}]\\
\nonumber
&=&d^{-2}\sum_P\Tr[T_{(12)}P^{\ot2}(V_{\rm stab}^2\ot V_{\rm stab}^{\dag2})]\\
\nonumber
&=&d^{-1}\Tr[T_{(12)}T_{(12)}(V_{\rm stab}^2\ot V_{\rm stab}^{\dag2})]\\
&=&d^{-1}\Tr[V_{\rm stab}^2]\Tr[V_{\rm stab}^{\dag2}]=d^{-1}g_2(2t)
\ea
As for the other 4-cycles:
\ba
\nonumber
\vec{q}_{T_{(ijk\ell)}}&=&\Tr[T_{(ijk\ell)}QV_{\rm stab}^{\ot2,2}]=d^{-2}\sum_P\Tr[PV_{\rm stab}PV_{\rm stab}PV_{\rm stab}^\dag PV_{\rm stab}^\dag]\\
\nonumber
&=&d^{-2}\sum_P\Tr[V_{\rm stab}PV_{\rm stab}PPV_{\rm stab}^\dag PV_{\rm stab}^\dag]\\
&=&d^{-2}\sum_P\Tr[\mathbb{I}]=d
\ea
Finally, one can compute the double swaps $T_{(12)(34)}$ and $T_{(ij)(k\ell)}$. Starting with $T_{(12)(34)}$:
\ba
\nonumber
\vec{q}_{T_{(12)(34)}}&=&\Tr[T_{(12)(34)}QV_{\rm stab}^{\ot2,2}]=d^{-2}\sum_P\Tr[PV_{\rm stab}PV_{\rm stab}]\Tr[PV_{\rm stab}^\dag PV_{\rm stab}^\dag]\\
\nonumber
&=&d^{-2}\sum_{\vec{x},\vec{z}}\sum_{i,j,k,\ell}e^{-i(E_i+E_j-E_k-E_\ell)t}\Tr[\mathbf{X}^{\vec{x}}\mathbf{Z}^{\vec{z}}\dyad{i}\mathbf{X}^{\vec{x}}\mathbf{Z}^{\vec{z}}\dyad{j}]\Tr[\mathbf{X}^{\vec{x}}\mathbf{Z}^{\vec{z}}\dyad{k} \mathbf{X}^{\vec{x}}\mathbf{Z}^{\vec{z}}\dyad{\ell}]\\
\nonumber
&=&d^{-2}\sum_{\vec{x},\vec{z}}\sum_{i,j,k,\ell}e^{-i(E_i+E_j-E_k-E_\ell)t}(-1)^{\vec{z}\cdot(i\oplus j\oplus k\oplus\ell)}\delta_{i\oplus x,j}\delta_{k\oplus x,\ell}\\
\nonumber
&=&d^{-1}\sum_{\vec{x}}\sum_{i,k}e^{-i(E_i+E_{i\oplus x}-E_k-E_{k\oplus x})t}\\
&=&d^{-1}\sum_{i,j,k}e^{-i(E_i+E_j-E_k-E_{i\oplus j\oplus k})t}=\tilde{g}_3^{\rm cb}(t)
\ea
similarly for the others $T_{(ij)(k\ell)}$ one finds:
\ba
\vec{q}_{T_{(ij)(k\ell)}}&=&\Tr[T_{(ij)(k\ell)}QV_{\rm stab}^{\ot2,2}]=d^{-2}\sum_P\Tr[PV_{\rm stab}PV_{\rm stab}^\dag]\Tr[PV_{\rm stab}PV_{\rm stab}^\dag]=\tilde{g}_3^{\rm cb}(t)
\ea

\subsection{Spectral form factors for the Toric Code Hamiltonian}

Our goal is to compute the Haar and Clifford spectral form factors for the operator $V_{\rm Tor}$. As the Toric code Hamiltonian is a stabilizer Hamiltonian we can exploit the results of App.~\ref{app:stab_ham_spectral_functions} to cut the computation short. As the Toric code Hamiltonian can be exactly diagonalized, we are going to write explicit and analytical spectral form factors. In order to compute these, we will repeatedly make use of the following facts. First, notice that each combination of products of vertex and facet operators corresponds to a Pauli string. It follows that the unitary operator V can be written as:
\ba
V_{\rm Tor}=\prod_{v,f=1}^{N^2}(\cos JtI+i\sin Jt A_v)(\cos JtI+i\sin Jt B_f)=\sum_{v,f=0}^{N^2}(\cos Jt)^{2N^2-v-f}(+i\sin Jt)^{v+f}\sum_{j,k=1}^{\binom{N^2}{v,f}}P_{vf}^{(j,k)}
\ea
where we collect the resulting Pauli strings according to the power $v+f$, corresponding to the total number of sine factors attached to that operator. Indeed, for each $v+f$ there are $\binom{N^2}{v}\binom{N^2}{f}$ different combinations of vertex and facet operators. Moreover, for notational compactness, we indicate:
\ba
\sum_{j,k=1}^{\binom{N^2}{v,f}}=\sum_{j=1}^{\binom{N^2}{v}}\sum_{k=1}^{\binom{N^2}{f}}
\ea
The most important property of the Pauli strings $P_{vf}^{(j,k)}$ that will be used in the computation of the spectral form factors is:
\ba
\nonumber
\tr{P_{vf}^{(j,k)}P_{v'f'}^{(j',k')}}&=&d\big(\delta_{v,v'}\delta_{f,f'}\delta_{j,j'}\delta_{k,k'}+\delta_{N^2-v,v'}\delta_{f,f'}\delta_{j,j'}\delta_{k,k'}\\
\label{eq:delta_toric}
&&+\delta_{v,v'}\delta_{N^2-f,f'}\delta_{j,j'}\delta_{k,k'}+\delta_{N^2-v,v'}\delta_{N^2-f,f'}\delta_{j,j'}\delta_{k,k'}\big)
\ea
which follows from Eq.~\eqref{eq:toric_prod_rule}. It can be explained heuristically noting that in order to obtain the identity operator inside the trace, given an operator $P_{vf}^{(j,k)}$, one has to have one of the following for the operator $P_{v'f'}^{(j',k')}$:
\begin{enumerate}
\item the stabilizer operators composing $P_{v'f'}^{(j',k')}$ are exactly the same as the ones composing $P_{vf}^{(j,k)}$. This corresponds to the first delta on the rhs of~\eqref{eq:delta_toric};
\item the stabilizer operators appearing in $P_{v'f'}^{(j',k')}$ are such that all the vertex (facet) operators are the same as in $P_{vf}^{(j,k)}$, while the facet (vertex) operators in $P_{v'f'}^{(j',k')}$ are exactly all the ones missing in $P_{vf}^{(j,k)}$, so that their product still gives the identity. This corresponds to the second and third Kronecker delta of Eq.~\eqref{eq:delta_toric};
\item finally, one could have that both the vertex and facet operators appearing in $P_{v'f'}^{(j',k')}$ are the ones missing in the expression of $P_{vf}^{(j,k)}$ in order to obtain the identity. This gives the fourth delta.
\end{enumerate}

We are now ready to compute the components of both $\vec{c}$ and $\vec{q}$ for the Toric code, that is, we are going to compute the regular spectral form factors $g_2^{\rm tor}(t)$, $g_3^{\rm tor}(t)$, $g_4^{\rm tor}(t)$ and  the Clifford spectral form factor $\tilde{g}_3^{\rm tor}(t)$.

Let us then start from the computation of $g_4(t)$, which is obtained when one chooses as permutation $\pi=I$:
\ba
\tr{IV_{\rm Tor}^{\ot2,2}}=|\tr{V_{\rm Tor}}|^4
\ea
One can compute the trace of $V_{\rm Tor}$ as:
\ba
\nonumber
\tr{V_{\rm Tor}}&=&\tr{\prod_{v,f=1}^{N^2}(\cos Jt I+i\sin Jt A_v)(\cos Jt I+i\sin Jt B_f)}\\
&=&\sum_{v,f=0}^{N^2}(\cos Jt)^{2N^2-v-f}(+i\sin Jt)^{v+f}\sum_{j,k=1}^{\binom{N^2}{v,f}}\tr{P_{vf}^{(j,k)}}
\ea
The only thing to note in order to perform the sum above is that only four terms, namely those for which $(v,f)=(N^2,N^2),(N^2,0),(0,N^2),(0,0)$, are different from zero. This is because the term with only cosine functions ($0,0$) is just the product of identities, and it is thus the identity. Similarly, the term with only sine functions ($N^2,N^2$) is the product of all vertex and facet operators, and is thus the identity. Finally, there are the two terms being the product of all facet and of all vertex operators, respectively ($N^2,0$) and $(0,N^2)$. Then one has:
\ba
\nonumber
&&\sum_{v,f=0}^{2N^2}(\cos Jt)^{2N^2-v-f}(+i\sin Jt)^{v+f}\sum_{j,k=1}^{\binom{N^2}{v,f}}\tr{P_{vf}^{(j,k)}}\\
\label{eq:trace_U_toric}
&&=\left[(\cos Jt)^{2N^2}+(+i\sin Jt)^{2N^2}+2(\cos Jt)^{N^2}(+i\sin Jt)^{N^2}\right]d=d\left[(\cos Jt)^{N^2}+(+i\sin Jt)^{N^2}\right]^2
\ea
It follows:
\ba
g_4^{\rm Tor}(t)=\tr{IV_{\rm Tor}^{\ot2,2}}=d^4\left|\left[(\cos Jt)^{N^2}+(+i\sin Jt)^{N^2}\right]^2\right|^4.
\ea
Let us now turn to the computation of the other spectral form factors $g_2(t)$ and $g_3(t)$, which can be obtained as:
\ba
g_3^{\rm Tor}(t)&=&\tr{V_{\rm Tor}^2}\tr{V_{\rm Tor}^\dag}^2\\
g_2^{\rm Tor}(t)&=&\tr{V_{\rm Tor}}\tr{V_{\rm Tor}^\dag}=\left|\tr{V_{\rm Tor}}\right|^2
\ea
As we have already computed $\tr{V_{\rm Tor}}$, the only other ingredient needed to compute these functions is $\Tr[V_{\rm tor}^2]$. This reads
\ba
\nonumber
\tr{V_{\rm Tor}^2}&=&\sum_{v,f,v',f'=0}^{N^2}(\cos Jt)^{4N^2-v-f-v'-f'}(+i\sin Jt)^{v+f+v'+f'}\sum_{j,k,j',k'=1}^{\binom{N^2}{v,f,v',f'}}\tr{P_{vf}^{(j,k)}P_{v'f'}^{(j',k')}}\\
\nonumber
&=&d\sum_{v,f,v',f'=0}^{N^2}(\cos Jt)^{4N^2-v-f-v'-f'}(+i\sin Jt)^{v+f+v'+f'}\times\\
&\times&\sum_{j,k,j',k'=1}^{\binom{N^2}{v,f,v',f'}}(\delta_{v,v'}\delta_{f,f'}+\delta_{N^2-v,v'}\delta_{f,f'}+\delta_{v,v'}\delta_{N^2-f,f'}+\delta_{N^2-v,v'}\delta_{N^2-f,f'})\delta_{j,j'}\delta_{k,k'}
\ea
At this point we need to compute the various combinations of deltas. Let us start from $\delta_{v,v'}\delta_{f,f'}$:
\ba
\nonumber
&&d\sum_{v,f,v',f'=0}^{N^2}(\cos Jt)^{4N^2-v-f-v'-f'}(+i\sin Jt)^{v+f+v'+f'}\sum_{j,k,j',k'=1}^{\binom{N^2}{v,f,v',f'}}\delta_{v,v'}\delta_{f,f'}\delta_{j,j'}\delta_{k,k'}\\
\nonumber
&=&d\sum_{v,f=0}^{N^2}(\cos Jt)^{4N^2-2v-2f}(+i\sin Jt)^{2v+2f}\sum_{j,k=1}^{\binom{N^2}{v,f}}\\
\nonumber
&=&d\sum_{v,f=0}^{N^2}\binom{N^2}{v}\binom{N^2}{f}(\cos^2 Jt)^{2N^2-v-f}(-\sin^2 Jt)^{v+f}\\
&=&d(\cos 2Jt)^{2N^2}
\ea
The computation for $\delta_{N^2-v,v'}\delta_{f,f'}$ and $\delta_{v,v'}\delta_{N^2-f,f'}$ is exactly the same, so that we need to compute only the former:
\ba
\nonumber
&&d\sum_{v,f,v',f'=0}^{N^2}(\cos Jt)^{4N^2-v-f-v'-f'}(+i\sin Jt)^{v+f+v'+f'}\sum_{j,k,j',k'=1}^{\binom{N^2}{v,f,v',f'}}\delta_{N^2-v,v'}\delta_{f,f'}\delta_{j,j'}\delta_{k,k'}\\
\nonumber
&=&d\sum_{v,f=0}^{N^2}(\cos Jt)^{3N^2-2f}(+i\sin Jt)^{N^2+2f}\sum_{j,k=1}^{\binom{N^2}{v,f}}\\
\nonumber
&=&d(+i\cos Jt\sin Jt)^{N^2}\sum_{v,f=0}^{N^2}\binom{N^2}{v}\binom{N^2}{f}(\cos^2 Jt)^{N^2-f}(-\sin^2 Jt)^{f}\\
\nonumber
&=&d(+i\sin 2Jt)^{N^2}(\cos 2Jt)^{N^2}
\ea
Last, we need to compute the term multiplying $\delta_{N^2-v,v'}\delta_{N^2-f,f'}$, obtaining:
\ba
\nonumber
&&d\sum_{v,f,v',f'=0}^{N^2}(\cos Jt)^{4N^2-v-f-v'-f'}(+i\sin Jt)^{v+f+v'+f'}\sum_{j,k,j',k'=1}^{\binom{N^2}{v,f,v',f'}}\delta_{N^2-v,v'}\delta_{N^2-f,f'}\delta_{j,j'}\delta_{k,k'}\\
\nonumber
&&d\sum_{v,f=0}^{N^2}(\cos Jt)^{2N^2}(+i\sin Jt)^{2N^2}\sum_{j,k=1}^{\binom{N^2}{v,f}}\\
&&d(+i\cos Jt\sin Jt)^{2N^2}\sum_{v,f=0}^{N^2}\binom{N^2}{v}\binom{N^2}{f}=d(+i\sin 2Jt)^{2N^2}
\ea
One can finally sum up all the previous results and write:
\ba
\nonumber
\tr{V_{\rm Tor}^2}&=&d\left[(\cos 2Jt)^{2N^2}+2(+i\cos 2Jt\sin 2Jt)^{N^2}+(+i\sin 2Jt)^{2N^2}\right]\\
&=&d\left[(\cos 2Jt)^{N^2}+(+i\sin 2Jt)^{N^2}\right]^2
\ea
It is worth noticing that this is the same as $\tr{V_{\rm Tor}}$ except for the argument of the trigonometric functions.

The function $g_2(t)$ and $g_3(t)$ follow immediately from Eq.~\eqref{eq:trace_U_toric}:
\ba
g_2^{\rm Tor}(t)&=&\left|\tr{V_{\rm Tor}}\right|^2=d^2\left|\left[(\cos Jt)^{N^2}+(+i\sin Jt)^{N^2}\right]^2\right|^2\\
g_3^{\rm Tor}(t)&=&\Tr[V_{\rm Tor}^2]\Tr[V_{\rm Tor}^\dag]^2=d^3\left[(\cos{2Jt})^{N^2}+(+i\sin{2Jt})^{N^2}\right]^2\left[(\cos Jt)^{N^2}+(-i\sin Jt)^{N^2}\right]^4
\ea

Let us then turn to the Clifford spectral form factor $\tilde{g}_3(t)$. Let us start with the permutation $\pi=I$. We have
\ba
\vec{q}_I=\tilde{g}^{\rm tor}_3(t)=d^{-2}\sum_P\tr{V_{\rm Tor}P}^2\tr{V_{\rm Tor}^\dag P}^2=d^{-2}\sum_P\left|\tr{V_{\rm Tor}P}\right|^4
\ea
Let us now compute $\tr{V_{\rm Tor}P}$:
\ba
\nonumber
\tr{V_{\rm Tor}P}&=&\tr{\prod_{v,f=1}^{N^2}(\cos JtI+i\sin JtA_\alpha)P}=\sum_{\alpha=0}^{2N^2}(+i\sin Jt)^\alpha(\cos Jt)^{2N^2-\alpha}\sum_{j=1}^{\binom{2N^2}{\alpha}}\tr{P_\alpha^{(j)}P}\\
&=&\sum_{\alpha=0}^{2N^2}(+i\sin Jt)^\alpha(\cos Jt)^{2N^2-\alpha}\sum_{j=1}^{\binom{2N^2}{\alpha}}d\delta_{P_\alpha^{(j)}P}
\ea
where we have made the substitution $\alpha=v+f$ and the double index $(j,k)$ has been simplified to only one index $j$ now running up to $2N^2$ instead of $N^2$. Plugging this expression back into the one for $\vec{q}_I$ one gets:
\ba
\nonumber
\tilde{g}^{\rm tor}_3(t)&=&d^{-2}\sum_P\tr{V_{\rm Tor}P}^2\tr{V_{\rm Tor}^\dag P}^2\\
\nonumber
&=&d^{-2}\sum_P\sum_{\alpha,\beta,\gamma,\delta=0}^{2N^2}(+i\sin Jt)^{\alpha+\beta}(-i\sin Jt)^{\gamma+\delta}(\cos Jt)^{8N^2-\alpha-\beta-\gamma-\delta}\times\\
\nonumber
&&\times\sum_{j=1}^{\binom{2N^2}{\alpha}}\sum_{k=1}^{\binom{2N^2}{\beta}}\sum_{\ell=1}^{\binom{2N^2}{\gamma}}\sum_{m=1}^{\binom{2N^2}{\delta}}\tr{P_{\alpha}^{(j)}P}\tr{P_{\beta}^{(k)}P}\tr{P_{\gamma}^{(\ell)}P}\tr{P_{\delta}^{(m)}P}\\
\nonumber
&=&d^{-1}\sum_{\alpha,\beta,\gamma,\delta=0}^{2N^2}(+i\sin Jt)^{\alpha+\beta}(-i\sin Jt)^{\gamma+\delta}(\cos Jt)^{8N^2-\alpha-\beta-\gamma-\delta}\times\\
\nonumber
&&\times\sum_{j=1}^{\binom{2N^2}{\alpha}}\sum_{k=1}^{\binom{2N^2}{\beta}}\sum_{\ell=1}^{\binom{2N^2}{\gamma}}\sum_{m=1}^{\binom{2N^2}{\delta}}\tr{P_{\beta}^{(k)}P_{\alpha}^{(j)}}\tr{P_{\gamma}^{(\ell)}P_{\alpha}^{(j)}}\tr{P_{\delta}^{(m)}P_{\alpha}^{(j)}}\\
\nonumber
&=&d^2\sum_{\alpha,\beta,\gamma,\delta=0}^{2N^2}(+i\sin Jt)^{\alpha+\beta}(-i\sin Jt)^{\gamma+\delta}(\cos Jt)^{8N^2-\alpha-\beta-\gamma-\delta}\times\\
\nonumber
&&\qquad\times\sum_{j=1}^{\binom{2N^2}{\alpha}}\sum_{k=1}^{\binom{2N^2}{\beta}}\sum_{\ell=1}^{\binom{2N^2}{\gamma}}\sum_{m=1}^{\binom{2N^2}{\delta}}(\delta_{\alpha,\beta}\delta_{j,k}+\delta_{2N^2-\alpha,\beta}\delta_{j,k})(\delta_{\alpha,\gamma}\delta_{j,\ell}+\delta_{2N^2-\alpha,\gamma}\delta_{j,\ell})(\delta_{\alpha,\delta}\delta_{j,m}+\delta_{2N^2-\alpha,\delta}\delta_{j,m})\\
\ea
The next step is to compute the expression above for each possible combination of Kronecker's deltas. To simplify things, let us first notice that in any of the combinations, one will have the product $\delta_{j,k}\delta_{j,\ell}\delta_{j,m}$, which will have the effect of canceling the sums over $k,\ell,m$. Thus, one is left to compute the remnant of the sum for each possible combination,i.e.:
\ba
\nonumber
&&d^2\sum_{\alpha,\beta,\gamma,\delta=0}^{2N^2}\binom{2N^2}{\alpha}(+i\sin Jt)^{\alpha+\beta}(-i\sin Jt)^{\gamma+\delta}(\cos Jt)^{8N^2-\alpha-\beta-\gamma-\delta}\times\\
&\times&(\delta_{\alpha,\beta}+\delta_{2N^2-\alpha,\beta})(\delta_{\alpha,\gamma}+\delta_{2N^2-\alpha,\gamma})(\delta_{\alpha,\delta}+\delta_{2N^2-\alpha,\delta})
\ea
Let us start from the first combination, $\delta_{\alpha,\beta}\delta_{\alpha,\gamma}\delta_{\alpha,\delta}$:
\ba
\nonumber
&&d^2\sum_{\alpha,\beta,\gamma,\delta=0}^{2N^2}\binom{2N^2}{\alpha}(+i\sin Jt)^{\alpha+\beta}(-i\sin Jt)^{\gamma+\delta}(\cos Jt)^{8N^2-\alpha-\beta-\gamma-\delta}\delta_{\alpha,\beta}\delta_{\alpha,\gamma}\delta_{\alpha,\delta}\\
\nonumber
&=&d^2\sum_{\alpha=0}^{2N^2}\binom{2N^2}{\alpha}(+i\sin Jt)^{2\alpha}(-i\sin Jt)^{2\alpha}(\cos Jt)^{8N^2-4\alpha}\\
\nonumber
&=&d^2\sum_{\alpha=0}^{2N^2}\binom{2N^2}{\alpha}(-\sin^2 Jt)^{\alpha}(-\sin^2 Jt)^{\alpha}(\cos^4 Jt)^{2N^2-\alpha}\\
&=&d^2\sum_{\alpha=0}^{2N^2}\binom{2N^2}{\alpha}(\sin^4 Jt)^{\alpha}(\cos^4 Jt)^{2N^2-\alpha}=d^2(\sin^4 Jt+\cos^4 Jt)^{2N^2}
\ea
Let us then move to the combination $\delta_{\alpha,\beta}\delta_{\alpha,\gamma}\delta_{2N^2-\alpha,\delta}$:
\ba
\nonumber
&&d^2\sum_{\alpha,\beta,\gamma,\delta=0}^{2N^2}\binom{2N^2}{\alpha}(+i\sin Jt)^{\alpha+\beta}(-i\sin Jt)^{\gamma+\delta}(\cos Jt)^{8N^2-\alpha-\beta-\gamma-\delta}\delta_{\alpha,\beta}\delta_{\alpha,\gamma}\delta_{2N^2-\alpha,\delta}\\
\nonumber
&=&d^2\sum_{\alpha=0}^{2N^2}\binom{2N^2}{\alpha}(+i\sin Jt)^{2\alpha}(-i\sin Jt)^{2N^2}(\cos Jt)^{6N^2-2\alpha}\\
\nonumber
&=&(-i\sin Jt\cos Jt)^{2N^2}d^2\sum_{\alpha=0}^{2N^2}\binom{2N^2}{\alpha}(-\sin^2 Jt)^{\alpha}(\cos^2 Jt)^{2N^2-\alpha}\\
&=&(-i\sin 2Jt)^{2N^2}d(\cos^2 Jt-\sin^2 Jt)^{2N^2}=d(-i\sin 2Jt)^{2N^2}(\cos 2Jt)^{2N^2}
\ea
Similarly for the combinations $\delta_{\alpha,\beta}\delta_{2N^2-\alpha,\gamma}\delta_{\alpha,\delta}$ and $\delta_{2N^2-\alpha,\beta}\delta_{\alpha,\gamma}\delta_{\alpha,\delta}$ one obtains respectively:
\ba
\nonumber
&&d^2\sum_{\alpha,\beta,\gamma,\delta=0}^{2N^2}\binom{2N^2}{\alpha}(+i\sin Jt)^{\alpha+\beta}(-i\sin Jt)^{\gamma+\delta}(\cos Jt)^{8N^2-\alpha-\beta-\gamma-\delta}\delta_{\alpha,\beta}\delta_{2N^2-\alpha,\gamma}\delta_{\alpha,\delta}=d(-i\sin 2Jt)^{2N^2}(\cos 2Jt)^{2N^2}\\
\nonumber
&&d^2\sum_{\alpha,\beta,\gamma,\delta=0}^{2N^2}\binom{2N^2}{\alpha}(+i\sin Jt)^{\alpha+\beta}(-i\sin Jt)^{\gamma+\delta}(\cos Jt)^{8N^2-\alpha-\beta-\gamma-\delta}\delta_{2N^2-\alpha,\beta}\delta_{\alpha,\gamma}\delta_{\alpha,\delta}=d(i\sin 2Jt)^{2N^2}(\cos 2Jt)^{2N^2}
\ea
We can then move to the combination $\delta_{\alpha,\beta}\delta_{2N^2-\alpha,\gamma}\delta_{2N^2-\alpha,\delta}$:
\ba
\nonumber
&&d^2\sum_{\alpha,\beta,\gamma,\delta=0}^{2N^2}\binom{2N^2}{\alpha}(+i\sin Jt)^{\alpha+\beta}(-i\sin Jt)^{\gamma+\delta}(\cos Jt)^{8N^2-\alpha-\beta-\gamma-\delta}\delta_{\alpha,\beta}\delta_{2N^2-\alpha,\gamma}\delta_{2N^2-\alpha,\delta}\\
\nonumber
&=&d^2\sum_{\alpha=0}^{2N^2}\binom{2N^2}{\alpha}(+i\sin Jt)^{2\alpha}(-i\sin Jt)^{4N^2-2\alpha}(\cos Jt)^{4N^2}\\
\nonumber
&=&d^2(\cos Jt)^{4N^2}\sum_{\alpha=0}^{2N^2}\binom{2N^2}{\alpha}(-\sin^2 Jt)^{\alpha}(-\sin^2 Jt)^{2N^2-\alpha}=d^2(\cos Jt)^{4N^2}(-2\sin^2 Jt)^{2N^2}\\
\nonumber
&=&d^2(\cos^2 Jt)^{2N^2}(-2\sin^2 Jt)^{2N^2}=d^2(-\cos Jt\sin Jt\sin 2Jt)=d(-\sin^2 2Jt)^{2N^2}=d(\sin^2 2Jt)^{2N^2}\\
\ea
Similarly for the combinations $\delta_{2N^2-\alpha,\beta}\delta_{\alpha,\gamma}\delta_{2N^2-\alpha,\delta}$ and $\delta_{2N^2-\alpha,\beta}\delta_{2N^2-\alpha,\gamma}\delta_{\alpha,\delta}$:
\ba
\nonumber
&&d^2\sum_{\alpha,\beta,\gamma,\delta=0}^{2N^2}\binom{2N^2}{\alpha}(+i\sin Jt)^{\alpha+\beta}(-i\sin Jt)^{\gamma+\delta}(\cos Jt)^{8N^2-\alpha-\beta-\gamma-\delta}\delta_{2N^2-\alpha,\beta}\delta_{\alpha,\gamma}\delta_{2N^2-\alpha,\delta}=d(\sin^2 2Jt)^{2N^2}\\\\
\nonumber
&&d^2\sum_{\alpha,\beta,\gamma,\delta=0}^{2N^2}\binom{2N^2}{\alpha}(+i\sin Jt)^{\alpha+\beta}(-i\sin Jt)^{\gamma+\delta}(\cos Jt)^{8N^2-\alpha-\beta-\gamma-\delta}\delta_{2N^2-\alpha,\beta}\delta_{2N^2-\alpha,\gamma}\delta_{\alpha,\delta}=d(\sin^2 2Jt)^{2N^2}\\
\ea
Finally, for the combination $\delta_{2N^2-\alpha,\beta}\delta_{2N^2-\alpha,\gamma}\delta_{2N^2-\alpha,\delta}$ one gets:
\ba
\nonumber
&&d^2\sum_{\alpha,\beta,\gamma,\delta=0}^{2N^2}\binom{2N^2}{\alpha}(+i\sin Jt)^{\alpha+\beta}(-i\sin Jt)^{\gamma+\delta}(\cos Jt)^{8N^2-\alpha-\beta-\gamma-\delta}\delta_{2N^2-\alpha,\beta}\delta_{2N^2-\alpha,\gamma}\delta_{2N^2-\alpha,\delta}\\
\nonumber
&=&d^2\sum_{\alpha=0}^{2N^2}\binom{2N^2}{\alpha}(+i\sin Jt)^{2N^2}(-i\sin Jt)^{4N^2-2\alpha}(\cos Jt)^{2N^2+2\alpha}\\
\nonumber
&=&d^2(+i\sin Jt\cos Jt)^{2N^2}\sum_{\alpha=0}^{2N^2}\binom{2N^2}{\alpha}(-\sin^2 Jt)^{2N^2-\alpha}(\cos^2 Jt)^{\alpha}\\
&=&d(+i\sin 2Jt)^{2N^2}(\cos 2Jt)^{2N^2}
\ea

One can then sum up all the obtained terms to get:
\ba
\tilde{g}^{\rm tor}_3(t)=d^2(\cos^4 Jt+\sin^4 Jt)^{2N^2}+3d(\sin(2Jt))^{2N^2}+4(-1)^{N^2}(\sin 4Jt)^{2N^2}
\ea

\begin{figure}[!ht]
\centering
\begin{subfigure}{0.49\textwidth}
\centering
\includegraphics[width=\linewidth]{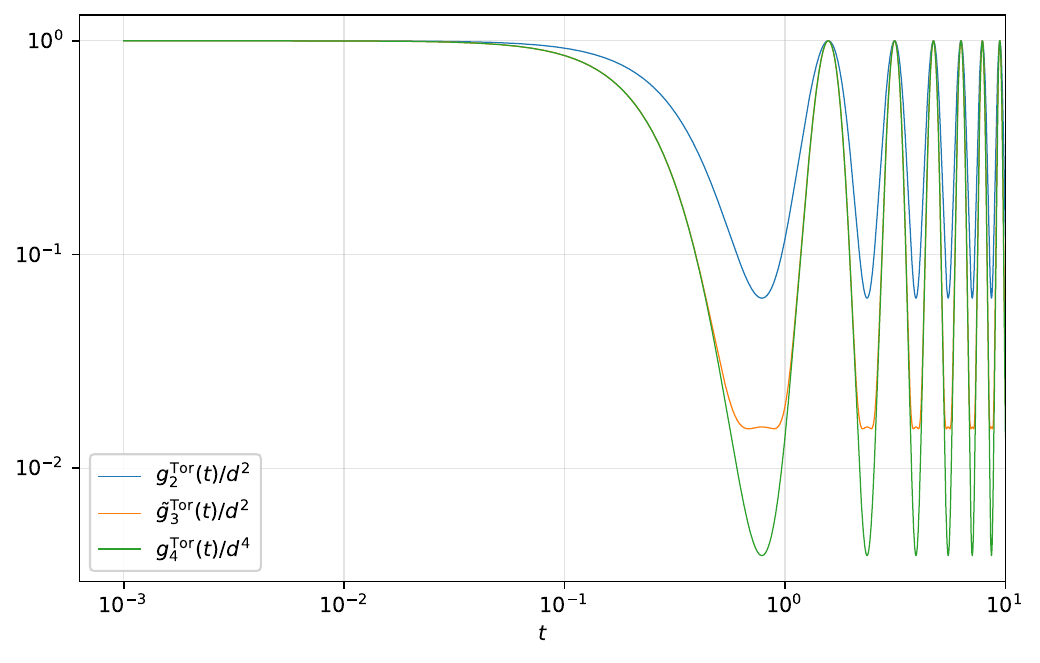}
\caption{Log-log plot of the Toric Code spectral form factors for $N=2$.}
\label{fig:g_tor_comparison_N2}
\end{subfigure}
\begin{subfigure}{0.49\textwidth}
\centering
\includegraphics[width=\linewidth]{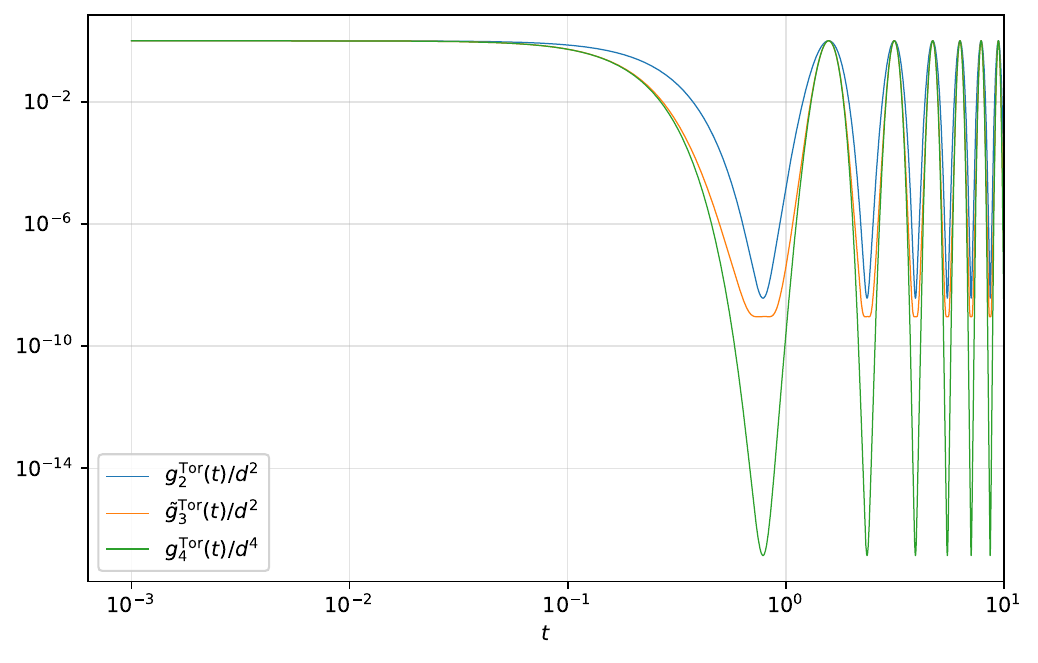}
\caption{Log-log plot of the Toric Code spectral form factors for $N=4$.}
\label{fig:g_tor_comparison_N4}
\end{subfigure}
\caption{Comparison between $g_2(t),\tilde{g}_3(t),g_4(t)$ for the Toric Code for lattice size $N=2$ and $N=4$}
\label{fig:g3_tilde_Tor}
\end{figure}

\section{Probes of chaos}
\subsection{\label{app:loschmidt}Loschmidt echo}
We can write the expressions for the Loschmidt echo of the second kind in matrix notation, which are more useful for practical calculations. We have:
\ba
&&\langle\mathcal{L}_2(t)\rangle_{\cal U}=\vec{c}\, W\cdot\vec{L}_2^{H}\\
&&\langle\mathcal{L}_2(t)\rangle_{\cal C}=\vec{q}\,W^{+}\vec{L}_2^{Q}+\vec{q}_\perp W^{-}\vec{L}_2^{Q^\perp}\\
&&\langle\mathcal{L}_2(t)\rangle_{\mathcal{C}_k}=\vec{t}\,\Xi^k\vec{L}_2^{Q}+\vec{t}\,\Gamma^{(k)}\vec{L}_2^H+\vec{b}\cdot\vec{L}_2^H
\ea
where we have defined:
\ba
(\vec{L}_2^H)_\pi&=&\tr{T_\pi T_{(14)(23)}A^{\ot2,2}}\\
(\vec{L}_2^Q)_\pi&=&\tr{T_\pi T_{(14)(23)}QA^{\ot2,2}}\\
(\vec{L}_2^{Q^\perp})_\pi&=&\tr{T_\pi T_{(14)(23)}Q^{\perp}A^{\ot2,2}}
\ea

Assuming the operator $A$ to be a Pauli string, one can compute the components of the vectors $\vec{L}_2^H,\vec{L}_2^Q$, summarized in Table~\ref{tab:loschmidt}. Further details on the computation of these components are found below.

\begin{table}[!th]
\begin{tabular}{|c|c|c|}
\hline
$T_\pi$&$\tr{T_\pi P^{\ot4}}$&$\tr{T_\pi QP^{\ot4}}$\\
\hline
$I$&$0$&$d^2$\\
$T_{(ij)}$&$0$&$d$\\
$T_{(ijk)}$&$0$&$1$\\
$T_{(ijk\ell)}$&$d$&$d$\\
$T_{(ij)(k\ell)}$&$d^2$&$d^2$\\
\hline
\end{tabular}
\caption{Value of the components of the vectors $\vec{L}_2^H,\vec{L}_2^{Q}$.}
\label{tab:loschmidt}
\end{table}

We can compute the components of the vector $\vec{L}_V^Q=V^{T}\vec{L}_2^{Q}$, obtaining:
\ba
&&(\vec{L}_V^Q)_+=\sqrt{\frac{2}{3}}(d-2)(d-1)\\
&&(\vec{L}_V^Q)_-=\sqrt{\frac{2}{3}}(d+2)(d+1)\\
&&(\vec{L}_V^Q)_1=\sqrt{2}(d^2-1)\\
&&(\vec{L}_V^Q)_2=0\\
&&(\vec{L}_V^Q)_3=0\\
&&(\vec{L}_V^Q)_4=-\sqrt{\frac{2}{3}}(d^2-1)
\ea
We can do the same operation on the Haar components, i.e. compute $\vec{L}_V^H=V^{T}\vec{L}_2^H$:
\ba
&&(\vec{L}_V^H)_+=\sqrt{\frac{3}{8}}d(d-2)\\
&&(\vec{L}_V^H)_-=\sqrt{\frac{3}{8}}d(d+2)\\
&&(\vec{L}_V^H)_1=\frac{3d^2}{2\sqrt{2}}\\
&&(\vec{L}_V^H)_2=0\\
&&(\vec{L}_V^H)_3=0\\
&&(\vec{L}_V^H)_4=-\sqrt{\frac{3}{8}}d^2
\ea
Let us now compute a couple of components of the vectors $\vec{L}_2^H$ and $\vec{L}_2^Q$ to show how the calculation is performed.
Let us consider for instance the case of the permutation $\pi=I$. We have:
\ba
(\vec{L}_2^H)_I=\tr{IP^{\ot 4}}=\tr{P}^4=0
\ea
because all Pauli strings different from the identity are traceless. The corresponding component of the vector $\vec{L}_2^Q$ reads:
\ba
(\vec{L}_2^Q)_I=d^{-2}\sum_{P'}\tr{PP'}^4=d^{2}\sum_{P'}\delta_{PP'}=d^2
\ea
One can proceed in a similar way for all the other permutations and obtain the values shown in Table~\ref{tab:loschmidt}.

The formula of the T-doped average is:
\ba
\nonumber
&&\langle\mathcal{L}_2\rangle_{\mathcal{C}_T}=\frac{\frac{3}{16}P_0(d)+\frac{3}{16}(-9+d^{2})B(t)\cos(4\theta)-\frac{1}{2}(4-5d^{2}+d^{4})K(\theta)}{3(-1+d^{2})^{2}\bigl(36 - 13 d^{2}+d^{4}\bigr)}
\ea
with
\ba
P_0(t)&=&c_0+c_1d+c_2d^2+c_3d^3+c_4d^4+56d^5-56d^6-16d^7\\
B(t)&=&-12g_2(2t)+2\left(g_3(t)+g_3^{*}(t)\right)d+\left(12 +g_2(2t)-4g_2(t)\right)d^{2}-d^{4}\\
K(\theta)&=&\Xi_+(t)\xi_+^k+\Xi_1(t)\xi_1^k+\Xi_-(t)\xi_-^k
\ea
where:
\ba
\nonumber
\Xi_+(t)&=&d(3+d)\Bigl[-g_3(t)+g_4(t)-g_3^{*}(t)-2g_2(2t)+dg_2(2t)+8g_2(t)-4dg_2(t)+(3d-d^{2})\tilde{g}_3-6 d^{2}+2d^{3}\Bigr]\\
\nonumber
\Xi_1(t)&=&4(-9+d^{2})\Bigl[g_2(2t)-4g_2(t)+g_4(t)+(3-\tilde{g}_3)d^{2}\Bigr]\\
\nonumber
\Xi_-(d)&=&d(3-d)\Bigl[-g_3(t)-g_4(t)-g_3^{*}(t)+2g_2(2t)+dg_2(2t)-8g_2(t)-4dg_2(t)+d(3+d)\,\widetilde{g}_3+6d^{2}+2d^{3}\Bigr].
\ea
and
\ba
c_0&=&96g_2(2t)-384 g_2(t)+24 g_3(t)+96 g_4(t)+24 g_3^*(t),\\
c_1 &=& -12 g_2(2t)-48 g_2(t)+8 g_3(t)+8 g_3^*(t),\\
c_2 &=& -192+36 g_2(2t)-176 g_2(t)+13 g_3(t)+40 g_4(t)+13 g_3^*(t),\\
c_3 &=& -52+24 g_2(2t)+54 g_2(t)-9 g_3(t)-9 g_3^*(t),\\
c_4 &=& 244+50 g_2(t)-8 g_3(t)-8 g_4(t)-8 g_3^*(t).
\ea

\subsection{OTOC\label{app:otoc}}
Also in this case to compute  the components of the vectors $\vec{O}_4^H,\vec{O}_4^Q,\vec{O}_4^{Q^\perp}$ one has to compute the traces of the form $\tr{T_\pi A^{\ot1,1}\ot B^{\ot1,1}}$ and $\tr{T_\pi QA^{\ot1,1}\ot B^{\ot1,1}}$. In particular, we assume $A$ and $B$ to be non-overlapping, i.e. different and commuting, Pauli strings. The values of these traces are reported in Table~\ref{tab:otoc}.

\begin{table}[!th]
\begin{tabular}{|c|c|c|}
\hline
$T_\pi$&$\tr{T_\pi P^{\ot2}\ot P'^{\ot2}}$&$\tr{T_\pi QP^{\ot2}\ot P'^{\ot2}}$\\
\hline
$I$&$0$&$0$\\
$T_{(ij)/\{(12),(34)\}}$&$0$&$0$\\
$T_{(12)},T_{(34)}$&$0$&$d$\\
$T_{(ijk)}$&$0$&$1$\\
$T_{(ijk\ell)/\{(1324),(1423)\}}$&$d$&$0$\\
$T_{\{(1324),(1423)\}}$&$d$&$d$\\
$T_{(ij)(k\ell)/\{(12)(34)\}}$&$0$&$0$\\
$T_{(12)(34)}$&$d^2$&$0$\\
\hline
\end{tabular}
\caption{}
\label{tab:otoc}
\end{table}
Also for the case of OTOCs we can compute the vectors $\vec{O}_V^H=V^{T}\vec{O_4^H}$ and $\vec{O}_V^Q=V^{T}\vec{O}_4^Q$. We obtain:
\ba
&&(\vec{O}_V^Q)_+=\sqrt{\frac{2}{3}}(d-2)\\
&&(\vec{O}_V^Q)_-=\sqrt{\frac{2}{3}}(d+2)\\
&&(\vec{O}_V^Q)_1=\sqrt{2}d\\
&&(\vec{O}_V^Q)_2=\sqrt{2}\\
&&(\vec{O}_V^Q)_3=\sqrt{\frac{2}{3}}\\
&&(\vec{O}_V^Q)_4=-\sqrt{\frac{2}{3}}d
\ea
for the vector $\vec{L}_V^Q$, while for the Haar case we get:
\ba
&&(\vec{O}_V^H)_+=\frac{d(d-6)}{2\sqrt{6}}\\
&&(\vec{O}_V^H)_-=-\frac{d(d-6)}{2\sqrt{6}}\\
&&(\vec{O}_V^H)_1=0\\
&&(\vec{O}_V^H)_2=-\frac{d^2}{2\sqrt{2}}\\
&&(\vec{O}_V^H)_3=-\frac{d^2}{2\sqrt{6}}\\
&&(\vec{O}_V^H)_4=0
\ea

Computing the components of the vectors $\vec{O}_4^H,\vec{O}_4^Q,\vec{O}_4^{Q^\perp}$ goes the same way as for the Loschmidt echo vectors, with the only difference that we replace $P^{\ot4}$ with $P^{\ot2}\otimes P^{'\ot2}$. Let us this time  consider the component $\pi=T_{(12)}$. We have:
\ba
\left(\vec{O}_4^H\right)_{T_{(12)}}=\tr{T_{(12)}P^{\ot2}\otimes P'^{\ot2}}=\tr{P^2}\tr{P'}^2=0.
\ea
The corresponding component of the vector $\vec{O}_4^Q$ reads:
\ba
\left(\vec{O}_4^Q\right)_{T_{(12)}}&=&\tr{T_{(12)}QP^{\ot2}\otimes P'^{\ot2}}=d^{-2}\sum_{P''}\tr{P''PP''P}\tr{P''P'}^2\\
&=&\sum_{P''}\delta_{P'P''}\tr{P''PP''P}=\tr{(PP')^2}=d
\ea
where in the last passage we use the assumption of $P,P'$ being non-overlapping, i.e. $\comm{P}{P'}=0$.

Finally, here we report the expression for the T-doped average:
\ba
\nonumber
&&\langle{\rm OTOC}_4\rangle_{\mathcal{C}_T}=\frac{-\frac{3}{16}\,\mathcal{P}_0(d)+
\frac{3}{16}B(d)\cos(4\theta)+\frac{1}{2}\,(d^2-1)\Bigl[
\Xi_+(t)\xi_+^{k}
+\Xi_1(t)\xi_1^{k}
+\Xi_-(t)\xi_-^{k}
\Bigr]}{3\,(1-d^{2})^{2}\,\bigl(36-13d^{2}+d^{4}\bigr)}\\
\ea
with
\ba
P_0(t)&=&c_0+ c_1 d+ c_2 d^2+ c_3 d^3+ c_4 d^4+ c_5 d^5+ 8 d^7,\\
B(t)&=&d(-9 + d^{2})
\Bigl(4g_2(2t)-d\bigl(g_3(t)+g_3^{*}(t)+4d- 2 g_2(t)d\bigr)\Bigr).\\
\Xi_+(d)&=&(3+d)d(2+d)\bigl(g_3(t)+g_3^{*}(t)\bigr)-d(3+d)(-4 + d^{2})g_2(2t)\\
&&-d(3+d)(2+d)\Bigl(g_4(t)-4 g_2(t)(d-2)-(\tilde{g}_3-2d)(-3+d)d\Bigr).\\
\Xi_1(t)
&=&(3+d)(-3+d)\left[5\bigl(g_3(t)+g_3^{*}(t)\bigr)-9 dg_2(2t)+d\Bigl(6 g_2(t)-4g_4(t)+(-7+4\tilde{g}_3)d^{2}\Bigr)\right]\\
\nonumber
\Xi_-(t)&=&-(-3+d)(-2+d)d\bigl(g_3(t)+g_3^{*}(t)\bigr)\\
&&+d(-3+d)(-2+d)\Bigl[g_2(2t)(2+d)-g_4(t)-4g_2(t)(2+d)+d(3 + d)\bigl(\tilde{g}_3 +2d\bigr)\Bigr].
\ea
where
\ba
c_0 &=& 120\bigl(g_3(t)+g_3^{*}(t)\bigr),\\
c_1 &=& -28\,g_2(2t) + 400\,g_2(t) - 64\,g_4(t) - 576,\\
c_2 &=& -151\bigl(g_3(t)+g_3^{*}(t)\bigr),\\
c_3 &=& 748 + 4\,g_2(2t) - 498\,g_2(t) + 80\,g_4(t),\\
c_4 &=& 39\bigl(g_3(t)+g_3^{*}(t)\bigr),\\
c_5 &=& -148 - 8\,g_2(2t) + 82\,g_2(t) - 16\,g_4(t).
\ea
\subsection{\label{app:entropy}Entanglement entropy}

Another context in which the isospectral twirling reveals useful is the evolution of entanglement~\cite{Page1993,ZanardiZalkaFaoro2000,Zanardi2001,WangGhoseSandersHu2004,VidmarRigol2017,KumariGhose2019,MezeiStanford2017,PhysRevA.98.032119,YouGu2018,rz86-47h3,Iannotti2025entanglement}. In fact, an important measure of the entanglement present in a quantum system is the set of $\alpha$-R\'enyi entropies.
They have found wide use in diverse field, such as condensed matter~\cite{Vidal2003,Srednicki1993}, quantum field theory~\cite{CalabreseCardy2004}, and quantum holography~\cite{RyuTakayanagi2006,doi:10.1142/S0219887824400103}. 

Given an initial pure state $\psi$, one can define the partial state over a bipartition of the system as:
\ba
\psi_A=\Tr_B\left[\dyad{\psi}\right]=\sum_{i=1}^d\lambda_i\dyad{i}
\ea
The more mixed the partial state, the more the entanglement between the bipartition. One can then define the $\alpha$-R\'enyi entropy as:
\ba
S_\alpha(\psi_A)=(1-\alpha)^{-1}\log\sum_i\lambda_i^\alpha=(1-\alpha)^{-1}\log\Tr[\psi_A^\alpha]
\ea
Let us focus on the 2-R\'enyi entropy. For a system evolving under a unitary dynamics $V$, one can upper bound the entanglement of the system by computing the reduced state with respect to an arbitrary bipartition of the system, and then evaluating the purity $\rm{Pur}(\psi_A)=\tr{\psi_A^2}$. First, one can exploit the swap trick to write the purity as:
\ba
\tr{\psi_A^2}=\tr{T_{(12)}^{(A)}V^{\ot2}\psi^{\ot2}V^{\dag\ot2}}=\tr{T_{(13)(24)}(V^{\ot2,2})(\psi^{\ot2}\ot T_{(12)}^{(A)})}
\ea
where the operator $T_{(12)}^{(A)}=d_A^{-1}\sum_{P_A}(P_A\ot\mathbb{I}_B)^{\ot2}$ is a swap acting on the two copies of partition $A$.
One can then compute the average over a group and lower bound exploiting the Jensen's inequality:
\ba
\langle S_2\rangle_{\cal G}&=&\left\langle-\log\tr{T_{(13)(24)}(V^{\ot2,2})(\psi^{\ot2}\ot T_{(12)}^{(A)})}\right\rangle_{\cal G}\\
\nonumber
&\geq&-\log\left\langle\tr{T_{(13)(24)}(V^{\ot2,2})(\psi^{\ot2}\ot T_{(12)}^{(A)})}\right\rangle_{\cal G}=-\log\tr{T_{(13)(24)}\mathcal{R}_{\cal G}^{(4)}(V)(\psi^{\ot2}\ot T_{(12)}^{(A)})}
\ea

\begin{figure}[!th]
\begin{subfigure}{0.49\textwidth}
\centering
\includegraphics[width=\textwidth]{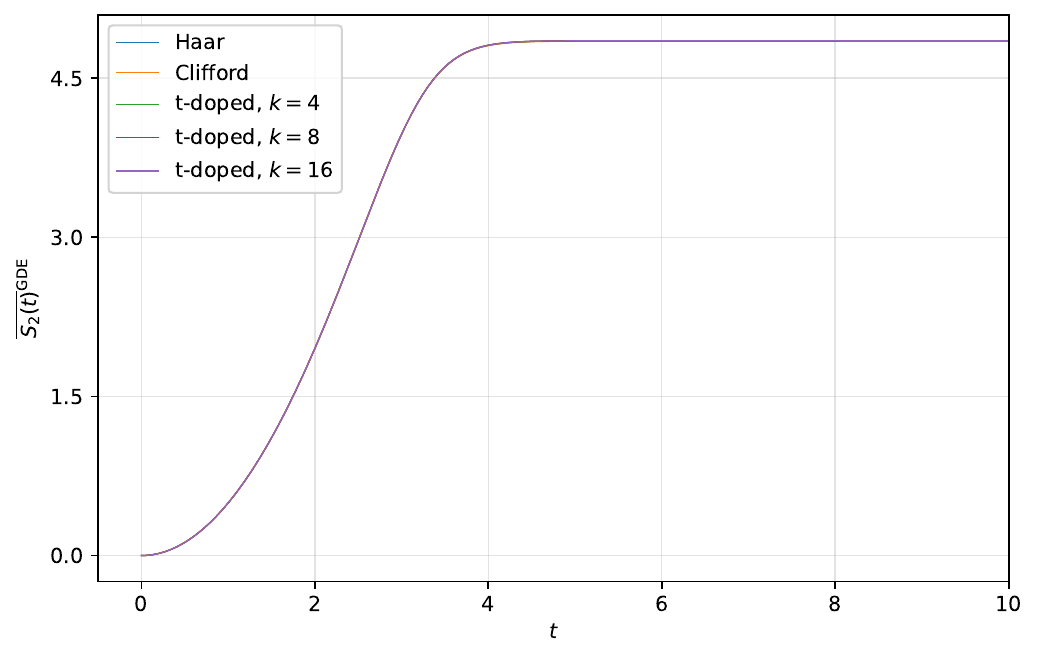}
\caption{Plot of $\overline{\langle S_2(t)\rangle}^{GDE}$.}
\label{fig:entropy_GDE}
\end{subfigure}
\begin{subfigure}{0.49\textwidth}
\centering
\includegraphics[width=\textwidth]{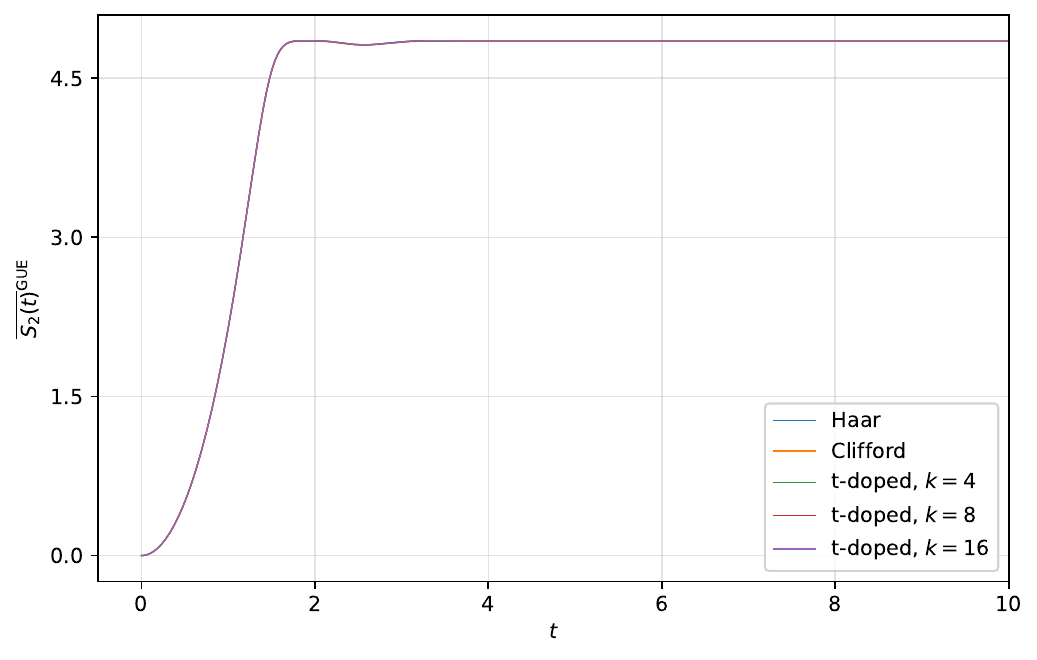}
\caption{Plot of $\overline{\langle S_2(t)\rangle}^{GUE}$.}
\label{fig:entropy_GUE}
\end{subfigure}
\caption{Plot of the bound on the 2-R\'enyi entropy averaged over the GDE(~\ref{fig:entropy_GDE}) and the GUE(~\ref{fig:entropy_GUE}) for $d=2^{16}$. One notices that averaging over the Clifford or unitary group makes little difference, as the R\'enyi entropies are measures of entanglement, which can be efficiently generated via Clifford operations.}
\label{fig:entropy}
\end{figure}

What is left to do is to compute the average purity, i.e. the average of the argument of the logarithm. For the averages over the Unitary and Clifford groups one gets:
\ba
\left\langle\rm{Pur}_{AB}\right\rangle_{\cal U}&=&
-\Bigg[g_2(2t)-4 g_2(t)
+ g_3(t)+g_4(t)+g_3^*(t)-g_2(2t)\sqrt{d}\\
\nonumber
&&-\sqrt{d}\Bigl(-4g_2(t)+g_3(t)+g_4(t)+g_3^*(t)+2(1+\sqrt{d})d^{2}(3+d)
\Bigr)\Bigg]/\left((1+\sqrt{d}) d^{2}(1+d)(3+d)\right)\\
\left\langle\rm{Pur}_{AB}\right\rangle_{\cal C}&=&\frac{g_2(2t)(\sqrt{d}-1)+\tilde{g}_3(\sqrt{d}-1)d+2d(1+d)\bigl(1+\sqrt{d}+d\bigr)}{(1+\sqrt{d})d(1+d)(2+d)}
\ea

We can see the results in the plots in Fig.~\ref{fig:entropy}. As one can see, averaging over the Clifford or unitary group makes little difference when it comes to the 2-R\'enyi entropy. This is indeed consistent with the fact that the Clifford group is able to generate arbitrarily high levels of entanglement. Thus, adding t-gates and injecting non-stabilizerness makes no effect when entanglement entropy is concerned. Finally, the asymptotic values and the equilibration times are the same as the ones obtained in~\cite{10.21468/SciPostPhys.10.3.076}.

\begin{figure}[!ht]
\begin{subfigure}{0.33\textwidth}
\centering
\includegraphics[width=\linewidth]{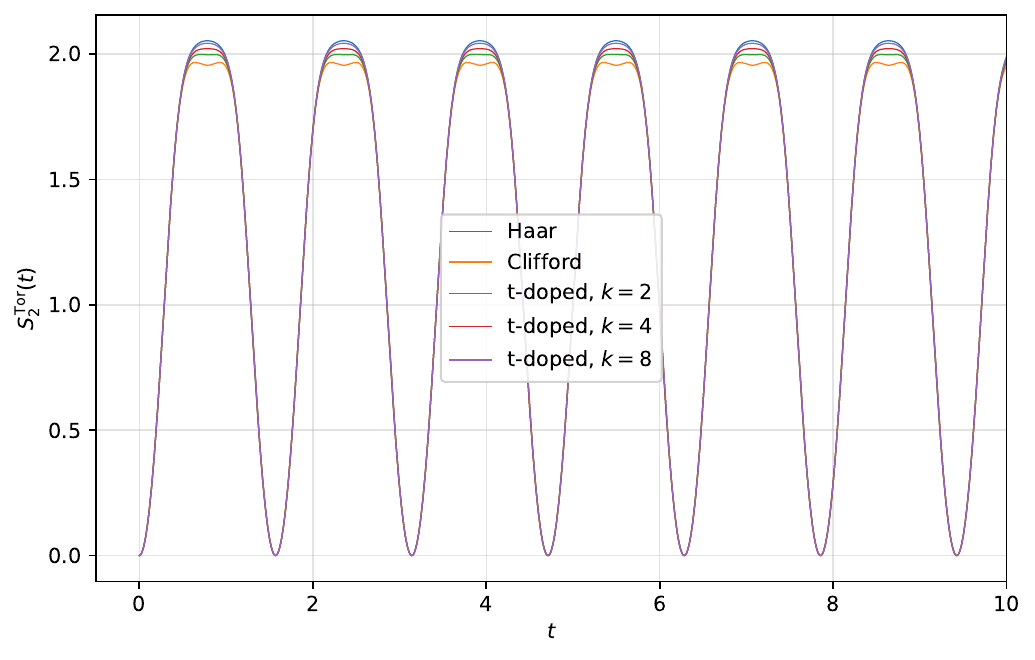}
\caption{$N=2$}
\label{fig:entropy_Tor_N2}
\end{subfigure}
\begin{subfigure}{0.33\textwidth}
\centering
\includegraphics[width=\linewidth]{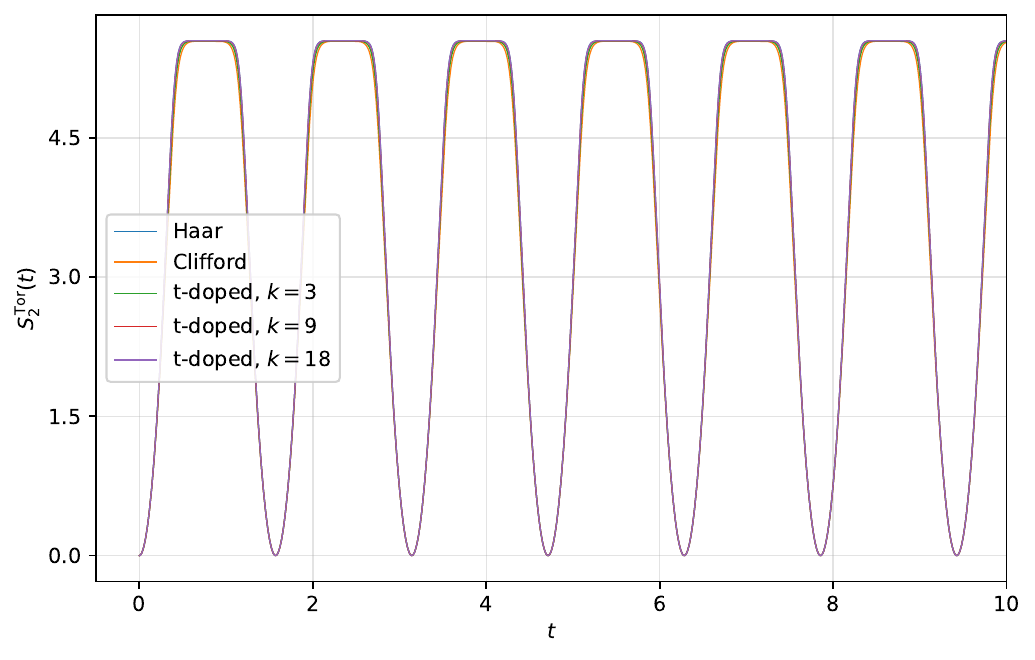}
\caption{$N=3$}
\label{fig:entropy_Tor_N3}
\end{subfigure}
\begin{subfigure}{0.33\textwidth}
\centering
\includegraphics[width=\linewidth]{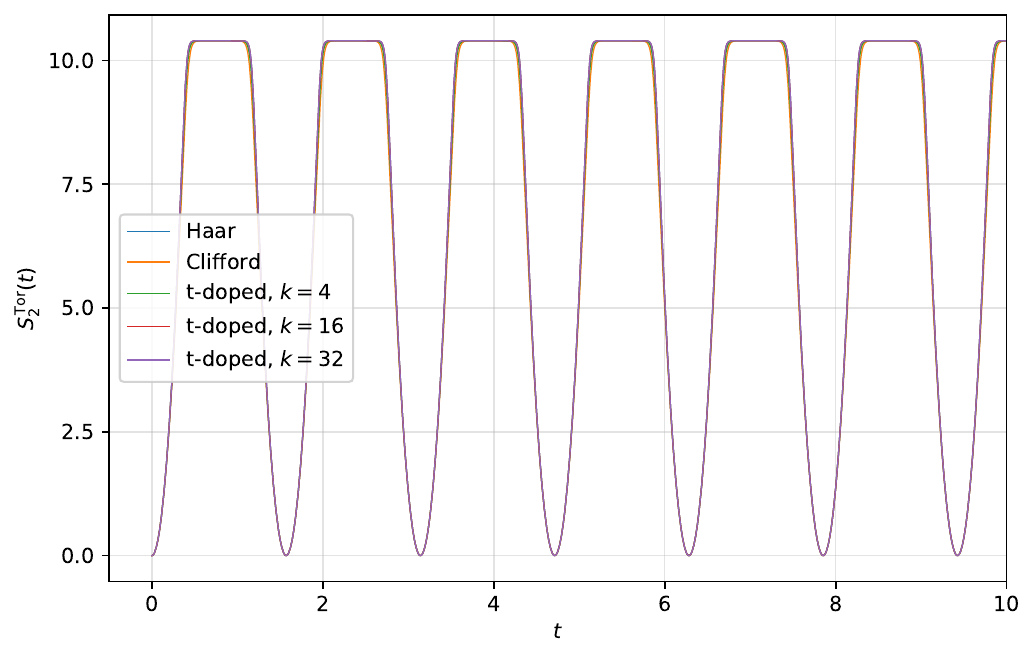}
\caption{$N=4$}
\label{fig:entropy_Tor_N4}
\end{subfigure}
\caption{Plot of the entanglement entropy $S_2$ for the Toric code for different lattice size.}
\label{fig:entropy_Tor}
\end{figure}

Let us set $\rm{Pur}_{AB}=\tr{T_{(13)(24)}(V^{\ot2,2})(\psi^{\ot2}\ot T_{(12)}^{(A)})}$. For the average value under isospectral twirling, we can use the usual matrix notation and obtain:
\ba
\left\langle\rm{Pur}_{AB}\right\rangle_{\cal U}&=&\vec{c}\,W\vec{p}_H\\
\left\langle\rm{Pur}_{AB}\right\rangle_{\cal C}&=&\vec{q}\,W^+\vec{p}_Q+\vec{q}_\perp\, W^-\vec{p}_{Q^\perp}\\
\left\langle\rm{Pur}_{AB}\right\rangle_{{\cal C}_k}&=&\vec{t}\,\Xi^k\vec{p}_Q+\vec{t}\,\Gamma^{(k)}\vec{p}_H+\vec{b}\cdot\vec{p}_H
\ea
where we have defined:
\ba
\left(\vec{p}_H\right)_\pi&=&\tr{T_{(12)(34)}T_\pi(\psi^{\ot2}\otimes T_{(12)}^{(A)})}\\
\left(\vec{p}_Q\right)_\pi&=&\tr{T_{(12)(34)}QT_\pi(\psi^{\ot2}\otimes T_{(12)}^{(A)})}\\
\left(\vec{p}_{Q^\perp}\right)_\pi&=&\tr{T_{(12)(34)}Q^\perp T_\pi(\psi^{\ot2}\otimes T_{(12)}^{(A)})}.\\
\ea

Once again, one has to evaluate the components of the vectors $\vec{p}_H,\vec{p}_Q,\vec{p}_{Q^\perp}$ and to compute traces of the form $\tr{T_\pi(\psi^{\ot2}\ot T_{(12)}^{(A)})}$ and $\tr{T_\pi Q(\psi^{\ot2}\ot T_{(12)}^{(A)})}$, whose values are reported in Table~\ref{tab:2renyi}. See below for details on the calculations.

\begin{table}[!th]
\begin{tabular}{|c|c|c|}
\hline
$T_\pi$&$\tr{T_\pi(\psi^{\ot2}\ot T_{(12)}^{(A)})}$&$\tr{T_\pi Q(\psi^{\ot2}\ot T_{(12)}^{(A)})}$\\
\hline
$I$&$dd_B$&$\tr{\psi_A^2}$\\
$T_{(ij)/\{(12),(34)\}}$&$d_B$&$d^{-1}\tr{\psi_A^2}$\\
$T_{(12)}$&$dd_B$&$\tr{\psi_B^2}$\\
$T_{(34)}$&$dd_A$&$\tr{\psi_B^2}$\\
$T_{(ijk)}^{(1,2)}$&$d_A$&$d^{-1}\tr{\psi_B^2}$\\
$T_{(ijk)}^{(3,4)}$&$d_B\tr{\psi^2}$&$d^{-1}\tr{\psi_B^2}$\\
$T_{(ijk\ell)/\{(1324),(1423)\}}$&$d_A\tr{\psi^2}$&$d^{-1}\tr{\psi_A^2}$\\
$T_{\{(1324),(1423)\}}$&$\Tr[\psi_B^2]$&$\tr{\psi_B^2}$\\
$T_{(ij)(k\ell)/\{(12)(34)\}}$&$\tr{\psi_A^2}$&$\Tr[\psi_A^2]$\\
$T_{(12)(34)}$&$dd_A\tr{\psi^2}$&$\tr{\psi_A^2}$\\
\hline
\end{tabular}
\caption{Components of the vectors $\vec{p}_H,\vec{p}_Q,\vec{p}_{Q^\perp}$.}
\label{tab:2renyi}
\end{table}
As usual, we can compute the vectors $\vec{p}_V^Q=V^{T}\vec{p}_Q$ and $\vec{p}_V^H=V^{T}\vec{p}_H$, for the case where $\psi(0)=\psi_A\ot\psi_B$ and $\psi=\dyad{\psi}$ obtaining:
\ba
&&(\vec{p}_V^Q)_+=0\\
&&(\vec{p}_V^Q)_-=\frac{2\sqrt{\tfrac{2}{3}}\,(2+d)}{d}\\
&&(\vec{p}_V^Q)_1=\frac{\sqrt{2}\,(d-1)}{d}\\
&&(\vec{p}_V^Q)_2=-\frac{\sqrt{2}\,(d-1)}{d}\\
&&(\vec{p}_V^Q)_3=-\frac{\sqrt{\tfrac{2}{3}}\,(d-1)}{d}\\
&&(\vec{p}_V^Q)_4=-\frac{\sqrt{\tfrac{2}{3}}\,(d-1)}{d}
\ea
for the Clifford case and:
\ba
&&(\vec{p}_V^H)_+=0\\
&&(\vec{p}_V^H)_-=\sqrt{\tfrac{2}{3}}\!\left(1 + \sqrt{d}\,(4+d)\right)\\
&&(\vec{p}_V^H)_1=\frac{1 + (d-2)\sqrt{d}}{\sqrt{2}}\\
&&(\vec{p}_V^H)_2=-\frac{1 + (d-2)\sqrt{d}}{\sqrt{2}}\\
&&(\vec{p}_V^H)_3=-\frac{1 + (d-2)\sqrt{d}}{\sqrt{6}}\\
&&(\vec{p}_V^H)_4=-\frac{1 + (d-2)\sqrt{d}}{\sqrt{6}}
\ea
for the Haar case.

In order to compute the components of the vectors $\vec{p}_H,\vec{p}_Q,\vec{p}_{Q^\perp}$ we need to evaluate traces involving the swap operator acting on the two copies of one part of the system. Let us once again show a couple of examples, one for the vector $\vec{p}_H$ and one for the vector $\vec{p}_Q$.

First of all, let us remind that the swap operator can be written as:
\ba
T_{(12)}=d^{-1}\sum_PP^{\ot2}
\ea
This means that we can write the components of $\vec{p}_H$ and $\vec{p}_Q$ as:
\ba
\tr{T_\pi\left(\psi^{\ot2}\otimes T_{(12)}^{(A)}\right)}&=&d_A^{-1}\sum_{P_A}\tr{T_\pi\left(\psi^{\ot2}\otimes (P_A\otimes\mathbb{I}_B)^{\ot2}\right)}\\
\tr{T_\pi Q\left(\psi^{\ot2}\otimes T_{(12)}^{(A)}\right)}&=&d_A^{-1}\sum_{P_A}\tr{T_\pi Q\left(\psi^{\ot2}\otimes (P_A\otimes\mathbb{I}_B)^{\ot2}\right)}
\ea 
Let us now consider as an example the case $\pi=I$. We obtain:
\ba
\left(\vec{p}_H\right)_I=d_A^{-1}\sum_{P_A}\tr{\psi}^2\tr{P_A\otimes\mathbb{I}_B}^2=d_Ad_B^2=dd_B
\ea
where we have used $d=d_Ad_B$. Similarly for the vector $\vec{p}_Q$ we find:
\ba
\nonumber
&&\left(\vec{p}_Q\right)_I=d_A^{-1}d^{-2}\sum_{P,P_A}\tr{P\psi}^2\tr{P(P_A\otimes\mathbb{I}_B)}^2=d_A^{-1}\sum_{P_A}\tr{(P_A\otimes\mathbb{I}_B)\psi}^2\\
\nonumber
&&=d_A^{-1}\sum_{P_A}\Tr_A\left[\Tr_B\left[(P_A\otimes\mathbb{I}_B)^{\ot 2}\psi^{\ot2}\right]\right]=d_A^{-1}\sum_{P_A}\tr{P_A^{\ot2}\psi_A^{\ot2}}=\tr{T_{(12)}^{(A)}\psi_A^{\ot2}}=\tr{\psi_A^2}=\rm{Pur}[\psi_A]\\
\ea

Finally, the expression for the T-doped average is:
\ba
\langle\rm{Pur}\rangle_{\mathcal{C}_T}=\frac{A(t)+d(3+d)(1+\sqrt{d}+d)\cos(4\theta)B(d,t)-(1-d^2)C(t)\xi_1^k}{(1-\sqrt{d})(1+\sqrt{d})^2(d-2)d^2(1+d)^2(2+d)(3+d)}
\ea
with
\ba
A(t)&=&\sum_{j=0}^{16} d^{j/2} F_j(t)\\
B(t)&=&\frac{1}{2} g_2(2t)
- \frac{1}{8} d\, g_3(t)
- \frac{1}{8} d\, g_3^*(t)
- \frac{1}{2} d^{2}
+ \frac{1}{4} g_2(t)\, d^{2},\\
C(t)&=&\sum_{j=0}^{11} d^{j/2}H_j(t)
\ea
where
\ba
\nonumber&&
F_0 = -4 g_2(2t) + 16 g_2(t) - g_3(t) - 4 g_4(t) - g_3^*(t),\quad
F_1 = 4 g_2(2t) - 16 g_2(t) + g_3(t) + 4 g_4(t) + g_3^*(t),\\
\nonumber&&
F_2 = -\frac{9}{2} g_2(2t) - 6 g_2(t) - 2 g_3(t) - 2 g_3^*(t),\quad
F_3 = \frac{3}{2} g_2(2t) + 6 g_2(t) - g_3(t) - g_3^*(t),\\
\nonumber&&
F_4 = 5 g_2(2t) - 16 g_2(t) + \frac{11}{8} g_3(t)
      + 5 g_4(t) + \frac{11}{8} g_3^*(t),\\
\nonumber&&
F_5=24-\frac{9}{2}g_2(2t)+22 g_2(t)
      - \frac{13}{8} g_3(t)-5 g_4(t)-\frac{13}{8}g_3^*(t),\\
\nonumber&&
F_6 = \frac{7}{2} g_2(2t) + \frac{29}{4} g_2(t)
      + \frac{5}{2} g_3(t) + \frac{5}{2} g_3^*(t)
      + \frac{57}{2},\quad
F_7=\frac{52}{8}-3g_2(2t)-\frac{27}{4} g_2(t)
      + \frac{9}{8}g_3(t)+\frac{9}{8}g_3^*(t),\\
\nonumber&&
F_8 = 8 - 3 g_2(2t) - g_2(t) + \frac{1}{8} g_3(t)
      - g_4(t) + \frac{1}{8} g_3^*(t),\quad
F_9 = -\frac{61}{2} - \frac{25}{4} g_2(t)
      + g_3(t) + g_4(t) + g_3^*(t),\\
\nonumber&&
F_{10} = -\frac{67}{2} - g_2(2t) - \frac{9}{4} g_2(t),\quad
F_{11} = -7,\quad
F_{12} = -8,\quad
F_{13} = 7,\quad
F_{14} = 7,\quad
F_{15} = 2,\quad
F_{16} = 2.
\ea
As for the coefficients of $C(t)$ we have:
\ba
&&\nonumber
H_0 = g_3(t) + 4 g_4(t) + g_3^*(t),\quad
H_1 = - g_3(t) - 4 g_4(t) - g_3^*(t),\\
&&\nonumber
H_2 = \tfrac{4}{3} g_3(t) - \tfrac{2}{3} g_4(t) + \tfrac{4}{3} g_3^*(t),\quad
H_3 = \tfrac{5}{3} g_3(t) + \tfrac{2}{3} g_4(t) + \tfrac{5}{3} g_3^*(t),\\
&&\nonumber
H_4 = 12 + \tfrac{1}{3} g_3(t) - \tfrac{2}{3} g_4(t) + \tfrac{1}{3} g_3^*(t) - 4 \tilde{g}_3(t),\quad
H_5 = -12 + \tfrac{2}{3} g_3(t) + \tfrac{2}{3} g_4(t) + \tfrac{2}{3} g_3^*(t) + 4 \tilde{g}_3(t),\\
&&\nonumber
H_6 = -1 + \tfrac{2}{3} \tilde{g}_3(t),\quad
H_7 = 1 - \tfrac{2}{3} \tilde{g}_3(t),\quad
H_8 = -\tfrac{2}{3} + \tfrac{2}{3} \tilde{g}_3(t),\quad
H_9 = \tfrac{11}{3} - \tfrac{2}{3} \tilde{g}_3(t),\quad
H_{10} = \tfrac{1}{3},\quad
H_{11} = \tfrac{2}{3}.
\ea

\subsection{Tripartite Mutual Information\label{app:tripartite_info}}
Thus, in order to compute the components of the vectors $\vec{I}_{3,C^2}^H,\vec{I}_{3,C^2}^Q,\vec{I}_{3,C^2}^{Q^\perp}$ needs to compute traces of the form $\tr{T_\pi T_{(12)}^{(C)\ot2}}$, $\tr{T_\pi T_{(12)}^{(C)}\ot T_{(12)}^{(D)}}$, $\tr{T_\pi QT_{(12)}^{(C)\ot2}}$, and $\tr{T_\pi QT_{(12)}^{(C)}\ot T_{(12)}^{(D)}}$, whose values are reported in Table~\ref{tab:TMI1} and Table~\ref{tab:TMI2}

\begin{table}[!th]
\begin{tabular}{|c|c|c|}
\hline
$T_\pi$&$\tr{T_\pi T_{(12)}^{(C)\ot2}}$&$\tr{T_\pi QT_{(12)}^{(C)\ot2}}$\\
\hline
$I$&$d^2d_D^2$&$d^2$\\
$T_{(ij)/\{(12),(34)\}}$&$dd_D^2$&$d$\\
$T_{(12)}$&$d^3$&$d$\\
$T_{(34)}$&$d^3$&$d$\\
$T_{(ijk)}^{(1,2)}$&$d^2$&$1$\\
$T_{(ijk)}^{(3,4)}$&$d^2$&$1$\\
$T_{(ijk\ell)/\{(1324),(1423)\}}$&$dd_C^2$&$d$\\
$T_{\{(1324),(1423)\}}$&$d$&$d$\\
$T_{(ij)(k\ell)/\{(12)(34)\}}$&$d^2$&$d^2$\\
$T_{(12)(34)}$&$d^2d_C^2$&$d^2$\\
\hline
\end{tabular}
\caption{}
\label{tab:TMI1}
\end{table}

\begin{table}[!th]
\begin{tabular}{|c|c|c|}
\hline
$T_\pi$&$\tr{T_\pi T_{(12)}^{(C)}\ot T_{(12)}^{(D)}}$&$\tr{T_\pi QT_{(12)}^{(C)}\ot T_{(12)}^{(D)}}$\\
\hline
$I$&$d^3$&$d$\\
$T_{(ij)/\{(12),(34)\}}$&$d^2$&$1$\\
$T_{(12)}$&$d^2d_C^2$&$d^2$\\
$T_{(34)}$&$d^2d_D^2$&$d^2$\\
$T_{(ijk)}^{(1,2)}$&$dd_D^2$&$d$\\
$T_{(ijk)}^{(3,4)}$&$dd_C^2$&$d$\\
$T_{(ijk\ell)/\{(1324),(1423)\}}$&$d^2$&$1$\\
$T_{\{(1324),(1423)\}}$&$d^2$&$d^2$\\
$T_{(ij)(k\ell)/\{(12)(34)\}}$&$d$&$d$\\
$T_{(12)(34)}$&$d^3$&$d$\\
\hline
\end{tabular}
\caption{}
\label{tab:TMI2}
\end{table}
Let us write down the averages over the various groups in matrix form:
\ba
\left\langle I_{3_{(2)}}^{C^2}\right\rangle_{\cal U}&=&\vec{c} W\vec{I}_{3,C^2}^H\\
\left\langle I_{3_{(2)}}^{C^2}\right\rangle_{\cal C}&=&\vec{q} W^+\vec{I}_{3,C^2}^Q+\vec{q}_{\perp} W^-\vec{I}_{3,C^2}^{Q^\perp}\\
\left\langle I_{3_{(2)}}^{C^2}\right\rangle_{\mathcal{C}_k}&=&\vec{t}\, \Xi^k\vec{I}_{3,C^2}^Q+\vec{t} \Gamma^{(k)}\vec{I}_{3,C^2}^H+\vec{b}\cdot\vec{I}_{3,C^2}^{H}
\ea
and similarly for $\left\langle I_{3_{(2)}}^{CD}\right\rangle_{\cal G}$, where we have defined:
\ba
\vec{I}_{3,C^2}^H&=&\tr{T_{(13)(24)}T_\pi T_{(12)}^{(C)\ot2}}\\
\vec{I}_{3,C^2}^Q&=&\tr{T_{(13)(24)}QT_\pi T_{(12)}^{(C)\ot2}}\\
\vec{I}_{3,C^2}^{Q^\perp}&=&\tr{T_{(13)(24)}Q\perp T_\pi T_{(12)}^{(C)\ot2}}\\
\ea
and analogously for the $CD$ versions.

Let us show the calculation of one component of the vectors $\vec{I}_{3,C^2}^H,\vec{I}_{3,C^2}^Q$, as the one for $\vec{I}_{3,CD}^H,\vec{I}_{3,CD}^Q$ goes in the same way. Let us consider the component corresponding to the identity permutation. For the vector $\vec{I}_{3,C^2}^H$ we have:
\ba
\nonumber
\left(\vec{I}_{3,C^2}^H\right)_I&=&\tr{T_{(12)}^{(C)\ot2}}=d_C^{-2}\sum_{P_C,P'_C}\tr{(P_C\otimes\mathbb{I}_D)^{\ot2}\otimes(P'_C\otimes\mathbb{I}_D)^{\ot2}}\\
&=&d_C^{-2}\sum_{P_C,P'_C}\tr{(P_C\otimes\mathbb{I}_D)}^2\tr{(P'_C\otimes\mathbb{I}_D)}^2=d_C^{-2}d^4=d_D^2d^2
\ea
Let us then turn to the corresponding component of the vector $\vec{I}_{3,C^2}^Q$:
\ba
\left(\vec{I}_{3,C^2}^Q\right)&=&\tr{QT_{(12)}^{(C)\ot2}}=d^{-2}d_C^{-2}\sum_P\sum_{P_C,P'_C}\tr{P(P_C\otimes\mathbb{I}_D)}^2\tr{P(P'_C\otimes\mathbb{I}_D)}^2\\
&=&d_C^{-2}\sum_{P_C,P'_C}\tr{(P_C\otimes\mathbb{I}_D)(P'_C\otimes\mathbb{I}_D)}^2=\sum_{P_C}d^2d_C^{-2}=d^2
\ea
The computation for other permutation goes in a similar way.

Once again we can compute the corresponding vectors $\vec{I}_{C^2,V}^Q=V^{T}\vec{I}_{3,C^2}^Q$ and similarly the vectors $\vec{I}_{C^2,V}^H$, $\vec{I}_{CD,V}^Q$, $\vec{I}_{CD,V}^Q$, obtaining:
\ba
&&(\vec{I}_{C^2,V}^Q)_+=\sqrt{\tfrac{2}{3}}\,(d-2)(d-1)\\
&&(\vec{I}_{C^2,V}^Q)_-=\sqrt{\tfrac{2}{3}}\,(d+1)(d+2)\\
&&(\vec{I}_{C^2,V}^Q)_1=\sqrt{2}(d^{2}-1)\\
&&(\vec{I}_{C^2,V}^Q)_2=0\\
&&(\vec{I}_{C^2,V}^Q)_3=0\\
&&(\vec{I}_{C^2,V}^Q)_4=-\sqrt{\tfrac{2}{3}}\,(d^{2}-1)
\ea
and
\ba
&&(\vec{I}_{CD,V}^Q)_+=-\sqrt{\tfrac{2}{3}}\,(d-2)(d-1)\\
&&(\vec{I}_{CD,V}^Q)_-=\sqrt{\tfrac{2}{3}}\,(d+1)(d+2)\\
&&(\vec{I}_{CD,V}^Q)_1=0\\
&&(\vec{I}_{CD,V}^Q)_2=-\sqrt{2}\,(d^{2}-1)\\
&&(\vec{I}_{CD,V}^Q)_3=-\sqrt{\tfrac{2}{3}}\,(d^{2}-1)\\
&&(\vec{I}_{CD,V}^Q)_4=0
\ea
for the Clifford case, while for the Haar case one gets:
\ba
&&(\vec{I}_{C^2,V}^H)_+=\frac{(d-1)d}{\sqrt{6}}\\
&&(\vec{I}_{C^2,V}^H)_-=\frac{d\big(1+d(9+2d)\big)}{\sqrt{6}}\\
&&(\vec{I}_{C^2,V}^H)_1=\frac{(d-1)d^{2}}{\sqrt{2}}\\
&&(\vec{I}_{C^2,V}^H)_2=-\frac{(d-1)^{2}d}{\sqrt{2}}\\
&&(\vec{I}_{C^2,V}^H)_3=-\frac{(d-1)^{2}d}{\sqrt{6}}\\
&&(\vec{I}_{C^2,V}^H)_4=-\frac{(d-1)d^{2}}{\sqrt{6}}
\ea
and
\ba
&&(\vec{I}_{CD,V}^H)_+=-\frac{(d-1)d}{\sqrt{6}}\\
&&(\vec{I}_{CD,V}^H)_-=\frac{d\bigl(1+d(9+2d)\bigr)}{\sqrt{6}}\\
&&(\vec{I}_{CD,V}^H)_1=\frac{(d-1)^{2}d}{\sqrt{2}}\\
&&(\vec{I}_{CD,V}^H)_2=-\frac{(d-1)d^{2}}{\sqrt{2}}\\
&&(\vec{I}_{CD,V}^H)_3=-\frac{(d-1)d^{2}}{\sqrt{6}}\\
&&(\vec{I}_{CD,V}^H)_4=-\frac{(d-1)^{2}d}{\sqrt{6}}
\ea

The expressions for the T-doped averages are:
\ba
\langle I_{3,C^2}\rangle_{\mathcal{C}_T}&=&
\frac{-\frac{3}{16}\mathcal{P}_{CC}(t)+
\frac{3}{16}B_{CC}(t)\cos(4\theta)+
\frac{1}{2}(d-1)^{2}\Big(
\Xi^{+}_{CC}(t)\xi_{+}^{k}
+\Xi^{1}_{CC}(t)\xi_{1}^{k}
+\Xi^{-}_{CC}(t)\xi_{-}^{k}
\Big)}{3d(d^{2}-1)(d^{2}-4)(d^{2}-9)}.
\\
\nonumber
\langle I_{3,CD}\rangle_{\mathcal{C}_T}&=&\frac{-\frac{3}{16}\,\mathcal{P}_{CD}(t)+\frac{3}{16}\,B_{CD}(t)\,\cos(4\theta)+
\frac{1}{2}\,(d-1)(d+1)\Big(\Xi^{+}_{CD}(t)\,\xi_{+}^{\,k}+\Xi^{1}_{CD}(t)\xi_{1}^{\,k}+\Xi^{-}_{CD}(t)\,\xi_{-}^{\,k}\Big)}{3\,(-3+d)(-2+d)(-1+d)d(1+d)^{2}\,(2+d)\,(3+d)}.\\
\ea
where
\ba
\nonumber
\Xi^{1}_{CC}(t)&=&2(d-3)(d+3)\Big[\big(g_{3}(t)+g_{3}^{*}(t)\big)+d\Big(-3g_{2}(2t)+6g_{2}(t)-2g_{4}(t)+\big(-5+2\tilde{g}_{3}(t)\big)d^{2}\Big)\Big].\\
\nonumber
\Xi^{+}_{CC}(t)&=&(d+3)\Big[d(d+2)\big(g_{3}(t)+g_{3}^{*}(t)\big)-d(d^{2}-4)g_{2}(2t)-d(d+2)\Big(g_{4}(t)-4g_{2}(t)(-2 + d)-\big(\tilde{g}_{3}(t)-2d\big)(d-3)d\Big)\Big]\\
\nonumber
\Xi^{-}_{CC}(t)&=&(d-3)(d-2)\,d\Big[-\big(g_{3}(t)+g_{3}^{*}(t)\big)+(d+2)\,g_{2}(2t)-g_{4}(t)-4(d+2)g_{2}(t)+d(d+3)\big(\tilde{g}_{3}(t)+2d\big)\Big].
\ea
and
\ba
\Xi_{CD}^{1}(t)&=&\Big(72g_{4}(t)+(18g_{3}(t)-72 g_{4}(t) + 18 g_{3}^{*}(t))d+(216+36g_{3}(t)-8g_{4}(t)-72\tilde{g}_{3}(t)+36 g_{3}^{*}(t))d^{2}\\
&&+ (-216 - 2 g_{3}(t) + 8 g_{4}(t) + 72 \tilde{g}_{3}(t) - 2 g_{3}^{*}(t))\,d^{3}
+ (-6 - 4 g_{3}(t) + 8 \tilde{g}_{3}(t) - 4 g_{3}^{*}(t))d^{4}\\
\nonumber
&&+(60-8\tilde{g}_{3}(t))d^{5}-2d^{6}-4d^{7}\Big)\\
\nonumber
&&+(3+d)\Big(-4g_{2}(t)(-3+d)(8 + d(-8 + d + 2 d^{2}))+2g_{2}(2t)(-3+d)( -4+d(4+d+2d^{2}))\Big).\\
\nonumber
\Xi^{+}_{CD}(t)&=&\Big(6g_{3}(t)d - 6 g_{4}(t)d+6 g_{3}^{*}(t)d-g_{3}(t)d^{2} + g_{4}(t)d^{2} - 18\,\tilde{g}_{3}(t)\,d^{2} - g_{3}^{*}(t)d^{2}
+ 36d^{3} - 4 g_{3}(t)d^{3}\\
\nonumber
&&+ 4 g_{4}(t)d^{3} + 9\tilde{g}_{3}(t)d^{3} - 4 g_{3}^{*}(t)d^{3}\\
\nonumber
&&- 18d^{4} - g_{3}(t)d^{4} + g_{4}(t)d^{4} + 11\tilde{g}_{3}(t)\,d^{4} - g_{3}^{*}(t)d^{4}
- 22d^{5} - \tilde{g}_{3}(t)\,d^{5} + 2\,d^{6} - \tilde{g}_{3}(t)d^{6} + 2\,d^{7}
\Big)\\
\nonumber
&&+ (3+d)\,(-2+d)\,(-1+d)\,d\,(2+d)\,\big(\,g_{2}(2t) - 4\,g_{2}(t)\,\big).\\
\nonumber
\Xi^{-}_{CD}(t)&=&(-3+d)(-2+d)(-1+d)d(2+d)\Big(g_{2}(2t)-4g_{2}(t)-g_{3}(t)-g_{4}(t)-g_{3}^{*}(t)+d(3+d)\big(\tilde{g}_{3}(t)+2d\big)\Big).\\
\ea
The other coefficients are worth:
\ba
\nonumber
\mathcal{P}_{CC}(t)&=&
-48\big(g_{3}(t)+g_{3}^{*}(t)\big)
+ d\Big(96\big(g_{3}(t)+g_{3}^{*}(t)\big) - 152\,g_{2}(2t) - 544\,g_{2}(t) + 64\,g_{4}(t)\Big)\\
\nonumber
&&+ d^{2}\Big(2\big(g_{3}(t)+g_{3}^{*}(t)\big) + 160\,g_{2}(2t) + 1088\,g_{2}(t) - 128\,g_{4}(t)\Big)\\
\nonumber
&&+ d^{3}\Big(-64\big(g_{3}(t)+g_{3}^{*}(t)\big) - 72\,g_{2}(2t) - 484\,g_{2}(t) + 48\,g_{4}(t) + 1368\Big)\\
\nonumber
&&+ d^{4}\Big(30\big(g_{3}(t)+g_{3}^{*}(t)\big) - 192\,g_{2}(t) + 32\,g_{4}(t) - 1440\Big)\\
&&+ d^{5}\Big(100\,g_{2}(t) - 16\,g_{4}(t) - 296\Big)
+ 448\,d^{6}
+ 16\,d^{7}
- 32\,d^{8}\\
\nonumber
\mathcal{P}_{CD}(t)&=&
2\Big[
-96\big(g_{2}(2t)-4g_{2}(t)+g_{4}(t)\big)
+ 8\big(12 g_{2}(2t)-48 g_{2}(t)-5 g_{3}(t)+12 g_{4}(t)-5 g_{3}^{*}(t)\big)d\\
\nonumber
&&+4\big(39 g_{2}(2t)-84 g_{2}(t)-8 g_{3}(t)+30 g_{4}(t)-8 g_{3}^{*}(t)\big)\,d^{2}\\
\nonumber
&&+ \big(576-120 g_{2}(2t)+480 g_{2}(t)+49 g_{3}(t)-120 g_{4}(t)+49 g_{3}^{*}(t)\big)\,d^{3}\\
\nonumber
&&- 2\big(18+50 g_{2}(2t)+41 g_{2}(t)-20 g_{3}(t)+12 g_{4}(t)-20 g_{3}^{*}(t)\big)\,d^{4}\\
\nonumber
&&+ \big(-784+24 g_{2}(2t)-96 g_{2}(t)- g_{3}(t)+24 g_{4}(t)- g_{3}^{*}(t)\big)\,d^{5}\\
&&+ 2\big(38+4 g_{2}(2t)+9 g_{2}(t)-4 g_{3}(t)-4 g_{3}^{*}(t)\big)\,d^{6}
+ 224\,d^{7} - 8\,d^{8} - 16\,d^{9}\Big]\\
\nonumber
B_{CC}(t)&=&-72g_{2}(2t)d+18\big(g_{3}(t)+g_{3}^{*}(t)\big)d^{2}+ \big(8g_{2}(2t)-36g_{2}(t) + 72\big)d^{3}\\
&&- 2\big(g_{3}(t)+g_{3}^{*}(t)\big)d^{4}+\big(-8+4g_{2}(t)\big)d^{5}\\
B_{CD}(t)&=&2\,d^{2}(d^{2}-9)\Big(4\,g_{2}(2t)-d\big(g_{3}(t)+g_{3}^{*}(t)+4d-2\,g_{2}(t)d\big)\Big).
\ea

\subsection{\label{app:coherence}Coherence}

Quantum coherence is one of the most striking features of quantum theory, being responsible for all the observed interference phenomena. Besides its foundational role~\cite{Zeh1970,Zurek1981,Zurek1982,JoosZeh1985}, quantum coherence is also a precious resource in quantum information processing~\cite{ZanardiStyliarisVenuti2017_CGP,ZanardiStyliarisVenuti2017_Unital,ZanardiVenuti2018_Grassmannian,StreltsovAdessoPlenio2017_Colloquium} and thermodynamics~\cite{Korzekwa2016_WorkExtraction,Campaioli2018_QuantumBatteries,AlickiFannes2013_EntanglementBoost,Andolina2019_ExtractableWork,PhysRevA.95.053838,PhysRevA.97.053811,GarciaPintos2020_Fluctuations,PhysRevLett.127.028901}. Moreover, it serves as a signature of quantum chaos~\cite{StyliarisAnandZanardi2020_arXiv,StyliarisZanardi2019_PRL,PhysRevResearch.3.023214} and quantum phase transitions~\cite{Styliaris2019_PRB,ZanardiPaunkovic2006_PRE}.

As quantum coherence depends on the basis used to describe the system, so do all its measures~\cite{ChitambarHsieh2016,Streltsov2015,Du2015,MarvianSpekkens2016,WinterYang2016}. Examples are given by the relative entropy of coherence and the $\mathcal{C}_{\ell_1}$ norm of coherence, which are both good measures of coherence in a resource theoretic sense. Here we will however use the $\mathcal{C}_{\ell_2}$ norm of coherence, which is not a good measure in resource theoretic sense~\cite{PhysRevLett.113.140401}. Nonetheless, it presents a greater advantage: it can be cast as an expectation value, and thus it is possible to apply the isospectral twirling technique. From now on we indicate with $\mathcal{C}_B(\psi)$ the $\ell_2$ norm of coherence in the basis $B$. This is defined as:
\ba
\mathcal{C}_B(\psi)=\tr{\psi^2}-\tr{\mathcal{D}_B(\psi)^2}=\pur(\psi)-\pur(\mathcal{D}_B(\psi))
\ea
with $\mathcal{D}_B(\cdot)$ being the dephasing channel over the basis $B$:
\ba
\mathcal{D}_B(\psi)=\sum_i\Pi_i\psi\Pi_i.
\ea
where $\Pi_i=\dyad{i}$ being projectors onto the basis states $\{\ket{i}\}$ of the basis $B$.

\begin{figure}
\begin{subfigure}{0.49\textwidth}
\centering
\includegraphics[width=\textwidth]{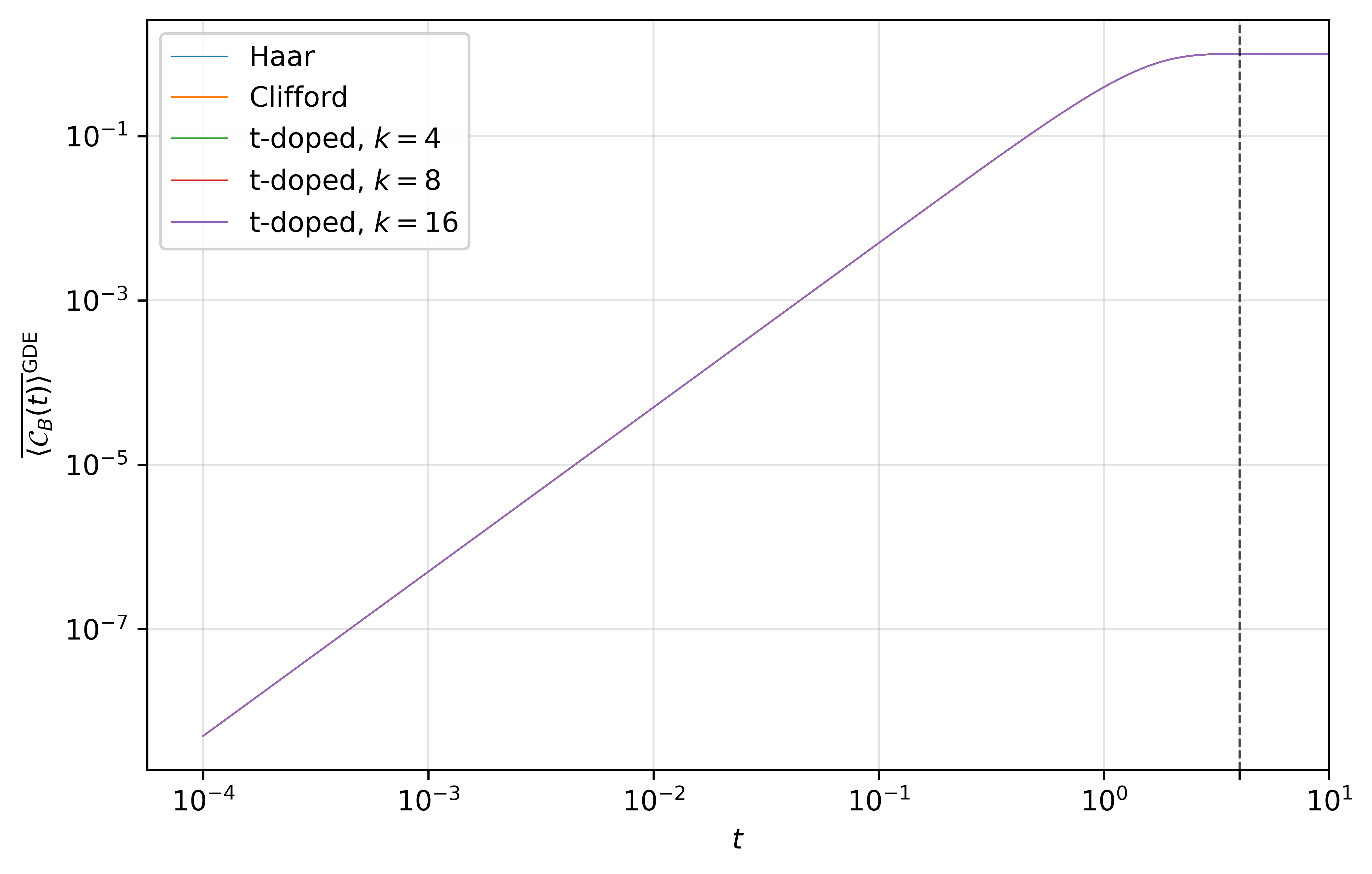}
\caption{Plot of $\overline{\langle\mathcal{C}_B(t)\rangle}^{GDE}$}
\label{fig:coherence_GDE}
\end{subfigure}
\begin{subfigure}{0.49\textwidth}
\centering
\includegraphics[width=\textwidth]{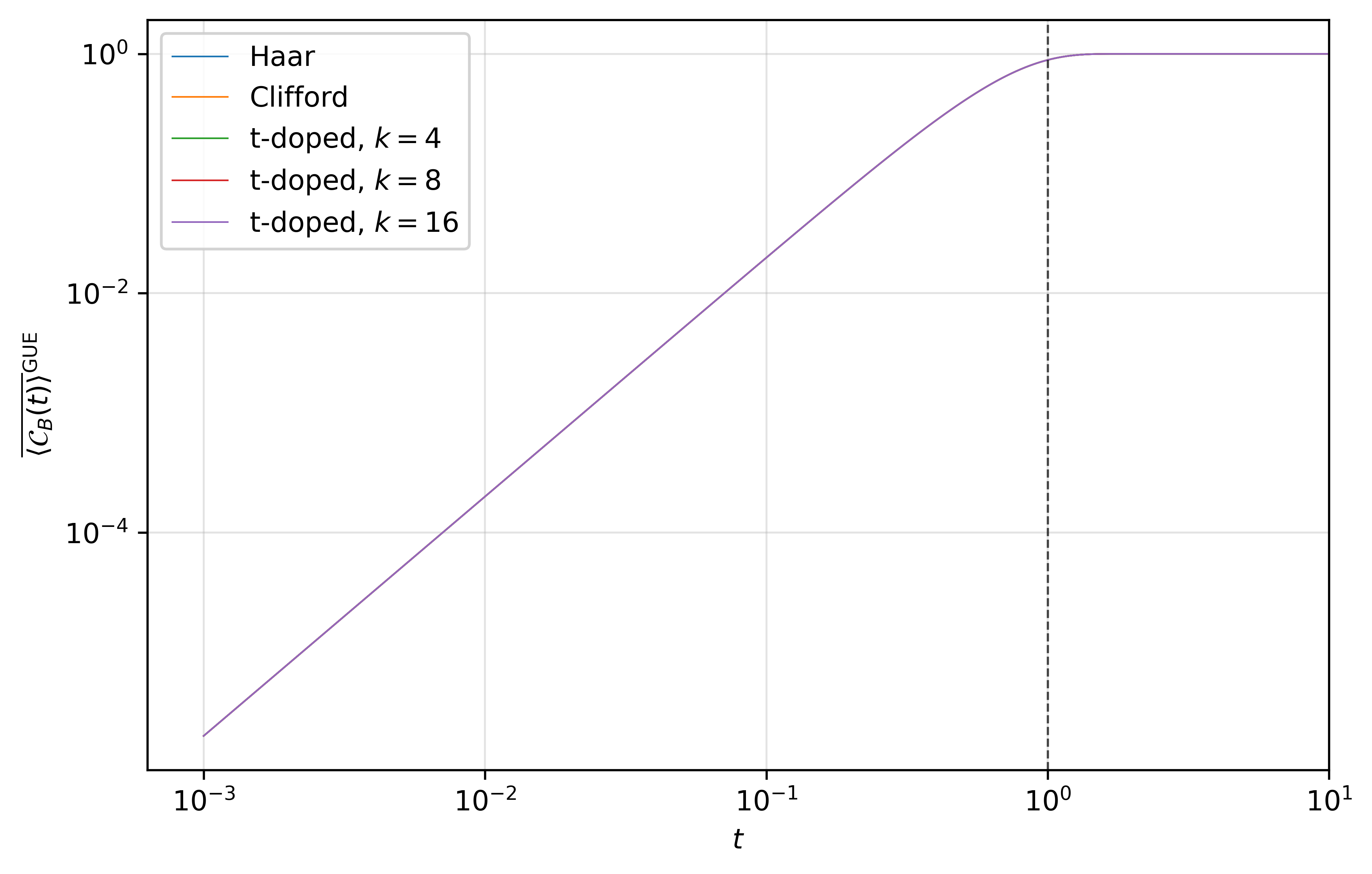}
\caption{Plot of $\overline{\langle\mathcal{C}_B(t)\rangle}^{GUE}$}
\label{fig:coherence_GUE}
\end{subfigure}
\caption{Plot of the 2 norm of coherence averaged over the GDE(~\ref{fig:coherence_GDE}) and the GUE(~\ref{fig:coherence_GUE})for $d=2^{16}$. One notices how averaging over different groups has no effect on coherence. This is explained by the fact that the Clifford group is large enough not to observe any difference in the decoherence process. Notice also that the equilibration time is $\mathcal{O}(\sqrt{\log d})$ for the GDE, while it is $\mathcal{O}(1)$ for the GUE, as indicated by the vertical dashed lines.}
\label{fig:coherence}
\end{figure}

Assuming a pure initial state $\psi=\dyad{\psi}$, we can let it evolve as $U\psi U^\dag$, and recast the coherence norm $\mathcal{C}_B$ in terms of the isospectral twirling as:
\begin{equation}
    \begin{split}
       \mathcal{C}_B(\psi_U)&=1-\sum_{i,j}\Tr[\Pi_i V\psi V^\dag\Pi_i \Pi_j V\psi V^\dag \Pi_j]=1-\sum_{i}\Tr[\Pi_i V \psi V^\dag \Pi_i V\psi V^\dag ]\\
       &=1-\Tr[T_{(13)(24)}(D_B\otimes \psi^{\otimes 2})V^{\otimes 2,2}]\,,
    \end{split}
\end{equation}
with $D_B:=\sum_i \Pi_i ^{\otimes 2}$.
Finally taking the average, we obtain:
\ba
\exv{\mathcal{C}_B(\psi_V)}{\mathcal G}&=&1-\Tr[\mathcal{R}^{(4)}_{\cal G}(V)T_{(13)(24)}(D_B\otimes \psi^{\otimes 2})].
\ea
The results for the Haar and Clifford averages are:
\ba
\label{eq:coherence_haar_average}
\langle\mathcal{C}_B\rangle_{\cal U}&=&\frac{g_2(2t)-4g_2(t)+g_3(t)+g_4(t)+g_3^{*}(t)+2d^{2}(3+d)}{d^{2}(1+d)(3+d)}\\
\label{eq:coherence_clifford_average}
\langle\mathcal{C}_B\rangle_{\cal C}&=&\frac{g_2(2t)+d\left(2+\tilde{g}_3(t)+2d\right)}{d(1+d)(2+d)}
\ea
Once again we observe that the Clifford average consistently does depend only on the Clifford spectral form factor $\tilde{g}_3(t)$, in contrast with the Haar average.
We can observe the results in Fig.~\ref{fig:coherence}. One can immediately observe that when the coherence is averaged over the GDE, the initial value of the Clifford average is $\mathcal{O}(d^{-2})$, in contrast with the Haar value $\mathcal{O}(d^{-1})$. This is explained by looking at Eqs.(~\ref{eq:coherence_haar_average},~\ref{eq:coherence_clifford_average}), and considering that $g_4(0)=d^4$, while $\tilde{g}_3(0)=d^2$. This difference is however washed away in the long time limit because of the dominant term of $\mathcal{O}(d^{-1})$ present in both expressions. The physical interpretation of this is that in presence of a stabilizer Hamiltonian, the eigenvectors are all stabilizer states. Thus, under any Clifford operation, stabilizer states are mapped into stabilizer states, and this does slow down the decoherence. In the long time limit this is however still sufficient to completely decohere the state.
The same effect is not observed when the average is taken over the GUE.

\begin{figure}[!ht]
\begin{subfigure}{0.33\textwidth}
\centering
\includegraphics[width=\linewidth]{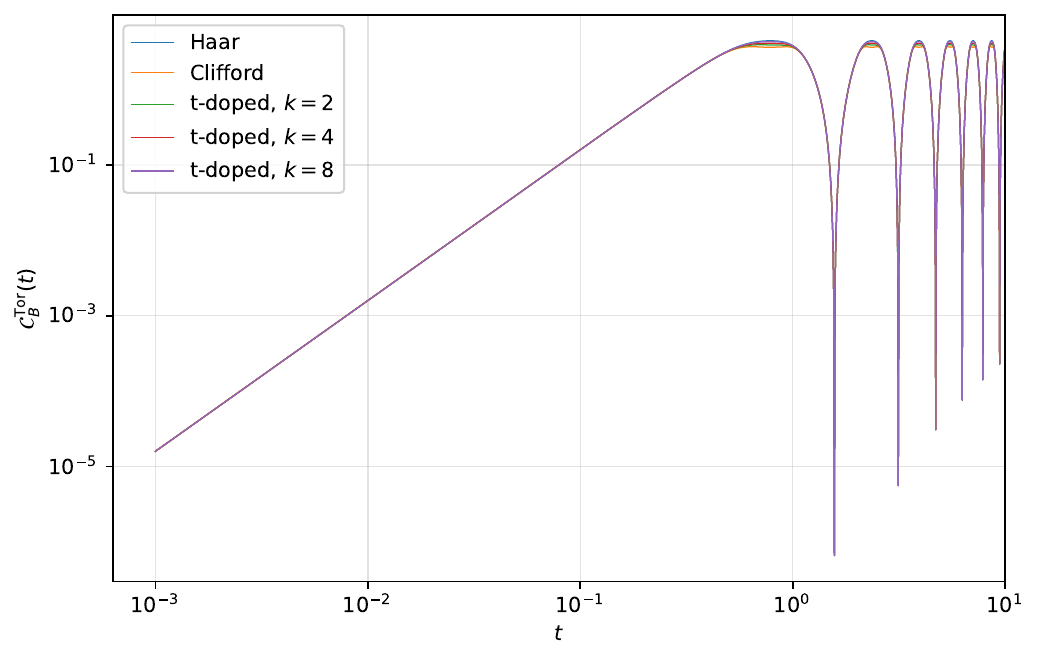}
\caption{$N=2$}
\label{fig:coherence_Tor_N2}
\end{subfigure}
\begin{subfigure}{0.33\textwidth}
\centering
\includegraphics[width=\linewidth]{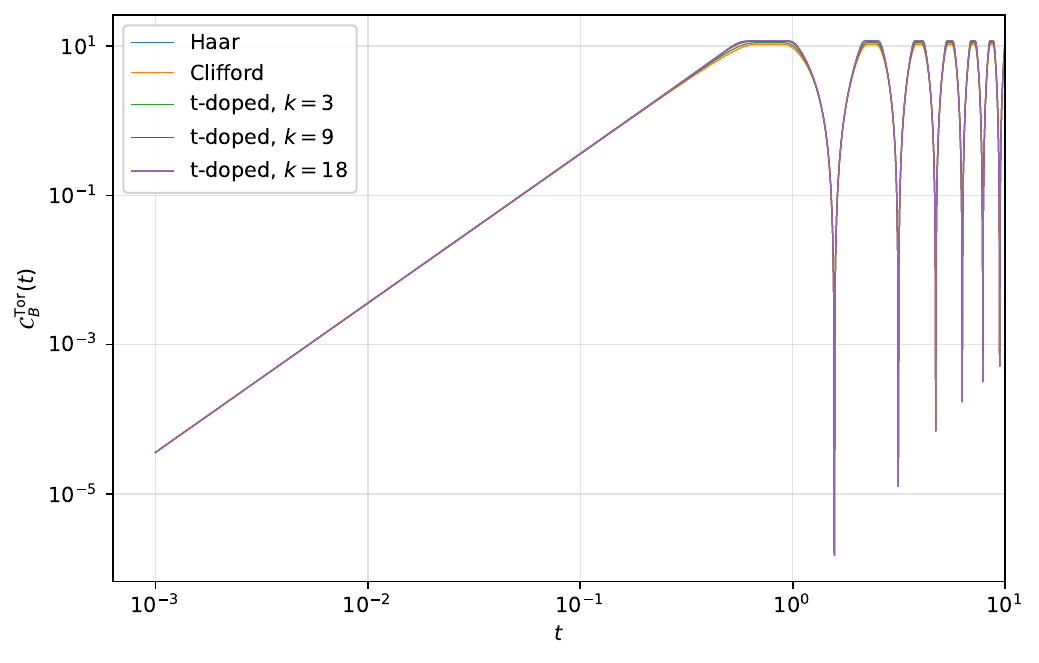}
\caption{$N=3$}
\label{fig:coherence_Tor_N3}
\end{subfigure}
\begin{subfigure}{0.33\textwidth}
\centering
\includegraphics[width=\linewidth]{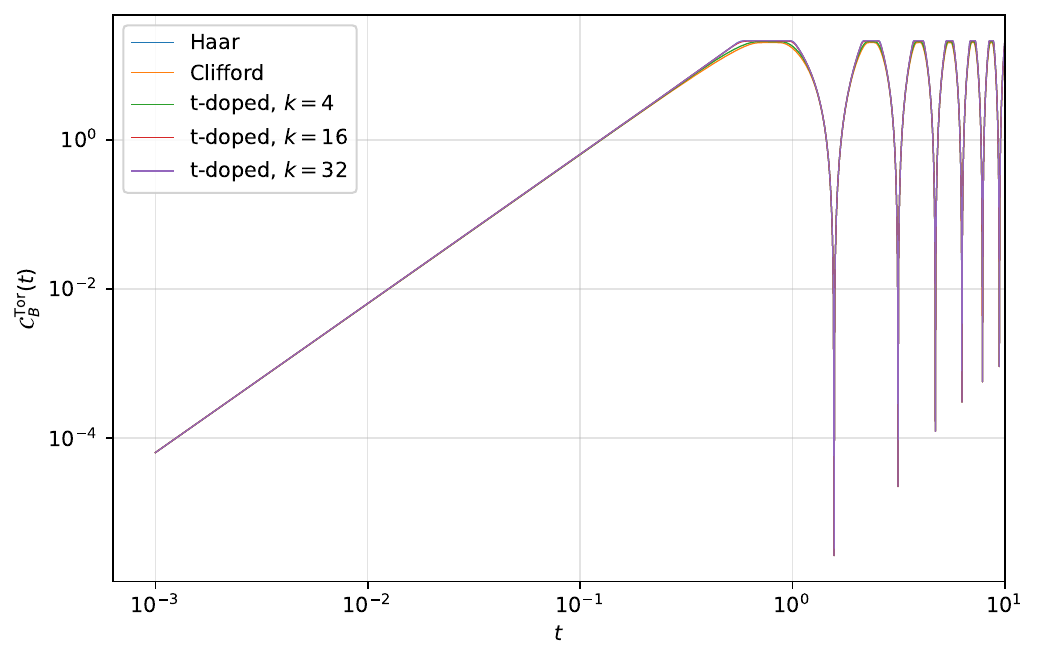}
\caption{$N=4$}
\label{fig:coherence_Tor_N4}
\end{subfigure}
\caption{Coherence $\mathcal{C}_B$ for the Toric Code for different lattice size.}
\label{fig:coherence_Tor}
\end{figure}

Once again we can cast the averages in matrix form for all three scenarios. We have:
\ba
\langle\mathcal{C}_B(\psi_V)\rangle_{\cal U}&=&1-\vec{t}\,W\vec{D}_B^H\\
\langle\mathcal{C}_B(\psi_V)\rangle_{\cal C}&=&1-\left(\vec{q}W^+\vec{D}_B^Q+\vec{q}_{\perp}W^-(\vec{D}_B^{Q^\perp})\right)\\
\langle\mathcal{C}_B\rangle_{\mathcal{C}_T}&=&1-\left(\vec{t}\,\Xi^k\vec{D}_B^Q+\vec{t}\,\Gamma^{(k)}\vec{D}_B^H+\vec{b}\cdot\vec{D}_B^H\right)
\ea
with: 
\ba
\left(\vec{D}_B^H\right)_\sigma&=&\tr{T_\sigma T_{(13)(24)}(D_B\otimes \psi^{\otimes 2})}\\
\left(\vec{D}_B^Q\right)_\sigma&=&\tr{QT_\sigma T_{(13)(24)} (D_B \otimes \psi^{\otimes 2})}\\
\left(\vec{D}_B^{Q^\perp}\right)_\sigma&=&\tr{Q^\perp T_\sigma T_{(13)(24)} (D_B \otimes \psi^{\otimes 2})}
\ea

\begin{table}[!th]
\centering
\begin{tabular}{|c|c|c|}
\hline
$T_\pi$&$\tr{T_\pi (\mathcal{D}_B\ot\psi^{\ot2})}$&$\tr{T_\pi Q(\mathcal{D}_B\ot\psi^{\ot2})}$\\
\hline
$I$&$d$&$1$\\
$T_{(12)}$&$d$&$1$\\
$T_{(34)}$&$d$&$1$\\
$T_{(ij)}$&$1$&$d^{-1}$\\
$T_{(ijk)}^{(1,2)}$&$1$&$d^{-1}$\\
$T_{(ijk)}^{(3,4)}$&$1$&$d^{-1}$\\
$T_{(ijk\ell)/\{(1324),(1423)\}}$&$1$&$d^{-1}$\\
$T_{\{(1324),(1423)\}}$&$1$&$1$\\
$T_{(ij)(k\ell)}$&$1$&$1$\\
$T_{(12)(34)}$&$d$&$1$\\
\hline
\end{tabular}
\caption{Components of the vectors $\vec{D}_B^H,\vec{D}_B^Q,\vec{D}_B^{Q^\perp}$. The coefficients for the case with the projector $Q$ are computed assuming the dephasing is acting with respect to the computational basis.}
\label{tab:coherence}
\end{table}

Also in this case, like for all the other probes, we can compute the vectors $\vec{D}_V^Q=V^{T}\vec{D}_B^Q$ and $\vec{D}_V^H=V^{T}\vec{D}_B^H$, assuming a pure initial state, obtaining:
\ba
&&(\vec{D}_V^Q)_+=0\\
&&(\vec{D}_V^Q)_-=\frac{2\sqrt{\tfrac{2}{3}}\,(\,2+d\,)}{d}\\
&&(\vec{D}_V^Q)_1=\frac{\sqrt{2}(d-1)}{d}\\
&&(\vec{D}_V^Q)_2=-\frac{\sqrt{2}\,(\,d-1\,)}{d},\\
&&(\vec{D}_V^Q)_3=-\frac{\sqrt{\tfrac{2}{3}}\,(\,d-1\,)}{d}\\
&&(\vec{D}_V^Q)_4=-\frac{\sqrt{\tfrac{2}{3}}\,(\,d-1\,)}{d}
\ea
for the Clifford case and
\ba
&&(\vec{D}_V^H)_+=0\\
&&(\vec{D}_V^H)_-=\sqrt{\tfrac{2}{3}}\,(5+d)\\
&&(\vec{D}_V^H)_1=\frac{d-1}{\sqrt{2}}\\
&&(\vec{D}_V^H)_2=-\frac{d-1}{\sqrt{2}}\\
&&(\vec{D}_V^H)_3=-\frac{d-1}{\sqrt{6}}\\
&&(\vec{D}_V^H)_4=-\frac{d-1}{\sqrt{6}}
\ea
for the Haar case.

Finally, the expression for the T-doped average is:
\ba
\langle\mathcal{C}_B\rangle_{\mathcal{C}_T}=\frac{\left[
P_0(t)
+ P_1(t)\cos(4\theta)
+ \Xi_1(t)\xi_1^k
+ \Xi_-(t)\xi_-^k
\right]}{24(-2+d)(-1+d)d^{2}(1+d)^{2}(2+d)(3+d)}
\ea
with:
\ba
\nonumber
P_0(t)&=&24\bigl(4 g_2(2t)-16 g_2(t)+g_3(t)+4g_4(t)+g_3^*(t)\bigr)-12 g_2(2t)d+8\bigl(-6 g_2(t)+g_3(t)+g_3^*(t)\bigr)d\\
\nonumber
&&+\bigl(-192+36g_2(2t)-176 g_2(t)+13 g_3(t)+40 g_4(t)+13 g_3^*(t)\bigr)d^{2}\\
\nonumber
&&+\bigl(-52+24 g_2(2t)+54g_2(t)-9g_3(t)-9 g_3^*(t)\bigr)d^{3}\\
&&+2\bigl(122+25 g_2(t)-4g_3(t)-4 g_4(t)-4 g_3^*(t)\bigr)d^{4}+56d^{5}-56d^{6}-16d^{7}.\\
P_1(t)&=&3d(3+d)\left[4g_2(2t)-d\bigl(g_3(t)+g_3^*(t)+4d-2 g_2(t)d\bigr)\right]\\
\nonumber
\Xi_1(t)&=&8(1-d^{2})\Bigl[(3+d)\bigl(4g_2(2t)-16g_2(t)+g_3(t)+g_4(t)+g_3^*(t)\bigr)\\
\nonumber
&&-3g_2(2t)d-2\bigl(-3 g_2(t)+g_3(t)+g_4(t)+g_3^*(t)\bigr)d\\
&&+2\bigl(6 + g_2(2t)+2 g_2(t) - 2\widetilde g_{3}\bigr)d^{2}+(-5 + 2\widetilde g_{3})d^{3}-2d^{4}\Bigr]\\
\Xi_-(t)&=&(-2+d)d\Bigl[-g_3(t)-g_4(t)-g_3^*(t)+g_2(2t)(2+d)-4g_2(t)(2+d)+d(3+d)\bigl(\tilde g_{3}+2d\bigr)\Bigr]
\ea
\subsection{\label{app:WYD}Wigner-Yanase-Dyson Skew Information}

Another measure of coherence is given by the Wigner-Yanase-Dyson (WYD) skew information~\cite{WignerYanase1997,Lieb2002,Yanagi2010}.
The WYD skew information quantifies how hard it is to measure a certain observable $X$ on a certain state $\psi_t$. In other words, it provides a measure of the strictly quantum uncertainty associated with the measurement of an observable $X$ on a given state $\psi$. The WYD information has found applications in diverse fields, such as quantum speed limits~\cite{Luo2004,LuoZhang2004,Furuichi2008}, quantum metrology~\cite{Luo2003,GibiliscoIsola2005,LuoFuruichi2005}, entanglement detection~\cite{LuoEntanglement2005,FuruichiYanagiKuriyama2004,Furuichi2006}.

The WYD skew information is defined as:
\ba
\mathcal{I}_{\eta}(\psi,X)=\tr{X^2\psi_t}-\tr{X\psi_t^{(1-\eta)}X\psi_t^{(\eta)}}
\ea
As $\mathcal{I}_\eta$ is the expectation value of an observable, it comes as no surprise that it can be recast in form of an isospectral twirling. Using the usual swap-trick techniques, and taking the average, one obtains:
\ba
\label{eq:wyd}
\left\langle\mathcal{I}_{\eta}(\psi,X)\right\rangle_{\cal G}=\tr{T_{(12)}X^2\ot\psi\mathcal{R}_{\cal G}^{(2)}(V)}-\tr{T_{(1423)}X^{\ot2}\ot\psi^{(1-\eta)}\ot\psi^{(\eta)}\mathcal{R}_{\cal{G}}^{(4)}(V)}
\ea
One can see that the second term on the rhs of Eq.~\eqref{eq:wyd} depends on the fourth moment of the operator $V$, and thus will distinguish between the unitary, Clifford and T-doped scenario.  Further details on the computation can be found in App.~\ref{app:WYD}

An interesting case to study is the one where the state $\psi$ is assumed to be proportional to a projector onto a computational basis state, i.e. $\psi=\dyad{j}$ with $\ket{j}$ a computational basis state, while the observable $X$ is set to be a Pauli string stabilizing the state, i.e. a Pauli string made out of $\mathbb{I},Z$ operators alone.

The average of the second order term can be easily computed with the techniques in App.~\ref{app:second_moment}, obtaining:
\ba
\langle\mathcal{I}_{1/2,2,P}\rangle_{U}=1
\ea

One can compute the average of the fourth order terms as:
\ba
\left\langle\mathcal{I}_{1/2,4,P}\right\rangle_{\cal U}&=&\frac{g_{2}(2t)-4g_{2}(t)+g_{3}(t)+g_{4}(t)+g_{3}^{*}(t)+(d-1)d(d+3)}{(d-1)d(d+1)(d+3)}\\
\left\langle\mathcal{I}_{1/2,4,P}\right\rangle_{\mathcal{C}}&=&\frac{-2+g_{2}(2t)+d\left(-1+\tilde{g}_{3}(t)+d\right)}{(d-1)(d+1)(d+2)}
\ea
where as usual the T-doped average is reported in App.~\ref{app:WYD}.

\begin{figure}[!th]
\begin{subfigure}{0.49\textwidth}
\centering
\includegraphics[width=\textwidth]{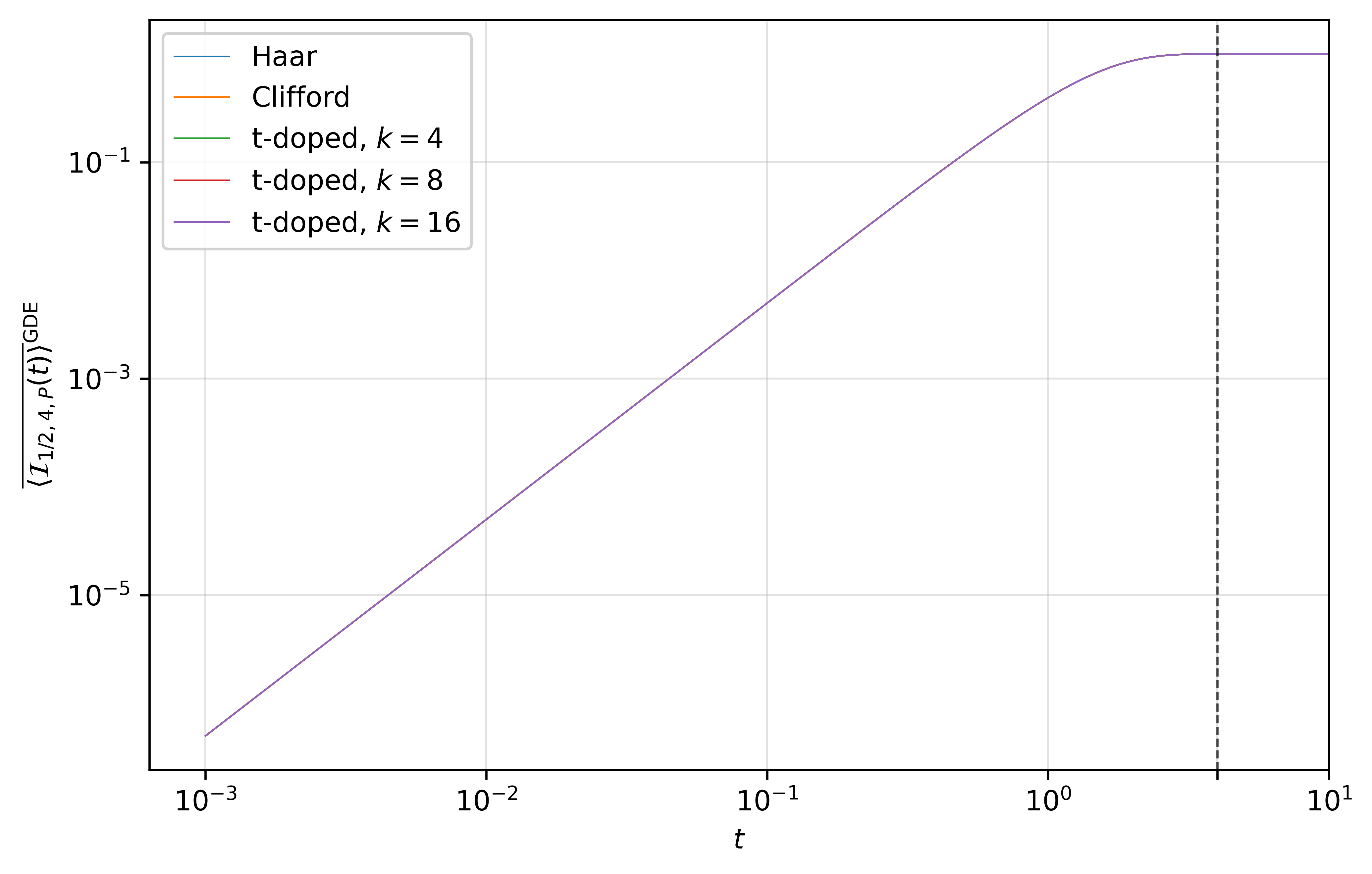}
\caption{Log-log plot of $\overline{\langle\mathcal{I}_{(1/2,4,P)}\rangle}^{\rm GDE}$.}
\label{fig:WYD_pauli_GDE}
\end{subfigure}
\begin{subfigure}{0.49\textwidth}
\centering
\includegraphics[width=\textwidth]{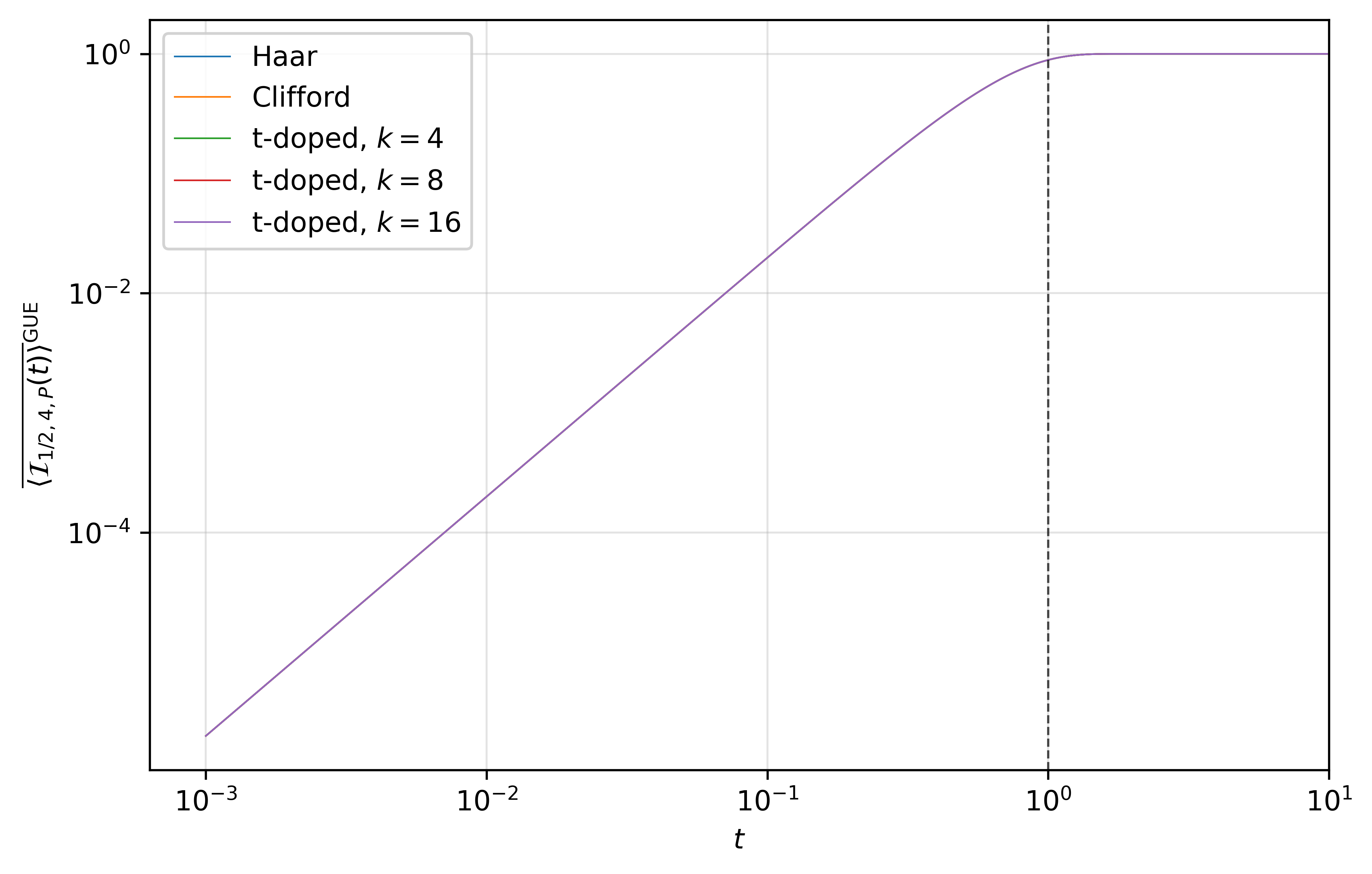}
\caption{Log-log plot of $\overline{\langle\mathcal{I}_{(1/2,4,P)}\rangle}^{\rm GUE}$.}
\label{fig:WYD_pauli_GUE}
\end{subfigure}
\caption{Plot of the WYD skew information for the case $X=P$ with $d=2^{16}$. Just as for the case of coherence, one does not observe any difference between the WYD averaged over the unitary of Clifford group. The explanation for this is the same as the one given for the norm of coherence. Similarly, also the equilibration times are the same, as indicated by the dashed vertical lines.}
\label{fig:WYD_pauli}
\end{figure}

The results are found in Fig.~\ref{fig:WYD_pauli}. One notices a similar behavior to the one of coherence norm. This is no surprise as the two measures are related, see below. Once again, the average over the GUE washes away the difference between the Clifford and Haar averages.

Another interesting setting can be obtained by noting that the WYD skew information can be set to be a good measure of coherence for mixed states by setting $\eta=1/2$ and by choosing $X$ to be a projector onto the basis $B=\{\ket{i}\}$ over which we want to compute the coherence. Then one can define:
\ba
\label{eq:wyd_coherence}
C(\psi)=\sum_i\mathcal{I}_{1/2}(\psi,\Pi_i)
\ea
with $\Pi_i=\dyad{i}$ being once again the projectors onto the basis states $\ket{i}$.
The average of the second order term is again easily calculated as:
\ba
\langle\mathcal{I}_{1/2,2,\Pi_i}\rangle_{U}=1
\ea
while the results of the averages of the second term of Eq.~\eqref{eq:wyd} for this case are:
\ba
\left\langle\mathcal{I}_{1/2,4,\Pi_i}\right\rangle_{\cal U}&=&\frac{g_{2}(2t)-4g_{2}(t)+g_{3}(t)+g_{4}(t)+g_{3}^{*}(t)+2d^{2}(3+d)}{d^{2}(1+d)(3+d)}\\
\left\langle\mathcal{I}_{1/2,4,\Pi_i}\right\rangle_{\mathcal{C}}&=&\frac{g_{2}(2t)+d\left(2+\tilde{g}_{3}(t)+2d\right)}{d(1+d)(2+d)}
\ea
the T-doped average being reported in the Appendix. It comes as no surprise that we obtain the same expressions as in Eqs.(~\ref{eq:coherence_haar_average},~\ref{eq:coherence_clifford_average}), showing the relation between the measures, and explaining the results similar to the case of coherence.

Setting $\mathcal{I}_{1/2,4,P}=\tr{T_{(1423)}\left(P^{\ot2}\ot\psi^{1/2}\ot\psi^{1/2}\right)\mathcal{R}_{\cal{G}}^{(4)}(V)}$, we can write the matrix expressions:
\ba
\left\langle\mathcal{I}_{1/2,4,P}\right\rangle_{\cal U}&=&\vec{c}\,W\vec{\cal I}_{1/2,4,P}^H\\
\left\langle\mathcal{I}_{1/2,4,P}\right\rangle_{\cal C}&=&\vec{q}\, W^+\vec{\cal I}_{1/2,4,P}^Q+\vec{q}_{\perp} W^-\vec{\cal I}_{1/2,4,P}^{Q^\perp}\\
\left\langle\mathcal{I}_{1/2,4,P}\right\rangle_{{\cal C}_k}&=&\vec{t}\,\Xi^k\vec{\cal I}_{1/2,4,P}^Q+\vec{t}\,\Gamma^{(k)}\vec{\cal I}_{1/2,4,P}^H+\vec{b}\cdot\vec{\cal I}_{1/2,4,P}^H\\
\ea
where:
\ba
\left(\vec{\cal I}_{1/2,4,P}^H\right)_\pi&=&\tr{T_\pi T_{(1423)}\left(P^{\ot2}\ot\psi^{1/2}\ot\psi^{1/2}\right)}\\
\left(\vec{\cal I}_{1/2,4,P}^Q\right)_\pi&=&\tr{QT_\pi T_{(1423)}\left(P^{\ot2}\ot\psi^{1/2}\ot\psi^{1/2}\right)}\\
\left(\vec{\cal I}_{1/2,4,P}^{Q^\perp}\right)_\pi&=&\tr{Q^\perp T_\pi T_{(1423)}\left(P^{\ot2}\ot\psi^{1/2}\ot\psi^{1/2}\right)}
\ea
and analogous versions can be written for the case where $X$ is a computational basis measurement.

Let us treat first the case of a Pauli measurement.
We have to compute traces of the form $\tr{T_\pi P^{\ot2}\ot\psi^{1/2}\ot\psi^{1/2}}$ and $\tr{T_\pi QP^{\ot2}\ot\psi^{1/2}\ot\psi^{1/2}}$ in order to compute the components of the vectors $\vec{\cal I}_{1/2,4,P}^H$ and $\vec{\cal I}_{1/2,4,P}^Q$. These are summarized in Table~\ref{tab:skew_info}.

The result of the T-doped average is:
\ba
\nonumber
\langle \mathcal{I}_{1/2,4,P}\rangle_{\mathcal{C}_T}=\frac{\frac{3}{16}{\mathcal P}(d,t)-\frac{3}{16}(d^{2}+d-6){\mathcal C}(d,t)\cos(4\theta)+\frac{1}{2}(d-1)(d+1)\Big[(3+d)\Xi_{1}(d,t)\xi_{1}^{k}+2(d-2)d\Xi_{-}(d,t)\xi_{-}^{k}\Big]}{3(d-2)d(d+2)(d+3)(d^{2}-1)^{2}}\\
\ea
where
\ba
\nonumber
\Xi_{1}(t)&=&8\Big(4g_{2}(2t)-16 g_{2}(t)+g_{3}(t)+4 g_{4}(t)+g_{3}^{*}(t)\Big)-8d\Big(\frac{3}{2} g_{2}(2t)+\big(-3 g_{2}(t)+g_{3}(t)+g_{4}(t)+g_{3}^{*}(t)\big)\Big)\\
&&+8d^{2}\Big(6+g_{2}(2t)+2g_{2}(t)-2\tilde g_{3}(t)\Big)+4\big(-5+2\tilde g_{3}(t)\big)d^{3}-8d^{4}\\
\Xi_{-}(t)&=&-\big(g_{3}(t)+g_{4}(t)+g_{3}^{*}(t)\big)+(2+d)\big(g_{2}(2t)-4 g_{2}(t)\big)+d(3+d)\big(\tilde g_{3}(t)+2d\big).
\ea
and
\ba
\nonumber
\mathcal {P}(t)
&=&(76g_{2}(2t)-256g_{2}(t)+64g_{4}(t)+40(g_{3}(t)+g_{3}^{*}(t)))\\
\nonumber
&&+d( -192 + 16\,g_{2}(2t)+48g_{2}(t) - 14(g_{3}(t)+g_{3}^{*}(t)) )\\
\nonumber
&&+d^{2}(116-73g_{2}(2t)+348g_{2}(t) - 80\,g_{4}(t) - 55\,(g_{3}(t)+g_{3}^{*}(t)) ) \\
\nonumber
&&+d^{3}( 288 - 23g_{2}(2t) - 50g_{2}(t) + 9(g_{3}(t)+g_{3}^{*}(t)) )\\
\nonumber
&&+d^{4}(-167+8g_{2}(2t) - 82g_{2}(t) + 16g_{4}(t) + 16(g_{3}(t)+g_{3}^{*}(t)))\\
&&-105d^{5}+40d^{6}+16d^{7}.\\
{\mathcal C}(t)&=& g_{2}(2t)(d-2)+d\Big( g_{3}(t)+g_{3}^{*}(t) - d\big(-2+2g_{2}(t)+d\big)\Big).
\ea

Let us now compute one component of the vectors $\vec{\cal I}_{1/2,4,P}^H$ and $\vec{\cal I}_{1/2,4,P}^Q$, where we are assuming $\psi=\dyad{j}$ with $\ket{j}$ a computational basis state and also $P$ to be a string containing only $\mathbb{I},Z$ operators. Let us consider the case $\pi=I$. We obtain:
\ba
\left(\vec{\cal I}_{1/2,4,P}^H\right)_I=\tr{P^{\ot2}\otimes\psi^{(1-\eta)}\otimes\psi^{(\eta)}}=0
\ea
because $X$ is a Pauli string. For the corresponding component of $\vec{\cal I}_{1/2,4,P}^Q$ we instead have:
\ba
\left(\vec{\cal I}_{1/2,4,P}^Q\right)_I&=&\tr{Q(P^{\ot2}\otimes\psi^{1/2}\otimes\psi^{1/2})}=d^{-2}\sum_P\tr{P'P}^2\tr{P'\psi^{1/2}}^2\\
&=&\tr{P\psi^{1/2}}^2=1
\ea

\begin{table}[!th]
\begin{tabular}{|c|c|c|}
\hline
$T_\pi$&$\tr{T_\pi \left(P'^{\ot2}\ot \psi^{1/2}\ot\psi^{1/2}\right)}$&$\tr{T_\pi \left(QP'^{\ot2}\ot\psi^{1/2}\ot\psi^{1/2}\right)}$\\
\hline
$I$&$0$&$1$\\
$T_{(ij)/\{(12),(34)\}}$&$0$&$d^{-1}$\\
$T_{(12)}$&$d$&$1$\\
$T_{(34)}$&$0$&$1$\\
$T_{(ijk)}^{(1,2)}$&$0$&$d^{-1}$\\
$T_{(ijk)}^{(3,4)}$&$1$&$d^{-1}$\\
$T_{(ijk\ell)/\{(1324),(1423)\}}$&$1$&$d^{-1}$\\
$T_{\{(1324),(1423)\}}$&$1$&$1$\\
$T_{(ij)(k\ell)/\{(12)(34)\}}$&$1$&$1$\\
$T_{(12)(34)}$&$d$&$1$\\
\hline
\end{tabular}
\caption{Traces for the WYD skew information for the case where the observable $X$ is a Pauli string $P'$ made out only of $\mathbb{I},Z$ operators and the initial state is a computational basis state.}
\label{tab:skew_info}
\end{table}

Let us also report the expression for the components $\vec{\mathcal{I}}_{1/2,4,P}^{Q,V}$ and $\vec{\mathcal{I}}_{1/2,4,P}^{H,V}$:
\ba
\left(\vec{\mathcal{I}}_{1/2,4,P}^{Q,V}\right)_+&=&0\\
\left(\vec{\mathcal{I}}_{1/2,4,P}^{Q,V}\right)_-&=&\frac{2\sqrt{2/3}\,(d+2)}{d}\\
\left(\vec{\mathcal{I}}_{1/2,4,P}^{Q,V}\right)_1&=&\frac{\sqrt{2}\,(d-1)}{d}\\
\left(\vec{\mathcal{I}}_{1/2,4,P}^{Q,V}\right)_2&=&-\frac{\sqrt{2}(d-1)}{d}\\
\left(\vec{\mathcal{I}}_{1/2,4,P}^{Q,V}\right)_3&=&-\frac{\sqrt{2/3}\,(d-1)}{d}\\
\left(\vec{\mathcal{I}}_{1/2,4,P}^{Q,V}\right)_4&=&-\frac{\sqrt{2/3}\,(d-1)}{d}
\ea
for the Clifford components, while for the Haar version one has:
\ba
\left(\vec{\mathcal{I}}_{1/2,4,P}^{H,V}\right)_+&=&0\\
\left(\vec{\mathcal{I}}_{1/2,4,P}^{H,V}\right)_-&=&\frac{d+6}{\sqrt{6}}\\
\left(\vec{\mathcal{I}}_{1/2,4,P}^{H,V}\right)_1&=&\frac{d}{2\sqrt{2}}\\
\left(\vec{\mathcal{I}}_{1/2,4,P}^{H,V}\right)_2&=&-\frac{d}{2\sqrt{2}}\\
\left(\vec{\mathcal{I}}_{1/2,4,P}^{H,V}\right)_3&=&-\frac{d}{2\sqrt{6}}\\
\left(\vec{\mathcal{I}}_{1/2,4,P}^{H,V}\right)_4&=&-\frac{d}{2\sqrt{6}}
\ea

Let us now move to the corresponding results and calculations when $X=\sum_i\Pi_i$. In this case, for the identity permutation one has:
\ba
\left(\vec{\cal I}_{\eta,4}^H\right)_I=\tr{\sum_i\Pi_i^{\ot2}\otimes\psi^{(1-\eta)}\otimes\psi^{(\eta)}}=d\tr{\psi^{(1-\eta)}}\tr{\psi^{(\eta)}}=d
\ea
for the Haar case, while for the Clifford case one gets:
\ba
\nonumber
\left(\vec{\cal I}_{\eta,4}^Q\right)_I&=&\tr{Q(\sum_i\Pi_i^{\ot2}\otimes\psi^{(1-\eta)}\otimes\psi^{(\eta)})}=d^{-2}\sum_P\sum_i\tr{P\Pi_i}^2\tr{P\psi^{(1-\eta)}}\tr{P\psi^{(\eta)}}=1
\ea
where we remind that we are assuming $\psi=\dyad{j}$ with $\ket{j}$ a computational basis state.

In Table~\ref{tab:wyd_coherence} we report the value of the components of the vectors for the case the observable $X=\Pi_i$.

\begin{table}[!th]
\begin{tabular}{|c|c|c|}
\hline
$T_\pi$&$\sum_i\tr{T_\pi \left(\Pi_i^{\ot2}\ot \psi^{1/2}\ot\psi^{1/2}\right)}$&$\sum_i\tr{T_\pi \left(Q\Pi_i^{\ot2}\ot\psi^{1/2}\ot\psi^{1/2}\right)}$\\
\hline
$I$&$d$&$1$\\
$T_{(ij)/\{(12),(34)\}}$&$1$&$d^{-1}$\\
$T_{(12)}$&$d$&$1$\\
$T_{(34)}$&$d$&$1$\\
$T_{(ijk)}^{(1,2)}$&$1$&$d^{-1}$\\
$T_{(ijk)}^{(3,4)}$&$1$&$d^{-1}$\\
$T_{(ijk\ell)/\{(1324),(1423)\}}$&$1$&$d^{-1}$\\
$T_{\{(1324),(1423)\}}$&$1$&$1$\\
$T_{(ij)(k\ell)/\{(12)(34)\}}$&$1$&$1$\\
$T_{(12)(34)}$&$d$&$1$\\
\hline
\end{tabular}
\caption{Value of the traces for the WYD skew information when the observable is a computational basis state and the initial state is also a computational basis state.}
\label{tab:wyd_coherence}
\end{table}

Also for the case where the observable is a computational basis state one can write the components of the vectors $\vec{\mathcal{I}}_{1/2,4,CB}^{Q,V}$ and $\vec{\mathcal{I}}_{1/2,4,CB}^{H,V}$:
\ba
\left(\vec{\mathcal{I}}_{1/2,4,CB}^{Q,V}\right)_+&=&0\\
\left(\vec{\mathcal{I}}_{1/2,4,CB}^{Q,V}\right)_-&=&\frac{2\sqrt{2/3}\,(d+2)}{d}\\
\left(\vec{\mathcal{I}}_{1/2,4,CB}^{Q,V}\right)_1&=&\frac{\sqrt{2}\,(d-1)}{d}\\
\left(\vec{\mathcal{I}}_{1/2,4,CB}^{Q,V}\right)_2&=&-\frac{\sqrt{2}\,(d-1)}{d}\\
\left(\vec{\mathcal{I}}_{1/2,4,CB}^{Q,V}\right)_3&=&-\frac{\sqrt{2/3}\,(d-1)}{d}\\
\left(\vec{\mathcal{I}}_{1/2,4,CB}^{Q,V}\right)_4&=&-\frac{\sqrt{2/3}\,(d-1)}{d}
\ea
and
\ba
\left(\vec{\mathcal{I}}_{1/2,4,CB}^{H,V}\right)_+&=&0\\
\left(\vec{\mathcal{I}}_{1/2,4,CB}^{H,V}\right)_-&=&\sqrt{\tfrac{2}{3}}\,(d+5)\\
\left(\vec{\mathcal{I}}_{1/2,4,CB}^{H,V}\right)_1&=&\frac{d-1}{\sqrt{2}}\\
\left(\vec{\mathcal{I}}_{1/2,4,CB}^{H,V}\right)_2&=&-\frac{d-1}{\sqrt{2}}\\
\left(\vec{\mathcal{I}}_{1/2,4,CB}^{H,V}\right)_3&=&-\frac{d-1}{\sqrt{6}}\\
\left(\vec{\mathcal{I}}_{1/2,4,CB}^{H,V}\right)_4&=&-\frac{d-1}{\sqrt{6}}
\ea
Finally, we report the expression for the T-doped average for the case of computational basis measurements:
\ba
\nonumber
\langle \mathcal{I}_{1/2,4,CB}\rangle_{\mathcal{C}_T}&=&
\frac{-\frac{3}{8}\mathcal{P}_{0}(t)+\frac{3}{8}d(3+d)B(t)\cos(4\theta)+(d-1)(d+1)\Big[(3+d)\Xi_{1}(d,t)\xi_{1}^{k}+(d-2)d\Xi_{-}(d,t)\xi_{-}^{k}\Big]}{3(d-2)(d-1)d^{2}(1+d)^{2}(d+2)(d+3)}\\
\ea
where
\ba
\Xi_{-}(d,t)&=&-(g_{3}(t)+g_{4}(t)+g_{3}^{*}(t))+(2+d)\left[g_{2}(2t)-4g_{2}(t)\right]+d(3+d)\left(\tilde g_{3}(t)+2d\right)\\
\nonumber
\Xi_{1}(t)&=&\big(4g_{2}(2t)-16g_{2}(t)+g_{3}(t)+4g_{4}(t)+g_{3}^{*}(t)\big)\\
\nonumber
&&-d\big(3g_{2}(2t)+2[-3g_{2}(t)+g_{3}(t)+g_{4}(t)+g_{3}^{*}(t)]\big)\\
&&+2d^{2}(6+g_{2}(2t)+2g_{2}(t)-2\tilde g_{3}(t))
+\big(-5+2\tilde g_{3}(t)\big)d^{3}-2d^{4}
\ea
and
\ba
\nonumber
\mathcal{P}_{0}(d,t)&=&-8\Big[4g_{2}(2t)-16g_{2}(t)+g_{3}(t)+4g_{4}(t)+g_{3}^{*}(t)\Big]\\
\nonumber
&&+(d-1)\Big[12 g_{2}(2t)+48 g_{2}(t)-8(g_{3}(t)+g_{3}^{*}(t))\Big]\\
\nonumber
&&+d^{2}\Big[36 g_{2}(2t)-176 g_{2}(t)+13(g_{3}(t)+g_{3}^{*}(t))+40 g_{4}(t)-192\Big]\\
\nonumber
&&+d^{3}\Big[24 g_{2}(2t)+54 g_{2}(t)-9(g_{3}(t)+g_{3}^{*}(t))-52\Big]\\
\nonumber
&&+d^{4}\Big[244 +50 g_{2}(t)-8(g_{3}(t)+g_{3}^{*}(t))-8 g_{4}(t)\Big]\\
&&+56 d^{5}-56 d^{6}-16 d^{7}\\
B(d,t)&=&4g_{2}(2t)-d\big[(g_{3}(t)+g_{3}^{*}(t))+4d-2g_{2}(t)d\big]
\ea

\end{document}